\documentclass[12pt,a4paper,oneside,openany]{book}
\usepackage{url,setspace,amsmath,amsfonts,amssymb,graphicx,fancyhdr}
\usepackage{natbib}
\usepackage{mathrsfs,bbm}

\renewcommand{\Re}{\ensuremath{\mathbb{R}}}
\newcommand{\vol}[1]{\ensuremath{\text{vol}\!\left(#1\right)}}
\newcommand{\N}{\ensuremath{\mathbb N}}

\newcommand{\E}[1]{\ensuremath{\mathbb{E}\!\left[#1\right]}}
\newcommand{\Eb}[1]{\ensuremath{\mathbb{E}\!\left\{#1\right\}}}

\newcommand{\norm}[1]{\ensuremath{\|#1\|}}
\newcommand{\gnorm}[1]{\ensuremath{\left\|#1\right\|}}

\newcommand{\abs}[1]{\ensuremath{\left|#1\right|}}
\newcommand{\sabs}[1]{\ensuremath{|#1|}}
\newcommand{\diam}[1]{\ensuremath{\text{diam}\!\left(#1\right)}}
\newcommand{\graph}[1]{\ensuremath{\text{graph}\! #1\,}}
\newcommand{\Hf}[2]{\ensuremath{\mathscr{H}^{#2}_{#1}}}
\newcommand{\Hfm}[1]{\ensuremath{\mathscr{H}^{#1}}}

\newcommand{\vel}{\ensuremath{\textbf{v}}}
\newcommand{\Vel}{\ensuremath{\textbf{V}}}
\newcommand{\svl}{\ensuremath{\textbf{u}}}
\newcommand{\vect}{\ensuremath{\textbf{r}}}
\newcommand{\wn}{\ensuremath{\textbf{k}}}
\newcommand{\mean}[1]{\langle #1 \rangle}
\newcommand{\gmean}[1]{\left< #1 \right>}
\newcommand{\eX}{\check{\mathbf{e}}_{x}}

\newcommand{\eZ}{\check{\mathbf{e}}_{z}}
\newcommand{\Vect}{\mathbf{ R}}
\newcommand{\brho}{\boldsymbol{\rho}}
\newcommand{\Bkappa}{\boldsymbol{\kappa}}

\newcommand{\LR}{\ensuremath{L^2_{\phi}(\Re)}}

\newcommand{\tr}{\ensuremath{\text{tr}\,}}
\newcommand{\R}[1]{\ensuremath{\mathcal{R}_e(#1)}}
\newcommand{\Cs}{C^{2}_{\varepsilon}}
\newcommand{\Tc}[2]{t_{#1}\!\left(\eX \; #2 \right)}
\newcommand{\Ts}[2]{t_{#1}\!\left[\eX \; #2 \right]}
\newcommand{\intDD}{\int_{D \times D}}
\newcommand{\Cons}{\frac{A_{0}\;k^{4}}{(2 \pi)^{2} l^{2} L^{2}}}
\newcommand{\cons}{\frac{A_{0}\;k^{2}}{L^{2}}}
\newcommand{\Heavy}[1]{ \Theta^{(2)}\!\left( \frac{1}{2}\,{\bf D} #1 \right)}

\newsymbol\vacio 203F
\DeclareMathAlphabet{\mathpzc}{OT1}{pzc}{m}{it}

\author{Dar\'{\i}o Gabriel P\'erez}
\title{Lightwave propagation through random media.}

\begin{document}
\pagestyle{fancy}
\renewcommand{\chaptermark}[1]{\markboth{#1}{}}
\renewcommand{\sectionmark}[1]{\markright{\thesection\ #1}}
\fancyhf{} 
\fancyhead[RO]{\bfseries\thepage}
\fancyhead[LO]{\bfseries\rightmark}
\renewcommand{\headrulewidth}{0.5pt}
\renewcommand{\footrulewidth}{0pt}
\addtolength{\headheight}{0.5pt} 
\fancypagestyle{plain}{%
\fancyhead{} 
\renewcommand{\headrulewidth}{0pt} 
}
\frontmatter
\newpage
\thispagestyle{empty}
\begin{center}
\textsc{Tesis Doctoral}

\vspace*{10em}
\textsc{\LARGE Propagaci\'on\vspace{1em} de Luz en Medios Turbulentos}
\vspace*{5em}

{\Large Dar\'{\i}o Gabriel P\'erez}
\vspace*{5em}
\begin{figure}[h]
\begin{center}
\includegraphics[width=0.15\textwidth]{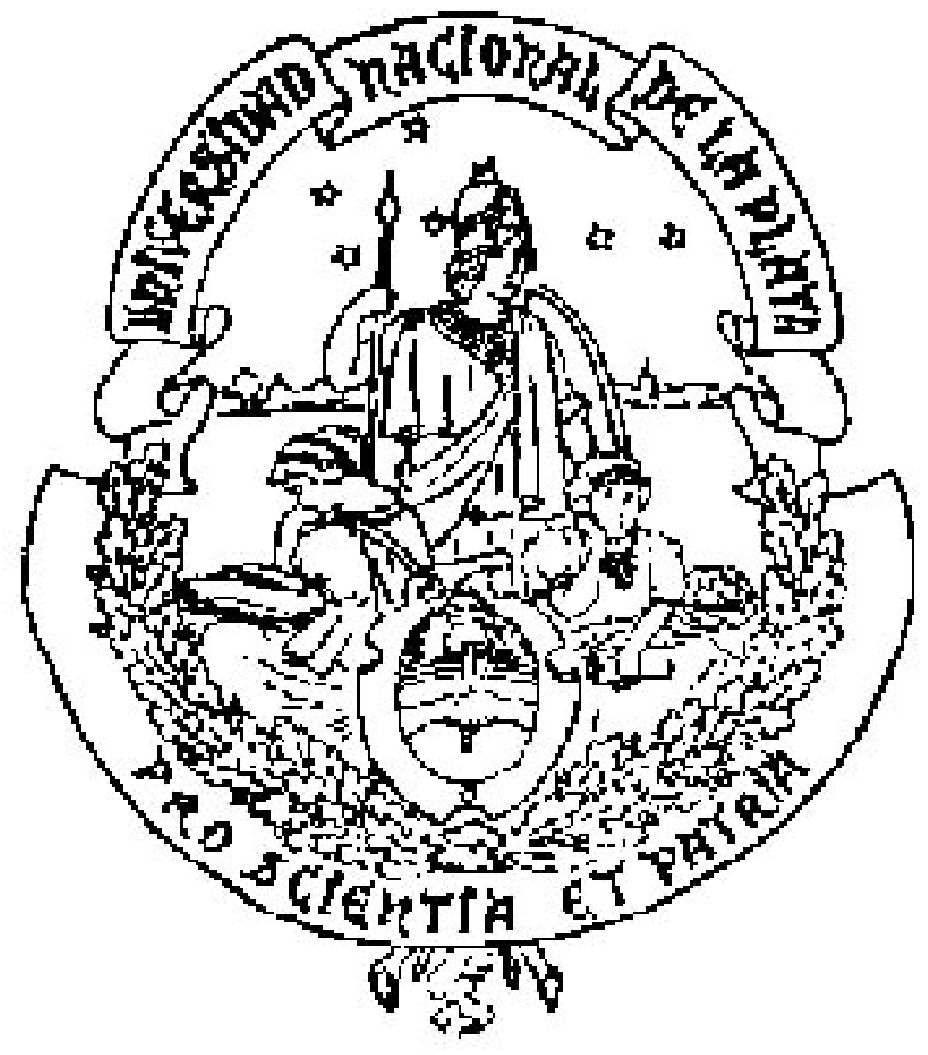}
\end{center}
\end{figure} 
 {\large

Universidad Nacional de La Plata

Facultad de Ciencias Exactas

Departamento de F\'{\i}sica}

\vspace*{7em}
La Plata, 23 de diciembre de 2002
\end{center}
\newpage
\thispagestyle{empty}
\begin{center}
\vspace*{5em}
\textsc{Tesis Doctoral}

\textit{presentada en la}

\textsc{Facultad de Ciencias Exactas}

\textsc{Universidad Nacional de La Plata}

\vspace*{5em}
\textit{Para obtener el t\'{\i}tulo de Doctor de la Facultad de Ciencias Exactas}

\vspace*{2em}
\textit{Por}

{\large Dar\'{\i}o Gabriel P\'erez}

\vspace*{2em}
\textit{Tema}

\textsc{\Large Propagaci\'on de Luz en Medios Turbulentos}

\vspace*{2em}
\textit{Director}

{\large Mario Garavaglia}

\end{center}
\newpage
\thispagestyle{empty}
\vspace*{15em}
\begin{center}
\textsc{A Selva}
\end{center}
\newpage
\thispagestyle{empty}
\begin{flushright}
\begin{quote}
\hspace*{15em}No cesaremos de explorar

\hspace*{15em}y al cabo de toda nuestra exploraci\'on 

\hspace*{15em}llegaremos al punto de partida

\hspace*{15em}y por primera vez conoceremos el lugar
\end{quote}
\vspace*{1em}
---T. S. Eliot
\end{flushright}

\vspace*{30em}
\begin{flushleft}
\begin{quote}
Nur aus dem Schweigen ward das Wort,

Nur aus dem Dunkel ward das Licht,

Nur aus dem Tod ward das Leben;

Hell ist der Flug des Falken,

In der Weite des Himmels.
\end{quote}
\vspace*{1em}
\hspace*{10em}---aus \textit{Die Erschaffung von \'Ea} 
\end{flushleft}

\chapter*{Acknowledgements}

For some reason the act of writting acknowledgements puts me in a contemplative state. Me, looking at an empty page. Not because of lack of people to whom I have to say thanks, but because I would like to find the best words. The dilemma is that, sometimes, these words seems weak to express oneself gratitude. Our `Thanks a lot' dissolves in the everyday use. For those appearing in these lines my thanks are more than what my poor writting is able to express.

\vspace*{1em}
To you, Mario, my most sincerly gratitude, respect and appreciation. Also, for all your support and encouragement over all this time. I hope to return you some of what I received from you.

\vspace*{1em}
To the Fundaci\'on Antorchas. I started my postgraduate studies with its support; now, I am also finishing with them. Since, the foundation financined my last year in Germany. Also, to all the people working there, whom have all my gratitude.

\vspace*{1em}
To Martina Z\"ahle, for all her pacience teaching me the mathematical tools used in part of this thesis. All those encounters helped me to consolidate the original idea displayed in the last chapters. Also, for all her effort turning my stay in Jena as best as possible.

\vspace*{1em}
Finally to my parents and friends. 

\vspace*{5em}
\begin{flushright}
December 23, 2002
\end{flushright}

\tableofcontents 
\contentsline {chapter}{\bibname}{\pageref{bib}}
\listoffigures

\mainmatter
\onehalfspacing
\chapter{The Turbulent Refractive Index: Dynamics and Stochastic Properties}
\label{chp:index}

The study of phenomena occurring in a turbulent fluid has been successively improved during the last 40 years. Specifically, the concentration of a substance advected by the turbulence has received most of the attention, for it covers a wide range of natural and engineering settings: heat transport, dye diffusion, microscopic organism movements, etc.. These substances are described by scalar fields with a negligible back-effect on the flow; thus, they are called \textit{passive scalar fields}. 

The turbulent refractive index also belongs to this class; this is not a novelty \citep{book:tatarskii}.  The temperature is a passive scalar field whenever it produces buoyancy forces smaller than the inertial stresses driving the flow, and a direct calculation shows that its fluctuations are proportional to those of the index.   

Our interest in lightwave propagation through turbulent media must start here then. That is, we have to comprehend the media before attempt a description of the propagation itself. In the forthcoming sections we will study the dynamics and stochastic properties of passive scalars, and eventually propose models for the refractive index. 

\section{Turbulence}
\subsection{The turbulent flow: Kolmogorov hypotheses}
\label{section:kolmogorov-h}

Above we have, without more precisions, referred to the turbulent media. From now on, we mean incompressible fluids in turbulent state; moreover, all our discussions will be targeting the atmosphere or experiments that resemble it. Of course, this section is intent to explain what `\textit{turbulent}' is.

Let us start from the beginning; as it is well known fluids are governed by the Navier-Stokes equation: 
\begin{equation}
\frac{\partial}{\partial t}\, \vel + (\vel \cdot \nabla) \vel -  \nu \bigtriangleup\! \vel =\frac{1}{\rho}\left(\,\mathbf{F} - \nabla p\right),
\label{eq:navier-stokes}\end{equation}
$\vel(\vect,t):\Re^3\times \Re_+ \rightarrow \Re^3$ is the velocity field, while $\nu$ is the viscosity of the fluid (with dimensions $[\nu]= L^2/T$), $\rho$ the density, $p$ the pressure and $\mathbf{F}$ the external force. It is worth noting that this equation is scale invariant. So it can be turned into the following adimensional equation,
\begin{equation}
\frac{\partial}{\partial\tilde{t}}\, \tilde{\vel} + (\tilde{\vel} \cdot \tilde{\nabla}) \tilde{\vel}- \left(\frac{\nu \tau}{l^2}\right) \tilde{\bigtriangleup} \tilde{\vel} =\frac{1}{\tilde{\rho}}\left(\,\tilde{\mathbf{F}} - \tilde{\nabla} \tilde{p}\right),
\label{eq:ad-navier-stokes}\end{equation}
with $l$ and $\tau$  the characteristic length and time of the system. The constant multiplying the first term at the right-hand side of the latter equation introduces the Reynolds number,
\begin{equation}
\R{l} = \frac{v_l\, l}{\nu},\label{def:reynolds-nr}
\end{equation}
$v_l$ is the velocity change on the scale length $l$. The Reynolds number is a scale dependent quantity, and its magnitude measures the flow regime: it compares the non-linear advection term $(\vel \cdot\nabla)\vel $ against the dissipation $-  \nu \bigtriangleup\! \vel$. While low Reynolds numbers, $\R{l}\ll 1$, correspond to regular and laminar flows and intermediate numbers, $1\lesssim \R{l}\lesssim 10^2$, exhibit complex patterns, higher Reynolds numbers, $\R{l}\gtrsim 10^4$, drive the flow to an apparent spatial disorder: parcels of fluids follow chaotic trajectories. In particular, when the Reynolds number tends to infinity  the flow exhibits a \textit{fully developed turbulence}. The non-linear advection is preponderant because the dissipative term goes to zero.

The equation (\ref{eq:navier-stokes}) induces the energy balance equation (per unit mass):
\begin{equation}
\begin{aligned}
\frac{d}{dt}\,u = \frac{d}{dt}\left(\frac{1}{2}\int_V\! d^3 r\; \norm{\vel}^2\right)&=\int_V\! d^3r\; \left[\,\left(\frac{\mathbf{F}}{\rho}\right)\cdot \vel - \nu \sum_{i,j} \frac{\partial v_i}{\partial x_j}  \left(\frac{\partial v_i}{\partial x_j}+\frac{\partial v_j}{\partial x_i}\right)\right]\\
&= \int_V\! d^3r\; \left[\,\left(\frac{\mathbf{F}}{\rho}\right)\cdot \vel - \frac{\nu}{2} \sum_{i,j} \left(\frac{\partial v_i}{\partial x_j}+\frac{\partial v_j}{\partial x_i}\right)^2\right].
\end{aligned}
\label{eq:energy-balance}\end{equation}
The balance is given here between the first term on the rightmost-hand side of this equation, which represents the energy injected per unit time into the system, and the energy dissipated by the viscous forces, that is, the second term. 

It was \citet{paper:kolmogorov} who first realized that from dimensional and reliable heuristic arguments the energy transfer could be explained. His success was to notice that the results of this analysis become universal laws in the statistical sense.  The turbulent velocity field should be thought a stochastic variable in the ensemble's sense of the statistical mechanics. It is independent on how the turbulence began: it does not matter the way the energy is injected. That is, the statistics of the chosen force has no effect over the statistics of the turbulence. 

Moreover, we will also assume that a fully developed turbulence is spatially isotropic, homogeneous, and stationary: for any linear transformation and translation the system looks the same.

In this section we will treat the turbulence development under the direct energy injection
. That is, the energy is injected by the largest disturbances of size $L$---the integral scale---, corresponding to the size of the bath, and then it is transferred towards the smallest scales. Finally a minimum scale $l_0$---the inner length---is reached, there the energy is dumped by the viscosity into heat (the magnitude of the inner length oscilates between $10^{-3}$m and $10^{-2}$m). 

The range of scales $l$ where the energy transfer happens without loss, the flux of energy from scale to scale is constant, is called \textit{inertial range}
\begin{equation*}
l_0\ll l \ll L.
\end{equation*}
This process can be thought as a cascade of energy that propagates through the scales via a succession of disturbances (\textit{eddies} which are portions of fluid with size $l$ and velocity $v_l$): big eddies break up smaller ones. These eddies are arranged in a hierarchy according to its size, from the bigger to the smallest, as follows: 
\begin{equation}
l_n\sim L\, \pi^n,\;n=0,1,\cdots
\label{prop:lenght-contraction}\end{equation}
with $\pi<1$ the contraction ratio of the eddy size from one generation to the other. 

Now, we can use this scheme to estimate some of the quantities involved in the generation of the turbulence. Thus, let $V_{l_n}$ be the volume occupied by the eddies of the $n$-th generation; their energy density is $u_{l_n}\sim v^2_{l_n}/2$. It is straightforward then that the accumulated total  energy by the eddies of size $l\sim l_n$ is,
\begin{equation}
E_l\sim  v_l^2 V_l.
\label{eq:volume-energy}\end{equation}
Knowing that the characteristic life-span of the disturbances is $\tau_l\sim l_n/v_l$. We obtain the following estimation for the energy transfer rate,
\begin{equation*}
\varepsilon \sim \frac{E_l}{\tau_l}\sim \frac{v^3_l\, V_l}{l}. 
\label{eq:energy-rate}\end{equation*}
If we now consider that the volume occupied by the eddies is independent of the scale, i.e., $V_{l}\sim$ const.. Then the energy flux per unit volume $\overline{\varepsilon}$ is also constant and,
\begin{equation}
\overline{\varepsilon}\, l\sim v^3_l.
\label{eq:scaling}\end{equation}
This scaling law is the fundamental result, as we shall see, of the whole chapter for it will be underneath every property we are about to show. 

For instance, let us try some examples: we have defined before the inner length as the scale where the dissipative term becomes noticeable with respect to the convective one, that is,
\begin{equation*}
\norm{(\vel\cdot\nabla)\vel}\sim v_{l_0}\left(\frac{v_{l_0}}{l_0}\right)\sim \overline{\varepsilon}^{2/3}\,l_0^{-1/3}\sim \norm{-\nu\bigtriangleup\! \vel}\sim \nu \left(\frac{v_{l_0}}{l^2_0}\right)\sim \overline{\varepsilon}^{1/3} \nu\, l_0^{-5/3},
\end{equation*}
and thus the inner scale is roughly
\begin{equation}
l_0 \sim \nu^{3/4} \overline{\varepsilon}^{-1/4}.
\end{equation}
It vanishes as $\nu\rightarrow 0$, and this results to be an ultraviolet cut-off. Below this cut-off the advection can be neglected and the velocity turns more regular.

The local Reynolds number at a scale $l$ can be calculated from (\ref{def:reynolds-nr}); it is $\R{l} \sim (l/l_0)^{4/3}$. Moreover, the system's Reynolds number may be taken as $\mathcal{R}_e:=\R{L}$. So, the condition $l\gg l_0$ is in agreement with the conditions for turbulence development: the inertial range grows as the system's Reynolds number do so. 

From equation (\ref{eq:scaling}) we can also check the occurrence of equilibrium between the injected and dissipated energy. Using the isotropy and homogeneity properties of the velocity field:
\begin{equation}
\gnorm{- \frac{\nu}{2} \sum_{i,j} \left(\frac{\partial v_i}{\partial x_j}+\frac{\partial v_j}{\partial x_i}\right)^2}_{l=l_0}\sim \nu \left(\frac{v_{l_0}}{l_0}\right)^2\sim \overline{\varepsilon}.
\label{prop:energy-dissipation}\end{equation}

Let us start looking at the stochastic properties of the velocity field. That is, we want to compute the $n$-point correlations of the turbulent velocity. The usual procedure to overtake   is as follows: first, we separate the (stochastic) fluctuations $\svl$ from the mean (averaged) flow $\mean{\vel}$,  so it can be written $\vel(\vect,t)= \mean{\vel}(\vect) + \svl(\vect,t)$; second, we derive from the Navier-Stokes equations the corresponding equations for the $n$-point correlation. Here the real problem arises: these equations are non-linear, for it is a closure problem. Their solutions are found only in approximation. The Kolmogorov's method is so successful because it allows us to override this second step. 

Assume the mean flow is zero and the random velocity field has the properties we have discussed at the beginning: homogeneity and isotropy. The existence of scaling laws for the $n$-correlation functions means the existence of exponents $\eta_n$ such that
\begin{equation*}
\exists\lim_{\lambda\rightarrow 0}\lim_{\nu\rightarrow 0} \;l^{-\eta_n}\!  \mean{\svl(l\, \vect_1,t)\cdots \svl(l\, \vect_n, t)}.
\end{equation*}

Because the energy transfer per unit volume is constant all over the inertial range and there is independence from the source of turbulence, the $n$-correlators of the stochastic velocity field will just depend from the scale. If we look back at equation (\ref{eq:scaling}), we have:
\begin{equation}
\mean{\gnorm{\svl(\vect +\vect')-\svl(\vect')}^n}= C_n (\overline{\varepsilon}\norm{\vect})^{n/3},
\end{equation}
and the constants $C_n$ are universal. 

This scaling method, although effective determining some fundamental behavior of the turbulent velocity field, is scarce explaining the way the transference of energy occurs between the different scales. This the task to be tackled in the next section.

\subsection{The energy cascade in isotropic turbulence}
\label{section:isotropic-cascade}

As before, we are dealing with an homogeneous and isotropic turbulence. The energy is then constant throughout space. Thus, when we consider the transport of turbulent energy, this will be in wavenumber rather than in the coordinate space. So, we can foresee a transfer from one range of eddy sizes to another: the cascade phenomenon. We used \citeauthor{book:mccomb}'s \citeyearpar{book:mccomb} book as the main guide for this section.

In order to have isotropy and homogeneity we will make the boundary of our system go to infinity.  We will deal here and thereafter with incompressible fluids, so going back to equation (\ref{eq:navier-stokes}) we set $\nabla \cdot \mathbf{F} =0$. Additionally, we can obtain another property from the Navier-Stokes equation, applying the divergence to both sides of it yields:
\begin{equation*}
\bigtriangleup (\rho^{-1}p) = -\sum_{j,k} \frac{\partial u_k }{\partial x_j}\frac{\partial u_j }{\partial x_k}=-\sum_{j,k} \frac{\partial }{\partial x_j}\,\frac{\partial  }{\partial x_k}\, u_k u_j = - (\nabla \otimes \nabla)\cdot (\svl \otimes \svl ),
\end{equation*}
where $\otimes$ is the tensor product. That is, each vector is understood as a column matrix, $\vect\in\Re\times\Re^3$, and the inner product acts column by column like above. This is a Poisson equation, and it can be solved calculating the Green's function: 
\begin{equation}
\bigtriangleup G(\vect,\vect')= \delta(\vect-\vect'), 
\end{equation}
with condition  $\check{\mathbf{n}}\cdot\nabla G \rightarrow 0$, as the boundary goes to infinity. The pressure can be written,
\begin{equation}
\rho^{-1} p(\vect,t) = - (\nabla\otimes\nabla)\cdot \int_{\Re^3} \!\!d^3r'\,G(\vect,\vect')\left(\svl(\vect',t)\otimes\svl(\vect',t)\right) .
\end{equation}
There, the superficial terms are zero according to the conditions imposed to the turbulence. 

The formation of a stationary isotropic turbulence requires the external force $\mathbf{F}$ to counter-effect the action of the viscous force, but for the present discussion we momentarily set it equal to zero.

Using the latter equation it takes some effort turning equation (\ref{eq:navier-stokes}) into:
\begin{align*}
\frac{\partial \svl}{\partial t} - \nu \bigtriangleup \svl = &- (\mathbf{1}\otimes\nabla)\cdot\left\{ (\svl\otimes\svl) -  \nabla\left[ (\mathbf{1}\otimes\nabla)\cdot \int_{\Re^3} \!\!d^3r'\, G(\vect,\vect')\, \left(\svl(\vect')\otimes\svl(\vect')\right)\right]\right\}\nonumber\\
=&- (\mathbf{1}\otimes\nabla)\cdot \mathbf{D}(\mathbf{1}\otimes\nabla)(\svl\otimes\svl),
\end{align*}
here $(\mathbf{1})_{j k}= \delta_{j k}$. The right-hand side of this equation can be changed into a symmetric form with the aid of the operator,
\begin{equation}
\mathbf{M}(\nabla)= -\frac{1}{2}\left[ (\mathbf{1}\otimes\nabla)\cdot \mathbf{D}(\mathbf{1}\otimes\nabla) + (\nabla\otimes\mathbf{1})\cdot \mathbf{D}(\nabla\otimes\mathbf{1}) \right],
\label{eq:matrix-operator}\end{equation}
so we finally find:
\begin{equation}
\frac{\partial \svl}{\partial t} - \nu \bigtriangleup \svl = \mathbf{M}(\nabla)(\svl\otimes\svl).
\label{eq:ns-simple}\end{equation}
This equation concentrates all the non-linear effects producing the advection on its left side, while the smoothing diffusive term is on the right-hand side. Also, for all the practical problems the non-linear term here is no more complex than the original one. 

Although possible, we would rather not build differential equations for the moments of the velocity field from equation (\ref{eq:ns-simple}); instead, it will be enough for us to recover an energy balance equation. Thus, we 
introduce the Fourier transform of the random velocity,
\begin{equation}
\svl(\vect,t)= \int_{\Re^3}\!\! d^3k\;\hat{\svl}(\mathbf{k},t) \exp(i \mathbf{k}\cdot\vect).
\label{eq:fourier-vel}\end{equation}
The continuity equation for incompressible fluids changes in the wavenumber space to
\begin{equation}
\wn\cdot \hat{\svl}=0,
\label{eq:orthogonality}\end{equation}
that is, the wavenumber vector is perpendicular to the velocity field. 

We can transform the Navier-Stokes equation into a wavespace equation, as usual, using the Fourier Analysis:
\begin{equation}
\left(\frac{\partial}{\partial t} + \nu k^2\right) \hat{\svl}(\wn) = \widehat{\textbf{M}}(\wn)\cdot \int_{\Re^3}\!\! d^3k'\,\hat{\svl}(\wn')\otimes\hat{\svl}(\wn-\wn') ,
\label{eq:spectral-ns}\end{equation}
where
\begin{equation}
\widehat{\textbf{M}}(\wn) = \frac{1}{2 i}\left[\wn\otimes\widehat{\textbf{D}}(\wn)+\widehat{\textbf{D}}(\wn)\otimes\wn\right]
\end{equation}
and
\begin{equation}
\widehat{\textbf{D}}(\wn) = \mathbb{I}- \frac{\wn\otimes\wn}{k^2}.
\label{def:proyection}\end{equation}
here $\mathbb{I}\in \Re^3\times\Re^3$ and we have dropped the time dependence on $\hat{\svl}$ to simplify things. These last two operators are, of course, Fourier transforms of its counterparts in (\ref{eq:matrix-operator}). It is straightforward from the equation (\ref{eq:fourier-vel}) that: $\wn\cdot\widehat{\textbf{D}}(\wn)=0$: it is a projection.

Moreover, the moments of $\hat{\svl}$  inherit some properties from the turbulent system initial setup. In fact, from
\begin{equation*}
\hat{\svl}(\wn)= \frac{1}{(2\pi)^3} \int_{\Re^3}\!\!d^3r\; \svl(\vect) \exp(-i \wn\cdot\vect), 
\end{equation*}
it follows that the two-point correlation in the $\wn$-space is related to the corresponding in the $\vect$-space by
\begin{multline}
\mean{\hat{\svl}(\wn)\otimes\hat{\svl}(\wn')}\\=\frac{1}{(2\pi)^6}\int_{\Re^3\times\Re^3}\!\!d^3r\, d^3r' \mean{\svl(\vect)\otimes\svl(\vect'-\vect)}\exp[-i(\wn+\wn')\cdot\vect]\exp(i \wn'\cdot\vect').
\label{eq:2-correlation}\end{multline}
The space correlation is invariant under translations due to the homogeneity of the turbulence; so, we have
\begin{equation}
\mean{\svl(\vect')\otimes\svl(\vect'-\vect)}=\mean{\svl(0)\otimes\svl(\vect)}.
\end{equation}
Equation (\ref{eq:2-correlation}) becomes,
\begin{align}
\mean{\hat{\svl}(\wn)\otimes\hat{\svl}(\wn')}&=\frac{1}{(2\pi)^6}\int_{\Re^3\times\Re^3}\!\!d^3r\, d^3r'\, \mean{\svl(0)\otimes\svl(\vect)}\times\nonumber\\
&\times\exp[-i(\wn+\wn')\cdot\vect]\exp(i \wn'\cdot\vect')=\nonumber\\
&= \delta(\wn+\wn')\;\left[ \frac{1}{(2\pi)^3}\int_{\Re^3}\!\!\! d^3r\; \mean{\svl(0)\otimes\svl(\vect)}\exp(i \wn'\cdot\vect')\right]=\nonumber\\
&= \delta(\wn+\wn')\;\left[ \frac{1}{(2\pi)^3}\int_{\Re^3}\!\!\! d^3r\; \mathbf{Q}(\vect)\exp(i \wn'\cdot\vect')\right]=\nonumber\\
&=\delta(\wn+\wn')\;\widehat{\mathbf{Q}}(\wn),
\end{align}
$\mathbf{Q}(\vect)$ is the isotropic correlation. Hence, the 2-point spectral correlation has a non-vanishing contribution only when $\wn+\wn'=0$. 

Also, we can prove, with the same arguments, that higher order correlations have the same property. That is,
\begin{equation}
\mean{\hat{\svl}(\wn_1)\otimes\hat{\svl}(\wn_2) \otimes\cdots\otimes\hat{\svl}(\wn_n)}= 0\quad \text{unless}\quad \wn_1+\wn_2+\cdots+\wn_n=0.
\label{eq:correlation-condition}\end{equation}

But, it is the isotropy which provides us with what can change all these tensor forms for the moments into 1-dimensional expressions. As we said, we are concerned with the energy transfer so just the Fourier transform of the second moment will be considered. A 2-tensor invariant under rotations and translations can only be expressed as follows \citep{book:batchelor},
\begin{equation*}
\widehat{\mathbf{Q}}(\wn)= B(k)\mathbb{I}+ A(k)\, \wn\otimes\wn
\end{equation*}
where the functions $A$ and $B$ are indeterminated but continous. If we multiply this equation by $\wn\cdot$, and make use of (\ref{eq:orthogonality}) then 
\begin{equation*}
\wn\cdot\widehat{\mathbf{Q}}(\wn)=0=B(k) \wn + A(k) k^2 \wn= [B(k) + A(k) k^2]\wn 
\end{equation*}
for all $\wn$. So,
\begin{equation*}
q(k)= B(k)= -k^2 A(k),
\end{equation*}
and finally it yields 
\begin{equation}
\widehat{\mathbf{Q}}(\wn)= q(k) \mathbb{I}-  \frac{q(k)}{k^2}(\wn \otimes\wn) = \widehat{\mathbf{D}}(\wn) q(k).
\label{def:spectral-correlation}\end{equation}

Now we can make some considerations about $q(k)$. Because $\tr \widehat{\mathbf{D}}(\wn) = 2$ from its definition. It is
\begin{equation*}
\tr \widehat{\mathbf{Q}}(\wn)  =\tr [ \widehat{\mathbf{D}}(\wn) q(k)] = 2 q(k).
\end{equation*}
This trace can also be linked to the energy $E$ per unit mass of fluid. The isotropic correlation is naturally related to the density of energy, and it gives the following:
\begin{equation*}
2E= 3\mean{u^2}=\left. \tr \mathbf{Q}(\vect)\right|_{\vect=0}=\tr \!\int_{\Re^3}\!\!d^3k\, \widehat{\mathbf{Q}}(\wn)=\tr \!\int^{\infty}_0\!\!k^2dk\, q(k)  \int\!d\Omega\; \widehat{\mathbf{D}}(\wn),
\end{equation*}
here $d\Omega$ is the solid angle. We have used definition (\ref{def:spectral-correlation}), and its Fourier relation with the isotropic correlation. The angle integration can easily be carried out, so
\begin{equation}
E= \frac{4\pi}{3} (\tr\mathbb{I}) \int^{\infty}_0 \!\!k^2 dk\; q(k) = \int^{\infty}_0\!\!4\pi k^2dk  \;q(k)=\int^{\infty}_0\!\! dk\, E(k).
\end{equation} 
We have thus defined $E(k)$, the \textit{wavenumber spectrum}, as the contribution to the total energy from harmonic components with wavevectors lying between $k$ and $k+dk$. The quantity $q(k)$ is the density of contributions in wavenumber space to the total energy; we will call it \textit{spectral density}.

It is now time  to calculate the dynamics of the spectral correlation. We will consider single-time moments. Henceforth, we $\otimes$-multiply equation (\ref{eq:spectral-ns}) by $\hat{\svl}(-\wn,t)$ to build the matrix,
\begin{multline*}
\gmean{\frac{\partial\hat{\svl}(\wn,t)}{\partial t} \otimes\hat{\svl}(-\wn,t)} + \nu k^2 \mean{\hat{\svl}(\wn,t)\otimes\,\hat{\svl}(-\wn,t)}= \nonumber\\
=\widehat{\textbf{M}}(\wn)\cdot \int_{\Re^3}\!\! d^3k'\,\mean{\hat{\svl}(\wn')\otimes\hat{\svl}(\wn-\wn')\otimes\hat{\svl}(-\wn,t)}.
\end{multline*}
We find a similar equation  when the change $\wn \rightarrow -\wn$ is made, that is,
\begin{multline*}
\gmean{ \hat{\svl}(\wn,t)\otimes\frac{\partial\hat{\svl}(-\wn,t)}{\partial t}} + \nu k^2 \mean{\hat{\svl}(\wn,t)\otimes\,\hat{\svl}(-\wn,t)}= \\
= \int_{\Re^3}\!\! d^3k'\, \mean{\hat{\svl}(\wn,t)\otimes\hat{\svl}(\wn')\otimes\hat{\svl}(-\wn-\wn')}\cdot\widehat{\textbf{M}}(-\wn).
\end{multline*}
Summing both equations, and using the property
\begin{equation*}
\gmean{ \frac{\partial\hat{\svl}(\wn,t)}{\partial t}\otimes\hat{\svl}(-\wn,t)} + \gmean{ \hat{\svl}(\wn,t)\otimes\frac{\partial\hat{\svl}(-\wn,t)}{\partial t}} = \frac{\partial}{\partial t}\mean{\hat{\svl}(\wn,t)\otimes\hat{\svl}(-\wn,t)},
\end{equation*}
we finally have:
\begin{align}
\left(\frac{d}{d t}+ 2 \nu k^2\right) \widehat{\mathbf{Q}}(\wn,t)= &\,\widehat{\mathbf{M}}(\wn)\cdot \int_{\Re^3}\!\!d^3k'\, \widehat{\mathbf{Q}}_3(\wn',\wn-\wn',t) \nonumber\\
+& \int_{\Re^3}\!\! d^3k'\; \widehat{\mathbf{Q}}_3(\wn',-\wn-\wn')\cdot \widehat{\mathbf{M}}(-\wn),
\label{eq:ns-spectral}\end{align}
here we have defined the 3-point spectral correlation $\widehat{\mathbf{Q}}_3(\wn',\wn-\wn',-\wn)=\mean{\hat{\svl}(\wn')\otimes\hat{\svl}(\wn-\wn')\otimes\hat{\svl}(-\wn,t)}\in \Re^3\times\Re^3\times\Re^3$. Taking the trace operator and multiplying by $2\pi k^2$ on both sides of (\ref{eq:ns-spectral}) we arrive to
\begin{equation}
\left(\frac{d}{d t}+ 2 \nu k^2\right) E(k,t) = T(k,t),
\label{eq:spectral-ns-1}\end{equation}
where the non-linear term on the right is given by,
\begin{align}
T(k,t)= & 2\pi k^2\, \tr\! \left\{\widehat{\mathbf{M}}(\wn)\cdot \int_{\Re^3} d^3k'\right.\times\nonumber\\
\times & \left[ \widehat{\mathbf{Q}}_3(\wn',\wn-\wn',-\wn,t) -\widehat{\mathbf{Q}}_3(\wn',-\wn-\wn',\wn,t)\right]\bigg\}.
\label{eq:advective-term}\end{align}

This term causes the advection of the spectral energy density: it redistributes the energy in the wavenumber space. Henceforth, it should satisfy 
\begin{equation}
\int^{\infty}_0\!\!dk\; T(k,t) =0.
\label{eq:spectral-advection}\end{equation}
To prove it let us first notice that from (\ref{eq:orthogonality}) and definition (\ref{def:proyection}) we have,
\begin{equation}
\widehat{\mathbf{D}}(\wn) \cdot\hat{\svl}(\wn)= \hat{\svl}(\wn).
\end{equation}
This property, together with equation (\ref{eq:orthogonality}), induces
\begin{equation}
\widehat{\mathbf{M}}(\wn)\cdot\widehat{\mathbf{Q}}_3(\wn',\mathbf{l},-\wn) = \widehat{\mathbf{D}}(\wn)\cdot \widehat{\mathbf{Q}}_3(\wn',\mathbf{l},-\wn)= \tr_{1,3}\widehat{\mathbf{Q}}_3(\wn',\mathbf{l},-\wn),
\end{equation}
where $\tr_{1,3}$ is the trace computed from the first and third arguments of the $3$-tensor  leaving the second free. 

Now, the integral (\ref{eq:spectral-advection}) can be put in the form:
\begin{align}
\int^{\infty}_0\!\! dk\, 2\,T(k,t)=& \int 4 \pi k^2\,dk \int_{\Re^3} d^3k'(2 i)^{-1}\times\nonumber\\
\times&\,\wn\cdot\left[ \tr_{1,3}\widehat{\mathbf{Q}}_3(\wn',\mathbf{l},-\wn) -  \tr_{1,3}\widehat{\mathbf{Q}}_3(\wn',\mathbf{l}-2\wn,\wn)\right].
\end{align}
It is also true $\mathbf{l}\cdot\tr_{1,3}\widehat{\mathbf{Q}}(\mathbf{l})=0$, for it is $\mathbf{l}\cdot\hat{\svl}(\mathbf{l})=0$. We can replace $\wn$ in the first term on the right-hand side above by $\wn-\mathbf{l}=\wn'$, because of condition (\ref{eq:correlation-condition}). Therefore, we find
\begin{align}
\int^{\infty}_0\!\! dk\, T(k,t)=& \int_{\Re^3}\!\! d^3k \int_{\Re^3}\!\! d^3k'(2 i)^{-1}\times\nonumber\\
\times&\,\left[ \wn'\cdot\tr_{1,3}\widehat{\mathbf{Q}}_3(\wn',\mathbf{l},-\wn) - \wn\cdot\ \tr_{1,3}\widehat{\mathbf{Q}}_3(\wn',\mathbf{l}-2\wn,\wn)\right].
\label{prop:integral-form}\end{align}
The definition of the $n$-point spectral correlation implies that they are all symmetric. Thus, the interchange $\wn'\leftrightarrow\wn$ shows that the above equation is antisymmetric. Therefore, we can conclude that equation (\ref{eq:spectral-advection}) holds. 

As we said, the advective term redistributes energy transferring it from one wavenumber to another. It has no influence over the total energy:
\begin{equation}
\frac{d E}{dt}  + \int^{\infty}_0\! 2\nu k^2 dk\, E(k,t) =0,
\end{equation}
the rate of decay of the total energy per unit mass is the dissipation rate $\overline{\varepsilon}=dE/dt.$ 

Henceforth, the advective non-linear term represents the collective action of all the modes over a specific one. Its general expression (\ref{eq:advective-term}) can be rewritten in an integral form; we use the property (\ref{prop:integral-form}) to put the spectral (2-point correlation) equation as follows
\begin{equation}
\frac{d}{d t} E(k,t)= \int^{\infty}_0\!\! dk'\,  S(k,k',\abs{\wn-\wn'},t) - 2 \nu k^2 E(k,t),
\label{eq:spectral-ns-2}\end{equation}
where $S$ satisfies the equation:
\begin{equation}
\int^{\infty}_{k_1}\!\! dk' \int^{k_2}_0\!\! dk\;  S(k,k',\abs{\wn-\wn'},t) =0,
\label{prop:advection-integral}\end{equation}
for arbitrary $k_1$ and $k_2$. 

Previously we introduced the Kolmogorov hypothesis for the configuration space assuming the process is also time-stationary, but equation (\ref{eq:spectral-ns-2}) does not posses that property. Let us consider for a moment that the advection is absent, then we have
\begin{equation*}
E(k,t)= E(k,t_0)\exp\!\left[-2\nu k^2 (t-t_0)\right]:
\end{equation*}
the greater the wavenumber the faster the energy density will decay. 

So, that is how the cascade happens: the non-linear term takes energy from the low wavenumbers, where there is net energy production, to compensate the net losses due to viscosity dissipation at  high wavenumbers. We hope that this transfer will lead the system at large times to a steady state. Because this situation is found in many real flows, the fully developed turbulence model is a representative class of turbulent phenomena. 

To consider a time-stationary state within this model we will introduce an artificial term. We will restore an external random force-like term $\mathbf{f}(\wn,t)$ into the spectral  equation. It should also satisfy, remember equation (\ref{eq:orthogonality}),
\begin{equation}
\wn\cdot \mathbf{f}(\wn,t)=0.
\end{equation}
It modifies equation (\ref{eq:spectral-ns-1}) as follows:
\begin{equation}
\left(\frac{d}{d t}+ 2 \nu k^2\right) E(k,t) = T(k,t)+2\pi k^2\mean{\mathbf{f}(\wn,t)\cdot\hat{\svl}(-\wn,t)}.
\label{eq:external-force}\end{equation}

Now, in order to find an explicit form for the rightmost term above we must characterize the random force. Lately, we have argued that the turbulent state should not be modified by external forces; then, suppose this force is Gaussian distributed, and its autocorrelation is given by
\begin{equation}
\mean{\mathbf{f}(\wn,t)\otimes\mathbf{f}(-\wn, t')}= \widehat{\mathbf{D}}(\wn)\, W(k) \delta(t-t').
\end{equation}
While the operator $\widehat{\mathbf{D}}$ is introduced to obtain an homogeneous, isotropic, and stationary force, the $\delta$-function makes it highly uncorrelated in time. Finally, $W$ has to be described: we will assume that the system response, for small time intervals $\abs{t-t'}$, is given by the Green function $g(k,t-t')$, such that
\begin{equation*}
\hat{\svl}(\wn,t) = \int_{\Re}\! dt'\, g(\wn,t-t')\, \mathbf{f}(\wn,t').
\end{equation*}
This kernel function also has the property
\begin{equation*}
g(\wn,t-t')=\left\{
\begin{aligned}
0 & \text{ for } t<t'\\
1 & \text{ for } t=t';
\end{aligned}\right.
\end{equation*}
so, it is causal and recovers the acceleration at equal times. We write then
\begin{equation}
2\pi k^2 \mean{\mathbf{f}(\wn,t)\cdot\hat{\svl}(-\wn,t)}= 2\pi k^2 \int_{\Re}\!\!dt'\; g(\wn,t-t')\,\tr\mean{\mathbf{f}(\wn,t)\otimes\mathbf{f}(-\wn, t')}= 4\pi k^2 W(k). 
\end{equation}
Henceforth, the equation (\ref{eq:external-force}) achieves its final form
\begin{equation}
\frac{d}{d t} E(k,t) = T(k,t)+ 4\pi k^2 W(k) - 2 \nu k^2 E(k,t).
\end{equation}

Stationary in time is found when the right-hand side of the latter equation is zero, under this circumstances it yields: 
\begin{equation}
\int^{\infty}_0\!\!\! dk'\;  S(k,k',\abs{\wn-\wn'},t) + 4\pi k^2 W(k) - 2 \nu k^2 E(k,t)=0.
\label{eq:energy-balance-stationary}\end{equation}
If we integrate this equation over the whole $\mathbf{k}$-space we obtain,
\begin{equation*}
\int^{\infty}_0\!\!\! 4\pi k^2 dk\; W(k) = \int^{\infty}_0\!\!\! 2\, \nu k^2 dk\; E(k) =-\overline{\varepsilon},
\end{equation*}
but a well-posed problem with separated input and dissipation ranges, it is what we have in the inertial range, implies the existence of a wave number $k^*$ such that the former is replaced by
\begin{equation}
\int^{k^*}_0\!\!\! 4\pi k^2 dk\; W(k) = \int^{\infty}_{k^*}\!\!\! 2\, \nu k^2 dk\; E(k) =-\overline{\varepsilon}.
\end{equation}
This means that the input term is peaked around $k=0$, and that the Reynolds number should not be too low.

Two energy-balance equations can be now drafted from (\ref{eq:energy-balance-stationary}): for the first we integrate from zero to $k^*$
\begin{equation}
\int^{k^*}_0\int^{\infty}_{k^*}\!\!\! dk'\,dk \; S(k,k',\abs{\wn-\wn'}) + \int^{k^*}_0\!\!\! 4\pi k^2 dk\;W(k) =0,
\end{equation}
here we have used property (\ref{prop:advection-integral}) to set the integration interval of the advective term; the second equation is obtained with the same argument but integrating from $k^*$ to infinity:
\begin{equation}
\int^{\infty}_{k^*}\int^{k^*}_0\!\!\! dk'\,dk \;  S(k,k',\abs{\wn-\wn'}) - \int^{\infty}_{k^*}\!\!\! 2\nu k^2dk\; E(k) =0.
\end{equation}

While the first equation tells us that the energy injected to the system from the low modes are transferred by the non-linear term to the higher modes, i.e., the inertial forces transfer energy from low to high wavenumbers. The second equation explains that the energy transferred is dissipated in the range $k^*<k'<\infty$.

\begin{figure}
\begin{center}
\includegraphics[width=0.65\textwidth]{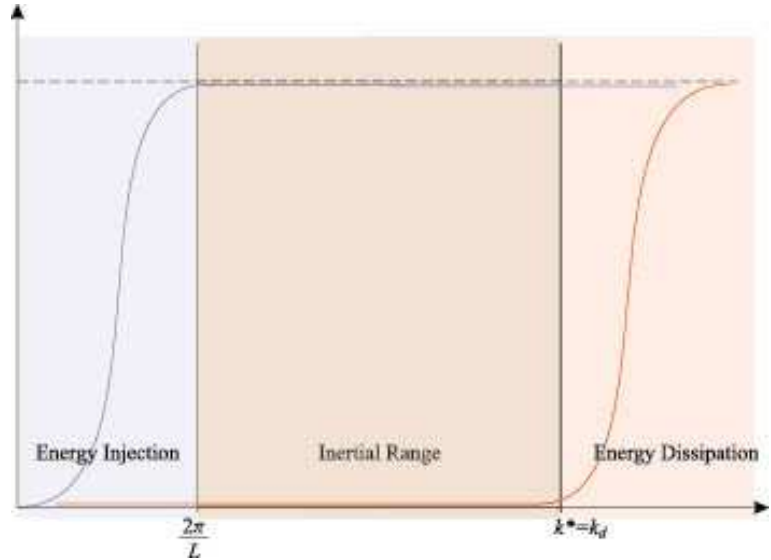}
\caption{Energy balance: three intervals are marked: below $2\pi/L$ (in blue) the energy is injected by the external forces, above $k^{\ast}$ (in red) the energy is dissipated by viscous forces, and finally the inertial range comes from the balance between the energy injection and dissipation---as it is shown in equation (\ref{eq:energy-balance-stationary}).
\label{fig:energy-balance}}
\end{center}
\end{figure}

We have finished characterizing the stochastic properties of the turbulence; also, we provided a model for the energy cascade. The inertial range is thus defined by those wavenumbers lesser than some $k^*$, where most of the advective term is concentrated. Therefore from the dimensional arguments we have used in Section \ref{section:kolmogorov-h}, we can take $k_d=k^*\sim 2\pi/l_0 \sim \overline{\varepsilon}^{1/4} \nu^{-3/4}$. We mentioned before that the injected energy $W$ should be concentrated around $k=0$; so, the limit $2\pi/L$ is the cut-off that sets the injection of energy. This is how the inertial range (see Figure \ref{fig:energy-balance}) is set within the wavenumber space:
\begin{equation}
\frac{2\pi}{L}\ll k \ll k_d.
\end{equation}

Now, the spectral density of energy becomes independent of the viscosity in this range as long as the Reynolds number is high. There is no other possibility for this function than to be
\begin{equation}
E(k)= \alpha\overline{\varepsilon}^{\,2/3} k^{-5/3},
\end{equation}
with $\alpha$ an adimensional constant. But, of course, this is the case of $\nu\rightarrow 0$ which is an idealization. If we introduce the dissipation via $k_d$ the distribution should be written,
\begin{equation}
E(k)= \alpha\overline{\varepsilon}^{\,2/3} k^{-5/3}F\left(\frac{k}{k_d}\right), 
\end{equation}
with $F(0)=1$. 

The theoretical form for $F$ is still under discussion. Moreover, there are strong clues that suggest that this functional relation extends all over the inertial range. It will be discussed next.

\subsection{The problem of the intermittency: Kolmogorov refined hypotheses}
\label{subsec:KRH}

The phenomenon of intermittent turbulence was first experimentally noticed by \citet{paper:batchelor}. They found that the energy was nonuniformly distributed throughout space in a fully developed turbulence. Some regions showed to be more active than others. The energy intermittence also implies that the dissipation behaves in the same way: this contradicts the assumptions that lead to equation (\ref{eq:scaling}). That is, the volume $V_l$ occupied by eddies of size $l$ is not constant, as we supposed after (\ref{eq:volume-energy}). Therefore, the global average should be replaced by local averages of the energy dissipation rate, for the former does not represent the behavior of $\varepsilon$. 

citet{paper:oboukhov} was the first to tackle this problem. His proposal was to divide the spatial domain into a collection of ensembles with characteristic dissipation $\varepsilon_r$---where $\varepsilon_r$ is the locally averaged dissipation over a spherical volume with diameter $r$ and center $\vect'$. Later \citet{paper:kolmogorov-2} used this proposal to rebuild his hypotheses. He added another hypothesis to shape the randomness of the energy dissipation rate: $\varepsilon_r(\vect',t)$ is a Gaussian variable having as variance of $\log\varepsilon_r$ the following
\begin{equation*}
\sigma^2_r(\vect',t)= A(\vect',t)+ 9\mu \log\left(\frac{L}{r}\right).
\end{equation*}
$\mu$ is a universal constant. Eventually he found the following scaling law for the $n$-order structure function:
\begin{equation}
\mean{\norm{\Delta \svl_{\vect}}^n}= K_n (\overline{\varepsilon} r)^{n/3} \left(\frac{L}{r}\right)^{\mu n(n-3)/3},
\label{law:scalar-intermittency}\end{equation}
where $\Delta\svl_{\vect}=\svl(\vect+\vect')-\svl(\vect')$, $K_n$ is an adimensional constant, and $\overline{\varepsilon}$ is, as before, the energy dissipation rate averaged over the entire bath of characteristic size $L$. Nevertheless, it is found that the universal constant $\mu$ depends on the order $n$. This is known as scale-similar theory of intermittency. It is worth nothing that it is unnecessary the log-normality to explain (\ref{law:scalar-intermittency}). 

To get a picture of how intermittence is created from the continous process of stretching and twisting of the advective term we will follow \citet{paper:frisch}. Let us restart from equation (\ref{prop:lenght-contraction}), assume that the average number of offspring of any eddy is $N$---an eddy of scale $l_n$ is supposed to give rise to $N$ eddies of scale $l_{n+1}$, irrespective of the value of $n$. These $N$ eddies occupy a fraction $\beta$ of their predecessor
. This fractional reduction in volume from one generation to the next is given by
\begin{equation}
\beta = \frac{N l_{n+1}^3}{l_n^3} = N\pi^3 \leq 1.
\label{def:volume-reduction}\end{equation}
Furthermore, let us suppose that the largest eddies fill all the space available to them $V_L\sim L^3$, the $n$-generation occupies just its \textit{active} fraction
\begin{equation}
V_{l_n}= \beta^n V_L.
\end{equation}
With the arguments we have given, the accumulated energy from the eddies of size $n$ is now
\begin{equation}
E_{l_n}\sim V_{l_n}v_{l_n}^2\sim V_L \beta^n v_{l_n}^2,
\end{equation}
while the globally averaged energy flux gives---remember that $\overline{\varepsilon}\sim\varepsilon_n$ because there is no energy losses within the inertial range---the following scaling law for the velocity
\begin{equation}
\overline{\varepsilon}\sim\frac{E_{l_n}/\tau_{l_n}}{V_L}\sim \frac{V_{l_n}v_{l_n}^3/l_n}{V_L}\sim \beta^n l_n^{-1} v_{l_n}^3.
\end{equation}
However, we want to express the above formulas in terms of the scale length. Let us assume that the splitting number $N$ is inversely  proportional to a power of the contraction ratio, i.e. $N=\pi^{-D}$. The volume reduction for the $n$-generation of eddies results
\begin{equation}
\beta^n=\left(N\pi^3\right)^n = \pi^{(3-D) n}= (\pi^n)^{3-D}=\left(\frac{l_n}{L}\right)^{3-D},
\end{equation}
here we have used definitions (\ref{def:volume-reduction}) and (\ref{prop:lenght-contraction}). 
Now, we can estimate the intermittent scaling law for the energy
\begin{equation}
E_{l_n}/V_L\sim \overline{\varepsilon}^{2/3} l_n^{2/3} \left(\frac{l_n}{L}\right)^{\frac{3-D}{3}},
\end{equation}
$D$ is identified as the Hausdorff fractal dimension by \citet{paper:mandelbrot}, and represents how much space is filled by the eddies so $3-D>0$. The functional expression for $2$-order structure function, or just structure function, should be 
\begin{equation}
\mean{\norm{\Delta \svl_{\vect}}^2}= K_n (\overline{\varepsilon} r)^{2/3} \left(\frac{r}{L}\right)^{\frac{3-D}{3}}, 
\label{def:singular}\end{equation}
and comparing against equation (\ref{law:scalar-intermittency}) it is $2\mu = 3-D$. 

Intermittency means that the probability for having large velocity fluctuations increases at small scale, $v_{l_n}\sim \overline{\varepsilon}^{1/3} l_n^{1/3} (L/l_n)^{(3-D)/3},$ but at the same time the amount of these eddies decreases with the scale. Furthermore, this implies the existence of a singularity in the Navier-Stokes  equation when $\nu\rightarrow 0$. It can be stated formally as follows: given $h=1/3$, these singularities are contained in a set $S_h$ with fractal dimension $D_h=D$. In the sense, $\vect' \in S_h$
\begin{equation}
\lim_{\vect\rightarrow 0}\frac{\Delta \svl_{\vect}(\vect')}{\norm{\vect}^h}\neq 0.
\end{equation}

Therefore, we have proven that the Frisch's $\beta$-model accounts for all these properties of the intermittence, and produces a 2-point structure function coherent with the experimental findings. Besides, we can estimate higher structure functions for the velocity field, 
\begin{equation}
\mean{\norm{\Delta \svl_{\vect}}^n} \propto r^{\zeta_n}
\label{eq:n-point-structure-velocity}\end{equation}
with $\zeta_n= h\, n + 3-D$, and $h= (D-2)/3$. The structure function exponent has a linear growing with $n$; actually, experimental tests \citep{paper:anselmet} shows a non-linear grow. The   statement affirming that the energy transfer per unit volume is constant through the different eddies scales is not valid \citep{paper:frisch-parisi}. 

To override this problem let us suppose that instead a single scale $h$ there is a range  of them   $h_{\min}\leq h\leq h_{\max}$ where (\ref{def:singular}) happens. Thus, we have a wider set of singularities, $\cup_{h\in[h_{\min},h_{max}]} S_h$.  Each subset has its own fractal dimension $D_h$. To calculate the moments of the velocity variation $\Delta \svl_{\vect}$ at a point $\vect'$ we have to look for the probability of finding such variation in one of the subsets $S_h$. This probability is proportional to the volume of $S_h$ thickened along the \textit{normal} direction by a length $r$. 

This enlarged set is defined for any given set $F\subset \Re^n$ as the $\lambda$-parallel body:
\begin{equation*}
F_{\lambda}=\{\vect\in\Re^n:\norm{\vect'-\vect}\leq \lambda,\quad \vect'\in F\}. 
\end{equation*}
We want to estimate its volume, let us pick up some examples first. For a single point set, $F=\{\vect\}$ it is obviously $\vol{F_{\lambda}}= 4\pi \lambda^3/3$. If we take $F$ as a segment of length $l$ it is $\vol{F_{\lambda}}\sim\pi l \lambda$, and for an extended \textit{flat} surface of area $A$ is $\vol{F_{\lambda}}\sim A\lambda^2$. In each case, $\vol{F_{\lambda}}\sim c \lambda^{3-s}$ with $s$ the dimension of $F$. This idea can be extended to fractal dimensions \citep{book:falconer}.  It follows that the $n$-order structure function is the average
\begin{equation}
\mean{\norm{\Delta\svl_{\vect}}^n} \propto \int \mu(dh)\, r^{n h + 3 - D_h}
\end{equation}
where $\mu(dh) r^{3-D_h}$ is the probability over the spectrum $[h_{\min},h_{\max}]$. In the short distance limit this integral can be estimated using a saddle point approximation, so
\begin{equation}
\mean{\norm{\Delta\svl_{\vect}}^n} \propto r^{\zeta_n},\text{ with}\quad\zeta_n=\min_h(n h +3-D_h).
\end{equation}
Since intermittency is related to singular velocity variations one should expect $h\leq h_{\max}=1/3$. Of course $0\leq D_h\leq 3$ and then $\zeta_n\leq n/3$. Also, we can see that each $\zeta_n$ depends on a specific value of $h$; therefore, the velocity moments pick out a particular subset $S_h$.

Besides the latter model, other alternatives to the $\beta$-model using fractal geometry has been proposed \citep{paper:benzi} to explain the energy transfer: the random $\beta$-model. It is assumed that the eddy splitting is not constant; from one generation to the other $N$ changes. That is, the contraction factors $\beta$ are independent random variables. 

These fractal models gives a better understanding of the phenomenon of advection and intermittence. But neither of them predict the values the anomalous dimension, and it must be found by experimental means.

\section{Passive Scalar Fields' Characterization}
\subsection{Scalar turbulence}

The turbulent flow transports and disperses any scalar by making parcels of fluid follow chaotic trajectories: the non-uniformity of the turbulence causes lines of constant scalar stretch and fold. This process drives the scalar concentration through smaller scales; eventually, the diffusivity $\kappa$ associated to the scalar (e.g. thermal diffusivity, molecular diffusivity, etc.) prevails over the advective mixing. There are so many parallels between this behavior and the one of velocity field that we address it as \textit{scalar turbulence} \citep{paper:shraiman}.

In a turbulent flow the scalar is controlled by two processes: transport, the physical translocation of the scalar via the combined action of fluid advection and diffusion; and mixing, the irreversible decay of fluctuations because of the scalar diffusion, $\kappa$, that tends to reduce the scalar field to uniformity. In a turbulent flow, both processes become independent of $\kappa$ as it goes to zero. This limit is the so called \textit{fully developed scalar turbulence}.

Any given scalar (concentration of the) quantity $\Theta$ put into a static fluid is subject to a diffusion equation, that is,
\begin{equation*}
\frac{\partial}{\partial t}\;\Theta (\vect, t) = \kappa\;\triangle \Theta (\vect, t).
\end{equation*}
The extension to flowing fluids is accomplished replacing the partial time derivative by the total derivative, so to speak:
\begin{equation}
\frac{D}{Dt}\;\Theta (\vect, t) =\frac{\partial}{\partial t}\;\Theta (\vect, t) + (\vel\cdot\nabla) \Theta (\vect, t)= \kappa\;\triangle \Theta (\vect, t),
\label{eq:scalar-flow}\end{equation}
with $\vel$ the \textit{random} velocity field which is naturally solenoidal, i.~e.,$\nabla \cdot \textbf{v}=0$. 

Now, the above equation is our starting point. One should consider introduce the Navier-Stokes equation to describe the velocity field, but we will not do so. In fact, most of the actual developments in scalar turbulence does not need a deep understanding of the velocity field: for all the purposes here will be enough to describe it as a given isotropic and homogeneous stochastic field. 

Because of the incompressible nature of the fluids equation (\ref{eq:scalar-flow}) can be rewritten:
\begin{equation}
\frac{\partial}{\partial t}\;\Theta + \nabla \cdot \left( \svl\, \Theta - \kappa\;\nabla \Theta \right) =0, 
\label{eq:scalar-flow-2}\end{equation}
here $\svl$ is the stochastic velocity field for the zero mean turbulent velocity field, and we also used $\nabla \cdot(\svl\,\Theta) = \svl\cdot\nabla\Theta + (\nabla\cdot\svl)\Theta$---with the last term vanished. We write, again, the concentration of the scalar as the sum of a mean and a random fluctuation,
\begin{equation}
\Theta(\vect, t) = \mean{\Theta}(\vect, t) + \vartheta(\vect, t),
\label{def:mean-stch}\end{equation}
with $\mean{\vartheta}=0$. In this case the mean of the scalar $\mean{\Theta}$ can not be neglected, because scalar turbulence is usually generated by maintaining a mean scalar gradient. Note that the homogeneous and isotropic velocity field does not guarantee the same properties on the scalar turbulence. 

Henceforth, using the latter definition, the averaged equation (\ref{eq:scalar-flow-2}) gives us,
\begin{equation}
\frac{\partial}{\partial t}\;\mean{\Theta} + \nabla \cdot 
\left( \mean{\svl\,\vartheta} - \kappa\;\nabla \mean{\Theta} \right) =0. 
\label{eq:averaged}\end{equation}
The first term within the divergence term represents the effect of velocity field in the transport of the concentration. We may assume that the bulk effect of this transport is proportional to the action of the mean concentration's gradient. That is,
\begin{equation}
\mean{\svl\,\vartheta} = - \kappa_T \nabla \mean{\Theta}, 
\label{def:eddy-diff}\end{equation}
$\kappa_T$ is defined as the \textit{eddy diffusivity}. This prescription for $\mean{\svl\,\vartheta}$ is widely used in engineering applications, for it `solves' the closure problem. It also expresses that anisotropies introduced by the large scales are maintained overall the scales range. Henceforth, the universality of the Kolmogorov treatment could be lost, because the turbulence can not be unbounded from the large scale forces. We will discuss the full effect of this behavior in forthcoming sections.

The integral $1/2 \int_V d^3r\; \vartheta^2$ is a good measure of the concentration of the scalar field. It is zero only when $\vartheta$ is zero, and therefore no turbulence is present. An \textit{energy balance} equation can be obtained for it: we just take the difference between eqs. (\ref{eq:scalar-flow-2}) and (\ref{eq:averaged}), and afterwards the average to have,
\begin{equation}
\frac{1}{2}\frac{\partial}{\partial t}\mean{\vartheta^2} + \frac{1}{2}\,\nabla \cdot \left(\mean{\svl\,\vartheta^2}-\kappa \nabla\mean{\vartheta^2}  \right) + \mean{\svl\,\vartheta} \cdot \nabla \mean{\Theta} + \kappa \,\mean{\left(\nabla \vartheta\right)^2} =0,
\end{equation}
integrating it over the whole bath, where the divergence terms are suppose to vanish, yields 
\begin{align}
\frac{d}{d t} \int_V d^3r\; \frac{\mean{\vartheta^2}}{2} =& \int_V\! d^3r\; \left[\mean{\vartheta \svl\,} \cdot \nabla \mean{\Theta} -\kappa \mean{\left(\nabla \vartheta \right)^2}\right]\nonumber\\
=& \int_V\!d^3r\; \left[\kappa_T \left(\nabla \mean{\Theta}\right)^2 -\kappa \mean{\left(\nabla \vartheta \right)^2}\right].
\label{eq:balance}\end{align}
We have used definition (\ref{def:eddy-diff}) to arrive to this last equation. The scalar will be stationary when $d \mean{\vartheta^2}/d t=0$. Then, there is balance between the injected energy from the scalar gradient and diffusive term:
\begin{equation}
\chi=\int_V\!\!d^3r\;  \kappa_T \left(\nabla \mean{\Theta}\right)^2= \int_V\!\!d^3r\; \kappa \mean{\left(\nabla \vartheta \right)^2}.
\label{eq:stationary-balance}\end{equation}
while the left-hand term represents the scalar concentration \textit{created} per unit time, the right-hand is the concentration \textit{destroyed} by molecular diffusion. 

But if we look for isotropic and homogeneous scalar turbulence the gradient of the mean field must be zero, and also $\nabla\cdot\mean{\svl\,\vartheta}=0$; therefore, the former model for energy injection should be left behind. Time-stationary turbulence needs an energy source to counterbalance the scalar dissipation, so an external `force' $f$ should be supplied at the right-side of (\ref{eq:scalar-flow-2}). Doing so, in equation (\ref{eq:balance}) we just make the replacement $\kappa_T \left(\nabla \mean{\Theta}\right)^2 \rightarrow \mean{f \vartheta}$ . 

Let us inspect briefly the decaying of the scalar concentration. It is appropriate to use the Fourier formalism we applied with the velocity field; from equation (\ref{eq:scalar-flow-2}) with the above prescriptions
\begin{equation*}
\left( \frac{\partial}{\partial t}  - \kappa\,\triangle \right)\vartheta(\vect,t) =-\nabla \cdot (\svl\, \vartheta)(\vect,t),
\end{equation*}
we get the following representation:
\begin{equation}
\left(\frac{d}{dt} + \kappa k^2 \right) \hat{\vartheta}(\wn,t)= -i \int_{\Re^3}\!\!d^3k'\; \wn\cdot\hat{\svl}(\wn-\wn',t)\, \hat{\vartheta}(\wn',t).
\end{equation}
Obtaining an evolution equation for the non-stationary spectrum is straightforward: multiply both sides of the above equation by $\hat{\vartheta}(-\wn,t)$, and then average. The result is the following balance equation for the spectral distribution of the concentration:
\begin{equation}
\left(\frac{d}{dt} + \kappa k^2 \right) F(k,t) = T_{\vartheta}(k,t),
\end{equation}
where it is defined $F(k,t)= 4\pi k^2 \mean{\vartheta(\wn,t)\,\vartheta(-\wn,t)}$ with 
\begin{equation}
\int^{\infty}_0\!\!dk\;  F(k,t)= \mean{\vartheta^2},
\label{eq:scalar-spectra}\end{equation}
and the scalar transfer spectrum
\begin{equation*}
T_{\vartheta}(k,t)= - 8\pi k^2 \int_{\Re^3}\!\! d^3k'\; i \wn\cdot\mean{\vartheta(\wn-\wn',t)\,\svl(\wn',t)\,\vartheta(-\wn,t)}.
\end{equation*}
Now, as in Section \ref{section:isotropic-cascade} it can be interpreted in the same way as the  energy-balance equation. The scalar transfer also possesses a conservative property:
\begin{equation*}
\int^{\infty}_0\!\!dk\; T_{\vartheta}(k,t)=0,
\end{equation*}
this is because of the isotropy we have imposed on the transport term $\mean{\svl\,\vartheta}$. Taking integrals on both sides of the spectral balance equation we obtain again,
\begin{equation}
\frac{1}{2}\,\frac{d}{dt}\mean{\vartheta^2}= -\chi=- \kappa \int^{\infty}_0\!k^2 dk\;  F(k,t).
\end{equation}

The idea that all the properties encountered studying the turbulent velocity can be mapped to the scalar turbulence was also employed to build an inertial convective (time-stationary) spectrum for the scalar variance. \citet{paper:oboukhov-2} and, independently, \citet{paper:corrsin} settled down the first steps towards a scaling law for scalar fields. Using the analogy between the scalar diffusion coefficient $\kappa$ and the viscosity coefficient they both used dimensional arguments to find an extra cut-off for the energy spectrum:
\begin{equation}
k_c = \overline{\varepsilon}^{1/4} \kappa^{-3/4}.
\end{equation}

Because we now have two ultraviolet cut-offs, $k_c$ and $k_d$, the inertial convective range should be that where both the viscosity and the diffusion coefficients go to zero, i.e.,
$2\pi/L\ll k \ll \min{\{k_c,k_d\}}.$ Afterwards, within this range the power spectrum for the scalar is,
\begin{equation}
F(k)= A \overline{\chi}\,\overline{\varepsilon}^{-1/3} k^{-5/3},
\end{equation}
where $A$ is known as the Oboukhov-Corrsin constant. 

Unlike the velocity field, there is an innate term $\mean{\svl\,\vartheta}$ injecting energy down to the small scales, and it can not be discarded at will. It causes one of the characteristic features of the scalar turbulence: the observation of ramp-and-cliff structures regardless the model introduced for the velocity field. This is the source of scalar intermittence and anisotropy seen in experiments and numerical simulations. On the contrary, the Kolmogorov-Oboukhov-Corrsin (KOC) theory assumes that the advective term restores universality, i.e., independence of large-scale injection mechanisms, and thus isotropy at the inertial-convective range. This is far from our final goal, which describes the full behavior of the scalars. But before inspect the anomalous scaling inherent to intermittence, we will introduce an isotropic model binding the scalar fields to the velocity scaling behavior.

\subsection{Kraichnan's model}
\label{section:kraichnan-model}

The \citeauthor{paper:kraichnan}'s model \citeyearpar{paper:kraichnan} for passive advection  assumes that both velocity and scalar fields are homogeneous and isotropic. So, we must introduce an \textit{external} force $f$---proportional to the diffusion $\kappa$---in order to compensate the energy dissipation. Equation (\ref{eq:scalar-flow-2}) is modified to yield,
\begin{equation}
\frac{\partial}{\partial t}\;\Theta + \nabla \cdot \left( \svl\, \Theta - \kappa\;\nabla \Theta \right) =f.
\label{eq:kraichnan}\end{equation}
Let the velocity field be a Gaussian process independent of the random force. This synthetic velocity has the following $2$-point correlation
\begin{equation}
\mean{\svl(\vect,t)\otimes\svl(\vect',t')}= \mathbf{D}(\vect-\vect')\,\delta(t-t') 
\end{equation}
with $\mathbf{D}(\vect)= \mathbf{D}(0)\mathbb{I} - \mathbf{d}(\vect)$ such that,
\begin{equation}
\mathbf{d}(\vect) =  D \left[ (d -1 + \zeta)\,\mathbb{I} - \zeta\frac{(\vect\otimes\vect)}{\norm{\vect}^2}\right] \norm{\vect}^{\zeta},
\label{eq:irregular-cov}\end{equation}
where  $0<\zeta<2$ is a fixed parameter, and $d(=3)$ is the dimension of the configuration space. More rigorous approaches require a regular covariance function, so it is often introduced an infrared cut-off $m$ to rewrite it as follows 
\begin{equation}
\mathbf{D}(\vect) = D_0 \int_{\Re^3}\!\! d^3 k\;\frac{\exp( i \wn\cdot\vect)}{(k^2+m^2)^{(d+\zeta)/2}}\left(\mathbb{I}- \frac{\wn\otimes\wn}{k^2}\right),
\label{eq:regular-cov}\end{equation}
where $D_0$ is related to the constant in (\ref{eq:irregular-cov}) by
\begin{equation}
D = \frac{\Gamma[(2-\zeta)/2]}{2^{2+\zeta}\pi^{3/2}\zeta(3+\zeta)\, \Gamma[(\zeta+3)/2]}D_0, 
\label{eq:diffusivity}\end{equation}
as a direct calculation shows.

This model for the stochastic velocity field is far from realistic. Besides the fact that it mimics the spatial dependence of the correlation function, the opposite happens with the time. The velocities are white noise in time: they are uncorrelated everywhere but $t=t'$. Hence, at any instant all the moments of the velocities are infinite! There is independence from the past, which contradicts the KOC model where the time-to-live of the scalar inhomogeneity is $\tau(r)\sim r^{2/3}$. 

Moreover, the KOC theory predicts the $\mean{ \norm{\Delta\svl}^2}\sim r^{2/3} \sim r^{4/3}/\tau(r)$ law. When compared against the Kraichnan's model $\mean{ \norm{\Delta\svl}^2}\sim r^{\zeta}\delta(t)$: we have $\zeta = 4/3$. It is said then that the velocity field changes very rapidly in time. Additionally, we observe that $\mathcal{O}(\norm{\svl})\sim \sqrt{t^{-1}}$.

Also, we will set the force to be Gaussian with zero mean, and having the following 2-point correlation function:
\begin{equation}
\mean{f(\vect,t)f(\vect',t')}= C(\vect-\vect')\,\delta(t-t'),
\end{equation}
where $C$ is invariant under rotations and its compact support set has the extension of the bath $L$. We are interested in the stationary regime which is found when (\ref{eq:stationary-balance}) holds: it requires
\begin{equation}
\overline{\chi}\equiv \kappa \mean{(\nabla \vartheta)^2} = \mean{f\,\vartheta} =\frac{1}{2}\, C(0).
\end{equation}

Now, with these prescriptions we will solve equation (\ref{eq:kraichnan}). From the homogeneous equation, which is a Fokker-Planck equation, we obtain the Green's function:
\begin{equation}
\left[\frac{\partial}{\partial t} +  \svl \cdot \nabla - \kappa\;\triangle \right] G_{\kappa}(\vect,t;\vect',t')=0
\end{equation}
with initial condition $G_{\kappa}(\vect,t_0;\vect',t_0)=\delta^3(\vect-\vect')$. Hence, the Kraichnan's equation has the solution, 
\begin{equation}
\vartheta(\vect,t)=\int^t_{t_0} dt'\int_{\Re^3}\! d^3r'\, G_{\kappa}(\vect,t;\vect',t') f(\vect',t') + \int_{\Re^3}\! d^3r'\, G_{\kappa}(\vect,t;\vect',t_0) \vartheta_0(\vect'),
\end{equation}
for $\vartheta(\vect,t_0)= \vartheta_0(\vect)$.

The resolvent is directly found from the Lagrangian trajectories of the fluid particles, that is,
\begin{equation*}
d\boldsymbol{\rho}(t)= \svl(\boldsymbol{\rho}(t),t)\, dt+\sqrt{2\kappa}\, d\mathbf{B}(t),\quad \text{with }x(t_0)=x_0
\end{equation*}
where $\mathbf{B}$ the isotropic Brownian motion---note that both terms are of order $\sqrt{t}$. This is the symbolic representation for a \textit{Stochastic Integral}, and we will delay its interpretation until Chapter \ref{chp:stochastic-calculus}. Instead, we are going to follow the original Kraichnan's paper. There, the resolvent is found to have the formal expression,
\begin{equation}
G_{\kappa}(\vect,t;\vect',t_0)= \left.\exp \left\{-\int^t_{t_0}dt' \left[\kappa\triangle - \svl\cdot\nabla\right](t')\right\}\right|_{(\vect,\vect')}.
\end{equation}
If we want to estimate the solution at time $t_0+\delta t$, we just need to approximate up to the first order in $\delta t$ of the Green's function, that is, 
\begin{align*}
G_{\kappa}(\vect,t_0+\delta t;\vect',t_0)=& \;1 + \int^{t_0+\delta t}_{t_0}\!\!dt'\, \left.(\svl(t')\cdot\nabla)\right|_{(\vect,\vect')}\, -\left.\kappa \triangle \delta t\right|_{(\vect,\vect')} + \\
+& \int^{t_0+\delta t}_{t_0}\!dt'\int^{t'}_{t_0}\!\!dt'' \left.(\svl(t')\cdot\nabla)(\svl(t'')\cdot\nabla)\right|_{(\vect,\vect')}\,+ \cdots.
\end{align*}
This procedure allows us to build a differential equation for the moments of the scalar. But here, the first order approximation let us only estimate the time evolution of the 2-point correlation:
\begin{align*}
\mean{\vartheta(\vect,t_0+\delta t)\vartheta(\vect',t_0+\delta t')}-\mean{\vartheta(\vect,t_0)\vartheta(\vect',t_0)}= \kappa \left(\triangle_{\vect}+\triangle_{\vect'}\right)\mean{\vartheta(\vect,t_0)\vartheta(\vect',t_0)}\,\delta t +\\
+ \int^{t_0+\delta t}_{t_0}dt'\int^{t'}_{t_0}dt'' \mean{(\svl(\vect,t')\cdot\nabla)(\svl(\vect,t'')\cdot\nabla)\,\vartheta(\vect,t_0)\vartheta(\vect',t_0)} +\left\{\vect\leftrightarrow\vect'\right\}+\\
+\int^{t_0+\delta t}_{t_0}dt'\int^{t_0+\delta t}_{t_0}dt''\mean{(\svl(\vect,t')\cdot\nabla)(\svl(\vect',t'')\cdot\nabla)\,\vartheta(\vect,t_0)\vartheta(\vect',t_0)}+\\
+\int^{t_0+\delta t}_{t_0}dt'\int^{t_0+\delta t}_{t_0}dt'' \mean{f(\vect,t')f(\vect',t'')}.
\end{align*}
The process $\vartheta(\vect,t_0)$ is markovian\footnote{
It is said that a process $X_t$ is markovian or possess the \textit{Markov property} if the future behavior of it given what has happened up to time $t$ is the same as the behavior obtained when starting the process at $X_t$ (a detailed description can be found at \citet[Chapter 7]{book:shiryayev}).
}, for it is statistically independent of $\svl(t)$ and $f(t)$ whenever $t>t_0$. Therefore, assuming the scalar is time-stationary the left-hand side is zero. Under the homogeneous hypothesis it is $F_2(\vect)=\mean{\vartheta(\vect)\vartheta(0)}$ and so we have:
\begin{equation}
\left(-\kappa\, \triangle - \nabla\cdot \mathbf{d}(\vect)\cdot\nabla\right) F_2(\vect)= C(\vect),
\label{eq:correlation-diff}\end{equation}
adding the isotropy condition implies the operator $\mathcal{M}^0_2= \nabla\cdot \mathbf{d}(\vect)\cdot\nabla$ can be rewritten as
\begin{equation*}
\mathcal{M}^0_2= -D(d-1)\frac{1}{r^{d-1}}\frac{d}{dr} r^{d+\zeta -1}\frac{d}{dr}.
\end{equation*}
Once we set the boundary conditions, the solution to (\ref{eq:correlation-diff}) is:
\begin{equation*}
F_2(r)= \int^{\infty}_r\!\! dr_1\int^{r_1}_0\!\!dr_0\,\frac{ r_0^{d-1} C(r_0)}{D(d-1)\, r_1^{d+\zeta-1}+ \kappa r_1^{d-1}}.
\end{equation*}
With a bath having finite extension $L$, the inviscid limit $\kappa\rightarrow 0$ yields
\begin{equation}
\left.F_2(r)\right|_{\kappa=0}= C L^{2-\zeta} - \frac{2\overline{\chi}}{ D d(d-1)(2-\zeta)}\, r^{2-\zeta} + \dots,
\label{eq:correlation-function-scalar}\end{equation}
the constant $C$ comes from the selected force correlation $C(\vect)$. Besides, the 2-point structure functions is independent of it, and is
\begin{equation}
S_2(r)= \mean{(\vartheta(\vect)-\vartheta(0))^2}= \frac{4\overline{\chi}}{D d(d-1)(2-\zeta)} r^{2-\zeta}.
\label{eq:structure-function-scalar}\end{equation}

Therefore, homogeneous and isotropic scalar turbulence imposes a universal law for the correlation function, since it only depends on the mean dissipation rate and the distance.

\subsection{Anomalous scaling, anisotropy and diffusion}

We surveyed with the Kraichnan's model the behavior of the isotropic and homogeneous scalar turbulence. We note that intermittence in the velocity field is directly translated to the scalar field, equation (\ref{eq:structure-function-scalar}). The anomalous scaling in scalar turbulence is just the separation from this inherited power spectra. 

In the present context universality of scalar fields should be understood as the existence of limiting correlation functions $\mean{\Pi_n\vartheta(\vect_n,t_n)}$ in a stationary regime when $\kappa, m,1/L\rightarrow 0$---as defined in the former section---are independent of the external force, i.e., the shape of the correlation function $C$. This condition let us build the following inertial range:
\begin{equation*}
\left(\frac{\kappa}{D}\right)^{1/\zeta}\ll \norm{\vect}\ll \min\{L,m^{-1}\},
\end{equation*}
with $D$ as in (\ref{eq:irregular-cov}).

\citet*{paper:gawedzki} determined that the correlators become independent of the diffusion and infrared cut-off, but there exits a dependence upon the external force. The integral scale $L$ contributes to the $n$-point structure functions 
\begin{equation}
S_{2n}(r)=\mean{[\vartheta(\vect)-\vartheta(0)]^{2n}}\sim A\norm{\vect}^{\zeta_n} = \gamma_{2n}\left( L/\norm{\vect}\right)^{\rho_{2n}}\norm{\vect}^{(2-\zeta)n},
\end{equation}
for $\norm{\vect}\ll L$ as $\kappa,m$ go to zero. The amplitudes $\gamma_{2n}$ depend on $\zeta$ and the function $C$, while the exponents $\rho_{2n}$ are just functions of $\zeta$ but not $C$. While for $n=1$ the intermittence is absent, $\rho_2=0$, for $n>1$ the $\rho$'s are positive, increasing, and convex function of $n$. 

They followed the arguments from \citet{paper:kraichnan-2} who noticed that an exact expression can be written for advection effects on scalar structure functions when velocity fields change rapidly in time, they are delta correlated. The second assumption is the need for the scalar field $\vartheta$ to be Gaussian, because it makes possible to break down the closure problem, that is, the dependence on higher moments of the scalar. 

Next, the velocity statistics is introduced through the two-particle eddy diffusivity 
\begin{equation*}
\eta(r)= \frac{1}{2}\int^t_{-\infty}\!\!dt'\, \mean{\delta_{\parallel}u(\vect,t)\delta_{\parallel}u(\vect,t')}
\end{equation*}
where $\delta_{\parallel}u(\vect,t)= [\svl(\vect',t)-\svl(\vect'+\vect,t)]\cdot \vect/r$. Inside the inertial range it has the form
\begin{equation}
\eta(r)=\eta_0 \left(\frac{r}{L}\right)^{\zeta(\eta)}\!,
\end{equation}
here we have changed the former exponent $\zeta$ into a functional of the diffusivity $\zeta(\eta)$. The final differential equation for the $n$-point structure functions is
\begin{equation*}
\frac{\partial S_{2n}(r)}{\partial t}- \frac{1}{r^{2}} \frac{\partial}{\partial r}\left(r^2\eta(r) \frac{\partial S_{2n}(r)}{\partial r}\right)=-2n\kappa \, \triangle S_2(0) \frac{S_{2n}(r)}{S_2(r)}.
\end{equation*}
The steady state makes the temporal derivative vanish, and the 2-point structure function is easily obtained:
\begin{equation}
S_2(r)= \frac{\kappa\,\triangle S_2(0)}{6(\eta_0/L^2)\zeta_2(\vartheta)}\, (r/L)^{\zeta_2(\vartheta)}
\label{eq:structure-2}\end{equation}
where $\kappa \,\triangle S_2(0)$ is the rate of dissipation of the scalar $\overline{\chi}$, and the scalar exponent is related to the eddy diffusivity exponent by
\begin{equation}
\zeta_2(\vartheta)= 2- \zeta(\eta),\quad\text{with }\; 0<\zeta(\eta)<2.
\label{eq:exp-relation}\end{equation}
These equations guarantee a precise functional relation for the exponents within the inertial-range, and again this exponents relation is found independent of the external source:
\begin{equation}
\zeta_{2n}(\vartheta)=\frac{1}{2}\sqrt{12 n \zeta_2(\vartheta) +[3-\zeta_2(\vartheta)]^2} -\frac{1}{2}[3-\zeta_2(\vartheta)]
\label{eq:n-exp-relation}\end{equation}
providing an asymptotic behavior as $n$ increases $\zeta_{2n}\sim \sqrt{3n\zeta_2}$. If we suppose that the above equation can continuously extended to $n=1/2$ we found then the exponent of $\mean{\abs{\Delta\vartheta}}$ to be
\begin{equation*}
\zeta_1(\vartheta)=\frac{1}{2}\sqrt{6 \zeta_2(\vartheta) +[3-\zeta_2(\vartheta)]^2} -\frac{1}{2}[3-\zeta_2(\vartheta)].
\end{equation*} 
\citet{paper:constantin} showed this exponent is also related to the fractal dimension\footnote{
The fractal dimension $\dim_H$ is formally known as \textit{Hausdorff dimension}. See the {\appendixname} A for a definition and some relevant properties.
} of the set of isoscalar surfaces contained in a sphere of radius the order an inertial-range scale, that is,
\begin{equation}
\dim_H \vartheta^{-1}(c) =3-\zeta_1(\vartheta).
\label{eq:fractal-scalar-dimension}\end{equation}
It resembles our discussion of the volume fraction $\beta$ filled by eddies with characteristic length $l_n$ in Section \ref{subsec:KRH}, but in this case the scalar exponent is exactly the co-dimension. Meaning the scalar field exponents are strongly determined by the scalar spatial distribution.

Furthermore, we can examine the significance of the exponents in the limiting cases. As $\zeta(\eta)\rightarrow 2$, all the structure function's powers go to zero as (\ref{eq:exp-relation}) and (\ref{eq:n-exp-relation}) shows. This is realized when the scales are near the inner length defined by the velocity field, where the viscous term in the Navier-Stokes equation becomes relevant. We note that $\dim_H \vartheta^{-1}(c)\rightarrow 3$ and the scalar fills all the space.

The opposite limit, $\zeta(\eta)\rightarrow 0$, makes the power spectra approach to $k^{-3}$. But at the same time the effective eddie diffusivity $\eta_0$ grows. Thus, we observe from both equations (\ref{eq:structure-2}) and (\ref{eq:structure-function-scalar})---with $D$ from (\ref{eq:diffusivity})---their coefficient going to zero. 

Because at small $r$ the difference $\abs{\Delta_{\vect} \svl}$ is at most of order 1, it is $0<\zeta_{2n}<2n$. Also, applying the H\"older inequality we can prove $\zeta_q/q\leq\zeta_p/p$, for any two $p,q>0$. So we found an upper bound to the scalar dimension, as defined in (\ref{eq:fractal-scalar-dimension}), $2\leq \dim_H \vartheta^{-1}(c)\leq 2 +\zeta(\eta)/2$. Henceforth, in the limit $\zeta(\eta)=0$ the scalar turbulence is contained in $2$-dimensional sheets. 

Additionally, if the velocity field $\svl$ changes slowly in time, but remains Gaussian, and has a long scaling range $0<\zeta_2(\svl)<2$---in the sense of (\ref{eq:n-point-structure-velocity})---it acts like a rapidly changing field because the large scales sweep fluid elements rapidly through the small scales. So we should have $\zeta(\eta)=\zeta_2(\svl)+1$, but it does not hold for any pair of values $\zeta(\eta),\zeta_2(\svl)$. As can be seen from the ranges covered by each power; moreover, it seems valid near the classical values: $\zeta(\eta)=4/3$ and  $\zeta_2(\svl)=1/3$. In any case we can affirm that there is not exist an invective relation between both variables.

Up to now, we have seen that the effect of the external source can not be separated from the phenomenon of intermittence. Besides, scalar turbulence is generated from the existence of a mean field of the scalar. So the prescription of an isotropic external source must be abandoned: the scalar fluctuations are excited by entwining of the mean external gradient of the passive scalar by turbulent flow.

Therefore, given the external gradient $\nabla \mean{\Theta}_0\neq 0$ equation (\ref{eq:scalar-flow-2}) yields,
\begin{equation}
\frac{\partial}{\partial t}\;\vartheta + \nabla \cdot \left( \svl\, \vartheta - \kappa\;\nabla \vartheta \right) = -\nabla \cdot \left( \svl\, \mean{\Theta}_0\right).
\end{equation}
The transference of the scalar concentration is done through the wavenumber space. So, let us consider the spectrum of the anisotropic scalar in the wavenumber interval $2\pi/L< k< \min\{k_c,k_d\}$. Because the molecular diffusivity is supposed to go to infinity, the two relevant terms are the advection and mean scalar gradient in the above equation. Both must be of the same order. 

Besides, the  eddy diffusivity, as we briefly introduce it before, measures the rate of transport of scalar concentration from one portion of fluid to another. It was first introduced by Heisenberg (1948) \citep[see][p. 75]{book:mccomb} and as a function of the wavenumber is written,
\begin{equation*}
\eta(k)\propto \int^{\infty}_k  \frac{\sqrt{k'E(k')}\,dk'}{k'}
\end{equation*}
If we suppose the energy is peaked around $k=0$, the most important contribution to the diffusivity is around $k$: thus, $\eta(k)\sim \sqrt{E(k)/k}\sim u(k)/k$. Where $u(k)=\mean{\norm{\svl}^2}^{1/2}_k$ is the magnitude of the stochastic velocity field at a given wavenumber for
\begin{equation*}
\mean{\norm{\svl}^2}_k = \int^{\infty}_k E(k')\,dk'.
\end{equation*}

Henceforth, given the eddy diffusivity we can replace the advective term by $\eta(k)\triangle \vartheta$ and obtain
\begin{equation*}
\svl\cdot\nabla \mean{\Theta}_0\sim \eta(k)\triangle \vartheta.
\end{equation*}
So we estimate the scalar spectrum with this equation and using definition (\ref{eq:scalar-spectra}), 
\begin{equation}
F(k)\sim k^{-3}\norm{\nabla \mean{\Theta}_0}^2.
\end{equation}

It was encountered that a renormalization procedure \citep{paper:elperin} provides this spectrum for the anisotropic source. Also, the isotropic case is undertaken with this formalism. Given the power law $k^{-(\zeta_2(u)+1)}$ for the velocity spectra it is found  
\begin{equation}
F(k)\propto k^{-2+\zeta_2(u)/2}
\end{equation}
whenever the quotient between the effective viscosity $\nu(k)$ is greater than the kinematic $\nu_0$. In particular this is condition is broken when $\zeta_2(u)\rightarrow 0$: it is $\nu(k)\searrow \nu_0$. 

We are seeing two processes in this discussion: the largest eddies grabbing energy from the anisotropic mean field, they significantly contribute to the structure function below certain scale $1/k_*$; above this characteristic length the external forces turns isotropic introducing the suitable power spectra. 

These anomalous exponents sensibly modify the probability distribution of the scalar fields. That is, they deviate from the Gaussian shape adding (exponential) tails. But because in this work we are just interested in the second moments the Gaussian approximation is enough.

Among all the passive scalars fields, the kind represented by the temperature requires more attention. Usually its turbulent state is reached through convection, but because temperature differences are the trigger for turbulent mixing the temperature field should not, at first, be passive. Whether the temperature is active or passive depends on the magnitude of the contribution from buoyancy forces to the total energy. It is despicable when the Rayleigh number is small:
\begin{equation}
\mathcal{R}_a= \frac{\alpha g \Delta T L^3}{\nu \kappa}\ll 1,
\label{eq:rayleighnr}\end{equation}
where $\alpha$ is the volume expansion coefficient, $g$ is the acceleration due to gravity, and $\Delta T$ is the temperature difference between bottom and top of the bath of size $L$. Meaning it does not affect the state of turbulence. It happens when this number becomes relevant that scaling  and gaussianity of the temperature field are observed \citep{paper:sano,paper:gollub,paper:ching} (for the soft convective turbulence $\mathcal{R}_a < 10^7$). But the scaling exponent behavior is more complex than in the scalar turbulence problem, and most of the arguments presented here are not applicable.

\section{The Turbulent Index of Refraction}

We are going to finish this chapter describing the behavior of the refractive index inside a turbulent flow. It has been proven long ago \citep{book:bean} that 
\begin{equation*}
n -1 = \frac{77.6}{T} \left(P + \frac{4810 e}{T}\right)\times 10^{-6},
\end{equation*}
where it is: $n$ the index of refraction, $T$ the absolute temperature, $P$ the air pressure, and $e$ the water vapor partial pressure (both in millibars). This equation is considered valid for frequencies ranging from 1MHz to around 30GHz or more. For light propagation it is assumed that the humidity term is negligible, hence 
\begin{equation*}
n -1 = \frac{77.6 P}{T} \times 10^{-6}.
\end{equation*}
Without buoyancy effects present we can assume the local pressure to be constant; thus,
\begin{equation}
\Delta n =\left( - \frac{77.6 P}{T^2} \times 10^{-6}\right) \Delta T.
\end{equation}
That is, deviations from the mean temperature field are proportional to those of the refractive index. Adiabatic corrections must be introduced to this formula when the system is the whole atmosphere \citep[see][p. 523]{book:ishimaru}, but they do not break the linear relation. Therefore, the deviations from the mean fields are proportional, and the refractive index inherits the passive scalar properties from the temperature field proved it is a scalar field.

As we introduced a synthetic turbulent velocity field to make insight into the behavior of the scalar turbulence. We will do the same for the lightwave propagation in turbulent media. That is, we will introduce a model for the refractive index such that most of the properties described in the former sections are present in it. 

Our assignment is to associate to the turbulent index of refraction a suited stochastic process. We have seen that the anomalies in the exponents of the $n$-point correlation functions drives the scalar fields apart from the Gaussian statistics at first. Nevertheless, all the theoretical models presented here induce a Gaussian behavior for the scalars because the stochastic velocity field is Gaussian. Moreover, because in this work we will only be interested in the first moments of the propagated light, the  Gaussian distribution is enough.

The family of Gaussian processes is wide, and each member of it is defined, as we shall see in the next section, by its $2$-point correlation function or covariance. Because we are specially interested in the properties of the atmospheric propagation, a fully developed turbulence, the covariance defined within the inertial range characterize almost all its properties. Although until now we have not explicitly given an expression for this covariance, we will show that the structure function is sufficient.

When the velocity field is homogeneous and isotropic, and the external source is isotropic the structure function for the stochastic refractive index, according to (\ref{eq:structure-2}), should have the form:
\begin{equation}
\mean{[\mu(\vect+\vect')-\mu(\vect')]^2}= A_2 \norm{\vect}^{\zeta}, \quad l_0\ll \norm{\vect}\ll L,
\label{prop:zeta-correlation}\end{equation}
$\mu$ is the stochastic component of the turbulent refractive index $n$, the constant $A_2$ is known as \textit{structure constant}\footnote{
Usually is noted as $C^2_\varepsilon$ when $\zeta=2/3$. Its range is around $10^{-14}$--$10^{-11}\text{m}^{-2/3}$ for low altitude measures and $10^{-18}$--$10^{-16}\text{m}^{-2/3}$ for the high altitude \citep{book:tatarskii,book:ishimaru}.
}, and $0<\zeta<2$ . Even the limit $\zeta\rightarrow 2$ must be taken in consideration because of the anisotropic behavior of the passive scalars

The structure function is in fact the covariance of the increments. The translation invariance of equation (\ref{prop:zeta-correlation}) implies that these increments are stationary in the statistical sense. That is, the probability distribution remains the same under translations. 

On the other hand, previous works make different prescriptions for the turbulent index from the ones we have given. They assume the process should be stationary and give other covariance functions in consequence, for example: the exponential correlation function,
\begin{equation}
\mean{\mu(\vect+\vect')\mu(\vect')}= \mean{\mu^2}\exp\left(-\norm{\vect}^2/l^2\right),
\label{eq:beckman}
\end{equation}
used by \citet{paper:beckman} and \citet[see][p. 358]{book:ishimaru}, to solve some propagation problems; or the uncoupling relation
\begin{equation}
\mean{\mu(z+z',\brho+\brho')\mu(z',\brho')}= \delta(z)A(\brho),
\label{eq:markovian}\end{equation}
that makes the process markovian in the propagation direction $z$, first suggested\footnote{
see the {\appendixname} B for a short description of this model.
}  by \citet{paper:klyatskin} and later mathematically formalized by \citet{paper:leland}.

But we have seen the stationary property for the stochastic refractive index is unnecessary; moreover, the $2$-point correlation function (\ref{prop:zeta-correlation}) only demands stationary increments. Our proposal will be to use the \textit{fractional Brownian motion} as a model for passive scalar fields \citep{paper:dario-garavaglia}.

\subsection{The synthetic refractive index: the fractional Brownian motion}
\label{section:turbulent-index}

Finally, in this section we will construct the synthetic index we are going to employ studying lightwave propagation. For such a task the refractive index must follow most of the properties described above. From all the possible Gaussian processes, the fractional Brownian motions seem the best suited to accomplish this as they present the following second moment of the increments
\begin{equation}
\E{(B^H(t)-B^H(s))^2}=\abs{t-s}^{2H},
\label{eq:structure-fBm}
\end{equation}
in the 1-dimensional case. On the other hand, the equation (\ref{eq:nfBm}) provides a similar expression for the tridimensional space, $\E{(B^H(\vect)-B^H(\vect'))^2}=\prod^3_{i=1}\abs{x_i-x'_i}^{2H}$. This is anisotropic or nonstationary in statistical terms; therefore, it can not fulfill our prescriptions for the Structure Function. Instead of this $n$-dimensional version we will use the change of variable property (\ref{prop:change}) to introduce the following Gaussian process 
\begin{equation}
\tilde{B}^H(\vect):= B^H\!\left(\norm{\vect}\right).
\label{eq:fBm}\end{equation}
We will call it \textit{isotropic fractional Brownian motion}. The variance of its increments is given by
\begin{equation*}
\mathbb{E}\left[\tilde{B}^H(\vect+\vect')-\tilde{B}^H(\vect')\right]^2= \left(\norm{\vect+\vect'}-\norm{\vect'}\right)^{2H},
\end{equation*}
and when $\norm{\vect}\gg\norm{\vect'}$ or $\norm{\vect}\ll\norm{\vect'}$ we can approximate this equation by
\begin{equation}
\mathbb{E}\left[\tilde{B}^H(\vect+\vect')-\tilde{B}^H(\vect')\right]^2\simeq \norm{\vect}^{2H},
\label{eq:increments-fBm}\end{equation}
and so the increments are \textit{locally stationary}.  

Now, the departure $\mu$ from the mean index of refraction $n_0$ is a small quantity. That is why the stochastic index also measures the behavior of the \textit{permitivity}, i.e., $\varepsilon(\vect)\simeq n_0^2 + 2\mu(\vect).$ It is a convention among the literature to substitute the turbulent index by the stochastic permitivity $\epsilon(\vect)=2\mu(\vect)$, and so we will do here. If we write the exponent  in equation (\ref{prop:zeta-correlation}) as $\zeta=2H$, from the former property (\ref{eq:increments-fBm}) under the conditions given, thus we define
\begin{equation}
\epsilon(\vect):= \alpha \tilde{B}^H\!\left(\vect/l\right)
\label{eq:fBm-permitivity}\end{equation}
with $\alpha$ an adimensional constant and $l$ some characteristic scale. When we consider a fully developed turbulence, isotropic and homogeneous, the Kolmogorov hypotheses sets $H=1/3$. We are considering departures from this ideal situation so $1/3\leq H < 1$ will be our working range.  

Let us compare the structure function of the permitivity against the structure function generated by this synthetic permitivity. Thus using equations (\ref{prop:zeta-correlation}),  (\ref{eq:increments-fBm}) and (\ref{eq:fBm-permitivity}) we have,
\begin{align}
S_2^{\varepsilon}(r)&= 4 A_2 \norm{\vect}^{2H}\nonumber\\
&=\alpha^2\E{\tilde{B}^H(l^{-1}(\vect+\vect'))-\tilde{B}^H(l^{-1}\vect')}^2\nonumber\\
&\simeq\alpha^2l^{-2H}\norm{\vect}^{2H},
\end{align}
we used the self-similarity property in the third line. The comparation allows determine the coefficient $\alpha= 2\,l^H\sqrt{A_2}$. Therefore, we must determine the characteristic length and the constant $A_2$ of the structure function. There are two physically distinguishable scenes for  setting the scale $l$, whether we are in the persistent, $1/3\leq H<1/2$, or anti-persistent, $1/2<H<1$, case. For the latter continuity conditions  for the limit $H\rightarrow 1$ \citep{paper:sirovich} set $l=l_0$ and $A_2= A\, l_0^{2/3-2H}$, the former just keep $l=L$ and $A_2 = A' L^{2/3-2H}$. 

Furthermore, using the probability density (\ref{def:gaussian}) with the conditions given above we also note that 
\begin{equation*}
\mathbb{E}\abs{\epsilon(\vect)-\epsilon(\vect')}\sim \norm{\vect-\vect'}^H;
\end{equation*}
thus, in our model we have $\zeta_2= 2H$ and $\zeta_1 = H$. 

Although, the relation between these two is linear, and thus does not coincide with the Kraichnan's model. We will proof next it is well defined; that is, it is consistent with other well known quantities. In particular, the fractal dimension associated to the isoscalar surfaces of the isotropic refractive index---contained within a sphere of radius $l_0$: that is, $\dim_H \epsilon^{-1}(c)$. 

This isoscalar surface can be expressed as the set
\begin{equation*}
\epsilon^{-1}(c) = \left\{\vect\in\Re^3: \tilde{B}^H(\vect/l_0) =\alpha^{-1} c\right\}
\end{equation*}
from definition (\ref{eq:fBm-permitivity}). If we take the plane set $P_{\alpha^{-1}c}=\{(\vect,\alpha^{-1}c): \vect\in\Re^{3}\}\subset \Re^3\times\Re$:
\begin{equation*}
\epsilon^{-1}(c) = \graph \,\tilde{B}^H \cap P_{\alpha^{-1}c}.
\end{equation*}
Moreover, using the properties (\ref{prop:cross-dim}), (\ref{prop:intersect-dim}), (\ref{prop:intersect2-dim}) and considering that
\begin{equation*}
\overline{\dim}_B P_{\alpha^{-1}c} = \overline{\lim_{\delta\rightarrow 0}}\;\frac{\log N_\delta\left(P_{\alpha^{-1}c}\right)}{-\log \delta} = \overline{\lim_{\delta\rightarrow 0}}\;\frac{\log \delta^{-3}}{-\log \delta}=3;
\end{equation*}
it is
\begin{equation}
\dim_H\left(\graph \,\tilde{B}^H \cap P_{\alpha^{-1}c}\right) =\dim_H \graph\,\tilde{B}^H - 1.
\label{eq:iso-dim}
\end{equation}
It is necessary, thus, to calculate the Hausdorff dimension associated to the set $\graph\,\tilde{B}^H$. We will accomplish this in the following paragraphs.

First note that the 1-dimensional fractional Brownian motion is \textit{$\lambda$-H\"older continous} \citep[p. 246]{book:falconer}, that is, for all $0<\lambda\leq H$,
\begin{equation*}
\abs{B^H(t)-B^H(s)}\leq M \abs{t-s}^{\lambda},\quad\text{for some }M. 
\end{equation*}
We observe that our isotropic fractional Brownian motion is also $\lambda$-H\"older continous. From the triangle inequality, $\big|\norm{\vect}-\norm{\vect'}\big|\leq \norm{\vect-\vect'}$, and the latter equation we thus have for any $0<\lambda\leq H$:
\begin{equation}
\abs{\tilde{B}^H(\vect)-\tilde{B}^H(\vect')}=\Big|B^H(\norm{\vect})-B^H(\norm{\vect'})\Big|\leq M \big|\norm{\vect}-\norm{\vect'}\big|^{\lambda} \leq M \norm{\vect-\vect'}^{\lambda},
\label{eq:lhausdorff}
\end{equation}
for some $M$. 

Let be $\tilde{B}^H:[0,2l_0]^3\rightarrow\Re$, that is, it will be restricted to a box-set containing the sphere $B_{l_0}$ of radius $l_0$. From all the coverings to the graph let us choose a box-covering $\{B_\delta\}$ such that $\diam{B_\delta}\leq \delta$---with side of length less than $\delta/2$.  It is,
\begin{equation*}
\Hf{s}{\delta}\!\left(\,\graph\,\tilde{B}^H\right)\leq \sum^{N_\delta} \left(\diam{B_\delta}\right)^s\leq N_\delta\, \delta^s,
\end{equation*}
where $N_\delta$ is the number of boxes touched by the graph of $\tilde{B}^H$. Over the domain we have at most $l_0\delta^{-3} + 1$ boxes. On the other hand, it is 
\begin{equation*}
\big|\tilde{B}^H(\delta/2,\delta/2,\delta/2)-\tilde{B}^H(0)\big|\leq M \delta^\lambda
\end{equation*}
from equation (\ref{eq:lhausdorff}). Therefore, we have at most $M \delta^\lambda/\delta + 2$ boxes piled at a given box on the domain; so,
\begin{equation*}
N_\delta\, \delta^s\leq M' \delta^{\lambda-4+s} + 2 \delta^s.
\end{equation*}
Now, as $\delta\rightarrow 0$ the Hausdorff measure remains bounded if and only if $s>4-\lambda$. So from property (\ref{prop:bound}) we have
\begin{equation}
\dim_H \graph\, \tilde{B}^H \leq 4-H.
\label{eq:upp-bound}
\end{equation}
Therefore, the fractal dimension can not exceed $4-H$. Next we will find a lower bound to the dimension. We will apply a flavor of the potential theory commented at the {\appendixname} A. That is, we will look at the integral
\begin{equation*}
\tfrac{1}{2}\;\E{\int\!\int \frac{\mu_{\omega}(dx)\mu_{\omega}(dy)}{\norm{x-y}^s}}
\end{equation*}
for the stochastic mass measure $\mu_\omega$. If we find this integral finite then the fractal dimension of the set we are studying will be greater than $s$. We choose the \textit{occupation measure}:
\begin{equation*}
\mu^H_\omega(B):=\int_{B_{l_0}}\! \chi_B(\vect,\tilde{B}^H(\vect,\omega))\, d^3r,
\end{equation*}
it counts how many points of the set $B$ are in the graph of $\tilde{B}^H$. For simplicity, let us take the sphere $B_{l_0}$ centered at the origin. We have
\begin{multline*}
\tfrac{1}{2}\;\E{\int\!\int \frac{\mu^H_{\omega}(dx)\mu^H_{\omega}(dy)}{\norm{x-y}^s}}\\
=\tfrac{1}{2} \E{\int\limits_{B_{l_0}\times B_{l_0}} \left(\norm{\vect-\vect'}^2 + \abs{\tilde{B}^H(\vect/l_0)-\tilde{B}^H(\vect'/l_0)}^2\right)^{-s/2} d^3r\,d^3r'}.
\end{multline*}
Because $\tilde{B}^H(\vect)-\tilde{B}^H(\vect')\sim \mathcal{N}(0,\abs{\norm{\vect}-\norm{\vect'}})$ and the triangle inequality, we have the following inequality,
\begin{equation*}
\mathbbm{p}_{\abs{\norm{\vect}-\norm{\vect'}}}(z) \leq \frac{\norm{\vect-\vect'}}{\abs{\norm{\vect}-\norm{\vect'}}}\mathbbm{p}_{\norm{\vect-\vect'}}(z),
\end{equation*}
for the probability densities
\begin{equation*}
\mathbbm{p}_{\abs{\norm{\vect}-\norm{\vect'}}}(z)=\frac{\exp\left(-z^2/\abs{\norm{\vect}-\norm{\vect'}}^{2H}\right)}{\sqrt{2\pi \abs{\norm{\vect}-\norm{\vect'}}^{2H}}}
\quad\text{and}\quad
\mathbbm{p}_{\norm{\vect-\vect'}}(z)=\frac{\exp\left(-z^2/\norm{\vect-\vect'}^{2H}\right)}{\sqrt{2\pi \norm{\vect-\vect'}^{2H}}}.
\end{equation*}
Therefore,
\begin{align*}
\tfrac{1}{2} \mathbb{E}\Bigg[\int\limits_{B_{l_0}\times B_{l_0}}\hspace*{-1.5ex} \Big(\norm{\vect-\vect'}^2 & + \abs{\tilde{B}^H(\vect/l_0)-\tilde{B}^H(\vect'/l_0)}^2\, d^3r\,d^3r'\Big)^{-s/2}\Bigg]\\
\leq (1/2)  &\int\limits_{B_{l_0}\times B_{l_0}}\!\int\limits_\Re \frac{\norm{\vect-\vect'}^H}{\abs{\norm{\vect}-\norm{\vect'}}^H}\frac{d^3r\,d^3r'dz}{(\norm{\vect-\vect'}^2+z^2)^{s/2}}\\
\leq  (l_0^6/2) &\int\limits_{0<\norm{\textbf{v}}\leq 1}\hspace*{-2ex} d^3v\hspace*{-1ex} \int\limits_{\norm{\textbf{u} + \textbf{v}}\leq 1}\hspace*{-2ex} d^3u \int_\Re dz \;\mathbbm{p}_u(z)\; \frac{u^H}{\abs{\norm{\textbf{u}+\textbf{v}}-v}^H (l_0^2 u^2 + z^2)^{s/2}}\\
\leq  l_0^6 2\pi^2 &\int_\Re\!dz\int^1_0\! v^2 dv \int^1_{-1}\!dx\int^{\sqrt{(vx)^2+(1-v^2)}-vx}_0\hspace*{-2ex}\frac{\mathbbm{p}_u(z)\,u^H u^2du }{\abs{\abs{u-v}-v}^H (l_0^2u^2+z^2)^{s/2}},
\end{align*}
in the last inequality we have used $\abs{u-v}\leq \norm{\textbf{v}+\textbf{u}}$. Since $0\leq\sqrt{(vx)^2+(1-v^2)}-vx\leq 2$ we change the former inequality to
\begin{align*}
\tfrac{1}{2} \mathbb{E}\Bigg[ \int\limits_{B_{l_0}\times B_{l_0}}\hspace*{-1.5ex} \Big(\norm{\vect-\vect'}^2 & + \abs{\tilde{B}^H(\vect/l_0)-\tilde{B}^H(\vect'/l_0)}^2\, d^3r\,d^3r'\Big)^{-s/2}\Bigg]\\
\leq &\;l_0^6 4\pi^2 \int_\Re\!dz\int^1_0\! v^2 dv \int^2_0\hspace*{-1ex}\frac{\mathbbm{p}_u(z)\,u^H u^2du }{\abs{\abs{u-v}-v}^H (l_0^2u^2+z^2)^{s/2}}\\
\leq &\; l_0^6 4\pi^2 \int^2_0\hspace*{-1ex} u^2du \left(\int_\Re\frac{\mathbbm{p}_u(z)\,dz}{(l_0^2u^2+z^2)^{s/2}}\right)\left(u^H\int^1_0 \frac{ v^2 dv }{\abs{\abs{u-v}-v}^H }\right)\\
\leq &\; \frac{l_0^6 12\pi^2}{(1-H)} \int^2_0 u^2 du \int_\Re \mathbbm{p}_u(z)(l_0^2 u^2 + z^2)^{-s/2} dz.
\end{align*}
Now, the probability density $\mathbbm{p}_u(z)$ is bounded in any closed interval of $u$. Let us subdivide $[0,2]$ into intervals of the form $[c^{n+1}\delta, c^n \delta)$ with $c<1$ and $c^{-k}\delta\geq 2$ for some $k,\delta$ such that $\cup^\infty_{-k} [c^{n+1}\delta,c^n \delta)\supset [0,2]$. We can rewrite the right-hand side of the above inequality as,
\begin{multline*}
\tfrac{1}{2} \mathbb{E}\Bigg[\int\limits_{B_{l_0}\times B_{l_0}}\hspace*{-1.5ex} \Big(\norm{\vect-\vect'}^2 + \abs{\tilde{B}^H(\vect/l_0)-\tilde{B}^H(\vect'/l_0)}^2\, d^3r\,d^3r'\Big)^{-s/2}\Bigg]\\
\leq\frac{12\pi^2}{(1-H)} \sum^\infty_{-k}\int^{c^n \delta}_{c^{n+1}\delta }u^2 du \int_\Re \mathbbm{p}_u(z)(l_0^2 u^2 + z^2)^{-s/2} dz.
\end{multline*}
Making the change of variables $u=c^n t$ we have:
\begin{multline*}
\tfrac{1}{2} \mathbb{E}\Bigg[\int\limits_{B_{l_0}\times B_{l_0}}\hspace*{-1.5ex} \Big(\norm{\vect-\vect'}^2 + \abs{\tilde{B}^H(\vect/l_0)-\tilde{B}^H(\vect'/l_0)}^2\, d^3r\,d^3r'\Big)^{-s/2}\Bigg]\\
\leq \frac{12\pi^2}{(1-H)} \sum^\infty_{-k} c^{3n}\int^{\delta}_{c\delta}t^2 dt \int_\Re \mathbbm{p}_{c^n t}(z)(l_0^2 c^{2n} t^2 + z^2)^{-s/2} dz.
\end{multline*}
Since the fractional Brownian motion is self-similar we have:
\begin{align*}
\int_\Re \mathbbm{p}_{c^n t}(z) (l_0^2 c^{2n} t^2 + z^2 )^{-s/2} dz &= \int_\Re \mathbbm{p}_t (z) (l_0^2 c^{2n} t^2 + c^{2nH} z^2 )^{-s/2} dz\\
&\leq M_0 \int_\Re (l_0^2 c^{2n} t^2 + c^{2nH} z^2)^{-s/2} dz\\
&\leq \overline{M}_0 t^{1-s} c^{n(1-s-H)}.
\end{align*}
Therefore, using this the former inequality turns to be
\begin{align*}
\tfrac{1}{2} \mathbb{E}\Bigg[\int\limits_{B_{l_0}\times B_{l_0}}\hspace*{-1.5ex} \Big(\norm{\vect-\vect'}^2 & + \abs{\tilde{B}^H(\vect/l_0)-\tilde{B}^H(\vect'/l_0)}^2\, d^3r\,d^3r'\Big)^{-s/2}\Bigg]\\
&\leq \overline{M}_0 \sum^\infty_{-k} c^{3n} \left(\int^c_{\delta c} t^{3-s} ds\right) c^{n(1-s-H)}\\
&\leq \overline{M}_0 \sum^\infty_{-k} c^{3n} \left(\frac{1-\delta^{4-s}}{4-s} c^{4-s}\right) c^{n(1-s-H)}\\
& \leq M_0(\delta,s,c) \sum^\infty_{-k} c^{n(4-s-H)}.
\end{align*}
So the $s$-potential will be bounded whenever $4-s-H>0$, that is, $s<4-H$ and,
\begin{equation}
4-H\leq \dim_H \graph \,\tilde{B}^H
\label{eq:low-bound}
\end{equation}
Then comparing this equation against (\ref{eq:upp-bound}) it is: $\dim_H \graph\, \tilde{B}^H=4-H$. Finally, from equation (\ref{eq:iso-dim}) we find that 
\begin{equation}
\dim_H \epsilon^{-1}(c)=3-H.
\label{eq:ifBm-dim}
\end{equation}
This expression exactly matches the one calculated by \citet{paper:constantin}. Since, it replicates the equation (\ref{eq:fractal-scalar-dimension}) with $\zeta_1 = H$ as we proved above. 

We have shown here that the isotropic fBm not only provides stationary increments and reproduces the structure function for the stochastic refractive index but also gives the right fractal dimension associated to the passive scalars. Moreover, the structure function (\ref{prop:zeta-correlation})  gives a covariance function $v$ which corresponds to a non-differentiable process. A simple calculation  
\begin{equation*}
\frac{\partial}{\partial x_i} S^\varepsilon_2 (\norm{\vect-\vect'})= 2H\,A_2  \norm{\vect-\vect'}^{2H-2} (x_i-x'_i),
\end{equation*}
shows this derivative is undefined whenever $\vect=\vect'$, and because
\begin{equation*}
S^\varepsilon_2 (\norm{\vect-\vect'}) =\mean{[\mu(\vect+\vect')-\mu(\vect')]^2}= v(\vect,\vect) + v(\vect',\vect')  -2 v(\vect,\vect'). 
\end{equation*}
The \citeauthor{book:cramer}'s Lemma (\citeyear{book:cramer}) proves the refractive index is non-dif-{\break}ferentiable.  This is the same with the isotropic fractional Brownian motion as it is proved in the Appendix A.

Therefore, these reasons are enough to use this model as source in the Optics' differential equations. In particular, we will use plenty of it in the last chapter.
\chapter{Classical Methods Applied to Turbulent Lightwave Propagation}
\label{chp:classic}

The problem of light passing through a hollow made on a surface is explained in every Optics treatise. In this chapter we are going to make a brief introduction to it; afterwards, we will show how it is related to the Feynman's Path Integrals formalism. It is this technique which had proven fundamental studying image formation in the case of light propagating through a turbulent medium. Further ahead we will describe such a problem, and eventually use it to study some characteristic properties of the refractive index.

\section{From the Green's Theorem to the Feynman's Path Integral}

\begin{figure}
\begin{center}
\includegraphics[width=0.65\textwidth]{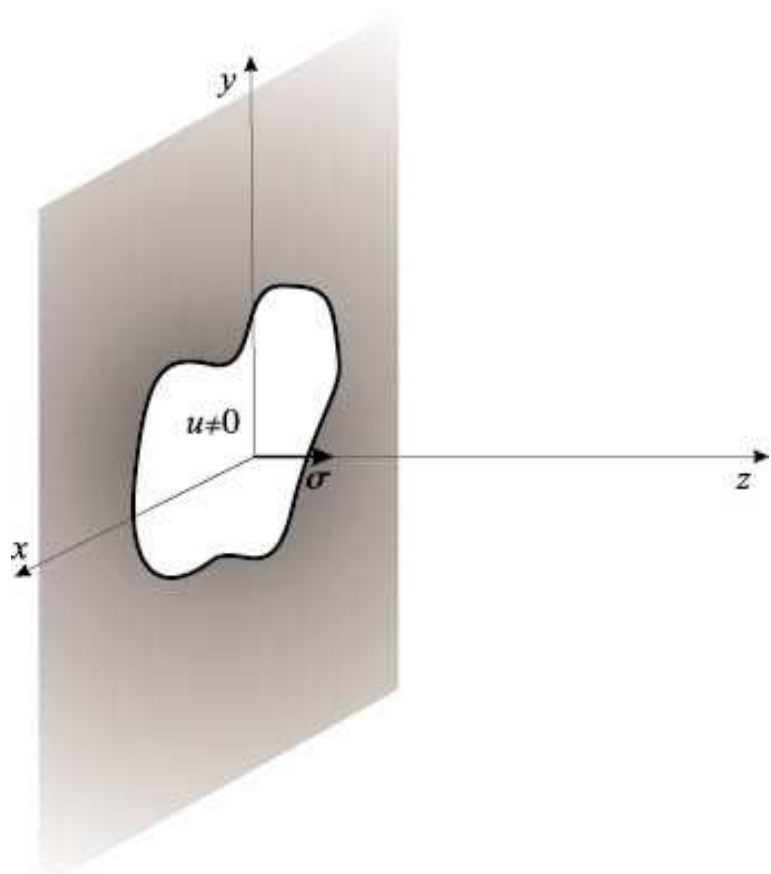}
\caption{The surface where the boundary condition $u$ is set is represented by its normal vector $\boldsymbol{\sigma}$. It is contained in the $(x,y)$-plane, while the $z$-axis is the direction of propagation.
\label{fig:propagation}}
\end{center}
\end{figure}

Assuming the polarization effects are negligible, the problem we have introduced is mathematically described as follows;
\begin{align}
\triangle G + k^2 G &= 0, \quad\text{and}\label{eq:wave-eq}\\
G &= 0 \text{ all over the surface } \boldsymbol{\sigma},\\
G &\rightarrow u \text{ as } r \rightarrow 0,\\
r \left( \nabla G \cdot \boldsymbol{\sigma} - i k G \right)& \rightarrow 0 \text{ as } r \rightarrow 0,
\end{align}
where $G$ is the Green's solution to the \textit{wave equation} (\ref{eq:wave-eq}), $u$ is the boundary condition given by the hollowed surface (Figure \ref{fig:propagation}).
 Therefore, the solution to the \textit{Kirchhoff-Huygens equation} is written in this context as 
\begin{equation}
4 \pi u_p = -\int_{\sigma}  \! d\boldsymbol{\sigma}\cdot(u \nabla G), 
\label{eq:KH}\end{equation}
where $p$ is the point where we evaluate the \textit{propagated} initial field. Now, let $\boldsymbol{\sigma}$ be a plane surface perpendicular to the $z$-axis, and by $u(\vect,z)$ design the propagated field at a distance $z$ from the initial (boundary condition) field $u:=u(\vect,0)$. So solution (\ref{eq:KH}) yields:
\begin{equation*}
u(\vect, z) = \frac{1}{4 \pi} \int_{\sigma} \!\!d^2 \rho\; u(\brho, 0) \left.\frac{\partial G}{\partial z}\right|_{z=0} .
\end{equation*}

Then, we just need to find the Green's function. Applying the \textit{image principle} to the free-space Green's function we build ours:
\begin{equation*}
G(\boldsymbol{{\rho}},z';\vect, z)= \frac{e^{i r_1 k}}{r_1}-\frac{e^{i r_2 k}}{r_2},
\end{equation*}
where $r^2_1= (\brho-\vect)^2+(z'-z)^2$ and $r^2_2= (\brho-\vect)^2+(z'+z)^2.$ Thus,
\begin{equation*}
-\left.\frac{\partial G}{\partial z}\right|_{z=0} = 2i k \frac{z e^{ikr'}}{r'^2} \left(1+\frac{i}{ k r'}\right),
\end{equation*}
where we have set $r'=r_1=r_2= \sqrt{(\brho-\vect)^2 + z^2}.$ Now with this expression at hand, we may evaluate the solution along the $z$-axis, that is,
\begin{equation}
u({\bf 0}, z)= \frac{k}{2i\pi}\int_{\sigma}\!d^2\rho\;\, u(\brho,0)\; \frac{z e^{ikr'}}{r'^2}\left(1+\frac{i}{2kr'}\right). 
\label{eq:original}\end{equation}
We can turn the pupil $\sigma$ into a function, and add it to the boundary condition $u$. Therefore, the integration is taken over the whole plane, which in turn can be expressed as the union of the sets $\{\norm{\brho}\leq z\}$ and $\{\norm{\brho} > z\}$. Also, we are going to introduce two mayor assumptions: that $u$ is symmetric around the $z$-axis, so  $d\sigma = 2\pi \rho\; d\rho$; and $kz\gg 1,$ i.e., the wavelength is smaller ($\sim 10^{-6}--10^{-9}$ m) than the distance to the origin. Under these conditions the integral over the first region has the following expression:
\begin{equation}
u_{\{\norm{\brho}\leq z\}}(z)= -ikz \int_{0}^z \! \rho d\rho\;u(\rho)\; \frac{e^{ikr'}}{r'^2}\left(1+\frac{i}{kr'}\right).
\label{eq:first-region}\end{equation}
Now, let us make the change of variables $k\rho=x$ and set $\epsilon= kz$; also,
\begin{equation}
kr'=\epsilon\left[1+\left(\frac{x}{\epsilon}\right)^2\right]^{1/2}.
\label{eq:change-of-variables}\end{equation}
Thus equation (\ref{eq:first-region}) is rewritten as
\begin{equation}
u_{\{\norm{\brho}\leq z\}}(z)= \frac{1}{i \epsilon}\int_0^{\epsilon}\!\!x dx \;u\!\left(\frac{x}{k}\right) \frac{\exp i\epsilon \left( 1+ x^2/\epsilon^2\right)^{\frac{1}{2}}}{(1+x^2/\epsilon^2)} \left[ 1+\frac{i}{\epsilon\left(1+x^2/\epsilon^2\right)^{\frac{1}{2}}} \right].
\end{equation}
We may replace by a Taylor series in $x/\epsilon$ all the expressions matching (\ref{eq:change-of-variables}), and finally keep the terms up to the second order. That is,
\begin{align*}
u_{\{\norm{\brho}\leq z\}}(z)&= \frac{e^{i\epsilon}}{i \epsilon}\int_0^{\epsilon}\!\! x dx\; u\!\left(\frac{x}{k}\right) \exp{i\left(\frac{x^2}{2\epsilon}+\cdots\!\right)}\!\!\left[1-\frac{x^2}{\epsilon^2} +\cdots\right] \!\!\left[ 1+\frac{i}{\epsilon}\left(1- \frac{x^2}{2\epsilon^2} + \cdots\!\right) \right]\nonumber\\
&\simeq \frac{e^{i\epsilon}}{i \epsilon}\int_0^{\epsilon}\!\! x dx\; u\!\left(\frac{x}{k}\right) \exp{i\left(\frac{x^2}{2\epsilon}\right)}= \frac{k}{2\pi i z}\;e^{i kz}\int_{\norm{\brho}\leq z} \!\! d^2\rho\;u(\rho) \exp{i\left(\frac{k}{2z}{\rho}^2\right)}. 
\end{align*}

Let us analyze the second case $\norm{\brho} > z$: its contribution to the solution can be split,
\begin{align*}
u_{\{\norm{\brho}> z\}}(z)&= -ikz \int_z^{\infty}\!\!\rho d\rho\; u(\rho)\; \frac{e^{ikr'}}{r'^2}\left(1+\frac{i}{kr'}\right) \nonumber\\
&= -ikz\int^{\infty}_z\!\!\rho d\rho\; u(\rho) \frac{e^{ikr'}}{r'^2}+ kz\int^{\infty}_z\!\!\rho d\rho\; u(\rho) \frac{e^{ikr'}}{kr'^3}.
\end{align*}
Again, with the change $k\rho=x$ both integrals are written as
\begin{equation*}
u_{\{\norm{\brho}> z\}}= -i\eta \int_\eta^{\infty}\!\!xdx\; u\left(\frac{x}{k}\right) \frac{e^{i\left(\eta^2+x^2\right)^{\frac{1}{2}}}}{\left(\eta^2+x^2\right)}+ \eta \int_\eta^{\infty}\!\!xdx\;  u\left(\frac{x}{k}\right) \frac{e^{i\left(\eta^2+x^2\right)^{\frac{1}{2}}}}{\left(\eta^2+x^2\right)^{\frac{3}{2}}},
\end{equation*}
where $\eta=kz$. Instead, another change can be used here $s=(\eta^2+x^2)^{\frac{1}{2}}$, and so $ds=xdx/(\eta^2+x^2)^{\frac{1}{2}}$,
\begin{equation}
u_{\{\norm{\brho}> z\}}= -i\eta \int_{\sqrt{2}\eta}^{\infty} u\left(k^{-1}(s^2-\eta^2)^{\frac{1}{2}}\right) \frac{e^{is}ds}{s} + \eta \int_{\sqrt{2}\eta}^{\infty} u\left(k^{-1}(s^2-\eta^2)^{\frac{1}{2}}\right) \frac{e^{is}ds}{s^2}.
\end{equation}
This contribution is bounded; moreover, it tends to zero as $\eta$ goes to infinity. With the condition $kz\gg 1$, we can make use of the mean value theorem and write,
\begin{equation}
u_{\{\norm{\brho}> z\}}\sim u(\eta/k) \eta \left\{ -i\int^{\infty}_{\sqrt{2}\eta}\frac{e^{is}ds}{s} +\int^{\infty}_{\sqrt{2}\eta}\frac{e^{is}ds}{s^2} \right\}.
\end{equation}
We observe that the integrals within the braces are compositions of \textit{Exponential-Integrals} and linear functions, 
\begin{align*}
\int^{\infty}_{\sqrt{2}\eta}\frac{e^{is}ds}{s} &= \text{Ei}(i \sqrt{2}\eta)\;\;\text{ and}\\
\int^{\infty}_{\sqrt{2}\eta}\frac{e^{is}ds}{s^2} &= i \text{Ei}(i \sqrt{2}\eta);
\end{align*}
therefore, it is $u_{\{\norm{\brho}> z\}}\sim 0.$ 

Thus, the solution is just the contribution of the initial field enclosed by a sphere of radius $z$:
\begin{equation*}
u(z)= \frac{k}{2\pi i z}\; e^{ikz} \int_{\norm{\brho}\leq z}\!\!\! d^2\rho\: u(\brho,0) \exp \left(i\frac{k}{2z}\rho^2\right).
\end{equation*}
Furthermore, this integral can be extended to the whole plane assuming $u(\brho,0)\sim u_0 \rho^{-2}$ as $\rho \rightarrow \infty.$ Remember, the solution we have build is  valid whenever $kz\gg 1$ is satisfied. 

Now, to evaluate the solution off the $z$-axis we just need to make the change,
\begin{equation}
u(\vect, z)= \frac{k}{2\pi i z}\; e^{ikz} \int_{\Re^2}\! d^2\rho\; u(\brho, 0) \exp\! \left[i\frac{k}{2z}(\brho-\vect)^2\right].
\label{eq:paraxial-aproximation}\end{equation}
This is the \textit{paraxial} or \textit{Fresnel approximation}, and it describes the free-space diffraction---from now on the term free-space will refer to a space free from inhomogeneities. Clearly, the phase term $e^{ikz}$ do not add information to the irradiance distribution---it is just the plane wave factor---so we will drop it off the paraxial approximation. Besides, setting $u(\vect,0)=e^{-ikz}\delta^{(2)}(\vect)$ as initial condition let us build 
\begin{equation*}
G(\vect,\brho;z)=\frac{k}{2\pi i z}\exp \left[i\frac{k}{2z}(\brho-\vect)^2\right],
\end{equation*}
that is, the Green function associated to the \textit{parabolic} or \textit{diffusion equation}:
\begin{equation*}
\left(2ik \frac{\partial}{\partial z} + \nabla_{\vect}^2\right) G(\vect,\brho;z) = \delta^{(2)} (\vect-\brho).
\end{equation*}
Also, the solution (\ref{eq:paraxial-aproximation}) has an alternative operator form, similar to the introduced in Section \ref{section:kraichnan-model},
\begin{equation}
u(\vect,z)= \exp\!\left(-i\frac{z}{2k}\nabla^2_{\vect}\right)u(\vect,0).
\label{prop:operator-form}\end{equation}
It can be deduced using the Fourier representation
\begin{equation}
\frac{1}{2\pi}\int_{\Re^2}\!\!d^2r\, \nabla^2_{\vect} u(\vect)\;e^{ i\Bkappa\cdot\vect}=\Bkappa^2 \hat{u}(\Bkappa).
\end{equation}
Hence,
\begin{equation*}
\begin{aligned}
u(\vect,z)&=\frac{1}{4\pi^2}\int_{\Re^2}\!\!d^2\kappa\int_{\Re^2}\!\!d^2\rho'  \exp\left(-\frac{iz}{2k}\nabla^2_{\brho'}\right) u(\brho')\; e^{i\Bkappa\cdot(\brho'-\vect)}\\
&=\frac{1}{4\pi^2} \sum^{\infty}_{n=0}\frac{1}{n!} \left(-\frac{iz}{2k}\right)^n\int_{\Re^2}\!\!d^2\kappa\; e^{-i\Bkappa\cdot\vect}\! \int_{\Re^2}\!\!d^2\rho'\;(\nabla^2_{\brho'})^n u(\brho')\; e^{i\Bkappa\cdot \brho'}\\
&=\frac{1}{2\pi} \sum^{\infty}_{n=0}\frac{1}{n!} \left(-\frac{iz}{2k}\right)^n\int_{\Re^2}\!\! d^2\kappa\; \Bkappa^{2n} \hat{u}(\Bkappa)\;e^{-i\Bkappa\cdot\vect}\\
&= \frac{1}{2\pi} \int_{\Re^2}\!\! d^2\kappa\;  \hat{u}(\Bkappa)\;\exp\! \left(-i\Bkappa\cdot\vect -i\frac{z}{2k}\Bkappa^2\right) \\
&= \frac{1}{4\pi^2}\int_{\Re^2} d^2\rho\; u(\brho) \int_{\Re^2}\!\! d^2\kappa\;  \exp\! \left[-i\Bkappa\cdot(\vect-\brho) -i\frac{z}{2k}\Bkappa^2\right]\\
&= \frac{1}{4\pi^2}\int_{\Re^2} d^2\rho\; u(\brho)\left(\frac{2\pi k}{iz}\right)\exp\! \left[i\frac{k}{2z}(\vect-\brho)^2\right]
\end{aligned}
\end{equation*}
\begin{equation*}
\!\!\!\!\!\!\!\!\!\!\!\!\!\!\!\!\!\!\!\!\!\!\!\!\!\!\!\!\!\!\!\!\!\!\!=\frac{k}{2\pi iz}\int_{\Re^2} d^2\rho\; u(\brho)\;\exp\! \left[i\frac{k}{2z}(\vect-\brho)^2\right], 
\end{equation*}
like we said.

This Green function will allows us to build a solution for 
\begin{equation}
\left[2ik \frac{\partial}{\partial z} + \nabla_{\vect}^2+k^2\epsilon(\vect,z)\right] U(\vect,z) = 0.
\label{eq:inhomogeneous-diffusion}\end{equation}
This differential equation is the main subject of the remaining of this chapter. It is obtained from changing the vacuum refractive index in equation (\ref{eq:wave-eq}) by an inhomogeneous one, like those we described in the last chapter; afterwards, the laplacian operator is approximated under the condition $kz\gg 1$ and the multiplicative term results from writing the solution as $e^{ikz}U$.

Now, we will follow \citet{paper:charnotskii} building the solution to (\ref{eq:inhomogeneous-diffusion}) from the free-space Green function. Suppose we want to find the scalar wave field at a point $(\Vect,L)$ given the boundary (initial) condition $u(\brho,0)=u_0(\brho)$. We will construct the field at that point by subdividing the interval $[0,L]$ in subintervals $[z_j,z_{j+1})$ of length $\Delta z=L/N$, with $N$ a large integer. Let us use equation (\ref{eq:inhomogeneous-diffusion}) to estimate how the field propagates from a point $z_j$ to the next $z_{j+1}$,
\begin{align}
u(\vect,z_{j+1})&= u(\vect,z_j)-\frac{\Delta z}{2ik}\left[\nabla^2_{\vect}+k^2\epsilon(\vect,z_j)\right]u(\vect,z_j) +\mathcal{O}(\Delta z^2)\nonumber\\
&=\left\{1 - \frac{\Delta z}{2ik}\left[\nabla^2_{\vect}+k^2\epsilon(\vect,z_j)\right]\right\}u(\vect,z_j) +\mathcal{O}(\Delta z^2)\nonumber\\
&= \exp\left\{ \frac{i\Delta z}{2k}\left[\nabla^2_{\vect}+k^2\epsilon(\vect,z_j)\right]\right\} u(\vect,z_j);
\label{eq:propagation-inhomogeneous}\end{align}
this last expression is exact. Remembering that $e^{A+B}=e^{[A,B]}e^A e^B$ we can detach both operators above; moreover, because both operators in above equation are of order one in $\Delta z$ the commutator is of $\mathcal{O}(\Delta z^2),$ and so its contribution can be neglected. Therefore, applying the homogeneous paraxial solution (\ref{eq:paraxial-aproximation}) and property (\ref{prop:operator-form}) we turn (\ref{eq:propagation-inhomogeneous}) into:
\begin{align} 
u(\vect,z_{j+1})&= \exp\!\left[i\frac{k \Delta z}{2}\;\epsilon(\vect,z_j)\right] \exp\!\left( i\frac{\Delta z}{2k}\nabla^2_{\vect}\right) u(\vect,z_j) +\mathcal{O}(\Delta z^2)\nonumber\\
&= \exp\!\left[i\frac{k \Delta z}{2}\;\epsilon(\vect,z_j)\right] \frac{k}{2\pi iz}\int_{\Re^2}\!\!d^2\rho\;  u(\brho, z_j)\exp\! \left[i\frac{k}{2z}(\vect-\brho)^2\right] +\mathcal{O}(\Delta z^2)\nonumber\\
&= \int_{\Re^2}\frac{k\,d^2\rho}{2\pi i\Delta z}\, u(\brho, z_j) \exp \frac{ik}{2}\!\left[\frac{(\vect-\brho)^2}{\Delta z}+ \Delta z\;\epsilon(\vect,z_j)\right] +\mathcal{O}(\Delta z^2).
\label{eq:propagation-path}\end{align}
We have build a recursive algorithm to estimate the field at the arriving point $z_j$ from its predecessor at $z_{j+1}$. The field at the point $(\Vect,L)$ should come after $N$ iterations from the initial field at $z_0=0$, that is,
\begin{equation}
\begin{aligned}
u(\Vect,L)&= \int_{\Re^2}\frac{k\;d^2\rho_{N-1}}{2\pi i\Delta z} \int_{\Re^2}\frac{k\;d^2\rho_{N-2}}{2\pi i\Delta z} \cdots \int_{\Re^2}\frac{k\;d^2\rho_0}{2\pi i\Delta z}\\
&\times \exp \left\{\frac{ik}{2\Delta z}\left[(\Vect-\brho_{N-1})^2 + (\brho_{N-1}-\brho_{N-2})^2 +\cdots+(\brho_1-\brho_0)^2 \right]\right.\\
&\left.+\frac{ik\Delta z}{2}\left[\epsilon(\Vect,z_{N-1})+\epsilon(\brho_{N-1},z_{N-2})+\cdots+\epsilon(\brho_1,0)\right]\right\} u_0(\brho_0)\\
&+\mathcal{O}\left[(\sqrt{N}\Delta z)^2\right].
\end{aligned}
\label{eq:path-sucession}\end{equation}

\begin{figure}
\begin{center}
\includegraphics[width=0.65\textwidth]{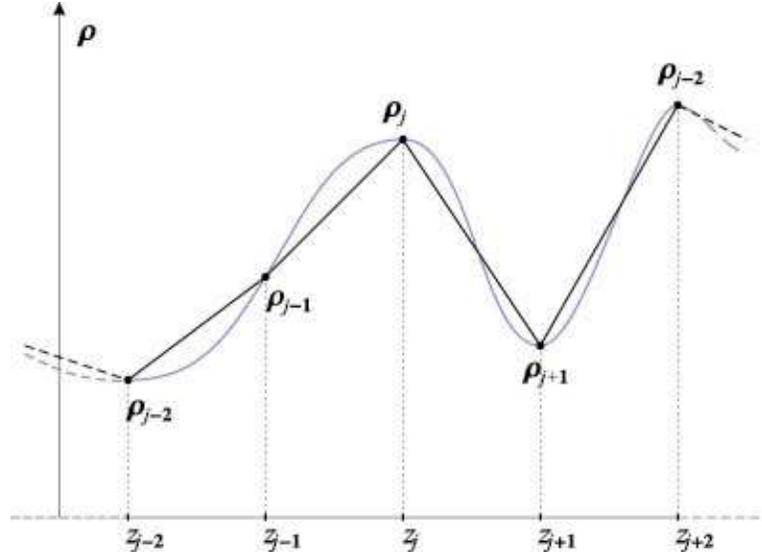}
\caption{The interpolating function provides a well defined derivative at each interval $[z_j,z_{j+1})$.
\label{fig:interpolating}}
\end{center}
\end{figure}

Afterwards, for any function given $\brho(z)$ the set $\{\brho(0)=\brho_0,\dots,\brho(z_j)=\brho_j,\dots,\brho(z_N=L)=\Vect\}$ will define an interpolating function (Figure \ref{fig:interpolating}), with constant derivatives within the subintervals $[z_j,z_{j+1})$,
which converges to it as $N\rightarrow \infty$.  Also, note that with $N$ growing we have $\sqrt{N}\Delta z\rightarrow 0,$ and then the terms inside the argument of the exponential in the latter equation behave:
\begin{align}
\lim_{N\rightarrow\infty}\frac{1}{\Delta z} \sum^{N-1}_{j=0} (\brho_{j+1}-
\brho_j)^2 &= \lim_{N\rightarrow\infty} \sum^{N-1}_{j=0} \left(\frac{\brho(z_{j+1})-
\brho(z_j)}{\Delta z}\right)^2\!\Delta z\nonumber\\
&= \int^L_0\!\!dz\; \left[\frac{d\brho(z)}{dz}\right]^2,\\
\lim_{N\rightarrow\infty}\Delta z \sum^{N}_{j=1} \epsilon(\brho_j,z_j) &= \int^L_0\!\! dz\; \epsilon(\brho(z),z).
\end{align}
Nevertheless, these convergences are not enough to prove the existence of a limit for the expression in (\ref{eq:path-sucession}) when $N\rightarrow \infty$. We may think that the following definition plays the role of a measure
\begin{equation}
\mathcal{D}^{2}\rho(z):=\lim_{N\rightarrow\infty} \prod^{N-1}_{k=0} \frac{k\;d^2\rho_j}{2\pi i\Delta z},
\label{def:pseudo-measure}\end{equation}
but it is not true. In fact, it is  $\exp \left\{(ik/2)\int^L_0\!\!dz\, [d\brho(z)/dz]^2\right\} \mathcal{D}^{2}\rho(z)$:
\begin{align}
\int^{\brho(L)=\Vect}\!\!\!\! \mathcal{D}^2\rho(z) \exp \left\{\frac{ik}{2}\int^L_0\!\!dz \left[\frac{d\brho(z)}{dz}\right]^2\right\}&= \lim_{N\rightarrow\infty} \int_{-\infty}^{\infty}\cdots \int_{-\infty}^{\infty}\prod^{N-1}_{j=0} \frac{kN\;d^2\rho_j}{2\pi i L}   \nonumber\\
\times \exp \left\{\frac{ikN}{2 L}\sum^{N-1}_{j=0} [\brho_{j+1}-\brho_j]^2\right\}&= \lim_{N\rightarrow\infty} \int_{-\infty}^{\infty}\cdots \int_{-\infty}^{\infty}\prod^{N-1}_{j=0}\frac{kN\;d^2\nu_j}{2\pi i L}\nonumber\\
\times\exp \left(\frac{ikN}{2 L}\sum^{N-1}_{j=0}\boldsymbol{\nu}_j^2\right)&= \lim_{N\rightarrow\infty} \prod^{N-1}_{j=0}\left[\int_{-\infty}^{\infty} \int_{-\infty}^{\infty} \frac{kN\;d^2\nu_j}{2\pi i L}\right.\nonumber\\
\left. \times\exp \left(\frac{ikN}{2 L}\boldsymbol{\nu}_j^2\right)\right]&= \lim_{N\rightarrow\infty} \prod^{N-1}_{j=0}\left[\int_0^{\infty} \frac{kN\;\nu_jd\nu_j}{i L}\right.\nonumber \\
\times\left.\exp \left(\frac{ikN}{2 L}\nu_j^2\right)\right]&= \lim_{N\rightarrow\infty} \prod^{N-1}_{j=0}(\,1)=1,
\end{align} 
here we have made the change $\brho_i=\Vect - \sum_{j=i}^N \boldsymbol{\nu}_j$---the notation for the integral on the left-hand side stresses the fact that the starting point is not fixed as one can see from the definition (\ref{def:pseudo-measure}).

We conclude that the limit to the expression at the right side of (\ref{eq:path-sucession}) exists, and we write it as:
\begin{equation}
\begin{aligned}
u(\Vect,L)&=\!\int^{\brho(L)=\Vect}\!\!\!\!\mathcal{D}^{2}\rho(z)\;u_0[\brho(0)]\\
&\times \exp\!\left\{ \frac{ik}{2}\int^L_0\!\!dz \left[\frac{d\brho(z)}{dz}\right]^2+\frac{ik}{2}\int^L_0\!\!dz \epsilon(\brho(z),z)\right\}. 
\end{aligned}
\label{eq:image-path-int}\end{equation}
This is the solution to the wave equation with a non-constant refractive index in the paraxial approximation, (\ref{eq:inhomogeneous-diffusion}). Even we can retrieve its associated Green function by making $u_0[\brho(0)]= \delta^{(2)}\left[\brho(0)-\Vect_0\right],$ i.e.,
\begin{equation*}
\begin{aligned}
G(\Vect,L;\Vect_0,0)&=\!\int^{\brho(L)=\Vect}\!\!\!\mathcal{D}^{2}\rho(z)\:\delta^{(2)}[\brho(0)-\Vect_0]\\
&\times \exp \left\{\frac{ik}{2}\int^L_0\!\! dz \left[\frac{d\brho(z)}{dz}\right]^2+ \frac{ik}{2}\int^L_0\!\! dz\;\epsilon(\brho(z),z)\right\}.
\end{aligned}\end{equation*}
Now, the \textit{reciprocity theorem} \citep{book:sommerfeld-vol-2} states: $G(\Vect,L;\Vect_0,0)=G(\Vect_0,0;\Vect,L).$ Applying it to the former equation yields,
\begin{equation*}
\begin{aligned}
G(\Vect_0,0;\Vect,L)&=\!\int_{\brho(0)=\Vect_0}\!\!\!\mathcal{D}^{2}\rho(z)\:\delta^{(2)}[\brho(L)-\Vect]\\
&\times \exp \left\{\frac{ik}{2}\int^L_0\!\! dz \left[\frac{d\brho(z)}{dz}\right]^2+ \frac{ik}{2}\int^L_0\!\! dz\;\epsilon(\brho(z),z)\right\}.
\end{aligned}
\end{equation*}
We observe here that the $\delta$-function alternatively replaces the conditions of the form $\brho(z)=\Vect,$ so symmetry imposes
\begin{equation}
\begin{aligned}
G(\Vect,L;\Vect_0,0)&=\!\int\mathcal{D}^{2}\rho(z)\:\delta^{(2)}[\brho(0)-\Vect_0]\;\delta^{(2)}[\brho(L)-\Vect]\nonumber\\
&\times \exp \left\{\frac{ik}{2}\int^L_0\!\! dz \left[\frac{d\brho(z)}{dz}\right]^2+ \frac{ik}{2}\int^L_0\!\! dz\;\epsilon(\brho(z),z)\right\}.
\end{aligned}
\label{eq:inhomogeneous-green-function}\end{equation}
In this context, the normalization can be rewritten in one of the following forms:
\begin{align}
1&=\int^{\brho(L)=\Vect}\!\!\!\! \mathcal{D}^2\rho(z) \exp \left\{\frac{ik}{2}\int^L_0\!\! dz  \left[\frac{d\brho(z)}{dz}\right]^2\right\}\nonumber\\
&= \int\mathcal{D}^{2}\rho(z)\:\delta^{(2)}[\brho(L)-\Vect]\exp \left\{\frac{ik}{2}\int^L_0\!\! dz  \left[\frac{d\brho(z)}{dz}\right]^2\right\}\\
&=\int\mathcal{D}^{2}\rho(z)\:\delta^{(2)}[\brho(0)-\Vect_0]\exp \left\{\frac{ik}{2}\int^L_0\!\! dz  \left[\frac{d\brho(z)}{dz}\right]^2\right\}.
\end{align}
Afterwards, the general solution to (\ref{eq:inhomogeneous-diffusion}) is 
\begin{equation}
\begin{aligned}
u(\Vect,L)&=\int d^2R_0\; u_0(\Vect_0)\int\mathcal{D}^{2}\rho(z)\:\delta^{(2)}[\brho(0)-\Vect_0]\;\delta^{(2)}[\brho(L)-\Vect]\\
&\times \exp \left\{\frac{ik}{2}\int^L_0\!\! dz \left[\frac{d\brho(z)}{dz}\right]^2+ \frac{ik}{2}\int^L_0\!\! dz\;\epsilon(\brho(z),z)\right\}.
\end{aligned}\label{eq:inhomogeneous-solution}
\end{equation}
Next, we are going to show how this construction help us determine the \textit{irradiance pattern} over a screen.  

\section{Image Formation using the Feynman's Path Integral Representation}

In the following we are going to change the representation space we are actually using by another  called \textit{velocity representation space.} This representation was introduced and used frequently by Russian scientists---it is just a functional change of variables. It has proven extremely useful in handling propagation problems \citep{paper:klyatskin-2,book:klyatskin-3}. Now, let us introduce it:
\begin{equation}
\left\{\begin{aligned}
\frac{d\brho}{dz}&=-\mathbf{v}(z)+\frac{[\brho(L)-\brho(0)]}{L},\\
\brho(L)&=\Vect,\;\brho(0)=\Vect_0\text{, and}\\
\mathbf{v}(0)&=\mathbf{v}(L)=\frac{[\brho(L)-\brho(0)]}{L}.
\end{aligned}\right.
\end{equation}
Also, this new variable requires its own \textit{normalization}, that is,
\begin{equation*}
\int\mathcal{D}^2 v(z)\; \exp\left\{\frac{ik}{2}\int^L_0\!\! dz\;\mathbf{v}^2(z)\right\}=1,
\end{equation*}
with $\mathcal{D}^2v(z)=\lim_{N\rightarrow\infty}\prod^{N-1}_{j=1}(kL\,d^2v_j/2\pi iN)$. 

This change transforms the inhomogeneous Green function (\ref{eq:inhomogeneous-green-function}) into;
\begin{equation}
\begin{aligned}
G(\Vect,L;\Vect_0,0)&=\exp\left\{\frac{ik}{2L}(\Vect-\Vect_0)^2\right\}\int\mathcal{D}^{2}v(z)\:\delta^{(2)}\!\left[\int^L_0\!\!\!dz\,\vel(z)\right]\\
\times \exp \left\{\frac{ik}{2}\right.&\left.\!\!\int^L_0\!\!\! dz\,\vel^2(z)+ \frac{ik}{2}\!\int^L_0\!\!\! dz\; \epsilon\!\left[\frac{z}{L}\Vect+\frac{(L-z)}{L}\Vect_0+\int^L_z\!\!\!d\eta\; \vel(\eta),z\right]\right\}.
\end{aligned}
\end{equation}
This representation allows us distinguish between the contribution made by the inhomogeneities from the free-space propagating wave. Now, let us calculate the free-space Green function to introduce the procedure we will afterwards follow in more complex situations. As always, we start dividing the interval $[0,L]$ in $N$ new subintervals, but this time we will also introduce the following Fourier transform,
\begin{align}
2\pi\; \delta^{(2)}\!\left[\int^L_0 \vel(\eta)d\eta\right]=\frac{1}{2\pi}\int_{\Re^2}\!\! d^2\kappa\; \exp\!\left\{-i\Bkappa\cdot\int^L_0\!\!d\eta\; \vel(\eta)\right\}.
\end{align}
We have then,
\begin{multline*}
\exp\left\{-\frac{ik}{2L}(\Vect-\Vect_0)^2\right\}G_0(\Vect,L;\Vect_0,0)=\\
=\lim_{N\rightarrow\infty} \left(\frac{kL}{2 \pi iN}\right)^{N-1}\!\! \int_{\Re^2}\!\!\cdots\int_{\Re^2}\! \prod^{N-1}_{j=1} d^2v_j\exp\left\{\frac{ik}{2}\sum^N_{j=0}\mathbf{v}^2_j\left(\frac{L}{N}\right)\right\}\,\delta^{(2)}\!\left[\sum^{N-1}_{j=0} \vel_j \left(\frac{L}{N}\right)\right]\\
=\lim_{N\rightarrow\infty} \left(\frac{kL}{2\pi iN}\right)^{N-1}\frac{1}{4\pi^2} \int_{\Re^2}\!\!d^2\kappa\,\int_{\Re^2}\!\!\cdots\int_{\Re^2}\!\prod^{N-1}_{j=1} d^2v_j \; \exp\left\{\sum^{N-1}_{j=0}\frac{ikL}{2N}\vel_j^2-\frac{iL}{N}\Bkappa\cdot\vel_j\right\}\\
\!\!\!\!\!\!\!\!\!\!=\lim_{N\rightarrow\infty}\left(\frac{kL}{2\pi iN}\right)^{N-1}\frac{1}{4\pi^2}\int_{\Re^2}\!\! d^2\kappa \prod^{N-1}_{j=1}\left\{\int_{\Re^2}\! d^2v_j \exp\!\left[\frac{ikL}{2N}\vel_j^2-\frac{iL}{N}\Bkappa\cdot\vel_j\right]\right\}\times\\
\!\!\!\!\!\!\!\!\!\!\times\exp\!\left[\frac{ikL}{2N}\vel_0^2-\frac{iL}{N}\Bkappa\cdot\vel_0\right]=\lim_{N\rightarrow\infty} \left(\frac{ikL}{2\pi iN}\right)^{N-1}\frac{1}{4\pi^2}\exp\left[\frac{kL}{2 N}\vel_0^2\right]\times\\
\times\int_{\Re^2}\!\!d^2\kappa\,\exp\!\left[-\frac{iL}{N}\Bkappa\cdot\vel_0\right]\prod^{N-1}_{j=1} \left\{\frac{2\pi iN}{kL}\exp\!\left[-i\frac{L}{2kN}\Bkappa^2 \right] \right\}\\
=\lim_{N\rightarrow\infty}\exp\!\left[\frac{kL}{2 N}\vel_0^2\right]\frac{1}{4\pi^2}\int_{\Re^2}\!\!d^2\kappa\,\exp\!\left[-i\frac{L(N-1)}{2kN}\Bkappa^2 -i\frac{L}{N}\Bkappa\cdot\vel_0\right]\\
= \frac{k}{2\pi iL}\lim_{N\rightarrow\infty}\frac{1}{\left(1-\frac{1}{N}\right)}\exp\!\left[ \frac{ikL}{2(N-1)}\right]=\frac{k}{2 \pi i L}.
\end{multline*}
So it is appropriate to write the Green function as:
\begin{equation}
G(\Vect,L;\Vect_0,0)=g(\Vect_0,\Vect,L)\;G_0(\Vect,L;\Vect_0,0),
\label{eq:inhomogeneous-green-2}\end{equation}
where $G_0$ is the free-space Green function and 
\begin{multline}
g(\Vect_0,\Vect,L):= \frac{2\pi i L}{k}\int\mathcal{D}^{2}v(z)\:\delta^{(2)}\!\left[\int^L_0\!\!dz\;\vel(z)\right]\times\\
\times \exp \left\{\frac{ik}{2}\int^L_0\!\!dz\;\vel^2(z) + \frac{ik}{2}\int^L_0\!\!dz\; \epsilon\!\left[\frac{ z}{L}\Vect+\frac{(L-z)}{L}\Vect_0+\int^L_z\!\!d\eta\;\vel(\eta),z\right]\right\};
\label{eq:inhomogeneous-term}
\end{multline}
therefore, we have concentrated all the random features of the medium within this last function---it is set equal to $1$ when perturbations are absent.

Before going further, we have to describe the environment in Optics where approximation (\ref{eq:inhomogeneous-diffusion}) holds. As in Section \ref{section:turbulent-index}, we write the permitivity as a constant term---we will assume equal to one---plus another term $\epsilon$ containing all the information coming from the medium. There, we consider the propagation of quasi-monochromatic lightwave radiation with frequency $\omega$, that is,
\begin{equation*}
\left\{\triangle + [1+\epsilon(\vect,z)] k^2\right\} \mathcal{E}(\vect,z)=0,
\end{equation*}
where $\mathcal{E}(\vect,z)$ is the scalar electromagnetic field---there is no polarization here. Supposing that backscattering is negligible, and thus the propagation has a preferred direction, let us say $\mathcal{E}(\vect,z)=E(\vect,z)\,e^{i(kz-\omega t)}$, the latter equation changes to
\begin{equation}
\left[\triangle + k^2 \epsilon(\vect,z) \right] E(\vect,z)=0.
\label{eq:inhomogeneous-wave}\end{equation}
Hence, under the condition $kz\gg 1$ we retrieve equation (\ref{eq:inhomogeneous-diffusion}). Thereafter this is a \textit{stochastic parabolic equation}, and it can be solved when $\epsilon$ is a markovian\footnote{
See Chapter \ref{chp:stochastic-calculus} for a detailed description.
} process along the propagation axis as it was shown by \citet{paper:rytov}. Moreover, its solution coincides with the deterministic solution we have just shown. \citet{paper:charnotskii-2} extensively discusses the applications of the path integral formulation to this problem. In the Appendix B we summarize how he characterizes the cohabitation of two regimes: \textit{weak} and \textit{strong}. Basically the technical differences between both regimes are the following. The weak regime cumulates a variety of methods grouped under the tag of Rytov's formalism---whose central idea is to solve equation (\ref{eq:inhomogeneous-wave}), or an equivalent form, using a \textit{Taylor}-like series expansion---in practical terms this distinction provides a way to select the best method for solving a particular problem. While the strong regime has, until now, one tool: the path-integral approach. Nevertheless, it is this approach the only one covering both regimes.

Finally, let us build the irradiance distribution from the solution to the inhomogeneous wave equation (\ref{eq:inhomogeneous-wave}). According to definition (\ref{eq:inhomogeneous-green-2}) and equation (\ref{eq:inhomogeneous-solution}) we have:
\begin{equation*}
\begin{aligned}
E(\Vect, L)&=\int_{\Re^2}\!\! d^2R_0\; E(\Vect_0, 0)\;G(\Vect,L;\Vect_0,0)\\
&= \int_{\Re^2}\!\! d^2R_0\; E(\Vect_0, 0)\;G_0(\Vect,L;\Vect_0,0)\,g(\Vect_0,\Vect,L),
\end{aligned}
\end{equation*}
and so the intensity function is 
\begin{align}
I(\Vect,L)&= E^*(\Vect,L)E(\Vect,L)\nonumber\\
&=\int_{\Re^2}\!\int_{\Re^2}\!\! d^2R'_0\, d^2R_0\;E^*(\Vect'_0, 0)E(\Vect_0,0)\; G^*(\Vect,L;\Vect'_0,0)\,G(\Vect,L;\Vect_0,0).
\end{align}
Now suppose the coherence time $\omega^{-1}$ is much smaller than the characteristic time scale $T$ of the detector, i.e., $\omega^{-1} \ll T$. Thus the irradiance pattern observed is time averaged;  furthermore, it can be assumed ergodic. So, if the source is spatially incoherent, the \textit{mutual intensity} is written \citep{book:goodman} 
\begin{equation}
\overline{E^*(\Vect'_0, 0)E(\Vect_0,0)}=A(\Vect_0)\,\delta^{(2)}(\Vect'_0-\Vect_0),
\label{prop:incoherent-source}\end{equation}
the overbar means the ensemble average. Besides, the function $A(\vect)$ has dimensions \textbf{irradiance}$\times$\textbf{area}, that is, is an intensity distribution; moreover, in most of the cases it is proportional to the initial irradiance distribution at the pupil:
\begin{equation}
A(\Vect)= \frac{\lambda^2}{\pi} I_0(\Vect).
\label{prop:irradiance-distribution}\end{equation}
Nevertheless, for the current discussion we will keep the distribution function $A$, and so we use the property (\ref{prop:incoherent-source}) to write the intensity function as
\begin{equation}
I(\Vect,L)= \int_{\Re^2}\!\! d^2R_0\;A(\Vect_0)\; \abs{G(\Vect,L;\Vect_0,0)}^2,
\label{eq:irradiance}\end{equation}
with $G$ as in (\ref{eq:inhomogeneous-green-2}). 

In the next section we will apply these results to the case of image formation in self-image system.

\section{Intensity Distribution of Self-image Systems into Turbulent Media}

Here we will describe light propagation through a Lau-like arrangement, that is, two Ronchi grids out of phase half a period within a turbulent medium. Also, we will inspect how degradation produced by the turbulence can be estimated in terms of the spacing between two parallel grids, the number of lines per millimeter, and $C^{2}_{\epsilon}$, the structure constant of the medium \citep{paper:perez-garavagia}.

\begin{figure}
\begin{center}
\includegraphics[width=0.65\textwidth]{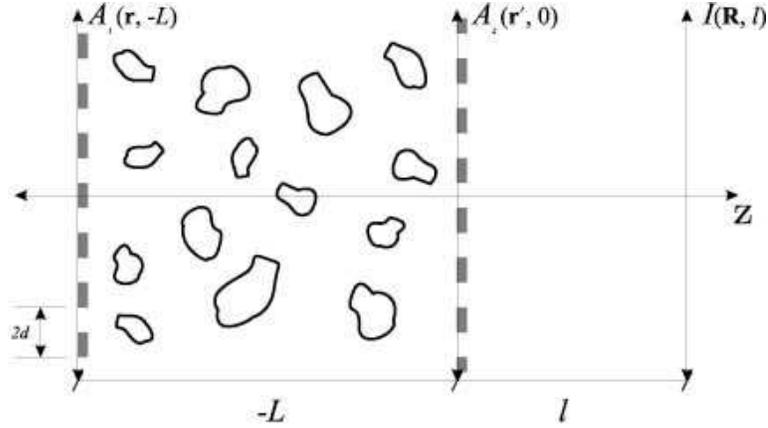}
\caption{The optical system employed in this work; two grids separated a distance $L$ with equal amplitude tramittance functions ($A_{1}$ is the outgoing intensity distribution and $A_{2}$ the tramittance function, these both have a period $2d$) and $I$ is the intensity distribution at a screen located at $z=l$.
\label{fig:grids}}
\end{center}
\end{figure}

Now, we introduce the optical system for the present discussion: as it is sketched in Figure \ref{fig:grids}, it consists of a Lau system of grids---we have no lens here---separated by a distance $L$ and half a period out of phase. That is, each grid can be thought as the negative image of the other. Finally, between the grids there is a turbulent medium with structure constant $\Cs$, roughly homogeneous in the plane perpendicular to the $z$ direction. At $z=l$  there is a screen where we want to observe the system behavior. We suppose that the medium between the second grid and the screen plane is free from turbulence. 

In this problem there are only two optical elements: the grids. Both of them have the same tramittance function, it can be modeled as a family of square wave functions 
\begin{equation}
t_{\alpha}(x)= \frac{1}{2} + \frac{1}{\pi} \sum_{n=0}^{\infty} 
\frac{\;(-1)^{n+1}}{(2n+1)}\, \cos\!\left[\frac{(2n+1)\pi}{d} (x + \alpha)\right],\label{eq:transmition}
\end{equation}
the parameter \(\alpha\) describes the relative phase and $2d$ is the grid period. There is a difference between both grids when we turn them into mathematical objects for our problem: the light passing through the first grid coming from a spatial incoherent source can be expressed as a delta correlated coherent function with intensity distribution $A_1({\bf r})$ given by
\begin{equation}
A_{1}({\bf r})= A_{0}\,t_{0}(\check{{\bf e}}_{x} \cdot {\bf r}) S_{D \times D}(\mathbf{r}),
\end{equation} \\
where
\begin{equation}
S_{D \times D}(\vect)= \left\{ \begin{array}{ll}
	1 & \text{if}\quad  \vect \in \left[-\frac{D}{2},\frac{D}{2}\right] \times \left[-\frac{D}{2},\frac{D}{2}\right] \\
	0 & \text{otherwise} \end{array} \right.,
\end{equation} 
$A_{0}$ is the maximum value for the irradiance distribution  and $D$ is the size of the rectangular grid; while the tramittance $A_2$ of the second grid is modeled using the same functional relationship, but with $A_0=1$ and setting the phase $\alpha=d$. 

The irradiance function $I(\Vect)$, as we showed arriving to equation (\ref{eq:irradiance}), can be expressed as follows:
\begin{equation}
I(\Vect)=\int_{\Re^2}\!\!d^2r \; A_1(\vect)|G_{\textit{tot}}
(\Vect,l;\vect,-L)|^{2}.
\label{eq:intensity}
\end{equation} 
This time the \textit{total Green function} $G_{\textit{tot}}(\Vect,l;\vect,-L)$ comes from combining the free-space Green function $G_{0}(\Vect,l;\vect',0)$, which corresponds to the zone between the second grid and the screen, and the turbulent Green function $G(\vect',0;\Vect,-L)$:
\begin{equation}
G_{\textit{tot}}(\Vect,l;\vect,-L)=\int_{\Re^2}\!\!d^2r'\;G_{0}(\Vect,l;\vect',0)\,A_2(\vect')\,G(\vect',0;\vect,-L). 
\label{eq:greenmelt} 
\end{equation}   
Remember that the irradiance distribution $A_1$ and the tramittance function $A_2$ were built from delta correlated fields as was discussed in the last section. Now, combining (\ref{eq:intensity}), (\ref{eq:greenmelt}) and the turbulent Green function definition (\ref{eq:inhomogeneous-green-2})  yields:
\begin{equation}
\begin{aligned}
I(\Vect) =&\Cons\;\intDD\!\!\!d^{2}r
\;\Tc{0}{\cdot \vect}\;\int_{\Re^2}\!\!\intDD\!\!\!d^{2}R'd^{2}r'\;\Ts{d}{\cdot\left(\Vect'-\frac{\vect'}{2}\right)} \\
 \times &\,\Ts{d}{\cdot\left(\Vect'+\frac{\vect'}{2}\right)}\exp\!\left[-\dot{\imath}\frac{k}{f} \vect'\cdot \Vect' \right]\exp\!\left[-\dot{\imath}k\left(\frac{\Vect}{l}+\frac{\vect}{L} \right)\cdot\vect'\right] \\
\times &\, g\!\left(\Vect'-\frac{\vect'}{2},\vect,L\right)g^{\ast}\!\left(\Vect'+\frac{\vect'}{2},\vect,L\right),
\end{aligned}
\label{eq:melt}
\end{equation}
here $f$ stands for the relation $1/f=1/l+1/L$. This equation is similar to that found by \citet{paper:charnotskii3} but here we have not lost the phase term $\exp(-\dot{\imath}\frac{k}{f}\vect'\cdot \Vect')$, just because we are working with grids instead of lenses.

\subsection{The non-turbulent case}

Before treating the turbulent problem we are going to consider light propagation in the absence of turbulence. Our goal here is to inspect the role of each grid in the image formation process. Let us assume that $g(\vect,\vect',|z-z'|)\equiv 1$, thus equation (\ref{eq:melt}) takes the form:
\begin{equation}
I(\Vect) =\Cons \intDD\!\!\!d^2r\;\Tc{0}{\cdot\vect}\intDD\!\!\!d^{2}r'\; C_{t}(\vect')\; \exp\!\left[ \dot{\imath}k \left( \frac{\Vect}{l}+ \frac{\vect}{L} \right) \cdot\vect' \right], 
\label{eq:noturb}
\end{equation}
where  
\begin{equation}
C_{t}(\vect')=\intDD\!d^{2}R'\;\Ts{d}{\cdot\left(\Vect'-\frac{\vect'}{2}\right)}\!\Ts{d}{\cdot\left(\Vect'+\frac{\vect'}{2}\right)} \exp\left(-\dot{\imath}\frac{k}{f}\vect'\cdot \Vect'\right)
\label{def:complex-correlation}\end{equation}
is a complex correlation function. 

Now, let us carefully inspect the former equation. Noticing that
\begin{multline}
\intDD\!\!\!d^2r\, \Tc{0}{\cdot\vect}\exp\left[ -i \vect\cdot\left( -\frac{k\vect'}{L} \right)\right]=\\
2\pi\mathcal{F}\left\{\Tc{0}{\cdot\vect}\Heavy{ + \vect} \Heavy{- \vect}\right\}\left(-\frac{k}{L}\,\vect'\right)=2\pi\;\;\hat{t}_{{0_{{}_{{}_{{}_{\!\!\!\!\!\!\!\!\!\!\!D \times D}}}}}}\!\!\!\left(-\frac{k}{L}\,\vect'\right),
\end{multline}
where $\Theta^{(2)}({\bf r})= \Theta(r_x) \Theta(r_y)$ is the two dimensional Heaviside function and  ${\bf D}= D \check{{\bf e}}_{x} + D \check{{\bf e}}_{y} $. We shall rewrite equation (\ref{eq:noturb}) as,
\begin{multline*}
2\pi I(\Vect) =\\
=\int_{\Re^2}\!\! d^{2}r'\left[ \left(\frac{A_{0}\;k^{4}}{l^{2} L^{2}}\right)\;\hat{t}_{{0_{{}_{{}_{{}_{\!\!\!\!\!\!\!\!\!\!\!D \times D}}}}}}\!\!\!\left(-\frac{k}{L}\,\vect'\right) C_{t}(\vect')\Heavy{ + \vect'} \Heavy{- \vect'}\right]
e^{ i k\vect' \cdot\Vect/l};
\end{multline*}
therefore, with the change of variables $\Bkappa=k\,\vect'/l$ the function between brackets results to be the Fourier transform of the irradiance $I$. That is,
\begin{equation}
\hat{I}(\Bkappa)=\cons\;\;\;
\hat{t}_{{0_{{}_{{}_{{}_{\!\!\!\!\!\!\!\!\!\!\!D \times D}}}}}}\!\!\left( -\frac{l}{L}\,\Bkappa \right) C_{t} \left( \frac{l}{k}\,\Bkappa \right) \Heavy{ + \frac{l}{k}\, \Bkappa} \Heavy{- \frac{l}{k}\, \Bkappa}.
\label{fourier}
\end{equation} 
Let us inspect each term in this transform. From the definition (\ref{eq:transmition}) we observe that the Fourier transform for the transmission function \;$\hat{t}_{{0_{{}_{{}_{{}_{\!\!\!\!\!\!\!\!\!\!\!D \times D}}}}}}$\!\! rests on the decomposition 
\begin{equation}
\cos\!\left[\frac{(2n+1)\pi}{d} x \right] = \frac{1}{2}\left\{ \exp\left[i\frac{(2n+1)\pi}{d} x \right] +\exp\left[-i\frac{(2n+1)\pi}{d} x \right]\right\}.
\end{equation}
As far as we are concerned with the effect produced by the grids, we are going to neglect the low spatial frequencies related to finite size effects, that is, we will take the limit $1/|\Bkappa|D\rightarrow 0$ each time we calculate the resultant irradiance---equivalent to the condition $D\gg\lambda$.  Otherwise, the optical system acts as a filter transmitting only a discrete numerable set of frequencies. 

Afterwards, using the dimensionless variable $\boldsymbol{\eta}/\lambda=\Bkappa$ we finally have,
\begin{align}
\hat{t}_{{0_{{}_{{}_{{}_{\!\!\!\!\!\!\!\!\!\!\!D \times D}}}}}}\!\!\left( -\frac{l}{\lambda L}\,\boldsymbol{\eta} \right) &= \frac{\lambda^2}{2\pi}\int\!\!\int^{\frac{D}{2\lambda}}_{-\frac{D}{2\lambda}} d^2x\; \Ts{0}{\cdot (\lambda\mathbf{x}) }\; \exp\!\left[i \mathbf{x}\cdot\left(\frac{l}{L}\,\boldsymbol{\eta}\right)\right]\nonumber\\
&= \hat{t}_0^{(0)}(\boldsymbol{\eta}) + \sum^{\infty}_{n=-\infty}\hat{t}_0^{(n)}(\boldsymbol{\eta}).
\end{align}
The last series comes from the limit $D/\lambda \rightarrow \infty$, its terms are:
\begin{align}
\hat{t}_0^{(0)}(\boldsymbol{\eta})&= \frac{\pi\lambda^2L^2}{l^2}\;\delta^{(2)}(\boldsymbol{\eta}),\label{eq:tramittance:t0}\\
\hat{t}_0^{(n)}(\boldsymbol{\eta})&= \frac{\lambda^2L^2}{l^2}\;\frac{(-1)^{n+1}}{(2n+1)}\;\delta^{(2)}\!\!\left[\boldsymbol{\eta}+ \frac{(2n+1)\pi \lambda L}{d\,l} \eX \right].
\label{eq:tramittance:t1}
\end{align}

Similarly, the complex correlation can be evaluated. First, we multiplicate the tramittance functions $t_d$ to obtain:
\begin{multline}
\Ts{d}{\cdot\left(\Vect-\frac{\vect}{2}\right)}\!\Ts{d}{\cdot\left(\Vect+\frac{\vect}{2}\right)} =\frac{1}{4} \\
+\frac{1}{2\pi}\sum^{\infty}_{n=-\infty}\frac{(-1)^{n+1}}{(2n+1)}\cos\!\left[\frac{(2n+1)\pi}{2d}\;\eX\cdot\vect\right]\exp\!\left[i\frac{(2n+1)\pi}{d}\;\eX\cdot\Vect\right]\\
+\frac{1}{4\pi^2}\sum^{\infty}_{m,n=-\infty}\frac{(-1)^{n+m}}{(2n+1)(2m+1)}\;\exp\!\left[i\frac{(m-n)\pi}{d}\;\eX\cdot\vect\right]\exp\!\left[i\frac{2(n+m+1)\pi}{d}\;\eX\cdot\Vect\right]\\
=M_0 + \sum^{\infty}_{n=-\infty} M^{(n)}_1 +\sum^{\infty}_{m,n=-\infty}M^{(m,n)}_2.
\end{multline}
These terms are directly Fourier-transformed because of the definition (\ref{def:complex-correlation}) under the condition we have given above, i.e.:
\begin{align}
\hat{M}_0(\boldsymbol{\eta})&= \frac{\pi^2 \lambda^2 f^2}{ l^2}\;\delta^{(2)}(\boldsymbol{\eta}),\label{eq:correlation:m0}\\
\hat{M}_1^{(n)}(\boldsymbol{\eta})&= \frac{2\pi \lambda^2 f^2}{l^2}\frac{(-1)^n}{(2n+1)}\cos\!\!\left[\frac{(2n+1) l}{ 4 d }\;\eX\cdot\boldsymbol{\eta}\right]\delta^{(2)}\!\!\left[\boldsymbol{\eta}-\frac{(2n+1)\pi \lambda f}{d\, l}\;\eX\right]\nonumber\\
&=\frac{2\pi \lambda^2 f^2}{l^2}\frac{(-1)^n}{(2n+1)}\cos\!\!\left[\frac{(2n+1)^2\pi \lambda f}{ 4 d^2 }\right]\delta^{(2)}\!\!\left[\boldsymbol{\eta}-\frac{(2n+1)\pi \lambda f}{d\, l}\;\eX\right]\label{eq:correlation:m1}\\
\hat{M}_2^{(m,n)}(\boldsymbol{\eta})&=\frac{\lambda^2 f^2}{l^2}\frac{(-1)^{m+n}}{(2n+1)(2m+1)}\exp\!\!\left[i\frac{(m-n) l}{ d}\;\eX\cdot\boldsymbol{\eta}\right]\delta^{(2)}\!\!\left[\boldsymbol{\eta}-\frac{2(m+n+1)\pi \lambda f}{d\, l}\;\eX\right]\nonumber\\
&=\frac{\lambda^2 f^2}{l^2}\frac{(-1)^{m+n}}{(2n+1)(2m+1)}\exp\left[i\frac{2(m-n)(m+n+1) \pi \lambda f}{ d^2}\right]\nonumber\\
\times&\;\delta^{(2)}\!\!\left[\boldsymbol{\eta}-\frac{2(m+n+1)\pi \lambda f}{d\, l}\;\eX\right].\label{eq:correlation:m2}
\end{align}
Finally, the Fourier transform of the irradiance distribution (\ref{fourier}) is obtained multiplying the tramittance and complex correlation transforms. Nevertheless, we should be cautious. The approximation we suggested  induce the constant terms to produce more delta functions than the Fourier integral is able to handle. Therefore, we will redefine those problematic terms. Let us start with $y$-axis, because all tramittance functions depends on the $x$-axis we can write,
\begin{align*}
\sqrt{I}_y(R_y)&=\frac{k^2\sqrt{A_0}}{lL} \int_{\left[-\frac{D}{2},\frac{D}{2}\right]^3}\!dr_y dR_y' dr_y' \exp\!\left[ik\left(\frac{R_y}{l}+\frac{r_y}{L}-\frac{R'_y}{f}\right)r'_y\right]\\
&=\frac{k^2\sqrt{A_0}}{lL}\int_{\left[-\frac{D}{2},\frac{D}{2}\right]^2} dr_y dR_y'\left\{\frac{\lambda}{2\pi} \int_{-\frac{\pi D}{\lambda}}^{\frac{\pi D}{\lambda}}d\zeta \exp\!\left[i\left(\frac{R_y}{l}+\frac{r_y}{L}-\frac{R'_y}{f}\right)\zeta\right]\right\}\\
&=\frac{k^2\sqrt{A_0}}{lL}\int_{\left[-\frac{D}{2},\frac{D}{2}\right]^2} dr_y dR_y'\;\frac{\lambda}{\sqrt{2\pi}}\;\delta\!\left(\frac{R_y}{l}+\frac{r_y}{L}-\frac{R'_y}{f}\right)\\
&=\frac{k\sqrt{2\pi A_0}}{lL}\int_{-\frac{D}{2}}^{\frac{D}{2}} dr_y\int_{\Re} dR_y'\;\delta\!\left(\frac{R_y}{l}+\frac{r_y}{L}-\frac{R'_y}{f}\right) \Theta\left(\frac{D}{2}-R'_y\right)\Theta\left(\frac{D}{2}+R'_y\right)\\
&=\frac{k f\sqrt{2\pi A_0}}{lL}\int_{-\frac{D}{2}}^{\frac{D}{2}} dr_y\;\Theta\left(\frac{D}{2}-\frac{f}{l}R_y-\frac{f}{L}r_y\right)\Theta\left(\frac{D}{2}+\frac{f}{l}R_y+\frac{f}{L}r_y\right).
\end{align*}
We observe from above that eliminating the effects from the edges corresponds to the condition $-D/2<R_y<D/2$, there we have the constant intensity distribution:
\begin{equation*}
\sqrt{I}_y(R_y)=\frac{k fD\sqrt{2\pi A_0}}{lL}.
\end{equation*}
Furthermore, a constant term will appear when multiplying both first terms in the tramittance and correlation functions, (\ref{eq:tramittance:t0}) and (\ref{eq:correlation:m0}),  
\begin{equation*}
I_e(\Vect)= 2I_0\left(\frac{ 2\pi D f}{l L}\right)^2
\end{equation*}
---this is the contribution from the edges to the irradiance. 

Also, it is worth noting that the products $\hat{t}^{(0)}_0 \hat{M}_1$ and $\hat{M}_0 \hat{t}^{(n)}_0$ are zero because  $(2n+1)\neq 0$ for all $n\in\mathbb{Z}$, but the remaining cross product by \textit{constants} contribute to the irradiance distribution on the $x$-axis:
\begin{align}
\widehat{\left(\sqrt{I_x}\right)}^{(0)}(\eta_x)&=\frac{2d\sqrt{2\pi I_0}}{l}\,\hat{t}^{(0)}_0\!\! \sum^{\infty}_{m,n=-\infty}\!\!\hat{M}_2^{(m,n)}\nonumber\\
&=\frac{d\sqrt{I_0}}{l}\,\delta(\eta_x)\sum^{\infty}_{n,m=-\infty}\frac{(-1)^{n+m}}{(2n+1)(2m+1)}\, \delta\!\left[\frac{2(m+n+1)\pi\lambda f}{d\,l}\right]\nonumber\\
&=\frac{d\sqrt{I_0}}{l}\,\delta(\eta_x)\sum^{\infty}_{n=-\infty}\frac{(-1)^n}{(2n+1)}\int^{\infty}_{-\infty} \frac{dw}{\left(2w+\frac{2\pi\lambda f}{d\,l}\right)}\nonumber\\
\times&\exp\left[i \pi(2k+1)\left(\frac{d\,l}{2\pi\lambda f}\right) w\right]\delta\left[w+\frac{2(n+1)\pi\lambda f}{d\,l}\right]\nonumber\\
&=\frac{d\sqrt{I_0}}{l}\,\delta(\eta_x)\sum^{\infty}_{n=-\infty}\frac{1}{(2n+1)^2}=\frac{d\sqrt{I_0}}{l}\,\left(\frac{\pi^2}{4}\right)\delta(\eta_x).
\label{eq:constant}
\end{align}
We changed the sum in $m$ to the integral, in $w= (2\pi\lambda f/d\,l) m$, because of the condition $d/\lambda\gg 1$, where $d$ is defined through $N(2d)=D$---$N$ is the number of lines of the grid.  

We find two others non-zero contributions to the irradiance, following the same procedure as above:
\begin{align}
\widehat{\left(\sqrt{I_x}\right)}^{(1)}(\eta_x)&=\frac{2d\sqrt{2\pi I_0}}{l}\,\sum^{\infty}_{m,n=-\infty}\hat{t}_0^{(n)}\hat{M}_1^{(m)}= \frac{2d\sqrt{I_0}}{l}\sum^{\infty}_{n=-\infty} \frac{(-1)^{n}}{(2n+1)}\,\delta\!\!\left[\eta_x+ \frac{(2n+1)\pi \lambda L}{d\,l}\right]\nonumber\\
\times\sum^{\infty}_{m=-\infty}&\frac{(-1)^{m}}{(2m+1)}\  \cos\!\!\left[\frac{(2m+1)^2\pi \lambda f}{ 4 d^2 }\right]\delta\!\!\left[\frac{(2n+1)\pi \lambda L}{d\,l}+\frac{(2m+1)\pi \lambda f}{d\, l}\right]\nonumber\\
&= \frac{2d\sqrt{I_0}}{l}\sum^{\infty}_{n=-\infty} \frac{(-1)^{n}}{(2n+1)}\,\delta\!\!\left[\eta_x+ \frac{(2n+1)\pi \lambda L}{d\,l}\right]\nonumber\\
\times\int^{\infty}_{-\infty}&\frac{dw}{(2w+\frac{2\pi\lambda f}{d\,l})}\exp\!\!\left[i\pi(2k+1)\left(\frac{d\,l}{2\pi\lambda f}\right) w \right]\cos\!\!\left[\left(\frac{l^2}{ 16\pi\lambda f }\right)\left(2w+\frac{2\pi\lambda f}{d\,l}\right )^2\right]\nonumber\\
\times&\delta\!\!\left[w+\frac{\pi \lambda f}{d\, l}+\frac{(2n+1)\pi \lambda L}{d\,l}\right]\nonumber\\
&= \frac{2d\sqrt{I_0}}{l}\sum^{\infty}_{n=-\infty} \frac{(-1)^{n+1}}{(2n+1)^2}\,\delta\!\!\left[\eta_x+ \frac{(2n+1)\pi \lambda L}{d\,l}\right]\cos\!\!\left[\frac{(2n+1)^2 \pi \lambda L^2}{4 d^2 f}\right]\nonumber\\
\times&\exp\left\{-i\frac{\pi(2k+1)}{2}\left[\frac{L}{f}(2n+1)+1\right]\right\}.
\label{eq:series-first}
\end{align}
The irradiance is real function, so the exponential in the latter equation should be real; it only happens when 
\begin{equation}
l= \frac{L}{2s} 
\label{eq:pairrel},
\end{equation}
for $s\in\mathbb{Z}^{+}$, any other choice in the quotient will make the series vanish. Finally, we write equation (\ref{eq:series-first}) as
\begin{multline}
\widehat{\left(\sqrt{I_x}\right)}^{(1)}(\eta_x)=\frac{(-1)^{s}(4s)d\sqrt{I_0}}{l}\\
\times\sum^{\infty}_{n=-\infty} \frac{1 }{(2n+1)^2}\cos\!\!\left[\frac{(2n+1)^2(2s+1)\pi \lambda L}{ 4 d^2 }\right]\delta\!\!\left[\eta_x+\frac{2s(2n+1)\pi \lambda}{d}\right].
\label{eq:TM}
\end{multline}

And the second non-zero term is,
\begin{multline}
\widehat{\left(\sqrt{I_x}\right)}^{(2)}(\eta_x)=\frac{d\sqrt{I_0}}{\pi l}\sum^{\infty}_{n,m,m'=-\infty}\!\!\!\hat{t}_0^{(n)}\hat{M}_2^{(m,m')}\\
=\frac{d\sqrt{I_0}}{\pi l}\sum^{\infty}_{n,m=-\infty}\frac{(-1)^{m+n+1}}{(2n+1)(2m+1)}\,\delta\!\!\left[\eta_x+ \frac{(2n+1)\pi \lambda L}{d\,l} \right]\\
\times\sum^{\infty}_{m'=-\infty}\frac{(-1)^{m'}}{(2m'+1)}\exp\!\!\left[i\frac{2(m-m')(m+m'+1)\pi\lambda f}{d^2}\right]\\
\times\delta\!\!\left[\frac{(2n+1)\pi \lambda L}{d\,l}\right.-\left.\frac{2(m+m'+1)\pi \lambda f}{d\, l}\right]\\
=\frac{d\sqrt{I_0}}{\pi l}\sum^{\infty}_{n,m=-\infty}\frac{(-1)^{m+n+1}}{(2n+1)(2m+1)}\;\delta\!\!\left[\eta_x+ \frac{(2n+1)\pi \lambda L}{d\,l} \right]\\
\times\int^{\infty}_{-\infty}\frac{dw}{\left(2w+\frac{2\pi\lambda f}{d\,l}\right)}\exp\!\left\{i\!\left(\frac{l^2}{2\pi\lambda f}\right)\right.\left.\!\left[\left(\frac{2\pi\lambda f}{dl}\right)m-w\right]\!\!\left[\left(\frac{2\pi\lambda f}{dl}\right)(m+1)+w\right]\right\}\\
\times \exp\!\!\left[i\pi(2k+1)\left(\frac{d\,l}{2\pi\lambda f}\right)\!w\right] \delta\!\left\{w-\left[\frac{(2n+1)\pi\lambda L}{d\,l}\;-\!\!\frac{2(m+1)\pi\lambda f}{d\,l}\right]\right\}\\
=\frac{d\sqrt{I_0}}{\pi l}\sum^{\infty}_{n,m=-\infty}\frac{(-1)^{m+n+1}}{(2n+1)(2m+1)}\;\delta\!\!\left[\eta_x+ \frac{(2n+1)\pi \lambda L}{d\,l} \right]\\
\times\frac{1}{\left[(2n+1)\frac{L}{f}-(2m+1)\right]}\;\exp\!\left\{i\pi(2k+1)\left[(2n+1)\frac{L}{2f}-(m+1)\right]\right\}\\
\times\exp\!\left\{i\left(\frac{\pi\lambda L}{d^2}\right)\right.\left.\!\!\left[(2m+1)-(2n+1)\frac{L}{2f}\right]\right\},
\label{eq:TM2}
\end{multline}
and these exponentials should give us a real number. It takes some algebra to rewrite them as
\begin{equation*}
\exp\left\{i\pi\left[ \left(\frac{\lambda L}{d^2}-1\right)\left(\frac{L}{2f}\right)+\frac{\lambda L}{d^2}\right]\right\}\in\mathbb{R}.
\end{equation*}
If we now assume that (\ref{eq:pairrel}) is fulfilled then it happens that only $\lambda L/d^2=2p+1$ makes this exponential real. On the other hand, it also makes the series in $m$ from the former equation to have a singular term. Nevertheless, we realize that the condition
\begin{equation*}
(2m+1)(2s+1)=(2n+1)
\end{equation*}
---according to (\ref{eq:pairrel}) it is $L/f=(2s+1)$---gives a zero term in the original series. Thus, we finally have the series
\begin{multline}
\widehat{\left(\sqrt{I_x}\right)}^{(2)}(\eta_x)=(-1)^p\frac{(2s)\lambda \sqrt{I_0}}{\pi d}\!\!\sum^{\infty}_{n=-\infty}\frac{(-1)^{n}}{(2n+1)}\,\delta\!\!\left[\eta_x+ \frac{2s(2n+1)\pi\lambda}{d} \right]\\
\times\!\!\!\!\!\!\!\!\!\!\!\! \sum^{\infty}_{{\substack{m=-\infty,\\(2m+1)\neq (2s+1)(2n+1)}}} \!\!\!\!\!\!\!\!\frac{1}{(2m+1)\left[(2m+1)-(2n+1)(2s+1)\right]}.
\label{eq:TM2-2}
\end{multline}
This series can be reduced: because 
\begin{multline*}
\sum^{\infty}_{{\substack{m=-\infty,\\(2m+1)\neq (2s+1)(2n+1)}}}\!\!\!\!\!\!\!\!\! \frac{1}{(2m+1)\left[(2m+1)-(2n+1)(2s+1)\right]}=\\
=\sum^{\infty}_{{\substack{m=0,\\(2m+1)\neq (2s+1)(2n+1)}}}\!\!\!\!\!\!\! \frac{2}{\left[(2m+1)^2-(2s+1)^2(2n+1)^2\right]}\\
=\!\!\!\!\!\!\!\sum^{\infty}_{\substack{k=0,\\k\neq (2s+1)(2n+1)}}\!\!\!\!\!\!\!\frac{2}{\left[k^2-(2s+1)^2(2n+1)^2\right]} - \sum^{\infty}_{k'=1}\frac{2}{\left[(2k')^2-(2s+1)^2(2n+1)^2\right]}\\
= \frac{3}{2(2s+1)^2(2n+1)^2}-\sum^{\infty}_{k'=1}\frac{2}{\left[(2k')^2-(2s+1)^2(2n+1)^2\right]},\end{multline*}
and knowing \citep{cd:gradshteyn}  that
\begin{equation*}
\sum^{\infty}_{k=0,\;k\neq m}\frac{1}{(m^2-k^2)}=-\frac{3}{4m^2},\quad\text{ and $m$ is an integer,}
\end{equation*}
we find 
\begin{multline*}
\sum^{\infty}_{k'=1}\frac{2}{\left[(2k')^2-(2s+1)^2(2n+1)^2\right]}\\
=-\frac{\pi}{2(2s+1)(2n+1)}\left\{\cot\left[\frac{\pi (2s+1)(2n+1)}{2}\right]-\frac{2}{\pi (2s+1)(2n+1)}\right\}\\
=\frac{1}{(2s+1)^2(2n+1)^2}.
\end{multline*}
Therefore, the equation (\ref{eq:TM2}) is  
\begin{equation}
\widehat{\left(\sqrt{I_x}\right)}^{(2)}(\eta_x)=(-1)^p\frac{(2s)\lambda \sqrt{I_0}}{(2s+1)^2\pi d}\,\sum^{\infty}_{n=-\infty}\!\!\frac{(-1)^{n}}{(2n+1)^3}\;\delta\!\!\left[\eta_x+ \frac{2s(2n+1)\pi\lambda}{d} \right],
\end{equation}
under the conditions $L=(2p+1)d^2/\lambda$ and $(2s)l=L$. 

Now we can change the relation (\ref{eq:pairrel}) to 
\begin{equation}
l=\frac{L}{(2s+1)}, 
\end{equation}
which automatically makes the equation (\ref{eq:TM}) vanish. That does not happen with (\ref{eq:TM2}); furthermore, it induces a relation of the type
\begin{equation}
L= \frac{d^2}{\lambda}2p 
\label{eq:setting}
\end{equation}
with $p\in\mathbb{Z}.$ It also forces the distance between the last grid and the screen to be
\begin{equation}
l= \frac{d^2}{\lambda}2q,\qquad q\in\mathbb{Z}.
\label{eq:setting-2}
\end{equation}
Observe that $q$ and $p$ must be simultaneously odd or even. Then the series (\ref{eq:TM2}) is written,
\begin{multline*}
\widehat{\left(\sqrt{I_x}\right)}^{(2)}(\eta_x)=(-1)^{s}(2s+1)\frac{\lambda \sqrt{I_0}}{\pi p\, d}\sum^{\infty}_{n=-\infty}\frac{(-1)^{n}}{(2n+1)}\delta\!\!\left[\eta_x+ \frac{(2n+1) (2s)\pi\lambda}{d} \right]\\
\times\sum^{\infty}_{m=-\infty}\!\! \frac{1}{(2m+1)\left[(2m+1)-2(s+1)(2n+1)\right]}.
\end{multline*}
Studying again the sum in $m$:
\begin{align}
\sum^{\infty}_{m=-\infty}&\!\! \frac{1}{(2m+1)\left[(2m+1)-2(s+1)(2n+1)\right]}= \sum^{\infty}_{m=0} \frac{2}{\left[(2m+1)^2-4(s+1)^2(2n+1)^2\right]}\nonumber\\
&=\sum^{\infty}_{\substack{k=0,\\k\neq 2(s+1)(2n+1)}}\!\!\!\!\!\!\!\frac{2}{\left[k^2-4(s+1)^2(2n+1)^2\right]} -\!\!\!\!\!\!\! \sum^{\infty}_{\substack{k'=0,\\k'\neq (s+1)(2n+1)}}\!\!\!\!\!\!\!\frac{2}{\left[(2k')^2-4(s+1)^2(2n+1)^2\right]}\nonumber\\
&= \frac{3}{2\left[4(s+1)^2(2n+1)^2\right]}-\frac{3}{2\left[4(s+1)^2(2n+1)^2\right]}=0\quad (\text{zero!})
\end{align}

Finally, we are ready to write the complete Fourier transform of the irradiance distribution due to the grids $I_g$: 
\begin{description}
\item[Case $L=(2s+1)l$:] All terms are zero but the constant (\ref{eq:constant}). Supposing the relations (\ref{eq:setting}) and (\ref{eq:setting-2}), imposed by equation (\ref{eq:TM2}) still applies, then 
\begin{equation}
\hat{I}_g(\eta_x)=\left[\frac{ \pi^2 (2s+1)^2}{4 \sqrt{2}  (s+1)p^2}\right] \left(\frac{\lambda}{d}\right)^2 N I_0 \delta(\eta_x).
\end{equation}

\item[Case $L=(2s)l$:] The full irradiance Fourier transform has two possible expressions. When $L=(2p+1)d^2/\lambda$,
\begin{multline}
\hat{I}_g(\eta_x)= \left[\frac{ 8\sqrt{2}\,\pi s}{(2s+1)}\right] \left(\frac{\lambda}{d}\right)^2 N I_0 \left\{ \left(\frac{\pi^2}{4}\right) \delta(\eta_x) \right.+\\
 + (-1)^{sp}\left(\frac{2s}{\sqrt{2}}\right) \left[\cos\frac{(s + p)\pi}{2}-\sin\frac{(s + p)\pi}{2}\right]\sum^{\infty}_{n=-\infty} \frac{1 }{(2n+1)^2}\,\delta\!\!\left[\eta_x+\frac{2s(2n+1)\pi \lambda}{d}\right] \\
 +(-1)^p \frac{1}{\pi (2s+1)^2}\left. \sum^{\infty}_{n=-\infty}\!\!\frac{(-1)^{n}}{(2n+1)^3}\;\delta\!\!\left[\eta_x+ \frac{2s(2n+1)\pi\lambda}{d} \right]\right\}.
\label{eq:firstnc}
\end{multline}
The other non-vanishing contribution has a simpler expression,
\begin{multline}
\hat{I}_g(\eta_x)= \left[\frac{ 8\sqrt{2}\,\pi s}{(2s+1)}\right] \left(\frac{d}{L}\right)^2 N I_0 \left\{ \left(\frac{\pi^2}{4}\right) \delta(\eta_x) \right.+\, (-1)^s(2s)\\
 \times\left.\sum^{\infty}_{n=-\infty} \frac{1 }{(2n+1)^2}\cos\!\!\left[\frac{(2n+1)^2(2s+1)\pi \lambda L}{ 4 d^2 }\right]\delta\!\!\left[\eta_x+\frac{2s(2n+1)\pi \lambda}{d}\right]\right\},\label{eq:secondnc}
\end{multline}
\end{description}
for any other $L$. Moreover, the above expression can be simplified by tuning $L$, we choose it in a similar fashion as in equation (\ref{eq:setting}):
\begin{equation}
L= \frac{2d^2}{\lambda}2p \quad\text{and}\quad l= \frac{2d^2}{\lambda}2q, 
\label{eq:maximal-relations}
\end{equation}
which maximizes all the series terms.

Now, we can recover the full irradiance distribution and calculate the visibility in each one of the exposed cases. Most of the situations will give us a constant intensity distribution, that is,
\begin{equation*}
I(\Vect)=\left\{
\begin{aligned}
(I_e+ I_g)(\Vect)&=\! \left[\frac{(2s+1)\pi}{(s+1)p}\right]^{\!2}\! \left(\frac{\lambda}{d}\right)^{\!\!2}\!\!\left[2N + \frac{(s+1)}{8\sqrt{\pi} }\right]N I_0,&\text{for}\;L&=(2s+1)l\\
I_e(\Vect)&=2\left[\frac{(2s+1)\pi}{(s+1)p}\right]^{\!2}\! \left(\frac{\lambda}{d}\right)^{\!\!2}N^2 I_0,&\text{for}\;L&\neq(2s)l.
\end{aligned}\right.
\end{equation*}
Thus, we obtain a non-constant irradiance distribution only with the condition $L=(2s)l$. We have seen there are two possible solutions: if $L=(2p+1)d^2/\lambda$ is
\begin{multline}
I_g(\Vect)= \left[\frac{ 8\sqrt{\pi} s}{(2s+1)}\right] \left(\frac{\lambda}{d}\right)^2 N I_0 \bigg\{ \frac{\pi^2}{4}+\\
+ (-1)^{sp}\left(\frac{2s}{\sqrt{2}}\right)\left[\cos\frac{(s + p)\pi}{2}-\sin\frac{(s + p)\pi}{2}\right] \sum^{\infty}_{n=0} \frac{1 }{(2n+1)^2} \, \cos\!\!\left[\frac{2s(2n+1)\pi }{d}R_x\right]\\
+ (-1)^p\frac{1}{\pi (2s+1)^2}\sum^{\infty}_{n=0}\frac{(-1)^{n}}{(2n+1)^3}\;\cos\!\!\left[\frac{2s(2n+1)\pi}{d} R_x\right]\bigg\},
\label{eq:firstt}
\end{multline}
or expressed in terms simple periodic functions as
\begin{multline*}
I_g(\Vect)= \left[\frac{ \pi^{5/2}2s }{(2s+1)}\right]\!\left(\frac{\lambda}{d}\right)^2\!\! N I_0\Bigg\{1+ (-1)^{sp}\left(\frac{2s}{\sqrt{2}}\right)\left[\cos\frac{(s + p)}{2}\pi-\sin\frac{(s+p )}{2}\pi\right]\times\\
\times\left(\frac{1}{2}-\abs{\frac{2sR_x}{d}}\right) + (-1)^p\frac{1}{2(2s+1)^2}\left(\frac{1}{4}-\abs{\frac{2sR_x}{d}}^2\right)\Bigg\},
\end{multline*}
for $-d/4s< R_x\leq d/4s$, and 
\begin{multline*}
I(\Vect)= \left[\frac{ \pi^{5/2}2s }{(2s+1)}\right]\!\left(\frac{\lambda}{d}\right)^2\!\! N I_0\Bigg\{1+ (-1)^{sp}\left(\frac{2s}{\sqrt{2}}\right)\left[\cos\frac{(s + p)\pi}{2}-\sin\frac{(s + p)\pi}{2}\right]\times\\
\times\left(\frac{1}{2}-\abs{\frac{2sR_x}{d}}\right) + (-1)^p\frac{1}{2(2s+1)^2}\left(\frac{1}{4}-\abs{1-\frac{2sR_x}{d}}^2\right)\Bigg\},
\end{multline*}
for $d/4s < R_x \leq d/2s$ or $  -d/2s < R_x\leq -d/4s$, and so on.  

\begin{figure}
\begin{center}
\includegraphics[width=0.80\textwidth]{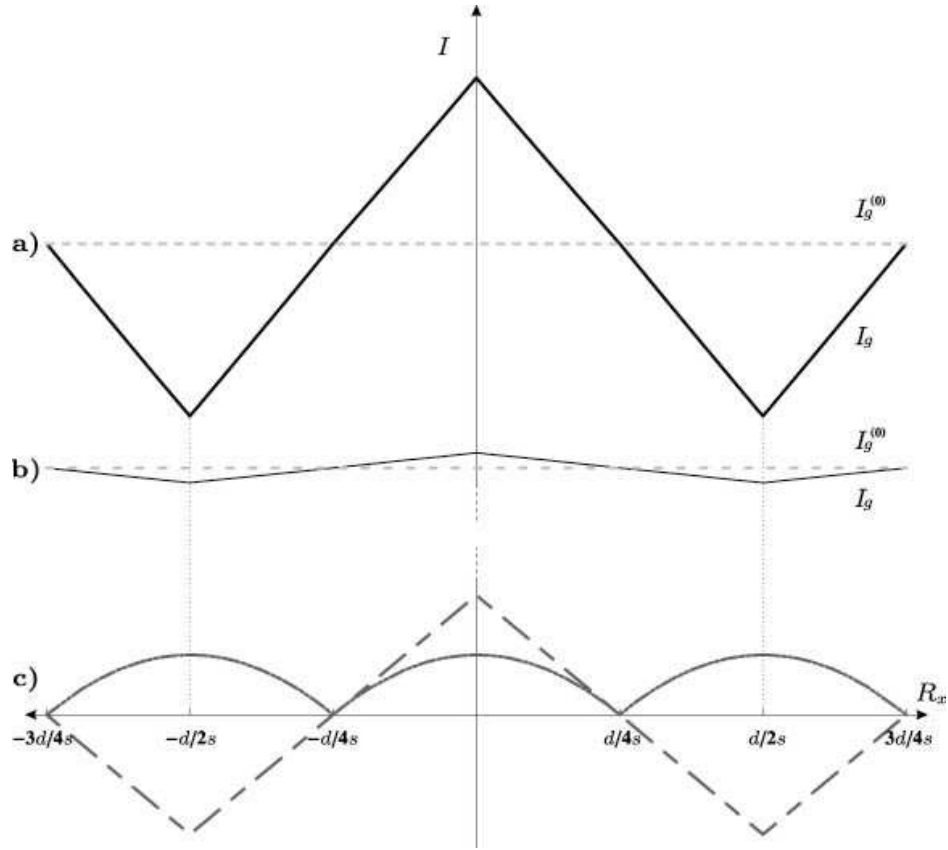}
\caption{When $L=(2s)l$ is satisfied we obtain non-constant irradiance patterns: \textbf{a)} The condition $L=(2p+1)d^2/\lambda$ gives an \textit{almost} triangular irradiance. \textbf{b)} A  triangular shape is found whenever $L= (2p)2d^2/\lambda$ and $l= (2q)2d^2/\lambda$, but the shape has less contrast---it is $1/16$ of the latter. \textbf{c)} The triangular shaped  and parabolic teethed functions contribute to \textbf{a)} but just the former to \textbf{b)}. The parabolic teethed function weights considerably less than the triangular so its contribution is almost negligible. 
\label{fig:irradiance-pattern}}
\end{center}
\end{figure}

On the other hand, when $L=(2s)l$ we choose the relations (\ref{eq:maximal-relations}) and thus
\begin{multline}
I_g(\Vect)= \left[\frac{ \sqrt{\pi} s}{(2s+1)2p^2}\right] \left(\frac{\lambda}{d}\right)^2 N I_0 \bigg\{ \frac{\pi^2}{4}+\\
+ (-1)^{s+p}\left(2s\right) \sum^{\infty}_{n=0} \frac{1 }{(2n+1)^2} \, \cos\!\!\left[\frac{2s(2n+1)\pi }{d}R_x\right]\bigg\}.
\label{eq:secondt}
\end{multline}
Again, we express it using periodic functions 
\begin{equation*}
I_g(\Vect)= \frac{\pi^{5/2}s}{8(2s+1)p^2}\!\left(\frac{\lambda}{d}\right)^2\!\! N I_0\left[1+ (-1)^{s+p}(2s)\left(\frac{1}{2}-\abs{\frac{2sR_x}{d}}\right)\right],
\end{equation*}
for $-d/2s< R_x\leq d/2s$. This irradiance pattern is a $16^\text{th}$ part of the latter (Figure \ref{fig:irradiance-pattern}).

The quality of an irradiance pattern---the contrast---produced by a system of grids is quantitatively measured the \textit{visibility} $\mathcal{V}$ defined by Michelson \citep{book:hech}:
\begin{equation}
{\mathcal V}=  \frac{I_\text{max}-I_\text{min}}{I_{\text{max}}+I_{\text{min}}},
\label{eq:visibility-eq:original}
\end{equation}
where $I_\text{max}$  and $I_\text{min}$ are two consecutive maximum and minimum. It is equal to zero for all distances but those described above. When $L$ is an even number of $d^2/\lambda$ we have
\begin{equation}
{\mathcal V}=  \frac{s}{\sqrt{2}\left[1 + \frac{(-1)^p}{2}(4s + 2)^{-2}\right]},
\label{eq:visibility-1}
\end{equation}
the with of the grid $d$, the wavelength $\lambda$ and the distance to the screen $l$ completely define $s$, and thus the visibility. Otherwise, when $L=(2p) 2d^2/\lambda$ the visibility is just
\begin{equation}
{\mathcal V}=  s.
\label{eq:visibility-2}
\end{equation}
Therefore, it only depends on the quotient between $L$ and $l$. The nearer the screen to the last grid the higher is the value of the visibility. 

\subsection{The turbulent case}

The statistical averages we will use here are understood as \textit{long exposure} time-averages \citep{paper:roddier}. Furthermore, the only relevant assumptions about the permitivity $\epsilon$ is being Gaussian process and markovian on the $z$-axis \citep{paper:tatarskii-2}. From equation (\ref{eq:melt}) and the definition (\ref{eq:inhomogeneous-term}) we must evaluate the following\footnote{
For any given Gaussian process $X$, we have $\langle\exp iX\rangle=\exp -\langle X^2 \rangle/2$.
}
\begin{multline}
\left< g\!\!\left(\Vect'-\frac{\vect'}{2},\vect,L\right) g^\ast\!\!\left(\Vect'+\frac{\vect'}{2},\vect,L\right)\right>=\frac{4\pi^2 L^2}{k^2} \int\!\!\!\int\mathcal{D}^2v_1(z)\,\mathcal{D}^2v_2(z)\\
\times\delta^{(2)}\!\left[\int^L_0\!\!dz\;\vel_1(z)\right]\delta^{(2)}\!\left[\int^L_0\!\!dz\;\vel_2(z)\right]\exp\left[\frac{i k}{2}\!\int^L_0\!\! dz \left(v^2_1(z)-v^2_2(z) \right)\right]\times\\
\times \exp -\frac{k^2}{8}\left<\left\{\int^L_0\!\!dz\; \epsilon\!\left[\frac{ z}{L}\vect+\frac{(L-z)}{L}\left(\Vect-\frac{\vect'}{2}\right)+\int^L_z\!\!d\eta\;\vel_1(\eta),z\right]+\right.\right.\\
\left.\left.-\epsilon\!\left[\frac{ z}{L}\vect+\frac{(L-z)}{L}\left(\Vect+\frac{\vect'}{2}\right)+\int^L_z\!\!d\eta\;\vel_2(\eta),z\right]\right\}^2\right>
\label{eq:mean-inohomo}
\end{multline}
Now, the averaged terms within the exponential can be rewritten using the markovian property, that is,
\begin{multline*}
\left<\epsilon\left({\textstyle \frac{z}{L}\vect+\frac{(L-z)}{L}\left(\Vect-\frac{\vect'}{2}\right)+\int^L_z\!\!d\eta\;\vel_1(\eta),z}\right) \epsilon\left({\textstyle \frac{z'}{L}\vect+\frac{(L-z')}{L}\left(\Vect-\frac{\vect'}{2}\right)+\int^L_{z'}\!\!d\eta\;\vel_1(\eta),z'}\right)\right>\\
=\left<\epsilon\left({\textstyle\frac{z}{L}\vect+\frac{(L-z)}{L}\left(\Vect+\frac{\vect'}{2}\right)+\int^L_z\!\!d\eta\;\vel_2(\eta),z}\right)\epsilon\left({\textstyle\frac{z'}{L}\vect+\frac{(L-z')}{L}\left(\Vect+\frac{\vect'}{2}\right)+\int^L_{z'}\!\!d\eta\;\vel_2(\eta),z'}\right)\right>\\
= \delta(z-z') A(0,z)
\end{multline*}
and 
\begin{multline*}
\left<\epsilon\left({\textstyle\frac{z}{L}\vect+\frac{(L-z)}{L}\left(\Vect-\frac{\vect'}{2}\right)+\int^L_z\!\!d\eta\;\vel_1(\eta),z}\right)\epsilon\left({\textstyle\frac{z'}{L}\vect+\frac{(L-z')}{L}\left(\Vect+\frac{\vect'}{2}\right)+\int^L_{z'}\!\!d\eta\;\vel_2(\eta),z'}\right)\right>\\
= \delta(z-z')\, A\!\left[\vect'\left(1-\frac{z}{L}\right)+\int^L_z\!d\eta \,\left(\vel_1(\eta)-\vel_2(\eta)\right),z\right]
\end{multline*}
where $A(\brho,z)$ is defined as in Appendix A, but with a $z$-axis dependence. Thus, we write the exponential term as
\begin{equation}
\exp - \frac{\pi k^2}{4}\int^L_0\!\!dz\, H\!\left[\vect'\left(1-\frac{z}{L}\right)+\int^L_z\!d\eta\,(\vel_2(\eta)-\vel_1(\eta)),z\right].
\label{eq:exponential}
\end{equation}
Therefore, we can introduce the linear change of variables: $\vel_1-\vel_2=\vel$  and $\vel_1+\vel_2=-\Vel$. Because $v^2_2(z)-v^2_1(z)=(\vel_2-\vel_1)\cdot(\vel_1+\vel_2)(z)=2\,\vel(z)\cdot\Vel(z)$, we turn (\ref{eq:mean-inohomo}) into
\begin{multline*}
\left< g\!\!\left(\Vect'-\frac{\vect'}{2},\vect,L\right) g^\ast\!\!\left(\Vect'+\frac{\vect'}{2},\vect,L\right)\right>\\
=\frac{4\pi^2 L^2}{k^2} \int\!\!\!\int\mathcal{D}^2v(z)\,\mathcal{D}^2V(z)\;\delta^{(2)}\!\left[\int^L_0\!\!dz\;\vel(z)\right]\delta^{(2)}\!\left[\int^L_0\!\!dz\;\Vel(z)\right]\times\\
\times \;\exp - \frac{\pi k^2}{4}\int^L_0\!\!dz\, H\!\left[\vect'\left(1-\frac{z}{L}\right)-\int^L_z\!d\eta\,\vel(\eta),z\right]\;\exp i k\!\int^L_0\!\! dz \,\left(\vel\cdot\Vel\right)(z).
\end{multline*}
We can group all the dependencies on $\Vel$ and integrate. As in the classical calculus it give us a delta function, that is,
\begin{equation}
\int\mathcal{D}^2V\;\exp i k \int^L_0\!\!dz\, \left(\vel\cdot\Vel\right)(z)= \delta_{\infty}(\vel)
\end{equation}
the  delta for \textit{functionals}. Moreover, when one of the extremes is fixed, as in our case, the path-integration of it is not equal to one:  
\begin{multline*}
\int\!\!\!\int\mathcal{D}^2v(z)\,\mathcal{D}^2V(z)\;\delta^{(2)}\!\left[\int^L_0\!\!dz\;\vel(z)\right]\delta^{(2)}\!\left[\int^L_0\!\!dz\;\Vel(z)\right]\exp i k\!\int^L_0\!\! dz \,\left(\vel\cdot\Vel\right)(z)\\
=\frac{1}{(4\pi^2)^2}\int_{\Re^2\times\Re^2}\!\!\!d^2\kappa\,d^2\kappa'\int\!\!\!\int\mathcal{D}^2v(z)\,\mathcal{D}^2V(z)\;\\
\times\exp -i\int^L_0\!\!dz\,\bigg[\Bkappa\cdot\vel(z)+\Bkappa'\cdot\Vel(z)- k\left(\vel\cdot\Vel\right)(z)\bigg]\\
=\frac{1}{(4\pi^2)^2}\int_{\Re^2\times\Re^2}\!\!\!d^2\kappa\,d^2\kappa'\lim_{N\rightarrow\infty} N_0^{N-1} \int_{\Re^2}\!\cdots\!\int_{\Re^2} \prod^{N-1}_{j=1}d^2v_j\,d^2V_j \\
\times\exp -i\sum^{N-1}_{j=0}\frac{L}{N}\bigg[\Bkappa\cdot\vel_j+\Bkappa'\cdot\Vel_j- k\,\vel_j\cdot\Vel_j\bigg]\\
=\frac{1}{(4\pi^2)^2}\int_{\Re^2\times\Re^2}\!\!\!d^2\kappa\,d^2\kappa'\lim_{N\rightarrow\infty} N_0^{N-1}\exp -i\frac{L}{N}\bigg[\Bkappa\cdot\vel_0+\Bkappa'\cdot\Vel_0- k\,\vel_0\cdot\Vel_0\bigg]\times\\
\times\left(\int_{\Re^2\times\Re^2}\!\! d^2v_1\,d^2V_1\,\exp -i\frac{L}{N}\bigg[\Bkappa\cdot\vel_1+\Bkappa'\cdot\Vel_1- k\,\vel_1\cdot\Vel_1\bigg]\right)^{N-1}.
\end{multline*}
The integral between parenthesis yields
\begin{equation*}
\int_{\Re^2\times\Re^2} d^2v_1\,d^2V_1\,\exp -i\frac{L}{N}\bigg[\Bkappa\cdot\vel_1+\Bkappa'\cdot\Vel_1- k\,\vel_1\cdot\Vel_1\bigg]= \frac{4\pi^2 L^2}{N^2} \exp -i \frac{L}{ k N} \Bkappa\cdot\Bkappa';
\end{equation*}
henceforth, $N_0\equiv N^2/4\pi^2 L^2$ and so it is
\begin{multline}
\int\!\!\!\int\mathcal{D}^2v(z)\,\mathcal{D}^2V(z)\;\delta^{(2)}\!\left[\int^L_0\!\!dz\;\vel(z)\right]\delta^{(2)}\!\left[\int^L_0\!\!dz\;\Vel(z)\right]\exp i k\!\int^L_0\!\! dz \,\left(\vel\cdot\Vel\right)(z)\\
=\frac{1}{(4\pi^2)^2}\int_{\Re^2\times\Re^2}\!\!\!d^2\kappa\,d^2 \exp -i \frac{L}{ k} \Bkappa\cdot\Bkappa'=\frac{k^2}{4\pi^2 L^2}.
\end{multline}
The latter property and the delta's definition allow equation (\ref{eq:mean-inohomo}) achieves its final form,
\begin{align}
\left< g\!\!\left(\Vect'-\frac{\vect'}{2},\vect,L\right) g^\ast\!\!\left(\Vect'+\frac{\vect'}{2},\vect,L\right)\right>&=\exp -\frac{\pi k^2}{4} \int^L_0\!\!dz\,H\!\left[\left(1-\frac{z}{L}\right)\vect',z\right]\nonumber\\
&=\exp -D(\vect',L)/2
\label{eq:structure-exponential}
\end{align}

We shall proceed to evaluate the mean irradiance
function.  We have shown the average adds a function dependant on the coordinate $\vect'$, then we arrive to an equation similar to (\ref{eq:noturb}) but with an extra term:
\begin{multline}
\langle I\rangle(\Vect) =\Cons \intDD\!\!\!d^2r\;\Tc{0}{\cdot\vect}\intDD\!\!\!d^{2}r'\; C_{t}(\vect')\;\exp -D(\vect',L)/2 \\
\times\exp \dot{\imath}k \left( \frac{\Vect}{l}+ \frac{\vect}{L} \right) \cdot\vect' .
\label{eq:turbulent}
\end{multline}
Its Fourier transform is now straightforward,
\begin{multline}
\widehat{\langle I \rangle}(\Bkappa)= \frac{1}{2\pi}\int_{\Re^2}\!\!d^2R\; \langle I\rangle(\Vect)\; \exp{-i\Bkappa\cdot\Vect} \\
=\Cons \int_{\Re^2}\!\!d^2R\intDD\!\!\!d^2r'\;\hat{t}_{{0_{{}_{{}_{{}_{\!\!\!\!\!\!\!\!\!\!\!D \times D}}}}}}\!\!\!\left(\frac{k}{L}\vect'\right) C_{t}(\vect')\;\exp -\frac{D(\vect',L)}{2} \exp \dot{\imath}  \left( \frac{k}{l}\vect'- \Bkappa \right) \cdot \Vect\\
=\cons\;\;\;\hat{t}_{{0_{{}_{{}_{{}_{\!\!\!\!\!\!\!\!\!\!\!D \times D}}}}}}\!\!\!\left(-\frac{l}{L}\Bkappa\right) C_{t}\left( \frac{l}{k} \Bkappa \right)\exp -\frac{D\left(\frac{l}{k}\Bkappa,L\right)}{2}\\
\times\Heavy{ + \frac{l}{k}\, \Bkappa} \Heavy{- \frac{l}{k}\, \Bkappa}.
\label{eq:last}
\end{multline}
Whether it is the case of equation (\ref{eq:firstnc}) or (\ref{eq:secondnc}) the exponential contribute to each term of them with
\begin{equation*}
\exp - D\!\big(Q(2n+1)\, d\, \eX,L\big)/2,
\end{equation*}
here $Q$ is an integer satisfying one of the conditions we have given. Assuming the structure constant $C^{2}_{\epsilon}(z)$ is roughly homogeneous we can write
\begin{equation}
D(r\eX,L)= \frac{\pi\, \Gamma(\mu+1)}{4\,\Gamma\!\left[(\mu+3)/2\right]^2}\sin\frac{\pi \mu}{2}\, k^2C^{2}_{\epsilon} L\, r^{^\mu+1}=D_\mu\, r^{\mu+1}.
\end{equation}
Henceforth, for $L=(2s) l$ is
\begin{multline}
\langle I_g\rangle(\Vect)= \left[\frac{ 8\sqrt{\pi} s}{(2s+1)}\right] \left(\frac{\lambda}{d}\right)^2 N I_0 \times\\
\times\bigg\{ \frac{\pi^2}{4}+ (-1)^{sp}\left(\frac{2s}{\sqrt{2}}\right)\left[\cos\frac{(s + p)\pi}{2}-\sin\frac{(s + p)\pi}{2}\right] \times\\
\times\sum^{\infty}_{n=0} \frac{\exp\! \left[- (D_\mu/2) Q^{\mu+1}(2n+1)^{\mu+1}d^{\mu+1}\right]}{(2n+1)^2} \, \cos\!\!\left[\frac{2s(2n+1)\pi }{d}R_x\right] + (-1)^p\frac{1}{\pi (2s+1)^2}\times\\
\times\sum^{\infty}_{n=0}\frac{(-1)^{n}\exp\! \left[- (D_\mu/2) Q^{\mu+1}(2n+1)^{\mu+1}d^{\mu+1}\right]}{(2n+1)^3}\;\cos\!\!\left[\frac{2s(2n+1)\pi}{d} R_x\right]\bigg\},
\label{eq:firstt-turbulent}
\end{multline}
with $Q=(2p+1)$ when $L=(2p+1)d^2/\lambda$, and when the relations (\ref{eq:maximal-relations}) are satisfied
\begin{multline}
I_g(\Vect)= \left[\frac{ \sqrt{\pi} s}{(2s+1)2p^2}\right] \left(\frac{\lambda}{d}\right)^2 N I_0 \times\\
\times \left\{ \frac{\pi^2}{4}+ (-1)^{s+p}\left(2s\right) \sum^{\infty}_{n=0} \frac{\exp\! \left[- (D_\mu/2) Q^{\mu+1}(2n+1)^{\mu+1}d^{\mu+1}\right] }{(2n+1)^2} \, \cos\!\!\left[\frac{2s(2n+1)\pi }{d}R_x\right]\right\},
\label{eq:secondt-turbulent}
\end{multline}
with $Q=4p$. 
\begin{figure}
\begin{center}
\includegraphics[width=0.7\textwidth]{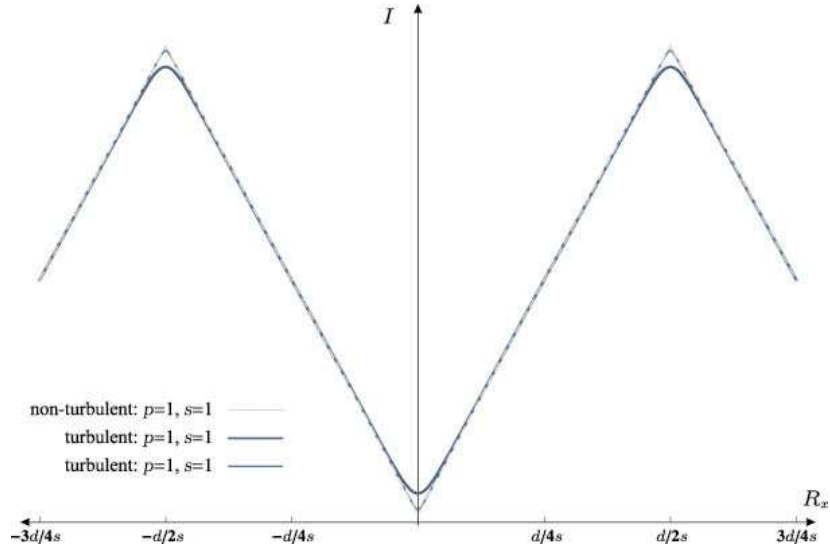}
\caption{The figure compares the irradiance patterns for two very different wavelengths, $400$nm (soft-ultraviolet) and $1.2\mu$m (red), given a fixed geometric configuration: $L= 0.976$m and $d=0.625\times 10^{-3}$m. The red-wavelength ($p=1$ to reach the distance $L$) function has been mirrored to compare against the other two.
\label{fig:comp-turbulent-irradiance-pattern}}
\end{center}
\end{figure}

Thus, only a finite number of terms contribute significantly to the
image formation. The exponential term in both series, (\ref{eq:firstt-turbulent}) and (\ref{eq:secondt-turbulent}), plays the role of a cutoff smoothing the original irradiance pattern. The integer $N \sim (10/D_\mu)^{\frac{1}{\mu+1}} (2Q d)^{-1}$ is a good measure of this cutoff---terms beyond that number add corrections of order less than $10^{-5}$ to the actual value. Also, it makes the irradiance extremely sensitive to changes in the wavelength and the structure constant. Figure \ref{fig:comp-turbulent-irradiance-pattern} shows the difference between the patterns generated by infrared and ultraviolet wavelengths for the same geometric arrangement. 

Afterwards, we can estimate the visibility. The visibility in the turbulent case is smaller than in the non-turbulent one because of the cutoff, and it turns smaller as the wavelength decreases: for $L=(2p+1)d^2/\lambda$,
\begin{equation*}
\mathcal{V}=0.69742,\quad\mathcal{V}_{1.2\mu\text{m}}=0.67865,\quad\text{and}\quad\mathcal{V}_{400\text{nm}}=0.63036;
\end{equation*}
and for $L=(2p) 2d^2/\lambda$
\begin{equation*}
\mathcal{V}=1,\quad\mathcal{V}_{1.2\mu\text{m}}=0.88163\quad\text{and}\quad\mathcal{V}_{400\text{nm}}=0.66679.
\end{equation*}
Amazingly, it is the second irradiance distribution pattern (\ref{eq:secondt-turbulent}), which has a flattened pattern, more sensitive to changes in the wavelength and turbulence behavior against what their patterns suggest.

\begin{figure}
\begin{center}
\includegraphics[width=0.80\textwidth]{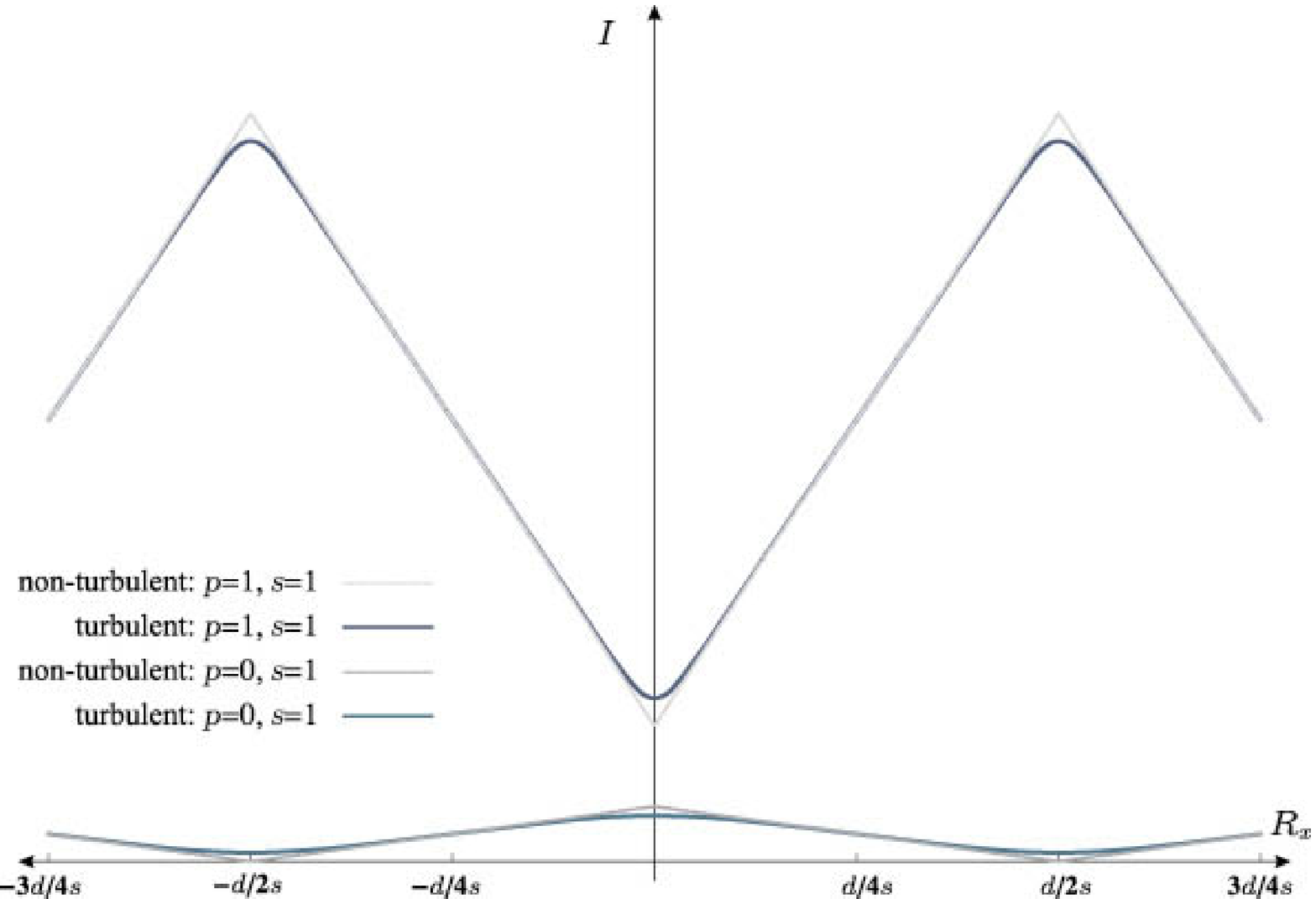}
\vspace{2em}
\includegraphics[width=0.80\textwidth]{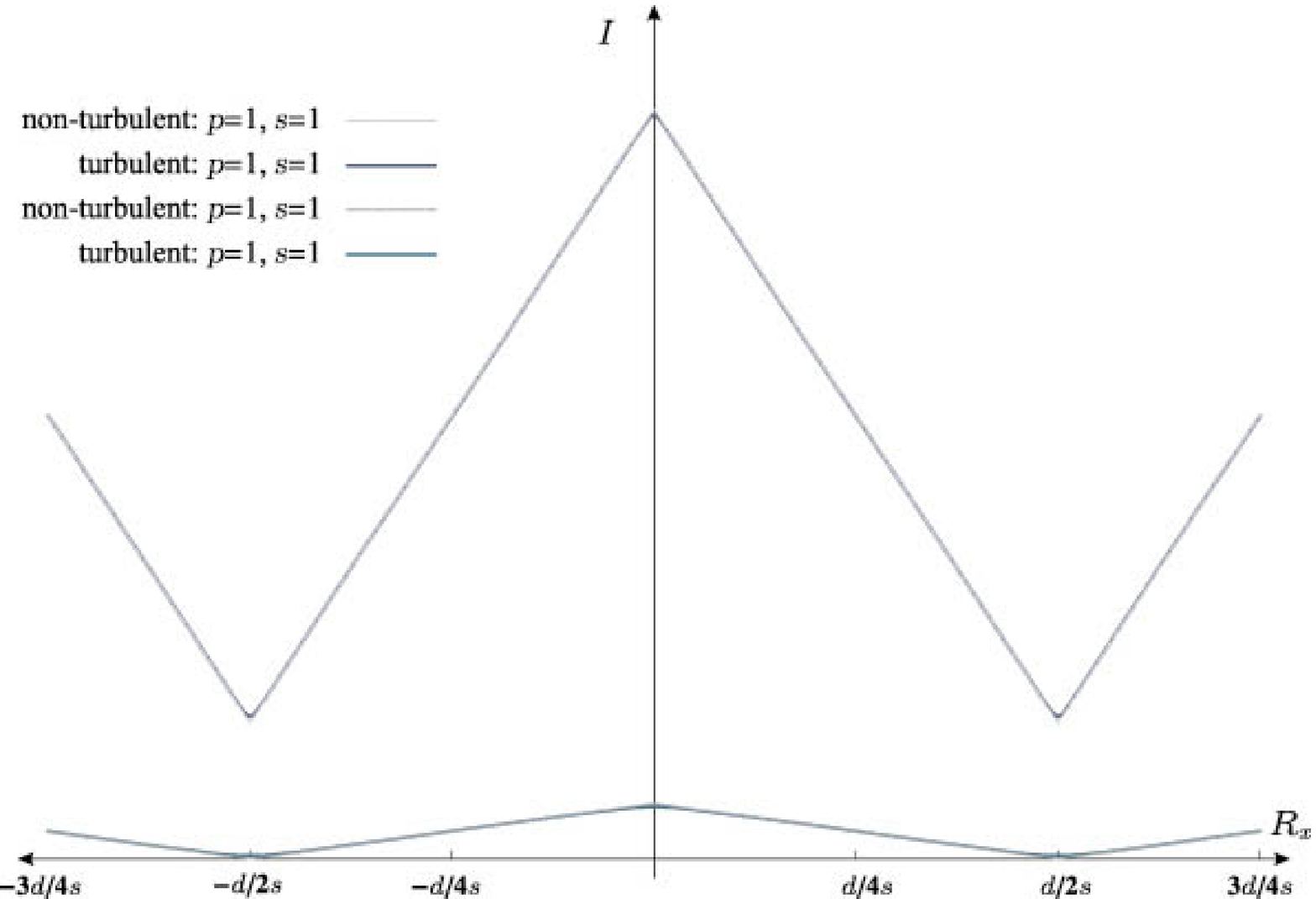}
\caption{The first graphic displays the irradiance patterns for a $\lambda=400$nm wavelength and the second for $\lambda=1.2\mu$m. The degradation is clearly observed in the first example, but hardly can be seen in the red wave length.
\label{fig:turbulent-irradiance-pattern}}
\end{center}
\end{figure}

The cutoff also depends on the geometry of the system. Two instances are relevant; as $d \rightarrow \infty$ the visibility goes to zero, otherwise if $d \rightarrow 0$ it takes the same value as in the non-turbulent case---equations (\ref{eq:visibility-1}) and (\ref{eq:visibility-2}). These results show us how the geometry influences  image formation in a turbulent media. The behavior of the visibility is in agreement with the results of \citet{paper:zavorotny} for an infinitely extended source as it vanishes when $d$ goes to infinity. Moreover, if $d$ is small enough the effects of the turbulent medium vanish and the visibility recovers the value it takes in the absence of turbulence.

Finally, here we have established the conditions for image formation in a Lau-like arrangement. 
For a visibility different from zero, the separation between grids, $L$, and the distance from the last of them to the screen, $l$, are related by the condition (\ref{eq:pairrel}). We observe the appearance of a characteristic length $d^2 / \lambda$, it is called Talbot distance and is widely present in grids systems. Only on integer multipliers of it we have found a non-zero visibility. In these situations we were able to express the degradation in terms of a few variables: the physical $C^{2}_{\epsilon}$ and $\lambda$, and the geometrical $L$ and $d$.

Also, the mean irradiance is exact: either it is useful in both strong and weak regimes. Equations like (\ref{eq:firstt-turbulent}) and (\ref{eq:secondt-turbulent}) provide us with a new way to calculate the structure constant of the medium at laboratory from a density section of an image. Indoor experiments carried out with laser beams through turbulent medium \citep{paper:consortini-1,paper:consortini-2} are based in measures of their wander and thus an statistical analysis. While ours just needs an interpolating Fourier polynomial. 

We have given an introduction to the classical methods in turbulent propagation based on a markovian model. In the forthcoming chapters we will introduce processes with memories to accurately resemble the model we introduced in the first chapter.
\chapter{Stochastic Calculus}
\label{chp:stochastic-calculus}

We have shown that defined the turbulent refractive index as a member of the family of fractional Brownian motions it is not differentiable. Furthermore, we usually find in Optics  derivatives of the refractive index within differential equations, but when the media is turbulent these equations are undefined in terms of the Classical Calculus.

For instance, let us suppose it is possible to define the derivative of a fractional or standard Brownian motion, the \textit{noise}: $\dot{B}^H$. Thus, the integral equation associated to $\dot{x}(t) = \dot{B}^H\!(t)\, x(t)$ is just
\begin{equation*}
x(t)=x_0 + \int^t_0 x(s)\,\dot{B}^H\!(s)ds\equiv x_0 + \int^t_0 x(s)\,d B^H\!(s),
\end{equation*}
for the last term above we have assumed that the change of variable formula is still valid. So, it is the existence of this kind of integrals what we should try to verify. If we attempt to define this integral as the limit of the Riemann series,
\begin{equation}
\sum^{n-1}_{i=0} x(t_i^n)\left(B^H(t_{i+1}^n)-B^H(t_i^n)\right),
\label{eq:riemann-stoch}\end{equation}
its existence can not be proven in general. 

Nevertheless, conditions over the argument function $x(t)$ for the existence of this type of integrals are now well established, and a Stochastic Calculus can be build from it. This calculus and how it can be used to solve stochastic differential equations will be described next.

\section{Introduction: White Noise and Brownian Motion}

In this section we will present the stochastic analysis for the standard Brownian motion, and set the notions that will be later extended to the more general fractional Brownian case.  

The theory of distributions had provided us with \textit{derivatives} for functions without them in the classical sense. Therefore, it is natural to propose the white noise as a distribution, but to do so we must also give the right abstract probability space. It was \citet{book:hida} who first used this idea as the building block for a stochastic analysis. Here we are going to build such  a space and show how it allows define integrals in the sense of (\ref{eq:riemann-stoch}).

Let $\mathcal{S}(\Re^d)$ be the Schwartz space of rapidly decreasing smooth ($C^{\infty}(\Re^d)$) real valued functions on $\Re^d$, and let us choose its dual $\mathcal{S}^{\ast}\!(\Re^d)$---the space of tempered distributions---as the probability space $\Omega$.  We represent with $\langle \omega,\phi\rangle=\omega(\phi)$  the action of the elements of the dual, $\omega\in \mathcal{S}^{\ast}\!(\Re^d)$, on the functions belonging to $\mathcal{S}(\Re^d)$. 

Of course, to properly define the probability space we have to attach a $\sigma$-algebra and a probability measure. The former is straightforward, we just use the family of Borel subsets $\mathcal{B}(\mathcal{S}(\Re^d))$, and associated to this algebra we need to prove the existence of a measure. The \textit{Bochner-Minlos theorem} \citep[for a proof see][Appendix A]{book:holden} shows that such a measure, $\mu$, exists; moreover, it has the following property: for all $\phi\in\mathcal{S}(\Re^d)$,
\begin{equation}
\E{\exp i \langle \cdot, \phi\rangle}:= \int_{ \mathcal{S}^{\ast}\!(\Re^d)}\!\!\!d\mu\, \exp i\langle\omega,\phi\rangle= \exp -\frac{1}{2}\norm{\phi}^2,
\end{equation}
where $\norm{\cdot}$ is the norm in $L^2(\Re^d)$. Therefore, we call the triplet $(\Omega,\mathcal{B}(\Omega),\mu)$ the \textit{1-dimensional white noise probability space}.

The probability measure is a Gaussian measure on $\mathcal{S}(\Re^d)$: we just have  to evaluate the finite dimensional measures. So, let us take a set of functions $\xi_1,\cdots,\xi_n\in\mathcal{S}(\Re^d)$ such that they are orthonormal in $L^2(\Re^d)$. Now, given a random variable $\omega$, we can project it into the finite random variable $(\langle\omega,\xi_1\rangle,\langle\omega,\xi_2\rangle,\dots,\langle\omega,\xi_n\rangle).$ For any  smooth function $f\in C^{\infty}(\Re^n)$ we have, 
\begin{align}
\E{f(\langle\cdot,\xi_1\rangle,\dots,\langle\cdot,\xi_n\rangle)}=& \frac{1}{\sqrt{(2\pi)^n}}\int_{\Re^n}\!\! d^n\! k\,\hat{f}(\wn)\, \E{e^{ i\langle\,\cdot\,,\sum_{\alpha} k_{\alpha} \xi_{\alpha}\rangle}}\nonumber \\
=& \frac{1}{\sqrt{(2\pi)^n}}\int_{\Re^n}\!\!d^n\! k\, \hat{f}(\wn)\, e^{-\frac{1}{2}\norm{\wn}^2} \nonumber\\
=& \frac{1}{(2\pi)^n}\int_{\Re^n}\!\!d^n x\, f(\textbf{x})\left[\int_{\Re^n}\!\!d^n\! k\, \exp\left(i\, \wn\cdot \textbf{x}-\frac{1}{2}\norm{\wn}^2\right)\right]\nonumber\\
=& \frac{1}{\sqrt{(2\pi)^n}}\int_{\Re^n}\!\!d^n x f(\textbf{x})\, e^{-\frac{1}{2}\norm{\textbf{x}}^2}
\label{eq:fdiminsional}
\end{align}
---we used above the properties of the Fourier transforms. Thus, we have found the $n$-dimensional Gaussian measure 
\begin{equation}
d \lambda^n(\textbf{x})=(2\pi)^{-n/2} e^{-\frac{1}{2}\norm{\textbf{x}}^2}dx_1\cdots dx_n.
\end{equation}

With the same procedure we can prove that if $\phi\in L^2(\Re^d)$ for any succession $\phi_n\in\mathcal{S}(\Re^d)$ such that $\phi_n\rightarrow\phi$ in $L^2(\Re^d)$, then $\exists \lim_{n\rightarrow\infty}\langle\omega,\phi_n\rangle:=\langle\omega,\phi\rangle$ in $L^2(\mu)$.

Let us introduce now the \textit{1-dimensional (d-parameter) smoothed white noise}. It is a map $\mathit{w}:\mathcal{S}(\Re^d)\times\mathcal{S}^{\ast}(\Re^d)\rightarrow \Re$ given by
\begin{equation}
w(\phi)=w(\phi,\omega)=\langle\omega,\phi\rangle;\quad \omega\in\mathcal{S}^{\ast}(\Re^d), \phi\in\mathcal{S}(\Re^d).
\label{eq:smoothed-wn}
\end{equation}

Now, we define the following process 
\begin{equation}
\tilde{B}(\textbf{x}):=\tilde{B}(x_1,\dots,x_d,\omega)=\langle\omega, \chi_{[0,x_1] \times\cdots\times [0,x_d]}\rangle, 
\label{def:Bm}\end{equation}
for $\textbf{x}=(x_1,\dots,x_d)\in \Re^d$, where $\chi$ is the index function: gives $1$ when $\textbf{x}$ is inside the box $[0,x_1] \times\cdots\times [0,x_d]$ and zero otherwise---when $x_i<0$ it is convention to assume $[0,x_i]$ represents $[x_i,0]$. This process has a continous version which turns to be a $d$-parameter Brownian motion.

It is evident from definition (\ref{def:Bm}) that this process is almost surely zero at $\textbf{x}=0$. Also, the process satisfy definitions (\ref{eq:mean-zero}) and (\ref{eq:Bm-covariance}) in their $d$-dimensional equivalent form, that is,
\begin{equation}
\E{\tilde{B}(\textbf{x})}=0\quad\text{ and }\quad\E{\tilde{B}(\textbf{x})\tilde{B}(\textbf{y})}= \prod^d_{i=1} \min\{x_i,y_i\}.
\label{eq:Bm-mean-properties}
\end{equation}
Checking these properties is straightforward, we choose $\textbf{x}^{(1)},\dots, \textbf{x}^{(n)}\in\Re^d$, and constants $c_1,\dots,c_n\in\Re$, so we build the index functions: $\chi^{(i)}=\chi^{(i)}_{[0,x_1] \times\cdots\times [0,x_d]}$; therefore, we compute the $n$-dimensional characteristic function,
\begin{align*}
\Eb{\exp\left[i\sum^n_{i=1} c_i \tilde{B}(\text{x}^{(i)})\right]}&=\Eb{\exp\left[i\langle\cdot, \sum^n_{i=i} c_i \chi^{(i)} \rangle\right]}\\
&=\exp\left(-\frac{1}{2}\norm{\sum^n_{i=1}c_i \chi^{(i)}}^2\right)\\
&=\exp\left(-\frac{1}{2}\sum^n_{i,j=1} c_i c_j \int_{\Re^d} \chi^{(i)}\chi^{(j)}d^d x\right)\\
&=\exp\left(-\frac{1}{2}\textbf{c}^T\textbf{V}\textbf{c}\right),
\end{align*}
where $\textbf{c}=(c_1,\dots,c_n)$ and $\textbf{V}$ is the symmetric nonnegative definite matrix defined by
\begin{equation*}
V_{i,j}=\int_{\Re^d} \chi^{(i)}(\textbf{x})\chi^{(j)}(\textbf{x})\,d^d x.
\end{equation*}
Therefore, $\tilde{B}$ is Gaussian with mean zero and covariance matrix given by $\textbf{V}$. It is better now, instead of directly evaluate the covariance, calculate the variance of its increments. So, making use of (\ref{eq:fdiminsional}):
\begin{align*}
\E{(\tilde{B}(\textbf{x})-\tilde{B}(\textbf{y}))^2}&= \E{\langle\cdot,\chi_{[0,\textbf{x}]}-\chi_{[0,\textbf{y}]}\rangle^2}\\
&= \norm{\chi_{[0,\textbf{x}]}-\chi_{[0,\textbf{y}]}}^2\,\E{\langle\cdot,\frac{\chi_{[0,\textbf{x}]}-\chi_{[0,\textbf{y}]}}{\norm{\chi_{[0,\textbf{x}]}-\chi_{[0,\textbf{y}]}}^2}\rangle^2}\\
&=\norm{\chi_{[0,\textbf{x}]}-\chi_{[0,\textbf{y}]}}^2\int_{\Re} u^2 d\lambda^1(u)\\
&=\norm{\chi_{[0,\textbf{x}]}-\chi_{[0,\textbf{y}]}}^2=\prod^d_{i=1} \abs{x_i-y_i}
\end{align*}
where $[0,\textbf{x}]=[0,x_1]\times\dots\times[0,x_d]$. Thus, the Brownian motion covariance follows from the variance we have found. Finally, the continous extension to the process comes from the application of the well-known Kolmogorov's continuity theorem, and makes the continous version $B(\textbf{x})$ the desired $d$-parameter Brownian motion.

With this definition of Brownian motion we can define the Wiener-It\^o integrals. We will simplify the following exposition setting $d=1$. Let $\phi\in L^2(\Re)$ be \textit{deterministic} with finite support set, let us say the interval $[a,b]$. Now, we build the succession:
\begin{equation}
\phi_n(t)= \sum^n_{i=1} \phi(t_i)\, \chi_{[t_i,t_{i+1})}(t),
\end{equation}
where $a\leq t_1<\dots<t_n<t_{n+1}=b$ is a partition such that $\max\abs{t_{i+1}-t_i}\rightarrow 0$ as $n\rightarrow\infty$. This family of functions belongs to $L^2(\Re)$ and converges to $\phi$ there. The requirement for $\{\phi_n\}_{n\in\N}$ being in $\mathcal{S}(\Re)$ is found making the edges of the function smooth in a neighborhood of the interval and approach to a step function as $n$ grows. Let us omit that step to simplify the exposition, therefore,
\begin{equation*}
\langle\omega,\phi_n\rangle= \sum^n_{i=1} \phi(t_i)\, (B(t_{i+1})-B(t_i))\xrightarrow{n\rightarrow\infty}\langle\omega,\phi\rangle
\end{equation*}
in $L^2(\mu)$---in mean square\footnote{
See footnote \ref{foot:conv} at Chapter \ref{chp:index}.
}. Thus, we can put
\begin{equation}
\int_{\Re} \phi(t)\, dB(t,\omega):=\langle\omega,\phi\rangle;\quad \omega\in\mathcal{S}^{\ast}(\Re),
\end{equation}

The same arguments can be used with the $d$-parameter Brownian motion to define the stochastic integral in the same way. Moreover, we can integrate by parts---provided the pathwise integral coincides with the $L^2$-stochastic integral---and get
\begin{equation*}
\begin{aligned}
w(\phi)=\int_{\Re^d} \phi(x)\, dB(x,\omega) =& (-1)^d \int_{\Re^d} \frac{\partial^d\phi}{\partial x_1\cdots\partial x_d }(x) B(x,\omega)\, d^dx\\
=& \left(B, (-1)^d\frac{\partial^d\phi}{\partial x_1\cdots\partial x_d }\right)=\left(\frac{\partial^d B}{\partial x_1\cdots\partial x_d}, \phi\right),
\end{aligned}
\end{equation*}
$(\cdot,\cdot)$ is the inner product in $L^2(\Re^d)$; thus, in the sense of distributions we write
\begin{equation}
w = \frac{\partial^d B}{\partial x_1\cdots\partial x_d}.
\end{equation}

Now, we would like to replace the deterministic function $\phi$ be a stochastic process $f(\omega,t)$. For most applications is enough to prove this replacement is possible for a closed set, say $T=[0,1]$, and we will do so. The extension, known as It\^o integral, is possible whenever the process has the following properties:
\begin{itemize}
\item[\textit{i)}] Given the set $\mathcal{F}_t=\{B(s): 0<s\leq t \}$, then $f$ is $\mathcal{F}_t$-measurable for any $t$ ($\mathcal{F}_0:=\{\Omega,\vacio\}$).

\item[\textit{ii)}] the map $(\omega,t)\rightarrow f(\omega,t)$ is  $\mathcal{B}(\Re)\times\mathcal{F}_T$-mensurable.

\item[\textit{iii)}] $\E{\int_T f^2(\omega,t) dt }< \infty$.
\end{itemize}
This conditions are enough to guarantee the existence of the following limit, \textit{It\^o integral},
\begin{equation}
\int_T f(\omega,t)\, dB(t):=\lim_{n\rightarrow\infty}\sum^{2^n-1}_{j=1} f(\omega, j 2^{-n})\left[B((j+1)2^{-n})-B(j 2^{-n})\right]\quad \text{in } L^2(\mu).
\label{eq:ito}\end{equation}
The choice of the step function $\sum^{2^n-1}_{j=1} f(\omega, j 2^{-n})\chi_{(j 2^{-n},(j+1) 2^{-n}]}$, which also converges in $L^2(\Omega\times\Re)$ to $f$, is crucial here. For it not only assures the limit in mean square but provides the \textit{isometry property},
\begin{equation}
\E{\abs{\int_T f(\omega,t)\, dB(\omega,t)}^2}=\E{\int_T f(\omega,t)^2 dt},
\end{equation}
and also
\begin{equation}
\E{\int_T f(\omega,t)\, dB(t)}=0.
\end{equation}
Alternatively it can be proved \citep{book:nualart} that the step function process
\begin{equation*}
\sum^{2^n-1}_{j=1} 2^n \left(\int^{j 2^{-n}}_{(j-1) 2^{-n}} f(\omega, s) ds\right)\chi_{(j 2^{-n},(j+1) 2^{-n}]}
\end{equation*}
is also $\mathcal{F}_{j 2^{-n}}$-adapted, converges to $f$ and gives the same limit integral (\ref{eq:ito}) with the properties above. Other approximations to the process can be built, but they do not obey the latter properties. 

No calculus can be built without a change-of-variable formula: the It\^o integrals provides one. Let $F:\Re\rightarrow\Re$ be a smooth function (or at least twice continuously differentiable). Also, suppose that $u$ and $v$ are measurable adapted processes such that $\int^t_0 u^2 ds <\infty$ and $\int^t_0 \abs{v} ds<\infty$ almost surely for every $t\in T$. For 
\begin{equation}
X(t)= X_0 + \int^t_0 u(s)\, dB(s) + \int^t_0 v(s) ds,
\label{def:martingale}\end{equation}
we have
\begin{equation}
\begin{aligned}
F(X(t))- F(X_0) &= \int^t_0 F'(X(s))\, u(s)\, dB(s) + \int^t_0 F'(X(s))\, v(s)\, ds\\
&+ \frac{1}{2}\int^t_0 F''(X(s))\, u^2(s)\, ds.
\end{aligned}
\end{equation}

This formula was obtained using the approximation by step functions we have previously commented. We may try guessing what happens if the point $t$, where we evaluate $f$ to build the former succession, is selected in a different way. For example, let us take the process (\ref{def:martingale}) and a partition $\pi_n=\{0=t_0<t_1<\dots<t_n=t\}$ of the interval $[0,t]$. The sums
\begin{equation*}
\sum^{n-1}_{j=0}\frac{1}{2}\left[X(t_j)+X(t_{j+1})\right]\left[B(t_{j+1})-B(t_j)\right]
\end{equation*}
converge to 
\begin{equation}
\int^t_0 X(s)\,dB(s) + \frac{1}{2}\int^t_0 u(s)\, ds.
\label{eq:strat-1}\end{equation}
This limit integral is called \textit{Stratonovich integral}. Now, comparing against the process $X(t)$ itself the second term in this integral looks like a derivative in the sense, `$d X(s)/dB(s)$'. So we could write it as
\begin{equation*}
\int^t_0 X(s)\,dB(s) + \frac{1}{2}\int^t_0 \frac{d X(s)}{dB(s)}\, ds.
\end{equation*}
Therefore, our next question is: can such an operator  be defined formally? The answer is yes. It appears when one tries to define the It\^o-Wiener integral for non-adapted processes. That is, let $F(\omega): \mathcal{S}^{\ast}(\Re)\rightarrow \Re$ be a process such that
\begin{equation}
F(\omega) =f(\langle\omega,\phi_1\rangle,\dots,\langle\omega,\phi_n\rangle),
\end{equation}
where $f\in C^{\infty}(\Re^n)$ has partial derivatives with polynomial growing, and the functions $\phi_1,\dots,\phi_n \in \mathcal{S}(\Re)$ are fixed. Thus, we define the Fr\'echet derivative, also known as Malliavin derivative, of $F$ as 
\begin{equation}
\begin{aligned}
 D_{\phi}F(\omega)=\lim_{\varepsilon\rightarrow 0} \frac{1}{\varepsilon}\left[f(w(\phi_1) +\varepsilon (\phi_1,\phi),\dots, w(\phi_n) +\varepsilon (\phi_n,\phi))\right.\\
 -\left.f(w(\phi_1),\dots,w(\phi_n))\right];
\end{aligned}
\label{eq:malliavin}
\end{equation}
moreover, if there exists a process $D_t F$ such that $D_{\phi}F= (D_.F,\phi)$---where $(\cdot,\cdot)$  is again the inner product in $L^2(\Re)$ or $L^2(\Re^d)$---we say it is differentiable. For $f=\text{id}: \Re\rightarrow\Re$ is
\begin{equation*}
 D_{\phi}\!\left[\int_{\Re} \phi_1(s)\, dB(s)\right]= (\phi_1,\phi),
\end{equation*}
and thus is $D_t\, w(\phi_1)=\phi_1(t)$. In general, the derivative is just the expression:
\begin{equation}
D_t F= \sum^n_{i=1} \frac{\partial f}{\partial x_i}(w(\phi_1),\dots,w(\phi_n)) \phi_i(t).
\end{equation}

This operator is closed and unbounded with values in $L^2(\Re\times\Omega)$ defined on the (dense) set $\mathbb{D}^{1,2}$ of smooth random variables with norm,
\begin{equation*}
\norm{F}^2_{1,2} =\E{\abs{F}^2}+ \E{\norm{D_{\cdot}F}^2_{L^2(\Re)}},
\end{equation*}
contained in $L^2(\Omega)$. We define the adjoint operator $\delta$ as an unbounded operator on $L^2(\Re\times\Omega)$ with values in $L^2(\Omega)$ such that:
\begin{itemize}
\item[\textit{i)}] Its domain, denoted by $\text{Dom}\,\delta$, is the set of processes $X\in L^2(\Re\times \Omega)$ with
\begin{equation*}
\abs{\E{\int_{\Re} D_t F\, X(t) dt}}\leq c \norm{F}_{1,2},
\end{equation*}
for all $F\in\mathbb{D}^{1,2}$, where $c$ is some constant depending on $X$.

\item[\textit{ii)}] If $X$ belongs to $\text{Dom}\,\delta$, then $\delta(X)$ is the element of $L^2(\Omega)$ characterized by 
\begin{equation}
\E{F\delta(X)}=\E{\int_{\Re} D_t F\, X(t) dt},\quad \forall F\in\mathbb{D}^{1,2}.
\end{equation}
\end{itemize}
This operator is called \textit{Skorohod stochastic integral} of the process $X$. It transforms square integrable processes into random variables. It is usually written as
\begin{equation}
\delta(X):= \int_{\Re} X(t)\,\delta B(t).
\end{equation}
This stochastic integral does not require adaptness for $X$; nevertheless, if it is adapted then it coincides with the It\^o integral. Moreover, The Skorohod integral is the right tool to understand stochastic integrals defined by Riemann sums.

Again, let us assume our parameter space is $T=[0,1]$. It is denoted by $\mathbb{L}_{1,2}$ the class of processes $X\in L^2(T\times\Omega)$ such that $X(t)\in \mathbb{D}_{1,2}$ for all $t$, and there exists a measurable version of the two-parameter process $D_s X(t)$ satisfying
\begin{equation*}
\E{\int_T\int_T (D_s X(t))^2\, ds\, dt}<\infty.
\end{equation*}
This space is a Hilbert space with norm $\norm{X}^2_{1,2}=\norm{X}^2_{L^2(T\times\Omega)}+\norm{DX}^2_{L^2(T^2\times\Omega)}$. It follows that $\mathbb{L}_{1,2}\subset \text{Dom}\,\delta$.

Now, for any process $X\in L^2(T\times\Omega)$ and any partition $\pi=\{t_0=0<t_1\leq \dots\leq t_{n-1}<t_n=1\}$ the step process 
\begin{equation*}
X^{\pi}(t)=\sum^{n-1}_{i=0}\frac{1}{t_{i+1}-t_i}\left(\int^{t_{i+1}}_{t_i}X(s)\,ds\right)\chi_{(t_i,t_{i+1}]}(t)
\end{equation*}
converges to the process $X$ in the norm of the space $L^2(T\times\Omega)$ as $\abs{\pi}=\text{max}_{i}\abs{t_{i+1}-t_i}$ tends to zero. Furthermore, it also holds in $\mathbb{L}_{1,2}$ whenever $X\in \mathbb{L}_{1,2}$. This means that the derivatives
\begin{equation*}
D_s\,X^{\pi}(t)=\sum^{n-1}_{i=0}\frac{1}{t_{i+1}-t_i}\left(\int^{t_{i+1}}_{t_i} d_s\,X(s)\,ds\right)\chi_{(t_i,t_{i+1}]}(t)\xrightarrow{\abs{\pi}\rightarrow 0} D_s X(t).
\end{equation*}
On the other hand, the Riemann sum associated to the preceding approximation is:
\begin{equation*}
S^{\pi}=\sum^{n-1}_{i=0}\frac{1}{t_{i+1}-t_i}\left(\int^{t_{i+1}}_{t_i}X(s)\,ds\right)(B(t_{i+1})-B(t_i)).
\end{equation*}
Thus, for any $X\in \mathbb{L}_{1,2}$ we find
\begin{equation}
\delta(X^{\pi})= S^{\pi}- \sum^{n-1}_{i=0}\frac{1}{t_{i+1}-t_i}\int^{t_{i+1}}_{t_i}\int^{t_{i+1}}_{t_i} D_s X(t)\,ds\,dt;
\end{equation}
moreover, it converges in $L^2(\Omega)$ to $\delta(X)$. Besides, this convergence does not guarantee the existence of the Riemann sum. Some conditions should be introduced to make the second term at the right-hand side converge. This summand is, in fact, an approximation of the trace of the kernel $D_s\,X_t$ in $T^2$. It is undefined for an arbitrary square integrable kernel. The set of functions where it exists has two properties: the mappings $s\rightarrow D_{t \vee s}X(t \wedge s)$ and $s\rightarrow D_{t \wedge s}X(t \vee s)$ are uniformly continuos with respect to $t$, and $\sup_{s,t} \E{\abs{D_s\,X(t)}^2}<\infty$. Then, we have the following limits (uniformly in $t$):
\begin{align*}
D_t^+ X(t)&=\lim_{\varepsilon\searrow 0} D_t X(t+\varepsilon)\\
D_t^- X(t)&=\lim_{\varepsilon\searrow 0} D_t X(t-\varepsilon),
\end{align*}
from it we construct the operator $\boldsymbol{\nabla}=D^+ + D^-$. With all these conditions at hand the Riemann sum converges to the Stratonovich integral and we have
\begin{equation}
\int_T X(t)\circ dB(t)= \int_T X(t)\,\delta B(t) + \frac{1}{2}\int_T (\boldsymbol{\nabla} X)_t\, dt.
\label{eq:strato-bm}
\end{equation}
Henceforth, we have accomplished a definition for the Riemann `like' approximating series, they are coherent with our previous views and our rough idea of derivative---see equation (\ref{eq:strat-1}). 

\section{Wiener-It\^o Chaos Expansion and Wick product}\label{Bm-wiener}

The chaos expansions allow us to write any given random variable as a series of smoothed white noise functionals. There are two versions: one based on terms of Hermite polynomials, the other using multiple It\^o integrals. Both version are, of course, related and eventually lead to the definition of a new product: the \textit{Wick product}. These three concepts are very important, for they provide a set of analytic tools---It\^o formula included---that will allow us to solve stochastic differential equations.

\subsection{Chaos expansion in terms of Hermite polynomials}

The \textit{Hermite polynomials} $H_n(x)$ are defined
\begin{equation}
H_n(x)= (-1)^n e^{x^2/2} \frac{d^n}{dx^n}(e^{-x^2/2});\quad n=0,1,\cdots.
\end{equation}
The first polynomials are then:
\begin{equation*}
H_0(x)=1, H_1(x)=x, H_2(x)=x^2-1, H_3(x)=x^3-3x,\,\text{etc.}.
\end{equation*}
Now, we define the \textit{Hermite functions}---a detailed description of their properties can be found in \citet{book:sundaram}:
\begin{equation}
\xi_n(x)= \left(2^{n-1} (n-1)! \sqrt{\pi}\right)^{- 1/2} e^{-x^2/2} H_{n-1}(x); \quad n=1,2,\cdots.
\label{def:hermite}\end{equation}
These functions belongs to $\mathcal{S}(\Re)$; moreover, they constitute an orthonormal basis for $L^2(\Re)$. We will use both, the Hermite polynomials and functions, to define a basis for $L^2(\mu)$.

Let $\delta=(\delta_1,\dots,\delta_d)\in \N^d$ denote $d$-dimensional multi-indices, then the family of tensor products
\begin{equation}
\xi_{\delta}:= \xi_{(\delta_1,\dots,\delta_d)}=\xi_{\delta_1}\otimes\cdots\otimes \xi_{\delta_d}
\end{equation}
is an orthonormal basis for $L^2(\Re^d)$. Let $\delta^{(j)}$ represent a given fixed order for the set of multi-indices, such that,
\begin{equation*}
i<j \Rightarrow \delta^{(i)}_1+\delta^{(i)}_2 +\cdots+ \delta^{(i)}_d \leq \delta^{(j)}_1+\delta^{(j)}_2 +\cdots+ \delta^{(j)}_d,
\end{equation*}
that is, an increasing order. Now we can define
\begin{equation*}
\eta_j:= \xi_{\delta^{(j)}}=\xi_{\delta^{(j)}_1}\otimes\cdots\otimes \xi_{\delta^{(j)}_d}; j=1,2,\cdots.
\end{equation*}
We will consider, in particular, the set $\mathcal{J}$ of all sequences $\alpha$ with only finitely many $\alpha_j\neq 0$. Therefore, for $\alpha\in \mathcal{J}$
\begin{equation}
\mathcal{H}_{\alpha}(\omega) = \prod^{\infty}_{i=1} H_{\alpha_i}(\langle\omega,\eta_i\rangle); \quad \omega\in \mathcal{S}^{\ast}(\Re^d).
\label{def:base-chaos}\end{equation}
These family of functions constitutes an orthogonal basis for $L^2(\mu)$, and 
\begin{equation*}
\norm{\mathcal{H}_{\alpha}}^2_{L^2(\mu)}=\alpha!:=\alpha_1!\,\alpha_2!\cdots.
\end{equation*}

Now, we are in conditions to formulate the \textit{Wiener-It\^o chaos expansion theorem}: every $f\in L^2(\mu)$ has a unique representation
\begin{equation}
f(\omega)=\sum_{\alpha\in\mathcal{J}} c_{\alpha} \mathcal{H}_{\alpha}(\omega), \quad \text{where}\quad c_{\alpha}\in\Re.
\label{eq:hermite-expansion}\end{equation}
Moreover, we have the isometry
\begin{equation}
\norm{f}^2_{L^2(\mu)}=\sum_{\alpha\in\mathcal{J}}\alpha! c_{\alpha}^2.
\end{equation}

Let us consider the $1$-dimensional smoothed white noise as it was defined in (\ref{eq:smoothed-wn}), it is
\begin{align}
w(\phi,\omega)=&\langle\omega,\phi\rangle=\langle\omega,\sum^{\infty}_{j=1}(\phi,\eta_j)\eta_j\rangle\nonumber\\
=& \sum^{\infty}_{j=1}(\phi,\eta_j)\langle\omega,\eta_j\rangle=\sum^{\infty}_{j=1}(\phi,\eta_j)\mathcal{H}_{\epsilon_j}(\omega),
\label{eq:smooth-wn-expansion}\end{align}
where $\epsilon_j=(0,0,\dots,1,\dots)$ with 1 on the entry number $j$, and 0 otherwise. This convergence is in $L^2(\mu)$. In this case, it is $\eta_{\epsilon_j}(t)=\xi_j(t)$. Also, we can calculate the expansion for the $1$-dimensional ($1$-parameter) Brownian motion defined in the preceding section. The expansion of the step function $\chi_{[0,t]}$, using the Hermite functions, is:
\begin{equation*}
\chi_{[0,t]}(s)=\sum^{\infty}_{i=0} (\chi_{[0,t]},\xi_i)\, \xi_i(s)=\sum^{\infty}_{i=0} \left(\int^t_0\xi_i(s)ds\right) \xi_i(s),
\end{equation*}
so the expansion for the Brownian motion is
\begin{align}
B(t,\omega)=& \langle\omega,\sum^{\infty}_{i=0}\left(\int^t_0\xi_i(s)ds\right) \xi_i\rangle=\sum^{\infty}_{i=0}\left(\int^t_0\xi_i(s)ds\right) \langle\omega,\xi_i\rangle\nonumber\\
=& \sum^{\infty}_{i=0}\left(\int^t_0\xi_i(s)ds\right) \mathcal{H}_{\epsilon_i}(\omega).
\label{eq:bm-expansion}\end{align}

\subsection{Chaos expansion in terms of It\^o integrals}

The latter expansion is equivalent to another one built using iterated It\^o integrals. This is defined as follows: Let $\Phi(t_1,\dots,t_n)$ be a symmetric function then its \textit{$n$-tuple It\^o integral} for $n\geq 1$ is
\begin{multline}
\int_{\Re^n}\!\!\! \Phi\,dB^{\otimes n}:= \\
n!\int^{\infty}_{-\infty}\int^{t_n}_{-\infty}\int^{t_{n-1}}_{-\infty}\!\!\!\!\cdots\int^{t_1}_{-\infty}\!\!\Phi(t_1,t_2,\dots,t_n)\,dB(t_1)\,dB(t_2)\cdots\,dB(t_n),
\end{multline}
each integrand in the iteration is adapted because of the integration limits of the preceding integrand. Using the It\^o isometry $n$ times whenever $\Phi\in L^2(\Re^n)$ we find
\begin{equation}
\E{\left(\int_{\Re^n}\Phi\,dB^{\otimes n}\right)^2}=n! \int_{\Re^n} \Phi(t_1,\dots,t_n)^2\,dt_1\cdots dt_n=n!\,\norm{\Phi}^2.
\end{equation}

Now, let $\alpha=(\alpha_1,\dots,\alpha_n)$ be a multi-index such that $n=\abs{\alpha}$. In 1951 It\^o found a fundamental result:
\begin{equation}
\int_{\Re^n} \xi_1^{\hat{\otimes}\alpha_1}\hat{\otimes}\cdots\hat{\otimes}\,\xi_k^{\hat{\otimes}\alpha_k}\,dB^{\otimes n}=\prod^k_{i=1} H_{\alpha_i}(\langle\omega,\xi_i\rangle),
\label{eq:ito-iterated}\end{equation}
where $\hat{\otimes}$ is the symmetrized tensor product, i.e., for $f,g:\Re\rightarrow\Re$ it is
\begin{equation*}
(f\otimes g)(x,y)=f(x)g(y)
\end{equation*}
and 
\begin{equation*}
(f\hat{\otimes} g)(x,y)= \frac{1}{2}\left[f\otimes g+ g\otimes f\right](x,y); \quad (x,y)\in \Re^2,
\end{equation*}
(the same applies to higher dimensions). Therefore, comparing equation (\ref{eq:ito-iterated}) with definition (\ref{def:base-chaos}) we have
\begin{equation}
\int_{\Re^n} \xi^{\hat{\otimes}\alpha}\,dB^{\otimes n}= \mathcal{H}_{\alpha}(\omega),
\end{equation}
here we have introduced the multi-index notation $\xi^{\hat{\otimes}\alpha}=\int_{\Re^n} \xi_1^{\hat{\otimes}\alpha_1}\hat{\otimes}\cdots\hat{\otimes}\,\xi_k^{\hat{\otimes}\alpha_k}$. If we assume now that $f\in L^2(\mu)$ has the chaos expansion (\ref{eq:hermite-expansion}); thus, we may rewrite $f$ using the latter equation as
\begin{equation*}
f(\omega)=\sum^{\infty}_{n=0}\sum_{\abs{\alpha}=n}\, c_{\alpha}\int_{\Re^n}\xi^{\hat{\otimes}\alpha}\,dB^{\otimes n}.
\end{equation*}
Henceforth,
\begin{align}
f(\omega)=\sum^{\infty}_{n=0}\int_{\Re^n} f_n\,dB^{\otimes n}&,&\text{with } f_n=\sum_{\abs{\alpha}=n}\,c_{\alpha}\,\xi^{\hat{\otimes}\alpha}\in \hat{L}^2(\Re^n),
\end{align}
where $\hat{L}^2(\Re^n)$ denotes the symmetric functions in $L^2(\Re^n)$. Moreover, the isometry relation reads
\begin{equation*}
\norm{f}^2_{L^2(\mu)}=\sum^{\infty}_{n=0}n!\norm{f_n}^2_{L^2(\Re^n)}.
\end{equation*}

\subsection{The Wick product}

The representation of stochastic processes by means of the chaos expansion representation provides a favorable setting to study stochastic differential equations. Until now, we have characterized these processes with function and distribution spaces, $\mathcal{S}(\Re^d)\subset L^2(\Re^d)\subset \mathcal{S}^{\ast}(\Re^d)$, but we will need to extend them a bit more. 

Again we impose a fixed order for the multi-index family $\delta=(\delta_1,\dots,\delta_d)\in \N^d$. Let us introduce the following notation: for $\alpha=(\alpha_1,\dots,\alpha_,\dots)\in \mathcal{J}$ and $\beta=(\beta_1,\dots,\beta_j,\dots)\in \Re^{\N}$ a finite sequence it is
\begin{equation*}
\alpha^{\beta}=\alpha_1^{\beta_1}\alpha_2^{\beta_2}\cdots\alpha_j^{\beta_j}\cdots\quad \text{where }\alpha^0_j=1.
\end{equation*}

It can be proven \citep{book:reed-simon} that:
\begin{itemize}
\item[\textit{i)}] For $\phi\in L^2(\Re^d)$, such that $\phi=\sum^{\infty}_{j=1} a_j\eta_j$, where the $a_j=(\phi,\eta_j)$ are the Fourier coefficients with respect to the multi-index Hermite functions. We have $\phi\in \mathcal{S}(\Re^d)$ if and only if
\begin{equation*}
\sum^{\infty}_{j=1}a^2_j(\delta^{(j)})^{\gamma}<\infty
\end{equation*}
for all $d$-dimensional multi-indices $\gamma=(\gamma_1,\dots,\gamma_d)$.

\item[\textit{ii)}] Also, the space $\mathcal{S}^{\ast}(\Re^d)$ can be identified with the space of all formal expansions 
\begin{equation*}
\Theta=\sum^{\infty}_{j=1}b_j\eta_j
\end{equation*}
such that
\begin{equation*}
\sum^{\infty}_{j=1}b^2_j(\delta^{(j)})^{-\gamma'}<\infty
\end{equation*}
for some $d$-dimensional multi-index $\gamma'=(\gamma_1',\dots,\gamma_d')$.
\end{itemize}
Similarly, we can define an analogue for the probability space $L^2(\mu)$: the \textit{Kondratiev spaces.} We will not give the more general version of these spaces, because it is not required in the present discussion. Therefore, let us define the quantity
\begin{equation}
(2\N)^{\gamma}:= \prod_j (2j)^{\gamma_j},
\end{equation}
where $\gamma=(\gamma_1,\dots,\gamma_j,\dots)\in \Re^{\N}$ has finite non-zero numbers. \textit{The stochastic test function spaces} $\mathcal{S}_{\rho}$ ($0\leq \rho\leq 1$ fixed) are the set of all the sums
\begin{equation}
f=\sum_{\alpha}c_{\alpha}\,\mathcal{H}_{\alpha}\in L^2(\mu);\quad c_{\alpha}\in \Re
\end{equation}
such that
\begin{equation}
\norm{f}^2_{\rho}:=\sum_{\alpha}c^2_{\alpha}(\alpha!)^{1+\rho}(2\N)^{k\alpha}<\infty\quad\text{for all }k\in \N. 
\end{equation}

On the other hand, \textit{the stochastic distribution spaces} $\mathcal{S}_{-\rho}$ consist of all formal expansions 
\begin{equation}
F=\sum_{\alpha}b_{\alpha} \mathcal{H}_{\alpha}\quad\text{with }b_{\alpha}\in \Re
\end{equation}
such that 
\begin{equation}
\norm{F}_{-\rho}:= \sum_{\alpha} b^2_{\alpha}(\alpha!)^{1-\rho}(2\N)^{-q \alpha}<\infty\quad\text{ for some }q\in\N.
\end{equation}
The seminorms $\norm{\cdot}_{\rho}$ gives a topology for $\mathcal{S}_{\rho}$, and the space $\mathcal{S}_{-\rho}$ can be thought to be the dual of the stochastic test function space by means of the inner product
\begin{equation*}
\langle F,f\rangle= \sum_{\alpha} b_{\alpha}c_{\alpha}\alpha!.
\end{equation*}
Note that for $\rho\in[0,1]$ we have
\begin{equation*}
\mathcal{S}_1\subset\mathcal{S}_{\rho}\subset\mathcal{S}_0\subset L^2(\mu)\subset\mathcal{S}_{-0}\subset\mathcal{S}_{-\rho}\subset\mathcal{S}_{-1}.
\end{equation*}
In particular if both $F$ and $G$ belong to $L^2(\mu)$, then $\langle F,G\rangle=\E{F G}$. The spaces $\mathcal{S}_0$ and $\mathcal{S}_{-0}$ are called \textit{Hida spaces}, and respectively denoted $\mathcal{S}$ and $\mathcal{S}^{\ast}$. 

Now, we can define the \textit{Wick product}: for two elements 
\begin{equation*}
F=\sum_{\alpha}a_{\alpha}\mathcal{H}_{\alpha}, G=\sum_{\alpha}b_{\alpha}\mathcal{H}_{\alpha}\in \mathcal{S}_{-1},
\end{equation*} 
we have
\begin{equation}
F\diamond G=\sum_{\alpha,\beta}a_{\alpha}b_{\beta}\mathcal{H}_{\alpha+\beta}.
\end{equation}
The product is independent of the base elements of $L^2(\mu)$. Moreover, the spaces $\mathcal{S}_1$, $\mathcal{S}_{-1}$ and $\mathcal{S}$, $\mathcal{S}_{\ast}$ are closed under the Wick product. In the sense $F,G\in\mathcal{A}\Rightarrow F\diamond G\in \mathcal{A}$ with $\mathcal{A}$ anyone of the former spaces. Of course, the three laws for products---associability, commutativity, and distributiveness---are obeyed.

The \textit{Wick powers} $F^{\diamond k};\quad k=0,1,2,\cdots$ of $F\in\mathcal{S}_{-1}$ are defined inductively as follows:
\begin{equation*}
\left\{\begin{aligned}
F^{\diamond 0}&=1\\
F^{\diamond k}&= F\diamond F^{\diamond (k-1)}\quad\text{for }k=1,2,\cdots.
\end{aligned}\right.
\end{equation*}
Moreover, given a polynomial $p(x)=\sum_{k=0}^N a_k x^k$ it is straightforward to define its Wick version,
\begin{equation*}
p^{\diamond}(F)=\sum_{k=0}^N a_k F^{\diamond k}.
\end{equation*}

It can be proven that for $F, G \in L^2(\mu)$ Gaussians, that is,
\begin{equation*}
F(\omega)=a_0+\sum_{k=1}^{\infty}a_k \mathcal{H}_{\epsilon_k}(\omega)\quad\text{and}\quad G(\omega)=b_0+\sum_{k=1}^{\infty}b_k \mathcal{H}_{\epsilon_k}(\omega)
\end{equation*}
with $\sum_{k=1}^{\infty}a_k^2,\sum_{k=1}^{\infty}b_k^2<\infty$, it is
\begin{equation}
(G\diamond F)(\omega)=(G F)(\omega)-\sum^{\infty}_{k=1} a_k b_k.
\label{eq:gaussian-relation}
\end{equation}
Where it had been used the property
\begin{equation*}
H_{\epsilon_j+\epsilon_k}=\left\{\begin{aligned}
H_{\epsilon_j}H_{\epsilon_k}\text{ for }k\neq j\\
H^2_{\epsilon_k}-1\text{ for }k=j
\end{aligned}\right..
\end{equation*}

Applying this formula to the smooth white noise expansion (\ref{eq:smooth-wn-expansion}) we find
\begin{equation}
w(\phi)\diamond w(\psi)= w(\phi)w(\psi)-(\phi,\psi);
\end{equation}
moreover, if $\psi=\phi$ and $\norm{\phi}=1$, then we have $w(\phi)^{\diamond 2}=H_2(w(\phi)),
$ and in general:
\begin{equation}
w(\phi)^{\diamond n}=H_n(w(\phi)).
\end{equation}

\subsection{Skorohod integration and Wick product}

The Skorohod integral can be written in terms of the chaos expansion. Let $Y(t)=Y(t,\omega)$ be a stochastic process such that $\E{Y(t)^2}<\infty$ for all $t$. We already know that this process can be expanded as
\begin{equation*}
Y(t)= \sum^{\infty}_{n=0}\int_{\Re^n}\! f_n(s_1,\dots,s_n,t)\, dB^{\otimes n}(s_1,\cdots,s_n),
\end{equation*}
where $f_n(\cdot,t)\in \hat{L}^2(\Re^n)$ for $n=0,1,2,\dots$ and for each $t$. We denote by $\hat{f}_n(s_1,\dots,s_{n+1})$ the symmetrization with respect to the $n+1$ variables. Thus, assume that
\begin{equation*}
\sum^{\infty}_{n=1}(n+1)!\norm{\hat{f}_n}_{L^2(\Re^{n+1})}<\infty.
\end{equation*}
We can define the \textit{Skorohod integral} of $Y(t)$ as
\begin{equation}
\int_{\Re} Y(t)\,\delta B(t)= \sum^{\infty}_{n=0}\int_{\Re^{n+1}} \hat{f}_n(s_1,\cdots,s_{n+1})\,dB^{\otimes (n+1)}(s_1,\cdots,s_{n+1}).
\end{equation}
It has the norm
\begin{equation*}
\gnorm{\int_{\Re} Y(t)\,\delta B(t)}_{L^2(\Re)}=\sum^{\infty}_{n=0}(n+1)! \norm{\hat{f}_n}_{L^2(\Re^{n+1})}.
\end{equation*}

On the other hand, we say $Z(t)=\sum_{\alpha} c_{\alpha}(t)\mathcal{H}_{\alpha}\in \mathcal{S}^{\ast}$ is $\mathcal{S}^{\ast}$-integrable if from its chaos expansion the expression 
\begin{equation*}
\int_{\Re} Z(t)\,dt=\sum_{\alpha} \left(\int_{\Re} c_{\alpha}(t)\,dt\right)\mathcal{H}_{\alpha}(\omega)
\end{equation*}
belongs to $\mathcal{S}^{\ast}$. Now, the process
\begin{equation*}
W(t)= \sum^{\infty}_{k=0} \xi_k(t)\,\mathcal{H}_{\epsilon_k}(\omega)\in\mathcal{S}^{\ast},
\end{equation*}
because the Hermite functions are bounded: $\xi_n(t)< n^{-1/12}$. From equation (\ref{eq:bm-expansion}) we have 
\begin{equation}
\int_{\Re} \chi_{(-\infty,0]} W(s)\,ds= \sum^{\infty}_{k=0}\left(\int^t_0 \xi_k(s)\,ds\right)\mathcal{H}_{\epsilon_k}(\omega)=B(t).
\end{equation}
Therefore, we have proven that $dB(t)/dt=W(t)$ is well defined in $\mathcal{S}^{\ast}$. Afterwards, a last fundamental theorem remains to be addressed \citep[\textbf{Theorem 2.5.9}]{book:holden}: assume that $Y(t)= \sum_{\alpha}c_{\alpha} \mathcal{H}_{\alpha}$ is a Skorohod integrable process, and let $a<b$ real numbers. Then $Y(t)\diamond W(t)$ is $\mathcal{S}^{\ast}$-integrable and 
\begin{equation}
\int^b_a Y(t)\,\delta B(t)= \int^b_a Y(t)\diamond W(t)\,dt.
\label{eq:noise-integral}
\end{equation}

\section{Stochastic Calculus for fractional Brownian motions}

In the past years different approaches have been given to produce a Stochastic Calculus for fractional Brownian motions: \citeauthor{paper:zaehle-1} (\citeyear{paper:zaehle-1},\citeyear{paper:zaehle-2}), \citet{paper:decreusefond}, and \citet{paper:follmer}. Basically, these approaches tackle the problem of constructing a calculus, but from two different starting points: one uses a pathwise definition of the integral while the other rests on the Malliavin Calculus as we sketched earlier in this chapter. In all these circumstances the processes have are persistent. We will follow \citet{paper:Hu} and \citet{paper:duncan} into the second approach. We will construct a Stochastic Analysis from a Chaos expansion.

Let $\phi:\Re_+\times\Re_+\rightarrow\Re$ be defined as follows
\begin{equation*}
\phi(s,z)= H(2H-1)\abs{s-z}^{2H-2},
\end{equation*}
for a fixed $H\in(1/2,1)$. Then we say $f\in\LR$ if it is measurable and 
\begin{equation}
\abs{f}^2_{\phi}:=\int_{\Re}\int_{\Re} f(s)f(z)\phi(s,z)\,\,ds\, dz <\infty.
\label{cond:squared}
\end{equation}
Afterwards, the inner product can be defined in $\LR$,
\begin{equation*}
(f,g)_{\phi}:=\int_{\Re}\int_{\Re} f(s)g(z)\phi(s,z)\,\,ds\, dz,\quad {\rm for\,all\,}f,g\in\LR;
\end{equation*}
therefore, {\LR} becomes a separable Hilbert space.

Again, we take $\mathcal{S}(\Re) \subset \LR$ to be the Schwartz space of rapidly decreasing smooth functions on $\Re$. Its dual $\Omega=\mathcal{S}'(\Re)$ is the probability space with the associated probability measure, $\mu_{\phi}$, found applying the \mbox{Bochner-Minlos} theorem,
\begin{equation*}
\E{e^{i\langle\cdot,f\rangle}}:=\int_{\Omega} e^{i\langle\omega,f\rangle} d\mu_{\phi}(\omega)=e^{-\frac{1}{2}\abs{f}_{\phi}^2},
\end{equation*}
where $\langle\omega,f\rangle$ is the usual pairing between elements in the dual and functions on $\Re$. Because of the latter construction  this probability measure can be shown  to induce properties like those in (\ref{eq:Bm-mean-properties}), i.e., 
\begin{equation}
\E{\langle\cdot,f\rangle}=0,\quad\text{and}\quad\E{\langle\cdot,f\rangle^2}=\abs{f}^2_{\phi}.
\label{eq:mean}
\end{equation}
Once more, the triplet $\left(\Omega, \mathcal{B}\!\left(\Omega\right),\mu_{\phi}\right)$ becomes a probability space---$\mathcal{B}\!\left(\Omega\right)$ is the Borel algebra on $\Omega$. It is usually called \textit{fractional white noise probability space}. 

Now, let $L^2(\mu_{\phi})=L^2(\Omega, \mathcal{B}\!\left(\Omega\right),\mu_{\phi})$ be the space of all the random variables $X:\Omega \rightarrow \Re$ such that
\begin{equation}
\norm{X}_{L^2(\mu_{\phi})}^2:=\mathbb{E}\abs{X}^2<\infty.
\end{equation}
Furthermore, the functions in \LR\, define a set of random variables of the form  $f(\omega)=\langle\omega,f\rangle$. It is included in  $L^2(\mu_{\phi})$; that is, the condition (\ref{cond:squared}) induces square measurable random variables because of equations (\ref{eq:mean}).

With the same arguments as before we have that: $\mathcal{S}(\Re)$ is dense in $\LR$; for any $f\in\LR$ the series $f_n\in\mathcal{\Re}$ are such that $f_n\rightarrow f$ in $\LR$; and so, the following limit 
\begin{equation}
\lim_{n\rightarrow\infty}\langle\omega,f_n\rangle:=\langle\omega,f\rangle
\label{eq:lim}
\end{equation}
exists in $L^2(\mu_{\phi})$.
 
We define now the fractional Brownian motion process as follows:
\begin{equation}
B^H(z):=B^H(z,\omega)=\langle\omega,\chi_{[0,z)}\rangle \in L^2(\mu_{\phi}).
\label{eq:fbm}
\end{equation}
For simplicity we will thought $B^H$ designates the $z$-continous version of the rightmost hand side term. As for the \textit{step function} $\chi_{[0,z)}:\Re\rightarrow [-1,1]$ again:
\begin{equation*}
\chi_{[0,z)}(s)=
\begin{cases}
1&\text{if }0\leq s\leq z\\
-1&\text{if }z<s\leq 0\\
0& \text{otherwise}
\end{cases}.
\end{equation*}
Because of property (\ref{eq:lim}) we have again that for any $f\in \LR$ definition (\ref{eq:fbm}) is equivalent to
\begin{equation}
\langle\omega,f\rangle=\int_{\Re} f(z)\, dB^H\!(z,\omega)
\label{eq:intfbm}.
\end{equation}
Under the same procedure we can verify for $f,g\in\LR$ that
\begin{equation}
\E{\langle \omega,f\rangle\langle \omega, g\rangle}= (f,g)_{\phi}
\label{eq:isometry}.
\end{equation}

For $f$ as above, we define the exponential function $\mathcal{E}:\LR\rightarrow L^2(\mu_{\phi})$ as
\begin{equation}
\mathcal{E}(f)= \exp\!\left(\int_{\Re} f\, dB^H-\frac{1}{2}\abs{f}^2_{\phi}\right).
\label{eq:basis}
\end{equation}
Thus, the Hilbert space {\LR} is naturally associated with the fBm process from the formulation as an abstract Wiener space. Let $\mathcal{E}$ be the \textit{linear span} of the exponentials, i.e.,
\begin{equation}
\mathcal{E}=\left\{\sum^n_{k=1} a_k \mathcal{E}(f_k); n\in\N, a_k\in\Re, f_k\in\LR {\rm\; for\, }k\in\{1,\dots,n\}\right\},
\end{equation}
is dense in $L^2(\mu_{\phi})$. 

Nevertheless, some tools we are going to introduce here require a more familiar functional expansion, and the Hermite functions (\ref{def:hermite}) will help us again. First, we note that we can map the orthonormal basis they form in $L^2(\Re)$ to an orthonormal one in {\LR} through the isometry map \citep[see \textbf{Lemma 2.1} in][]{paper:Hu} $\tilde{\xi}_n=\Gamma^{-1}_{\phi}\xi_n$ defined
\begin{equation*}
\Gamma_{\phi}f(s)=c_H\int_s^\infty(z-s)^{H-3/2} f(z)\, dz,
\end{equation*}
where
\begin{equation*}
c_H=\sqrt{\frac{H(2H-1)\,\Gamma\!\left(\frac{3}{2}-H\right)}{\Gamma\!\left(H-\frac{1}{2}\right)\Gamma(2-2H)}}.
\end{equation*}

From the identity \citep[p. 404]{paper:gripenberg}
\begin{equation*}
c_H^2\int_{-\infty}^{z\wedge s}(z-u)^{H-3/2}(s-u)^{H-3/2}\,ds=\phi(z,s)
\end{equation*}
we see that
\begin{equation*}
\int_{\Re}\tilde{\xi}_n(s)\, \phi(s,z)\,ds =c_H\int_{-\infty}^z(z-s)^{H-3/2} \xi_n(s)\,ds
\end{equation*}
---because the $\tilde{\xi}_n$'s are an orthonormal basis these integrals are also smooth.

Let $\mathcal{I}$ be the set of all finite multi-indices $\alpha=(\alpha_1,\cdots,\alpha_m)$ of nonnegative integers, we define
\begin{equation*}
\mathcal{H}_{\alpha}(\omega):= H_{\alpha_1}\!(\langle\omega,\tilde{\xi}_1\rangle)\cdots H_{\alpha_m}\!(\langle\omega,\tilde{\xi}_m\rangle).
\end{equation*} 
In particular, if we put $\alpha=\epsilon^i$ then, in the very same way as in Section \ref{Bm-wiener}, we get from (\ref{eq:intfbm}) and the definition of Hermite polynomials
\begin{equation*}
\mathcal{H}_{\epsilon^i}(\omega)= H_1\!(\langle\omega,\tilde{\xi}_i\rangle)=\langle\omega,\tilde{\xi}_i\rangle=\int_{\Re}\tilde{\xi}_i(s)\; dB^H(s).
\end{equation*}
These functionals are elements of $L^2(\mu_{\phi})$, and they form its basis \citep[\textbf{Theorem 6.9}]{paper:duncan}. That is, for $X\in L^2(\mu_{\phi})$ there are $c_{\alpha}\in\Re$ and $\alpha\in\mathcal{I}$, such that
\begin{equation}
X(\omega)= \sum_{\alpha\in\mathcal{I}} c_{\alpha}\mathcal{H}_{\alpha}(\omega),
\label{eq:1series}
\end{equation}
and also
\begin{equation}
\norm{X}^2_{L^2(\mu_{\phi})}=\sum_{\alpha\in\mathcal{I}}\alpha! c_{\alpha}^2.
\end{equation}
These coefficients are given by $c_{\alpha}=\E{X\,\mathcal{H}_{\alpha}}/\alpha!$. 

The existence of this property let us define a flavor of (fractional) Hida spaces: the \textit{fractional Hida test function space} $\mathcal{S}_H$ which is the set of all
\begin{equation*}
\begin{aligned}
\psi(\omega)&=\sum_{\alpha\in\mathcal{I}}a_{\alpha}\mathcal{H}_{\alpha}(\omega)\in L^2(\mu_{\phi}),\quad{\rm such\; that}\\
\norm{\psi}^2_{H,k}&=\sum_{\alpha\in\mathcal{I}}\alpha!\, a^2_{\alpha}(2\,\N)^{k\alpha}<\infty,\quad{\rm for\; all }\>k\in\N,
\end{aligned}
\end{equation*}
where 
\begin{equation*}
(2\,\N)^{\gamma}= \prod_j (2j)^{\gamma_j}\quad{\rm for\; any\; element }\>\gamma=(\gamma_1,\cdots,\gamma_m)\in\mathcal{I};
\end{equation*}
and the \textit{fractional Hida distribution space} $\mathcal{S}^{\ast}_H$, the set of all formal expansions
\begin{equation}
\begin{aligned}
Y(\omega)&=\sum_{\beta\in\mathcal{I}}b_{\beta}\mathcal{H}_{\beta}(\omega),\quad{\rm such\; that}\\
\norm{Y}^2_{H,-q}&=\sum_{\beta\in\mathcal{I}}\beta!\, a^2_{\beta}(2\,\N)^{-q\beta}<\infty,\quad{\rm for\; some }\>q\in\N.
\end{aligned}
\label{eq:hida-exp}\end{equation}
Using these definitions is not hard to see that $\mathcal{S}_H\subset L^2(\mu_{\phi})\subset\mathcal{S}_H^{\ast}$.

It is now time to show how the fractional white noise and integration with respect to $B^H$ is defined. Let us first calculate the expansion for the stochastic integral in (\ref{eq:intfbm}). For any $f\in L^2_{\phi}(\Re)$---any given deterministic function---we have from equations~(\ref{eq:1series}) and (\ref{eq:isometry}):
\begin{equation}
\int_{\Re} f(s)\> dB^H(s)= \sum^{\infty}_{i=1} (f,\tilde{\xi}_i)_{\phi}\,\mathcal{H}_{\epsilon^i}(\omega).
\label{eq:chaos-integral}
\end{equation}
When $f=\chi_{[0,z)}$ in the left hand side we recover (\ref{eq:fbm}) and the following relation holds
\begin{equation}
B^H(z)= \sum^{\infty}_{k=1}\left[\int_0^z\left(\int_{\Re}\tilde{\xi}_k(s)\phi(s,u)\,ds \right)du\right]\mathcal{H}_{\epsilon^k}(\omega)\;\in\mathcal{S}^{\ast}_H,
\end{equation}
if we check its norm
\begin{equation}
\begin{aligned}
\norm{B^H\!(z)}^2_{H,-q}&=\sum^{\infty}_{k=1}\left[\int_0^z\left(\int_{\Re}\tilde{\xi}_k(s)\phi(s,u)\,ds \right)du\right]^2(2k)^{-q}\leq \\
&\leq M^2\,z^2\,2^{-q}\sum^{\infty}_{k=1}k^{1/3-q}=M^2\,z^2\,2^{-q}\zeta\!\left(q-\frac{1}{3}\right),
\end{aligned}
\label{fBm-cont}\end{equation}
($\zeta$ is the Riemann's zeta function) because 
\begin{equation}
\abs{\int_{\Re}\tilde{\xi}_k(s)\phi(s,u)\, ds}=c_H\abs{\int_{(-\infty,u]}(u-s)^{H-3/2}\,\xi_k(s)\,ds}\leq M\,k^{1/6},
\label{eq:bound}
\end{equation}
here we have used the bound for the Hermite functions given by \citet[pp. 198--201]{book:szego}. Furthermore, when $q>4/3$ the former inequality also shows that $B^H$ is continuos and differentiable in $\mathcal{S}^{\ast}_H$. Its derivative 
\begin{equation}
\frac{d}{dz}B^H(z)= \sum^{\infty}_{k=1} \left(\int_{\Re}\tilde{\xi}_k(s)\phi(s,z)\,ds \right)\mathcal{H}_{\varepsilon^k}(\omega):=W^H(z)\in \mathcal{S}^{\ast}_H 
\label{eq:white-noise}
\end{equation}
is the formal definition of \textit{fractional white noise}. This noise is also continous in $\mathcal{S}^{\ast}_H$, when $z>s$
\begin{align}
\norm{W^H(z)-W^H(s)}^2_{H,-q}&=\sum^{\infty}_{k=1}\epsilon^i!\left|\int_{\Re}\tilde{\xi}_k(u)\phi(z,u)du-\int_{\Re}\tilde{\xi}_k(u)\phi(s,u)du \right|^2(2k)^{-q}\leq\nonumber\\
&\leq c_H \sum^{\infty}_{k=1} \left[\int^{z-s}_0\!\! [(z-s)-u]^{H-3/2}\abs{\xi_k(s+u)}du\right]^2(2k)^{-q}\leq\nonumber\\
&\leq 2^{-q} c_H^2  M^2\,\zeta\!\left(q+\frac{1}{6}\right) \left\{\int^{z-s}_0 [(z-s)-u]^{H-3/2}\right\}^2=\nonumber\\
&=\frac{2^{2-q} c_H^2  M^2}{(2H-1)^2}\,\zeta\!\left(q+\frac{1}{6}\right) (z-s)^{2H-1},
\label{fWn-cont}
\end{align}
the same holds when $z<s$.

Of course, this chaos expansion has its own Wick product. Let $X(\omega)=\sum_{\alpha\in\mathcal{I}} a_{\alpha} \mathcal{H}_{\alpha}(\omega)$ and $Y(\omega)=\sum_{\beta\in\mathcal{I}} b_{\beta} \mathcal{H}_{\beta}(\omega)$ be in $\mathcal{S}^{\ast}_H$, then
\begin{equation}
(X\diamond Y)(\omega)=\sum_{\alpha,\beta\in\mathcal{I}} a_{\alpha}b_{\beta}\mathcal{H}_{\alpha+\beta}(\omega)=\sum_{\gamma\in\mathcal{I}}\left(\sum_{\alpha+\beta=\gamma}a_{\alpha}b_{\beta}\right)\mathcal{H}_{\gamma}(\omega).
\label{eq:wick-prod}
\end{equation}
For example, let $f,g\in \LR$ then using equation (\ref{eq:chaos-integral}) we find
\begin{multline}
\left(\int_\Re f(s)\, dB^H(s)\right)\diamond \left(\int_\Re g(s)\, dB^H(s)\right)=\left(\sum^{\infty}_{i=1} (f,\tilde{\xi}_i)_{\phi}\,\mathcal{H}_{\epsilon^i}(\omega)\right)\diamond \left(\sum^{\infty}_{j=1} (g,\tilde{\xi}_j)_{\phi}\,\mathcal{H}_{\epsilon^j}(\omega)\right)\\
=\sum^{\infty}_{j,i=1} (f,\tilde{\xi}_i)_{\phi}(g,\tilde{\xi}_j)_{\phi}\,\mathcal{H}_{\epsilon^i+\epsilon^j}(\omega)\\
= \sum^{\infty}_{\substack{j,i=1\\i\neq j}} (f,\tilde{\xi}_i)_{\phi}(g,\tilde{\xi}_j)_{\phi}\,\langle\omega,\tilde{\xi}_i\rangle \langle\omega,\tilde{\xi}_j\rangle + \sum^{\infty}_{i=1} (f,\tilde{\xi}_i)_{\phi}(g,\tilde{\xi}_i)_{\phi}(\langle\omega,\tilde{\xi}_i\rangle^2-1)\\
=\left(\sum^{\infty}_{i=1} (f,\tilde{\xi}_i)_{\phi} \langle\omega,\tilde{\xi}_i\rangle\right)\cdot\left(\sum^{\infty}_{j=1} (g,\tilde{\xi}_j)_{\phi}\, \langle\omega,\tilde{\xi}_j\rangle\right)-\sum^{\infty}_{i=1} (f,\tilde{\xi}_i)_{\phi}(g,\tilde{\xi}_i)_{\phi}\\
=\left(\int_\Re f(s)\, dB^H(s)\right)\cdot\left(\int_\Re g(s)\, dB^H(s)\right) - (f,g)_\phi.
\label{eq:gaussian-integral}
\end{multline}
This property is a special case of a more general one for Gaussian variables, that is, for $X= a_0 + \sum^\infty_{i=0} a_i \mathcal{H}_{\epsilon^i}$ and $Y= b_0 + \sum^\infty_{i=0} b_j \mathcal{H}_{\epsilon^i}$ we have $X\diamond Y = X\cdot Y - \sum_{i=1}^\infty a_i b_i$  as was proved in (\ref{eq:gaussian-relation}) for the Brownian case. Afterwards, for $f=g=\tilde{\xi_i}$  proceeding inductively with the latter equation yields
\begin{equation*}
\langle\omega,\tilde{\xi}_i\rangle^{\diamond n}= \left(\mathcal{H}_{\epsilon^i}(\omega)\right)^{\diamond n}= \mathcal{H}_{n \epsilon_i}(\omega)=H_n(\langle\omega,\tilde{\xi}_i\rangle).
\end{equation*}

Now, as we extended polynomials into the Hida space for Brownian motions, we do the same here but with the power series. The \textit{Wick exponential} defined by the power series    
\begin{equation*}
\exp^{\diamond}(X)=\sum^{\infty}_{n=0}\frac{1}{n!}X^{\diamond n},
\end{equation*}
provided it converges in $\mathcal{S}^{\ast}_H$. It has the same algebraic properties as the usual exponential, e.g.:
\begin{equation*}
\exp^\diamond( X) \diamond \exp^\diamond(Y) = \exp^\diamond(X+Y).
\end{equation*}
This Wick exponential is the keystone of this section, for it provides a link between the two expansions given here. If we set $X=a\langle \omega,\tilde{\xi}_i\rangle$, it is 
\begin{align}
\exp^\diamond( a \langle \omega,\tilde{\xi}_i\rangle )&= \sum_{i=1}^{\infty} \frac{a^n}{n!} \langle\omega,\tilde{\xi}_i \rangle^{\diamond n}\nonumber\\
&= \sum_{i=1}^{\infty} \frac{a^n}{n!}\, H_n(\langle\omega,\tilde{\xi}_i\rangle)\nonumber\\
&= \exp\left(a\langle\omega,\tilde{\xi}_i\rangle - \frac{1}{2} a^2\right),
\end{align}
because of
\begin{equation*}
\exp\left(tx-\frac{1}{2}t^2\right)=\sum^\infty_{i=1} \frac{t^n}{n!} H_n(x).
\end{equation*}
Therefore, when $X=\langle\omega, f\rangle=\int_\Re f dB^H$
\begin{align*}
\exp^{\diamond}(\langle \omega,f\rangle) &=\exp^\diamond\left(\sum^\infty_{i=1} (f,\tilde{\xi}_i)_\phi \langle\omega,\tilde{\xi}_i\rangle\right)\\
&= \prod^\diamond_{i=1}\exp^\diamond\left( (f,\tilde{\xi}_i)_\phi \langle\omega,\tilde{\xi}_i\rangle\right)\\
&= \prod_{i=1}^\infty \exp\left((f,\tilde{\xi}_i)_\phi\langle\omega,\tilde{\xi}_i\rangle - \frac{1}{2} (f,\tilde{\xi}_i)_\phi^2\right)\\
&= \exp\left(\sum_{i=1}^\infty (f,\tilde{\xi}_i)_\phi\langle\omega,\tilde{\xi}_i\rangle - \frac{1}{2} \sum_{i=1}^\infty (f,\tilde{\xi}_i)_\phi^2\right)\\
&= \exp\left(\int_\Re f(s)\, dB^H(s) -\frac{1}{2}\abs{f}^2_\phi\right)
\end{align*}
and thus
\begin{equation}
\exp^{\diamond}(\langle \omega,f\rangle)=\mathcal{E}(f),
\label{eq:wick}
\end{equation}
as the right-hand side was defined in (\ref{eq:basis}) the relation between the expansions is settled.

It is appropriate to show now the behavior of the Wick product within an average. Let $X=\sum_{\alpha\in\mathcal{I}} a_{\alpha} \mathcal{H}_{\alpha}$ and $Y=\sum_{\beta\in\mathcal{I}} b_{\beta} \mathcal{H}_{\beta}$ have the usual chaos expansion thus
\begin{align}
\E{X\diamond Y}	&= \sum_{\gamma\in\mathcal{I}}\left(\sum_{\alpha+\beta=\gamma} a_{\alpha}b_{\beta}\right) \E{\mathcal{H}_\gamma}\nonumber\\
&=\sum_{\gamma\in\mathcal{I}}\left(\sum_{\alpha+\beta=\gamma} a_{\alpha}b_{\beta}\right) \E{\mathcal{H}_\gamma \cdot 1}\nonumber\\
&=\sum_{\gamma\in\mathcal{I}}\left(\sum_{\alpha+\beta=\gamma} a_{\alpha}b_{\beta}\right) \E{\mathcal{H}_\gamma \mathcal{H}_0}\nonumber\\
&=\sum_{\alpha+\beta=\gamma} a_{\alpha}b_{\beta}\, 0!=a_0 b_0=\E{X}\E{Y},
\label{prop:mean-fbm}
\end{align}
here we used the fact that the $\mathcal{H}$'s are an orthonormal basis.

Obviously the next step is to introduce the (fractional) Malliavin derivative for these processes or $\phi$-derivative, for $X\in L^2(\mu_{\phi})$ and $g\in L^2_{\phi}(\Re)$, the alternative version to (\ref{eq:malliavin}) reads
\begin{equation*}
D_{\Phi g}X(\omega)=\lim_{\delta\rightarrow 0} \frac{1}{\delta}\left\{X(\omega + \delta \int_{\Re}
(\Phi g)(u)\,du))-X(\omega)\right\},
\end{equation*}
where $(\Phi g)(z)=\int_{\Re}\phi(z,u)g(u)\,du$. Afterwards, if there exists a function $D^{\phi}(s) X$ such that
\begin{equation}
D_{\Phi g}X =\int_{\Re} (D^{\phi}_s X)\,g(s)\,ds,\> \forall g\in \LR,
\label{eq:malliavin-derivative}
\end{equation}
we say that $X$ is $\phi$-differentiable, and $D^{\phi}_s X$ is the $\phi$-differential. Let us point out some properties for the fractional Malliavin derivative, with $X$ defined be as always and $f,g:\Re\rightarrow\Re$ these are:
\begin{align}
D_{\Phi g}f(X)&=f'(X)D_{\Phi g}X,\\
D_{\Phi g}\langle\omega,f\rangle&=(f,g)_{\phi},\label{eq:dbrow1}\\
D^{\phi}_s \langle\omega,f\rangle&= \int_{\Re} \phi(u,s) f(u) du.
\label{eq:dbrow2}
\end{align}

Let us inspect another property for this operator. We can compute the second moment of $\mathcal{E}(f)\mathcal{E}(g)$. Because of $\E{\mathcal{E}(f)\diamond\mathcal{E}(g)}=1$,
\begin{align*}
1&=\E{\mathcal{E}(f)\mathcal{E}(g)} - \sum_{n=1} \frac{1}{n!} \E{(\langle\omega,f\rangle\langle\omega,g\rangle)^{\diamond n}}\\
&= \E{\mathcal{E}(f)\mathcal{E}(g)} - \sum_{n=1} \frac{1}{n!} (\E{\langle\omega,f\rangle\langle\omega,g\rangle})^{n}\\
&=\E{\mathcal{E}(f)\mathcal{E}(g)} - \sum_{n=1} \frac{1}{n!} (f,g)_{\phi}^{n}\\
&=\E{\mathcal{E}(f)\mathcal{E}(g)} - \exp(f,g)_{\phi} +1,
\end{align*}
we used property (\ref{eq:isometry}) in the last steps; so, $\E{\mathcal{E}(f)\mathcal{E}(g)} = \exp(f,g)_{\phi}$. We construct the following,
\begin{align*}
\E{(\mathcal{E}(h)\diamond\mathcal{E}(\delta f))(\mathcal{E}(h')\diamond\mathcal{E}(\varepsilon g))}&=\E{\mathcal{E}(h+\delta f)\mathcal{E}(h'+\varepsilon g)}\\
&= \exp (h+\delta f, h' +\varepsilon g)_\phi.
\end{align*}
Taking partial derivatives in $\delta$ and $\varepsilon$ afterwards yields
\begin{align*}
\E{\left(\mathcal{E}(h)\diamond\int_\Re f\,dB^H\right)\right.&\!\!\!\left.\left(\mathcal{E}(h')\diamond\int_\Re g\,dB^H\right)}\\
&=\exp(h,h')_\phi\left[(h,f)_\phi(h',g)_\phi+(f,g)_\phi\right]\\
&=\E{D_{\Phi_f}\mathcal{E}(h)D_{\Phi_g}\mathcal{E}(h')+ \mathcal{E}(h)\mathcal{E}(h')(f,g)_\phi}
\end{align*}
Henceforth, because any two $X, Y\in L^2(\mu_{\phi})$ can be decomposed by the span $\mathcal{E}$  we finally find,
\begin{multline}
\Eb{\left(X\diamond\!\int_{\Re}f(s)\,dB^H(s)\right)\!\left(Y\diamond\!\int_{\Re}g(s)\,dB^H(s)\right)}\\
=\E{(D_{\Phi f}X)( D_{\Phi g}Y) + XY(f,g)_{\phi}}.
\label{eq:cov}
\end{multline}
This equality will allow to change the integrator inside (\ref{eq:intfbm}) by a stochastic function $X: \Re \times \Omega \rightarrow \Re$ such that $\mathbb{E}\abs{X}^2_{\phi}<\infty$. That is, define the stochastic integral for fractional Brownian motion. 

The basic procedure consists of building a Riemann sum, replacing the standard product by the Wick one, 
\begin{equation}
S_n(X) := \sum_{i=0}^{n-1} X(z_i)\diamond (B^H(z_{i+1})-B^H(z_i)).
\label{eq:riemannsum}
\end{equation}
Observe that for any partition $\pi=\{z_0\leq z_1\leq\dots\leq z_{n-1}\}$,
\begin{align*}
\E{\sum_{i=0}^{n-1} X(z_i)\diamond (B^H(z_{i+1})-B^H(z_i))}&=\sum_{i=0}^{n-1}\E{X(z_i)\diamond (B^H(z_{i+1})-B^H(z_i))}\\
&=\sum_{i=0}^{n-1}\E{X(z_i)}\E{(B^H(z_{i+1})-B^H(z_i))}=0.
\end{align*}
Next, we compute the $L^2(\mu_\phi)$ norm of the former sum. Note that,
\begin{multline*}
\Eb{\left[X(z_i)\diamond(B^H(z_{i+1})-B^H(z_i))\right]\!\!\left[X(z_j)\diamond(B^H(z_{j+1})-B^H(z_j))\right]}\\
=\Eb{\int_{z_i}^{z_{i+1}} D^\phi_s F(z_i)\;ds\int_{z_j}^{z_{j+1}} D^\phi_u F(z_j)\;du + X(z_i)X(z_{j})\int^{z_{i+1}}_{z_i}\!\!\int^{z_{j+1}}_{z_j}\!\!\phi(s,u)\;duds}
\end{multline*}
is obtained from (\ref{eq:cov}); afterwards, 
\begin{multline*}
\E{(S_n(X))^2}=\sum^{n-1}_{i,j=0}\Eb{\int_{z_i}^{z_{i+1}}\!\! D^\phi_s F(z_i)\;ds\int_{z_j}^{z_{j+1}}\!\! D^\phi_u F(z_j)\;du\right.+ \\
\left.+ \,X(z_i)X(z_{j})\int^{z_{i+1}}_{z_i}\!\!\int^{z_{j+1}}_{z_j}\!\!\phi(s,u)\;duds}.
\end{multline*}
The continuity of $X$ and the existence of the trace of $D_s^\phi X$  \citep[\textbf{Theorem 3.9}]{paper:duncan} makes this sequence converge in $L^2(\mu_\phi)$ as $\abs{\pi}\rightarrow 0$, and it converges to 
\begin{equation*}
\E{\left(\int^L_0\!\!D^\phi_s X(s)\;ds \right)^2+ \abs{X}^2_\phi}.
\end{equation*}
In these conditions we say it is the \textit{fractional Brownian Stochastic Integral}:
\begin{equation}
\lim_{n\rightarrow\infty}S_n(X) := \int_0^L\!X(s)\> dB^H(s);
\label{eq:riemannsumlimit}
\end{equation}
moreover, the following equality holds
\begin{equation}
\int_0^L\!X(s)\> dB^H(s) =\int_0^L\!\! X(s) \diamond W^H(s)\;ds,
\label{eq:fBnoise-integral}
\end{equation}
while the integral on the left-hand side represents the limit (\ref{eq:riemannsumlimit}), the right-hand side is just the integral evaluated under the Hida expansion of the Wick product defined in (\ref{eq:hida-exp})--(\ref{eq:wick-prod}). 

Dropping the Wick product in definition (\ref{eq:riemannsum}) still produces a limit if the conditions given above are satisfied. This integral is the Stratonovich integral $\int^L_0 X(s)\circ dB^H(s)$ , because
\begin{align*}
\sum^{n-1}_{i=0} X(z_i)(B^H&(z_{i+1})-B^H(z_i))\\
&=\sum^{n-1}_{i=0} X(z_i)\diamond(B^H(z_{i+1})-B^H(z_i)) + \sum^{n-1}_{i=0} D_{\Phi \chi_{[z_i,z_{i+1})}}X(z_i)\\
&=\sum^{n-1}_{i=0} X(z_i)\diamond(B^H(z_{i+1})-B^H(z_i)) + \sum^{n-1}_{i=0} \int^{z_{i+1}}_{z_i}\!\!ds\; D^\phi_s X(z_i),
\end{align*}
we have
\begin{equation}
\int^L_0 X(s)\circ dB^H(s)= \int_0^L\!X(s)\> dB^H(s) + \int^L_0\!\! D^\phi_s X(s)\;ds.
\label{eq:stratonovich-int}
\end{equation}
This property is the counterpart from (\ref{eq:strato-bm}) in the Brownian motion case, as it should be if the analogy follows from (\ref{eq:noise-integral}) into (\ref{eq:fBnoise-integral}). But here both operators on the right-hand side can be evaluated without difficulty. We will finish this chapter with three theorems from \citet{paper:duncan} we will employ soon:
\begin{description}
\item[Theorem 4.2]\textit{Let $X(z)$ a stochastic process defined as above, and  $\sup_{z\in [0,L)} \mathbb{E}\abs{D^{\phi}_z X}^2_{\phi}<\infty$. Also, let $\eta(z)=\int_0^z X(s)\, dB^H(s)$. Then for $s,z\in [0,L)$}
\begin{equation}
D^{\phi}_s \eta(z) =\int_0^z\!D^{\phi}_s X(u)\, d B^H(u) + \int_0^z\! X(u) \phi(s,u)\, du.\label{thm:first}
\end{equation}

\item[Corollary 4.4]\textit{Let $\eta_z=\int_0^z f(s)\, dB^H(s)$ and  $F(z,x):\Re_+\times\Re\rightarrow \Re$, where $f\in L^2_\phi(\Re)$ is continous and $F$ has second continous derivatives. Then}
\begin{equation}
\begin{aligned}
F(z,\eta(z))&= F(0,0)+\int_0^z \frac{\partial F}{\partial s}(s,\eta(s))\, ds + \int_0^z\frac{\partial F}{\partial x} (s,\eta(z)) f(s)\, dB^H(s)\\
&+ \int_0^z\frac{\partial^2 F}{\partial x^2}(s,\eta(s)) \int_0^s \phi(s,s') f(s')\, ds'ds.
\end{aligned}
\label{eq:itof}
\end{equation}

\item[Theorem 6.11] \textit{If $X\in L^2(\mu_{\phi})$ then there exists a sequence $\{f_n \in L^2_{\phi}(\Re^n_+)\}_{n\in \N}$ such that $\sum^{\infty}_{n=1} \abs{f_n}^2_{\phi}< \infty$ and}
\begin{equation}
X= \E{X} + \sum^{\infty}_{n=1} \int_{\Re^n_+} f_n(s_1,\cdots,s_n)\> dB^H_{s_1}\cdots dB^H_{s_n}
\end{equation}\end{description}
where 
\begin{multline*}
\abs{f_n}^2_{\phi}=\!\! \int_{\Re^{2n}_+}\!\! f_n(s_1,\cdots,s_n)f_n(s'_1,\cdots,s'_n)\, \phi(s_1,s'_1)\cdots \phi(s_n,s'_n)\, ds_1\cdots ds_n\, ds'_1\cdots ds'_n .
\end{multline*}
$L^2_{\phi}(\Re^n_+)$ is the $n$-dimensional  space of symmetric functions. Given the base complete orthonormal base $\{\tilde{\xi}_n\}_{n\in\mathbb{N}}\subset L^2_\phi(\Re_+)$ then $L^2_{\phi}(\Re^n_+)$ is the completion of all function of the following form:
\begin{equation*}
f(s_1,\dots,s_n)= \sum_{1\leq k_1,\dots, k_n\leq k} a_{ k_1,\dots, k_n}\tilde{\xi}_{k_1}(s_1)\,\tilde{\xi}_{k_2}(s_2)\cdots\tilde{\xi}_{k_n}(s_n).
\end{equation*}
Associated to this functions we define the multiple integral as
\begin{equation*}
I_n(f)=\sum_{1\leq k_1,\dots, k_n\leq k} a_{ k_1,\dots, k_n} \int_\Re\tilde{\xi}_{k_1}(u)\, dB^H_u\diamond\,\int_\Re\tilde{\xi}_{k_2}(u)\, dB^H_u \diamond\cdots\diamond\int_\Re\tilde{\xi}_{k_n}(u)\, dB^H_u,
\end{equation*}
It is not difficult to prove \citep[\textbf{Lemma 6.6}]{paper:duncan} that given $f\in L^2_{\phi}(\Re^n_+)$ and $f\in L^2_{\phi}(\Re^m_+)$ it is
\begin{equation}
\E{I_n(f)I_m(g)}=\left\{\begin{aligned}
(f,g)_\phi &\quad\text{if }n=m\\
0\quad\quad &\quad \text{if }n\neq m\end{aligned}\right..
\label{eq:orthonorm-int}
\end{equation}
Moreover, for the iterated integral
\begin{multline*}
\int_{0\leq s_1\leq s_2<\cdots<s_n\leq t} f_n(s_1,\cdots,s_n)\> dB^H_{s_1}\cdots dB^H_{s_n}=\\
\int_0^t \left(\int_{0\leq s_1\leq s_2<\cdots<s_{n-1}\leq s_n}f_n(s_1,\cdots,s_{n-1},s_n)\> dB^H_{s_1}\cdots dB^H_{s_{n-1}}\right)dB^H_{s_n}
\end{multline*}
is $n!$ times $I_n(f)$.
\chapter{Stochastic Geometric Optics}\label{chp:stochastic}

Diverse experimental techniques have been devoted to the study of the optical properties of the turbulent atmosphere. Plenty of them are based on the analysis of the output of laser beams making their way through it. But also, controlled experiences had been developed for the laboratory, such as the experiments performed by \citet{consortini1,consortini2,consortini3}. These experiences apply Geometric Optics to interpret the data acquired. 

All these studies have their theoretical grounds on the precursor paper by \citet{paper:beckman}, who was able to find a nice relationship between the variance of the turbulent refractive index $\mu(\vect)$---being homogeneous and isotropic---and the variance of the laser beam wandering over a screen. As it was pointed out in Chapter \ref{chp:index}, he proposes (\ref{eq:beckman}) as covariance function because it gives meaning to the derivatives of the refractive index. Moreover, he pointed out that the Kolmogorov-like  structure functions ``\dots are mathematically fairly unmanageable''. The literature after him forgot this warning: modifications to his solution were given \citep[e.g.,][]{consortini1} but for the wrong covariance, the Kolmogorov structure function.

We do intent to show here, armed with our refractive index's model, that the ray-path equations are manageable. But this requires the Stochastic Calculus we have introduced in the last chapter. 

\section{Introduction}

Before start working in our approach we will briefly describe the differences between it and other works. In most of them the markovian model plays a central role. \citet{paper:leland} exaustively depicted by it. The markovian model provides the following covariance:
\begin{equation}
\E{\,\epsilon(\brho;z)\epsilon(\brho';z')}=\delta(z-z') A(\brho-\brho'),
\label{eq:markovian-cov}
\end{equation}
where $A$ is a differentiable function as defined in Appendix B. This covariance is associated to a process build from the Brownian motions' distribution space to a bounded linear operator $L$ on some Hilbert space $\mathcal{H}$; that is, $\epsilon = L(\dot{B}^{1/2})$. Since this operator can be described by using some kernel function whose coeficients are differentiable functions in $\rho$. Obviously, this model transfers all the discontinuities to the $z$-axis. 

For instance, let us illustrate the problem with the simple example: choose  $L(\dot{B}^{1/2})=\int^z_0 F(\brho;s) \dot{B}^{1/2}(s)\, ds$. Assuming $F$ is continuously differentiable in $\rho$, the following
\begin{equation}
\frac{\partial}{\partial x} L(\dot{B}^{1/2})(\brho;z) = \int^z_0 F_x(\brho;s)\, d B^{1/2}(s)
\label{eq:bm-derivative}
\end{equation}
is well-defined. On the other hand, the covariance of the original process is
\begin{equation*}
\E{ L(\dot{B}^{1/2})(\brho;z)\, L(\dot{B}^{1/2})(\brho';z')} = \int^{z\wedge z'}_0 F(\brho,s)F^{\ast}(\brho',s)\,ds;
\end{equation*}
therefore, differentiating the above by $\frac{\partial^2}{\partial x \partial x'}$ we find
\begin{equation*}
\frac{\partial^2}{\partial x \partial x'} \E{ L(\dot{B}^{1/2})(\brho;z)\, L(\dot{B}^{1/2})(\brho';z')} = \int^{z\wedge z'}_0 F_x(\brho,s)F^{\ast}_x(\brho',s)\,ds.
\end{equation*}
Henceforth, from equation (\ref{eq:bm-derivative}) we observe that
\begin{equation}
\frac{\partial^2 \E{ L(\dot{B}^{1/2})(\brho;z)\, L(\dot{B}^{1/2})(\brho';z')}}{\partial x \partial x'}=
 \E{ \frac{\partial}{\partial x}L(\dot{B}^{1/2})(\brho;z)\,\frac{\partial}{\partial x'} L(\dot{B}^{1/2})(\brho';z')}.
 \label{prop:commutative}
\end{equation}
It is this property the commonest property used in turbulent optics not regarding its original nature; that is, equation (\ref{eq:markovian-cov}) or the like. Moreover, we can also evaluate the fractal dimension of this type of processes. Let us use the \textit{Kolmogorv's criterion}\footnote{
Given a process $X(\vect)$, with $\vect$ in a closed domain $D$ in $\Re^d$. Assume that there exist positive constants $s, M$ and $\alpha_i, i=1,\dots,d$ with $\alpha_0^{-1} d=\sum^d_{i=1}\alpha_i^{-1}< 1$ satisfying
\begin{equation*}
\E{\abs{X(\vect)-X(\vect)}^s}\leq M \sum^d_{i=1}\abs{x_i-x'_i}^{\alpha_i}, \quad\text{for every } \vect,\vect'\in D.
\end{equation*}
Then it has a continuous modification $\tilde{X}$ such that
\begin{equation*}
\abs{\tilde{X}(\vect)-\tilde{X}(\vect)}\leq K(\omega) \sum^d_{i=1}\abs{x_i-x'_i}^{\beta_i}, \quad\text{for every } \vect,\vect'\in [0,1]^d,
\end{equation*}
holds for almost all $\omega$. The coefficients $\beta_i$ are arbitrary positive numbers less than $\alpha_i (\alpha_0 - d)/\alpha_0 s$. We call it a $(\beta_1,\dots,\beta_d)$-H\"older continuous process.
} \citep[\textbf{Theorem 1.4.1}, pg 31.]{book:kunita} for that. Let $n$ be an even integer, it is
\begin{multline*}
\Eb{\left[\int^z_0 F(\brho,s)\,dB^{1/2}_s-\int^{z'}_0 F(\brho',s)\,dB^{1/2}_s\right]^n}\\
=\Eb{\left[\int^z_0 (F(\brho,s)-F(\brho',s))\,dB^{1/2}_s + \int^z_{z'} F(\brho',s)\,dB^{1/2}_s\right]^n}.
\end{multline*}
Now, if we name $G_s\equiv G(\brho,\brho',s)=F(\brho,s)-F(\brho',s)$ and $H_s\equiv F(\brho',s)\chi_{[z',\infty)}(s)$ after applying the Newton's binomial theorem, then we will have a summatory with the following terms
\begin{equation*}
\binom{n}{j}\E{I_j(G)I_{n-j}(F)}, \quad\text{with}\quad I_n(f)=\left(\int^z_0\! f_s\, dB^{1/2}_s\right)^n,
\end{equation*}
these integrals can be turned into symmetric integrals as the ones shown in the latter chapter. We note from the orthonormal property of stochastic symmetric integrals---an equivalent to (\ref{eq:orthonorm-int}) for the Browinian case---that the only remaining are three:
\begin{align*}
\E{I_n(G)}=\E{I^2_{n/2}(G)}&=(n/2)! \left[\int^z_0\left(F(\brho,s)-F(\brho',s)\right)^2\,ds\right]^{n/2},\\
\E{I_n(F)}=\E{I^2_{n/2}(F)}&=(n/2)! \left[\int^z_{z'} F^2(\brho',s)\,ds\right]^{n/2},\\
\E{I_{n/2}(G)\,I_{n/2}(F)}&=(n/2)!\left[\int^z_{z'} (F(\brho,s)-F(\brho',s))F(\brho',s)\,ds\right]^{n/2}.
\end{align*}
It is not hard to find bounds to these,
\begin{align*}
(n/2)!\left[\int^z_0\left(F(\brho,s)-F(\brho',s)\right)^2\,ds\right]^{n/2}&\leq (n/2)! M_1 \norm{\brho-\brho'}^n ,\\
(n/2)! \left[\int^z_{z'} F^2(\brho',s)\,ds\right]^{n/2}&\leq(n/2)! M_2 \abs{z-z'}^{n/2} ,\\
(n/2)!\left[\int^z_{z'} (F(\brho,s)-F(\brho',s))F(\brho',s)\,ds\right]^{n/2}&\leq(n/2)!M_3 \norm{\brho-\brho'}^{n/2}\abs{z-z'}^{n/2}.
\end{align*}
Finally, using the property $\norm{\brho-\brho'}^n< 2^{n/2}(\abs{x-x'}^n+\abs{y-y'}^n)$ we have
\begin{equation}
\Eb{\left[L(\dot{B}^{1/2})(\brho;z)-L(\dot{B}^{1/2})(\brho';z')\right]^n}\leq C \left(\abs{x-x'}^n+\abs{y-y'}^n+\abs{z-z'}^{n/2}\right).
\end{equation}
Therefore, we observe that $\beta_{1,2}<(n-4)/n$ and $\beta_3<\tfrac{1}{2} (n-4)/n$. In particular,  $\min\{\beta_1,\beta_2,\beta_3\}=\beta_3<\tfrac{1}{2} (n-4)/n<1/2$. Using the H\"older continuity we observe this process gives a isoscalar fractal dimension less than $\inf\{3-\beta_3\}<2\tfrac{1}{2}$.    Moreover, the fact that 
\begin{equation*}
m^2\norm{\vect-\vect'}^2<\Eb{\left[\int^z_0 F(\brho,s)\,dB^{1/2}_s-\int^{z'}_0 F(\brho',s)\,dB^{1/2}_s\right]^2}\equiv\sigma^2,
\end{equation*}
provides us a bound for the \textit{potential theory}, and thus we will obtain---as we did in our first chapter---a isoscalar fractal dimension equal to $2\tfrac{1}{2}$. 

Therefore, not only this model does not match the covariance function but also does not provide the right dimension for the refractive index. It effectively allows some degree of differentiability but at the cost of eliminating some physical informatin from the refractive index covariance. Moreover, this markovian approach is not isotropic, and an isotropic version will inexorably lead to a non-differentiable process.

In particular, we may cite the work of \citet{consortini1}. They follow \citeauthor{paper:beckman}'s steps to evaluate the covariance of the displacements of a ray over a screen. Ending up with an equation of the form
\begin{equation*}
\Delta x = A \int^L_0\int^L_0 \frac{\partial\epsilon }{\partial x_i}\, dz\,dz',
\end{equation*}
where $A$ is some constant. Afterwards, the authors commutate the derivatives with the average. But they do not mention the markovian approximation as the cause of this, and soon after they replace the covariance function by the isotropic one. This violates the valid use of the commutation property (\ref{prop:commutative}); since an isotropic process does not provide derivatives, the above equation has a priori no meaning thus it is not true we can commute operators.
\begin{figure}
\begin{center}
\includegraphics[width=0.85\textwidth]{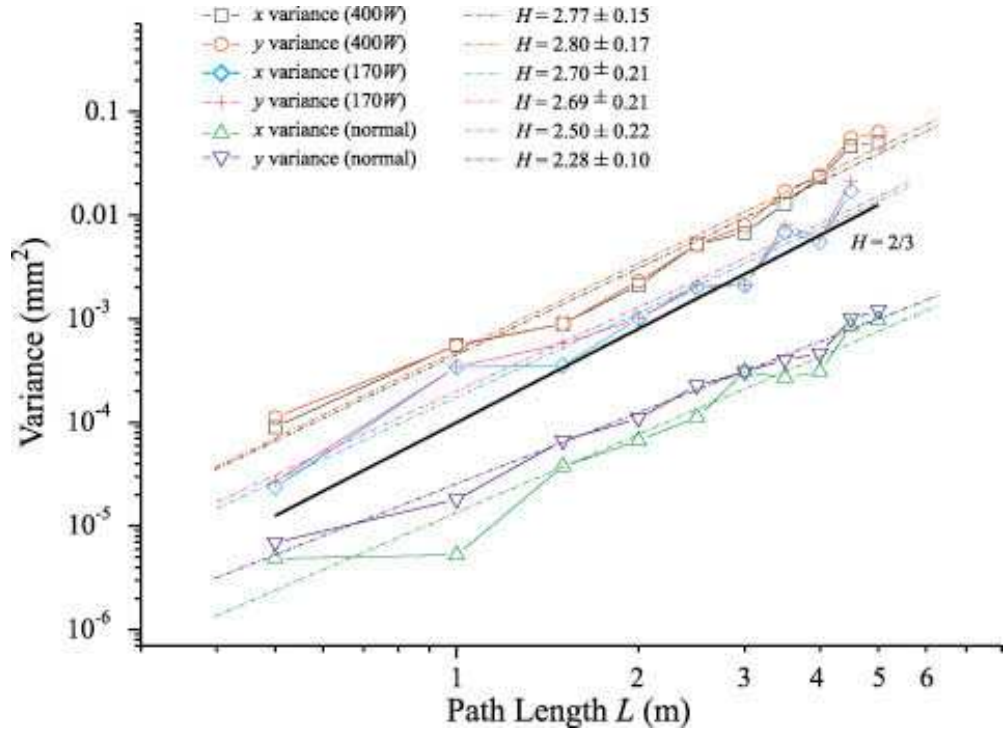}
\caption{The graphic shows the behavior of the $\log$ of the variance against the distance $L$. Interpolating lines can be calculated and the values of their tangents are shown.
\label{fig:consortini}}
\end{center}
\end{figure}

Moreover, we observe the markovian model is dependence on the characteristic lenght $L$ as $L^{1/2}$, then the former integral behaves as $L^{2\tfrac{1}{2}}$. The covariance of the displacements will grow proportional to $L^3$. This is a quality of the Brownian or markovian processes.

Finally, we observe in Figure \ref{fig:consortini} several plots of the logarithm of displacement covariance against the distance, based in the experimental data found in the work of \citet{consortini3}. In all cases the estimated power is below the theoretically estimated. Just in the higher cases the error is wide enough to cover the calculated value $\alpha=3$ and its value near it.  

Next, we will use the isotropic fractional Brownian model within the Geometric Optics to obtain an equation for the rays. We will show that under the correct framework a solvable stochastic equation exists and its result can be directly applied to the problem of a ray wandering over a screen.

\section{Stochastic Differential Equations in Geometric Optics}
\subsection{The ray-path equations}

As it is well-known, the Fermat's Extremal Principle is in the foundations of the Geometric Optics, that is, to find the ray trajectories we must find the variational solution to
\begin{equation}
\delta\!\left( \int n\, ds \right) = 0.
\label{fermat}
\end{equation}
We shall denote this solution by $q(\tau)$, and $\tau$ is a parameter with, in principle, no physical meaning. In Optics Treatises this parameter is usually replaced with one of the trajectory coordinates, which fulfills $dq_i/d\tau>0,$ and is thus called the propagation direction. But the election of this parameter can not be done at will \citep{synge}, since, for any parameterization chosen, the Optical Lagrangian
\begin{equation*}
L(q,\dot{q})=n(q)\norm{\dot{q}},
\end{equation*} 
($q,\dot{q} \in \Re^3$ are the position and velocity respectively\footnote{
$\Re^3$ is the configuration space. Usually is denoted by $Q$.
} ) is \textit{degenerated}. It is easy to show this property. Calculating the momentum,
\begin{equation}
p_i=\frac{\partial L}{\partial \dot{q}^i }= n(q)\frac{\dot{q}^i}{\norm{q}},
\label{eq:momentum}
\end{equation}
we see that the Lagrangian is rewritten as,
\begin{equation}
L(q,\dot{q})=\sum_i \frac{\partial L}{\partial \dot{q}^i}\, \dot{q}^i=\sum_i p_i \dot{q}^i.
\label{eq:lag}
\end{equation}
Since it is homogeneous in the velocities we can recalculate the momentum and find,
\begin{align*}
\frac{\partial L}{\partial \dot{q}^j} =\sum_i \left(\delta^{i j} \frac{\partial L}{\partial \dot{q}^i} + \frac{\partial^2 L}{\partial \dot{q}^i\partial\dot{q}^j} \dot{q}^j\right),\quad\text{then}\\
 0=\sum_i \frac{\partial^2 L}{\partial \dot{q}^i\partial\dot{q}^j}\, \dot{q}^j, \quad \forall\dot{q}^i.
\end{align*}
Therefore,  
\begin{equation*}
{\rm det}\left(\frac{\partial^2 L}{\partial \dot{q}^i\partial\dot{q}^j}\right)\equiv 0,
\end{equation*}
for any pair $(q,\dot{q})$: this matrix is singular.  As it is proved by \citet[\textbf{Theorem 7.3.3}]{book:marsden}, the solution is not univocally determined because the second order dynamics equation
\begin{equation*}
\ddot{q}^i = \left(\frac{\partial^2 L}{\partial \dot{q}^i\partial\dot{q}^j}\right)^{-1}\left[\frac{\partial L}{\partial q^i}+\frac{\partial^2 L}{\partial q^j\partial\dot{q}^i} \dot{q}^j\right],
\end{equation*}
obviously, can not be built. Nevertheless, equation (\ref{eq:momentum}) provides us more information, for it induces the following relation 
\begin{equation}
\norm{p}^2 = n^2(q),
\label{eq:condition}
\end{equation}
which indicates that the choice of coordinates and momenta is not free. 

The degeneracy of the Lagrangian should be worked out in the Hamiltonian framework because of the constraint we have just found. This problem of constrained Hamiltonians is known as Dirac's problem in the literature. The procedure is to reduce it to a Lagrangian problem: because given a set of constraint functions 
\begin{equation*}
\psi_1(q,\dot{q})= 0, \dots, \psi_k(q,\dot{q})= 0,\quad\text{for some}\quad(q,\dot{q})\in TQ,
\end{equation*}
associated to a Lagrangian $L$, there is a solution $q:[a,b]\rightarrow Q$ (critical point) if and only if $\exists\lambda_k:[a,b]\rightarrow \Re$ such that the following equation holds
\begin{equation}
\frac{d}{d\tau}\left(\frac{\partial L}{\partial\dot{q}^i}\right) -\frac{\partial L}{\partial q^i}=\sum^{k}\left\{\lambda_j\left[\frac{d}{d\tau}\left(\frac{\partial \psi_j}{\partial\dot{q}^i}\right) -\frac{\partial \psi_j}{\partial q^i}\right] + \dot{\lambda}_j\frac{\partial\psi_j}{\partial \dot{q}^i}\right\}
\label{eq:lagrange-multipliers}
\end{equation}
\citep[see][for a proof]{arnold}. Afterwards, we apply this theorem to  $\tilde{L}(p,q,\dot{p},\dot{q})=\theta(\dot{p},\dot{q})- H(p,q)$---where we have chosen the configuration space $P=T^\ast Q$, and $\theta$ is the canonical 1-form on $T^\ast Q$---with
\begin{equation}
\Psi_1(p,q)= 0, \dots, \Psi_k(p,q)= 0,
\label{eq:constraints-2}
\end{equation}
for $(p,q)\in P$. We thus find, from (\ref{eq:lagrange-multipliers}),
\begin{equation}
\left\{\begin{aligned}
\dot{q} &= \frac{\partial}{\partial p}\left(H + \sum^k \lambda_j \Psi_j \right)\\
-\dot{p}&= \frac{\partial}{\partial q}\left(H + \sum^k \lambda_j \Psi_j\right).
\end{aligned}\right.
\label{eq:dynamics-const}
\end{equation}
Also, we can calculate the dynamics equations for the constraints (\ref{eq:constraints-2}), that is,
\begin{equation}
\dot{\Psi}_i=\{\Psi_i,H\} + \sum^k \lambda_j\{\Psi_i,\Psi_j\},
\end{equation}
where $\{\cdot,\cdot\}$ is the Poisson bracket. The set of these equations is called \textit{compatibility condition set}: if $\{\Psi_i,\Psi_j\}\neq 0$ then the multipliers $\lambda_i$ are uniquely defined. Otherwise, if some $\{\Psi_i,\Psi_j\}$ are zero we have a new set of constraints, called \textit{secondary constraints}, that should be added to the original constraints. But, when we have $k=1$ and $\{\Psi_1,H\}=0$ then $\lambda_1$ is arbitrary.

Now going back to our problem, equation (\ref{eq:condition}) provides us with the constraint
\begin{equation*}
\Psi(p,q)=\frac{1}{2}\left[\norm{p}^2-n^2(q)\right], 
\end{equation*}
and the Hamiltonian $H$ obtained from the original Lagrangian is, combining equations (\ref{eq:momentum}) and (\ref{eq:lag}),
\begin{equation*}
H = \sum_i p_i \dot{q}^i-L = \sum_i p_i \dot{q}^i- \sum_i p_i \dot{q}^i\equiv 0.
\end{equation*}
We just need to build the new Hamiltonian, as (\ref{eq:dynamics-const}) suggests,
\begin{equation*}
\widetilde{H}(p,q):= H(p,q)+\lambda \Psi(p,q) = \lambda \Psi(p,q).
\end{equation*}
By doing so, we obtain the following dynamic equations
\begin{equation}
\left\{
\begin{aligned}
\dot{p}=& - \frac{\partial \widetilde{H}}{\partial q} = \lambda \frac{\partial \Psi}{\partial q}= \frac{\lambda}{2} \nabla_q n^2\\
\dot{q}=& \frac{\partial \widetilde{H}}{\partial p} = \lambda \frac{\partial \Psi}{\partial p}= \lambda p
\end{aligned}
\right.
\label{eq:hamilton}\end{equation}
and the constraint, 
\begin{equation}
0=\Psi(p,q)=\frac{1}{2}\left[\norm{p}^2-n^2(q)\right].
\label{eq:const}
\end{equation}
Finally, to ensure $\lambda$ is well defined we have to check the compatibility conditions. Because our original Hamiltonian is zero,  $\{H,\Psi\}=0$. The constraint is arbitrary; moreover, it is actually a smooth function on the constrained space that can be freely chosen. There are no \textit{secondary constraints} derived from the compatibility conditions so (\ref{eq:hamilton}) and (\ref{eq:const}) completely define our problem \citep{blago}.

Combining the pair (\ref{eq:hamilton}) of Hamiltonian equations yields to the following second order equation:
\begin{equation}
\frac{d}{d\tau}\left(\frac{1}{\lambda}\frac{d \dot{q}}{d\tau}\right)=\frac{\lambda}{2}\nabla_q n^2(q(\tau))\\
\label{eq:aceler}\end{equation}
with 
\begin{equation}
\norm{\dot{q}}^2= \lambda^2 n^2(q).
\label{eq:mod-const}
\end{equation}

We observe that with each selection we make for $\lambda$ the parameter $\tau$ is also set, i.e. if we choose $\lambda=n^{-1}$ then
\begin{equation}
\norm{\dot{q}}^2= 1\qquad{\rm and}\qquad ds=\norm{\dot{q}}d\tau=d\tau \label{eq:arclength}
\end{equation}
$\tau$ is then the arc-length. But selecting $\lambda=1$ gives us $ds=n d\tau$ and now the parameter is $\tau=\int ds/n.$

\subsection{Linearizing the trajectory equations}

The ray equations we have just found are evidently nonlinear, so in this section we are going to linearize them. But first, we must define the parameter $\tau$ and the refractive index. Let $n$ be the refractive index of the medium and $n_0$ its average, as it was defined in Section \ref{section:turbulent-index}, we write
\begin{equation}
n^2(q)= n_0^2 + \alpha\, \epsilon^\ast(q),
\label{index}
\end{equation}
we changed the stochastic permitivity $\epsilon(q)$ by $\alpha\,\epsilon^\ast$, where $\mathcal{O}(\epsilon^\ast)\sim 1$, so the strength of perturbation is due to $\alpha$. This term contains all the inhomogeneities of the media, thus when $\alpha=0$ the index is constant. Now, we suppose the solution to (\ref{eq:aceler}) can be expressed as power series on $\alpha$, i.e.,
\begin{equation}
q(\tau)= q_0 + \sum_{n=1}^\infty \alpha^n q_n(\tau). 
\label{eq:exp}
\end{equation}
Also, we should develop a series for the constraint function $\lambda$. Instead of using an undetermined constraint we will set its value beforehand: from all the possible parameterizations we choose  the arc-length (\ref{eq:arclength}). Now, we can rewrite equation (\ref{eq:aceler}) as follows
\begin{equation}
\frac{d^2 q}{d\tau^2}= \frac{1}{2}\left[\alpha\lambda^2 \nabla_q\epsilon^\ast + \frac{1}{\lambda^2}\left(\nabla_q\lambda^2 \cdot \dot{q}\right) \frac{d q}{d\tau}\right];
\label{eq:start}
\end{equation}
therefore, it is better to expand
\begin{equation}
\lambda^2(q)= \frac{1}{n^2(q)}= \frac{1}{n_0^2} +\sum^{\infty}_{n=1}\frac{(-1)^n {\epsilon^\ast}^n(q)}{n_0^{2n+2}}\; \alpha^n,
\end{equation}
in short we will write $\lambda^2_n := (-1)^n {\epsilon^\ast}^n(q)/n_0^{2n+2}$---note that $1/\lambda^2$ is exact.

Now, we must insert both power series in $\alpha$, the expansion (\ref{eq:exp}) and the latter for $\lambda^2$, into (\ref{eq:start}). We will obtain afterwards a family of differential equations from claiming the equality between the coefficients on the right and left for the same power. The second term on the right-hand side is tricky,
\begin{align*}
\frac{1}{\lambda^2}\left(\nabla_q\lambda^2 \cdot \dot{q}\right) =& (n^2_0+\alpha \epsilon^\ast)\left[-\left(\sum^{\infty}_{k=1}\frac{k\nabla\epsilon^\ast}{n^2_0}\;\lambda^2_{k-1} \alpha^k\right)\cdot\left(\sum^{\infty}_{n=0}\dot{q}_n\alpha^n\right)\right]\\
=&-(n^2_0+\alpha \epsilon^\ast)\sum^{\infty}_{n=1}\sum^n_{k=1}\frac{k\lambda^2_{k-1}}{n^2_0} (\nabla\epsilon^\ast\cdot \dot{q}_{n-k})\, \alpha^n \\
=&- \sum^{\infty}_{n=1}\sum^n_{k=1}k\lambda^2_{k-1} (\nabla\epsilon^\ast\cdot \dot{q}_{n-k})\alpha^n +\sum^{\infty}_{n=1}\sum^n_{k=1}k\lambda^2_{k} (\nabla\epsilon^\ast\cdot \dot{q}_{n-k})\,\alpha^{n+1}\\
=&-(\nabla\epsilon^\ast\cdot \dot{q}_0)\, \alpha+ \\
+&\sum^{\infty}_{n=2}\left[-\sum^n_{k=1}k\lambda^2_{k-1} (\nabla\epsilon^\ast\cdot \dot{q}_{n-k})+\sum^{n-1}_{k=1}k\lambda^2_k (\nabla\epsilon^\ast\cdot \dot{q}_{n-(k+1)})\right]\alpha^n\\
\end{align*}
\begin{align*}
\quad\quad\quad\quad=&-(\nabla\epsilon^\ast\cdot \dot{q}_0)\, \alpha +\nonumber\\
+&\sum^{\infty}_{n=2}\left[-\sum^n_{k=1}k\lambda^2_{k-1} (\nabla\epsilon^\ast\cdot \dot{q}_{n-k})+\sum^{n-1}_{k=2}(k-1)\lambda^2_{k-1} (\nabla\epsilon^\ast\cdot \dot{q}_{n-k})\right]\alpha^n\nonumber\\
=&- \sum^{\infty}_{n=1}\lambda^2_{k-1} (\nabla\epsilon^\ast\cdot \dot{q}_{n-k})\,\alpha^n.
\end{align*}
Thus, we finally have:
\begin{align}
\frac{d^2 q_0}{d\tau^2}&=0,\nonumber \\
\frac{d^2 q_1}{d\tau^2}&=\frac{1}{2n_0^2}\left[\nabla_q\epsilon^\ast - \left(\nabla_q\epsilon^\ast\cdot\frac{d q_0}{d \tau}\right)\frac{d q_0}{d\tau}\right],
\label{eq:linsol}
\end{align}
and\;when $n\geq 2:$
\begin{equation}
\frac{d^2 q_n}{d\tau^2}=\frac{1}{2}\left\{\lambda^2_n\; \nabla_q\epsilon^\ast - \sum^n_{m=1}\left[\sum^m_{k=1} \lambda^2_{k-1} \left(\nabla_q\epsilon^\ast\cdot\frac{d q_{m-k}}{d\tau}\right)\right]\frac{d q_{n-m}}{d\tau}\right\}.
\end{equation}
With the same criteria we obtain a constraint condition from (\ref{eq:mod-const}) for each differential equation above;
\begin{align}
\left(\frac{d q_0}{d\tau}\right)^2&=1,\label{eq:zero-const}\\
\frac{d q_1}{d\tau}\cdot\frac{d q_0}{d\tau}&=0,\label{eq:first-const} \\
\sum^{n}_{k=0} \frac{d q_k}{d\tau}\cdot\frac{d q_{n-k}}{d\tau}&=0,\qquad{\rm for\;all\;} n\geq 2,
\label{eq:constraints2}\end{align}
while the first constraint normalizes the zero-order solution, the second establishes it is orthogonal to the first-order solution.

We can readily find the solution for the zero-order equation in (\ref{eq:linsol}). The result is the linear relationship: $q_0(\tau)=\mathbf{a}\,\tau +\mathbf{b}$. Given that the initial condition to the problem is
\begin{equation}
q(0)=0,\label{bc}
\end{equation}
it implies that $\mathbf{b}=0$; also, using the constraint condition (\ref{eq:zero-const}) we obtain $\norm{\mathbf{a}}^2=1$, so we are free to choose the coordinate frame best suited to our purposes. Let us choose:
\begin{equation}
z\eZ:= q_0 = \tau \eZ,
\end{equation}
this will be our \textit{forward} direction of propagation. Now, we proceed to calculate the next differential equation: the first-order constraint condition (\ref{eq:first-const}) reads then
\begin{equation}
\frac{d q^z_1}{d z}=0.
\end{equation}
This and the initial condition (\ref{bc}) make the component along the $z$-axis null all over the ray trajectory. Of course, this constraint is compatible with its corresponding dynamical equation (\ref{eq:linsol}). Therefore, at first-order in $\alpha$ we just have a differential equation for the \textit{perpendicular (to the  direction of propagation) displacements}:
\begin{equation}
\frac{d^2}{d z^2}\,Q= \frac{\alpha}{2 n_0^2} \nabla_q\epsilon^\ast\left(z\eZ + Q + \dots\right),
\label{eq:non-linear}
\end{equation}
where $Q=\alpha q_1$. If we want to introduce the model we have previously introduced we just make the change $\epsilon^\ast\rightarrow \tilde{B}^H$, and so the parameter $\alpha=2\,l^H\sqrt{A_2}$. From the values given to structure constant  and the inner length, in the ideal case, we estimate $\alpha\sim 10^{-6}$. Afterwards, in order to examine the stochastic behavior of a wandering beam it will be enough to consider this first-order equation. 

With the tools we have used until now further analysis can not be done: the properties of the turbulent refractive index must be introduced in order to completely linearize the former equation.

\section{The Stochastic Volterra Equation}

As we already know, the gradient in equation (\ref{eq:non-linear}) should be given when looking for a solution; thus, we must provide a context to understand the previous equation. That is, a stochastic equation is not only determined by the type of process (the fractional Brownian motion in our case) attached to it, but also by the \textit{integro-differential} theory employed in defining its \textit{derivatives}. Moreover, there are distinctive stochastic integration methods whether $H>1/2$ or $H\leq 1/2$ \citep{decreusefond}. Here we are going to make use of the stochastic calculus exposed in the last chapter, so only the $H>1/2$ case will be considered. By doing so, either we are considering the inertial-diffusive range, in the following sense
\begin{equation*}
\zeta_n>\frac{1}{3},
\end{equation*}
or the anisotropic scalar situation $\zeta_n \rightarrow 1$ \citep{paper:elperin}. The physical interest about this particular situation comes from the many optical experiments where aspects regarding the creation of turbulence are neglected. Usually, heaters are used to create a turbulent medium but neither buoyancy or the temperature distribution are measured nor controlled, opposed to the conditions we have given through this work. Furthermore, the isotropic state of the index can be questioned.

Afterwards, because the turbulent refractive index oscillates around its mean value, it is expected that the light wanders around the \mbox{$z$-axis} over the screen (which corresponds to the case $\alpha=0$). So, the solution we are looking for should also have expectation zero. This is easily achieved by the formalism we introduced: the fractional It\^o integrals have expectation zero as it is seen from properties  (\ref{eq:fbm}), (\ref{prop:mean-fbm}) and their definition (\ref{eq:riemannsum}). Henceforth, using the model's definition (\ref{eq:fBm}) we can calculate the gradient of the refractive index:
\begin{align*}
\frac{\partial}{\partial x^i}\left[B^H(l_0^{-1}\norm{\vect})\right]&=\sum^\infty_{k=1} \frac{\partial}{\partial x^i}\left[\int_0^{l_0^{-1}\norm{\vect}}\left(\int_\Re \tilde{\xi}_k(s)\phi(s,u)\,ds\right)du\right]\mathcal{H}_{\epsilon^k}(\omega)\\
&= \sum^\infty_{k=1} \left.\left(\int_\Re \tilde{\xi}_k(s)\phi(s,u)\,ds\right)\right|_{s=l_0^{-1}\norm{\vect}}\frac{\partial}{\partial x_i}\left(l_0^{-1}\norm{\vect}\right) \mathcal{H}_{\epsilon^k}(\omega)\\
&= \frac{W^H\left(l_0^{-1}\norm{\vect}\right)}{l_0\norm{\vect}}\,x^i,
\end{align*}
for $ i=1,2$. This equation should be understood within $\mathcal{S}^*_H$---it has nothing to do with the usual concept of derivative: we have used the chain rule and the fractional white noise definition (\ref{eq:white-noise}).
 
The procedure to interpret equation (\ref{eq:non-linear}) requires to replace all the ordinary products containing stochastic variables by Wick products. If we do not follow this rule, the integrals should be interpreted as Stratonovich integrals. Thus, we observe from (\ref{eq:stratonovich-int}) that the mean value of the solution is non-zero, and we do not want that. Henceforth, 
\begin{equation}
\frac{d^2}{d z^2}\,Q= \frac{\alpha}{2\, l_0n_0^2}\ \left[\frac{W^H\!\!\left(l_0^{-1}\norm{z\eZ + Q}\right)}{\norm{z\eZ + Q}}\right] \diamond Q.
\label{volterra}
\end{equation}
Still, besides the changes, we have a non-linear \textit{stochastic} differential equation. Worse than that, we have a composition of two stochastic processes. We have to find a reasonable way to define it.  In the last chapter we explained that because any analytic function is expressed by a power series, it can be extended into the Hida space---whenever a stochastic process is an argument for it---by replacing the powers by Wick powers. We are going to extend this substitution rule. The representation for the noise in $\mathcal{S}^*_H$ is a series with analytic functions as components (\ref{eq:white-noise}); thus, it is valid \citep[private communication]{pc:oksendal}
\begin{equation*}
\int_{\Re} \phi(s,z)\,\tilde{\xi}_k(s)\, ds\rightarrow \int_{\Re} \phi^{\diamond}(s,Z)\,\tilde{\xi}_k(s)\,ds,
\end{equation*}
where $Z$ is some continuous stochastic process with $\E{Z}:=z_0\neq 0$, and $\phi^{\diamond}(s,\cdot)$ is the Wick representation of $\phi(s,\cdot)= H(2H-1) \abs{s- \cdot}^{2H-2}$. Now, we approximate $\norm{z\eZ+Q}\simeq z + Q^2/2z$ because $Q\sim \mathcal{O}(\alpha)$, and then evaluate the fractional white noise at $z +\alpha^2 Z(\omega)$:
\begin{multline*}
\phi^{\diamond}(s,z +\alpha^2 Z)\\
= H(2H-1)\abs{z +\alpha^2 Z-s}^{\diamond(2H-2)}=H(2H-1)\left[(z-s)+\alpha^2 Z\right]^{\diamond(2H-2)},
\end{multline*}
we have just took the positive part of the absolute value: it is enough for us examine this situation. If$\:\,\E{\alpha^2 Z}=\alpha^2 z_0$ then
\begin{multline*}
\left[(z-s)+\alpha^2 Z\right]^{\diamond(2H-2)}\\
=\left[(z-s)+ \alpha^2 z_0\right]^{2H-2} + \sum^{\infty}_{n=1}  \frac{\alpha^{2n}(2H-2)\cdots(2H-3-n)}{n!\left[(z-s)+ \alpha^2 z_0\right]^{n+2-2H}}\left(Z-z_0\right)^{\diamond n},
\end{multline*}
and all the terms in the series are of order higher or equal to 2 in $\alpha$. We just need to compare the first term against the deterministic coefficient in the white noise series expansion:
\begin{equation}
\phi(s,t+\alpha^2 z_0)-\phi(s,t)\sim z_0 (2H-2)  (t-s)^{(2H-3)}\alpha^2.
\end{equation}
This happens `coordinate' to `coordinate' in the fractional white noise decomposition, thus we have found
\begin{equation}
\frac{W^H\!(l_0^{-1}z)}{z} - \frac{W^H\!\!\left(l_0^{-1}\norm{z\eZ + Q}\right)}{\norm{z\eZ + Q}}
\sim \mathcal{O}(\alpha^2).
\label{eq:aprox}
\end{equation}
The first-order equation (\ref{volterra}) is unaffected by this replacement since these processes differ in $\alpha^2$. Finally, we arrived to the desired linear equation:
\begin{equation}
\frac{d^2}{d z^2}\,Q(z)= g \; \frac{W^H\!\!\left(l_0^{-1}z\right)\diamond Q(z)}{z},
\label{eq:diff-vol}
\end{equation}
we have set $g=\alpha/2\, l_0 n_0^2$ ($g\sim 10^{-3}$). 

\subsection{The stochastic Volterra equation and its solution}

The integral form of equation (\ref{eq:diff-vol}) is,
\begin{equation}
Q(z)= \dot{Q}_0 z + g \int^z_0\int^{s'}_0 \frac{W^H(l_0^{-1}s)}{s}\diamond Q(s)\: ds ds'.  
\label{eq:volterra}
\end{equation}
Let us set the following initial conditions $Q(0)=0$ and $\dot{Q}(0)\in \mathcal{S}_H^{\ast}$---the initial velocity is also uncertain. It can be simplified a bit more since 
\begin{align*}
\int^z_0\int^{s'}_0 \frac{W^H(l_0^{-1}s)}{s}\diamond Q(s)\:& ds\, ds'\\
&= \int_\Re\int_\Re \chi_{[0,z)}(s')\chi_{[0,s')}(s) \frac{W^H(l_0^{-1}s)}{s}\diamond Q(s)\: ds\, ds'\\
&=  \int_\Re\left(\int_\Re \chi_{[0,z)}(s')\chi_{[0,s')}(s)\, ds'\right)\frac{W^H(l_0^{-1}s)}{s}\diamond Q(s)\: ds\\
&= \int_\Re (z-s)\chi_{[0,z)} \frac{W^H(l_0^{-1}s)}{s}\diamond Q(s)\: ds\\
&=\int^z_0 \frac{(z-s)}{s} W^H(l_0^{-1}s)\diamond Q(s)\:ds.
\end{align*}
Thus we have a stochastic Volterra equation with (Fredholm) kernel:
\begin{equation*}
k^H(z,s):= g \frac{(z-s)}{s} \chi_{[0,z]}(s) W^H(l_0^{-1} s).
\end{equation*}
We will be interested in finding a solution on the (closed) interval $0\leq z\leq L$. The kernel is continous everywhere but $s=0$, and
\begin{equation}
\norm{k^H(z,s)}_{H,-q}\leq g \tilde{M}\; \chi_{[0,z]}(s)s^{-1}\abs{z-s},
\label{eq:notbounded}
\end{equation}
as can be seen from the bound (\ref{eq:bounda}).

Now we have to see what are the conditions that make equation (\ref{eq:volterra}) solvable. It should be, if we were able to apply a \textit{fixed-point theorem} to the above kernel. Therefore, proposing as \textit{ansatz} the usual resolvent for convoluted kernels, that is,
\begin{equation}
K_H(z,s)=\sum^{\infty}_{n=1} K^{(n)}_H(z,s),\label{eq:conv-kernel}
\end{equation}
such that 
\begin{align}
Q(z)&= \dot{Q}_0 z + \int^z_0 K_H(z,s)\diamond \left(\dot{Q}_0 s\right) ds\nonumber\\
&=\dot{Q}_0 \diamond\left[ z + \int^z_0 K_H(z,s) s \,ds\right]
\label{eq:solution}
\end{align}
with the $K^{(n)}_H$ given inductively by
\begin{align}
K^{(n+1)}_H(z,s)&= \int^z_s K^{(n)}_H(z,u)\diamond k^H(u,s)\; du,\quad {\rm with\quad }n\geq 1,\label{eq:iductive-ker}\\
K^{(1)}_H(z,s)&= k^H(z,s).
\label{eq:kernel-al}
\end{align}
It was found by \citet{book:holden} that  this is the unique solution for bounded kernels in the distribution Hida space. Their proof is based on the existence of a bound via the norm $\norm{\cdot}_{-1,-q}$. The same theorem can also be shown valid in the fractional Hida spaces with $\norm{\cdot}_{H,-q}$. But our kernel is unbounded, since the fractional white noise is continous and non-zero at $s=0$. 

\citet{grip} discuss this type of problematic kernels for normed spaces. Defined the space of continous functions $f:J\rightarrow K$  with norm $\norm{f}_{L^p(J)}=\left(\int_J \norm{f(s)}^p\,ds\right)^{1/p}$, where $J\subset\Re$ is not necessarily compact and $K$ is a Hilbert space---with $\norm{\cdot}$. Afterwards, they introduce a norm for the kernel $k$:
\begin{equation}
\norm{k}_{L^{p,p'}(J)}= \sup_{\substack{\norm{f}_{L^p(J)}\leq 1\\
\norm{g}_{L^{p'}(J)}\leq 1}} \int_J\int_J \norm{g(z)k(z,s)f(s)}\, dz\, ds,\quad\left(\frac{1}{p}+\frac{1}{p'}=1\right).
\label{def:ker-norm}
\end{equation}
Then, they proved that a resolvent solution exists whenever this norm is less than one (Corollary to \textbf{Theorem 3.9} in \citeauthor[p. 235]{grip}). This norm has also another property, using the H\"older inequality the following can be proved\footnote{
For a proof see Appendix C.
}:
\begin{equation*}
\norm{k}_{L^{p,p'}(J)}\leq \min\!\left\{\left[ \int_J\left(\int_J\norm{k(z,s)}^p ds\right)^\frac{p'}{p}\!\!\! dz\right]^\frac{1}{p'}\!\!\!\!,\left[\int_J\left(\int_J \norm{k(z,s)}^{p'}dz\right)^\frac{p}{p'}\!\!\! ds\right]^\frac{1}{p}\right\}.
\end{equation*}

The theorem and property above can be tracked back to the norm in the fractional Hida space. Hence, the same hypothesis applies for this stochastic Fredholm kernel defined $J=(0,L]$: $\norm{k^H}_{L^{p,p'}(J),-q}< 1$ for some $q>0$; moreover, 
\begin{multline}
\norm{k^H}_{L^{p,p'}(J),-q}
\leq \min\left\{\left[\int_J\left(\int_J\norm{k^H(z,s)}^p_{H,-q}ds\right)^\frac{p'}{p} dz\right]^\frac{1}{p'}\!\!,\right.\\
\left.\left[\int_J\left(\int_J\norm{k^H(z,s)}^{p'}_{H,-q}dz\right)^\frac{p}{p'} ds\right]^\frac{1}{p}\right\},
\label{cond:norm}\end{multline}
where $\norm{F}_{L^p(J),-q}=\left(\int_J \norm{F(s)}^p_{H,-q}\,ds\right)^{1/p}$. Then applying equation (\ref{eq:notbounded}) to the bounding condition (\ref{cond:norm}) we find\footnote{
Also in Appendix C.
} 
\begin{equation}
\norm{k^H}_{L^{p,p'}(J),-q}\leq g \tilde{M} p^{-1/p} \left(\frac{\pi p'}{\sin \pi p'}\right)^{1/p'}<1,
\label{eq:gral-bound}
\end{equation}
since $\tilde{M}$ is a small constant and $g\ll 1$. This guarantees the convergence of the proposed ansatz.

Unfortunately, the solution represented as a series of convoluted kernels, eqs. (\ref{eq:solution})---(\ref{eq:kernel-al}), is useless for calculations. Next, we will prove that a fractional chaos expansion exists for the solution. Let us take the second term in the Wick product of equation~(\ref{eq:solution}), it can be written 
\begin{align}
X(z)&= z+ \int^z_0 \left[\sum^{\infty}_{n=1} K_H^{(n)}(z,s)\right] s\; ds\nonumber\\
 & =z+ \sum^{\infty}_{n=1}\left[\int^z_0 K_H^{(n)}(z,s) s\; ds\right],
\end{align}
because it converges absolutely. The general term in this series can be written, using definition (\ref{eq:iductive-ker}),
\begin{align}
\int^z_0 K_H^{(n)}(z,s) &s\, ds \nonumber\\
&= g l_0^{1-H}\!\!\int^z_0\!\left[\int^z_{s_1} K_H^{(n-1)}(z,s_2)\frac{(s_2-s_1)(s_1-0)}{s_1}\,ds_2\right]\diamond  W^H(s_1)\,ds_1\nonumber\\
&= g l_0^{1-H}\!\!\int^z_0\!\int^z_{s_1} K_H^{(n-1)}(z,s_2)\frac{(s_2-s_1)(s_1-0)}{s_1}\,ds_2\,dB^H_{s_1}\nonumber\\
&= (g l_0^{1-H})^2\!\!\int^z_0\!\!\int^z_{s_1}\!\!\int^z_{s_2}\!\! K_H^{(n-2)}(z,s_3)\,\frac{(s_3-s_2)(s_2-s_1)(s_1-0)}{s_2s_1}\,ds_3\,dB^H_{s_2}dB^H_{s_1}\nonumber\\
&= (g l_0^{1-H})^n\!\!\int^z_0\cdots\int^z_{s_{n-1}}\int^z_{s_n}\frac{(z-s_n)(s_n-s_{n-1})\cdots(s_1-0)}{s_n s_{n-1}\cdots s_1}\,dB^H_{s_n}\cdots dB^H_{s_1}\nonumber\\
&= (g l_0^{1-H})^n\!\!\int_{\Re_+^n}(z-s_n)\prod^{n}_{i=1}\left[\frac{(s_i-s_{i-1})}{s_i} \chi_{[s_{i-1},z)}(s_i)\right]\,dB^H_{s_n}\cdots dB^H_{s_1}\nonumber\\
&= z(g z^H l_0^{1-H})^n\!\!\int_{\Re_+^n} f^{(n)}(s_n,\dots,s_1)\,dB^H_{s_n}\cdots dB^H_{s_1}.
\end{align}
Here we have used the self-similarity property (\ref{prop:selfsimilar}) to build the latter adimensional integrals, and defined 
\begin{equation}
f^{(n)}(s_n,\dots,s_1)=(1-s_n)\prod^{n}_{i=1}\left[\frac{(s_i-s_{i-1})}{s_i} \chi_{[s_{i-1},1)}(s_i)\right].\label{eq:chaos-term}
\end{equation}
with $s_0=0$. Now, we build the symmetrized form of the above function, that is,
\begin{equation*}
\hat{f}^{(n)}(s_n,\dots,s_1)=\frac{1}{n!}\sum_{\sigma\in\Pi}f^{(n)}(s_{\sigma_n},\dots,s_{\sigma_1}). \end{equation*}
Thus, it induces the following relation
\begin{equation}
\int_{\Re_+^n}\!\hat{f}^{(n)}(s_n,\dots,s_1)\,dB^H_{s_n}\cdots dB^H_{s_1}=\!\int_{\Re_+^n}\!\!f^{(n)}(s_n,\dots,s_1)\,dB^H_{s_n}\cdots dB^H_{s_1},
\end{equation}
because we can rename the each dummy variables of the $n!$  permutated terms to the normal order. Finally,
\begin{equation}
X(z)=z\left\{ 1+ \sum^{\infty}_{n=1}\int_{\Re^n_+}\left[\tilde{g}^n\, \hat{f}^{(n)}(s_n,\dots,s_1)\right]dB^H_{s_n}\cdots dB^H_{s_1}\right\},\label{eq:chaos-exp}
\end{equation}
where $\tilde{g}= l_0g (z/l_0)^H$. This will be nothing else but the fractional chaos expansion provided
\begin{equation}
\sum^{\infty}_{n=1}\tilde{g}^{2n} \sabs{\hat{f}^{(n)}}^2_{\phi}<\infty\label{eq:bounda}
\end{equation}
holds. In fact this condition express nothing else that the existence of the variance of the process,
\begin{equation}
\E{X^2(z)}= z^2 \left[1+ \sum^{\infty}_{n=1} \tilde{g}^{2n} \sabs{\hat{f}^{(n)}}^2_{\phi}\right]
\end{equation}
---we used property (\ref{eq:orthonorm-int}). The search of an upper bound for the succession of $\phi$-norms, given that the $\hat{f}^{(n)}$ are symmetric, is straightforward:
\begin{multline}
\sabs{\hat{f}^{(n)}}^2_{\phi}\\
= \int_{\Re^{2n}_+}\! \hat{f}^{(n)}(s_n,\dots, s_1)\hat{f}^{(n)}(s'_n,\dots,s'_1)\phi(s_n,s'_n)\cdots\phi(s_1,s'_1)\, ds_n\cdots ds_1\,ds'_n\cdots ds'_1\\
\leq \int^1_0\int^1_0\!\!\cdots\!\!\int^1_{s_{n-1}}\int^1_{s'_{n-1}}\phi(s_n,s'_n)\cdots\phi(s_1,s'_1)\, ds_n\cdots ds_1\,ds'_n\cdots ds'_1,
\label{eq:boundt}
\end{multline}
because of definition (\ref{eq:chaos-term}) and the fact $0<s_i-s_{i-1}\leq s_i$ (idem $0<s'_i-s'_{i-1}\leq s'_i$) the last inequality follows. Observing that
\begin{align}
\int^1_{s_{n-1}}\int^1_{s'_{n-1}}\phi(s_n,s'_n)&\; ds_n ds'_n \nonumber\\
&= H(2H-1)\int^1_{s_{n-1}}\int^1_{s'_{n-1}}\abs{s_n-s'_n}^{2H-2}ds_n\,ds'_n\nonumber\\
&= \frac{1}{2}\left[(1-s_{n-1})^{2H}+(1-s'_{n-1})^{2H}-\abs{s_n-s'_n}^{2H}\right]\leq 1,
\end{align}
we iteratively apply it in (\ref{eq:boundt}) to find: $\sabs{\hat{f}^{(n)}}^2_{\phi} \leq 1.$ Thus, the chaos expansion exists for all $ z\leq L$ whenever
\begin{equation*}
l_0 g \left( \frac{L}{l_0}\right)^H<1
\end{equation*}
is satisfied. From the definition of $g$ and the magnitude of the quantities\footnote{
See pages \pageref{eq:energy-balance} and \pageref{prop:zeta-correlation}.
} utilized here we have: 
\begin{equation}
L\ll l_0^{1-1/3H} (C^2_\varepsilon)^{-1/2H}.
\end{equation}
So, the condition above is always fulfilled.

\section{Ray-light Statistics: a Test Case}

In this section we will use the stochastic ray-equation solution to study the behavior of the displacements with respect to the characteristic variables of the system: $C^2_\varepsilon$, $l_0$ and $L$. We note that both coordinates of displacement are independent, and they also hold the same (non-coupled) differential equation. It is enough to consider a 1-dimensional case then. The parameter election (\ref{eq:arclength}), we have used in our treatment, also defines the meaning of the \textit{transversal} velocities, for they are the angles of deviation. Being the velocities continuous we can set,
\begin{equation*}
\dot{Q}_0:=\lim_{\epsilon \rightarrow 0}\dot{Q}(\epsilon) = \left.\theta\right|_{\epsilon=0}\in \mathcal{S}^*_H.
\end{equation*}
Since our solution is dependent of the initial refractive angle $\theta$, its behavior at the boundary, $\epsilon\rightarrow 0$, should be given.  This boundary is just the interface between turbulent and resting air. Henceforth, we will also model the initial angle as a fractional Brownian motion,
\begin{equation}
\theta(\epsilon)= c \int_{\Re_+}\chi_{[0,\epsilon)}(s)\, dB^H_s=c\, B^H(\epsilon),
\label{eq:initialangle}\end{equation}
the constant $c$ is adimensional and measures the strength of the noise. The length $\epsilon$ works as a kind of correlation distance, as it goes to zero we are examining the properties of the interface's short-range correlation. 

Besides, any stochastic process can be put in terms of the spans described in the past chapter, and these depend on the construction of stochastic integrals by step functions. In any case, even if the former model needs to be corrected---maybe the interface introduces long-range correlations---the next results are useful; since, they are the building blocks for more complex stochastic processes---see p. \pageref{eq:1series}.

Now, using the chaos expansion (\ref{eq:chaos-exp}) and the initial conditions given here, the solution (\ref{eq:solution}) is written :
\begin{align}
Q(z)&= \theta(\epsilon) \diamond X(z)\nonumber\\
&=z c \,B^H(\epsilon)\diamond \left( 1+ \sum^{\infty}_{n=1}\, \tilde{g}^n \int_{\Re^n_+}\!\! \hat{f}^{(n)}(s_n,\dots,s_1)\,dB^H_{s_n}\cdots dB^H_{s_1}\right)\!.
\end{align}
From the Wick product property (\ref{prop:mean-fbm}) we see
\begin{multline}
\E{Q(z)}\\=z c\,\E{B^H(\epsilon)}\E{1+ \sum^{\infty}_{n=1}\,\tilde{g}^n \int_{\Re^n_+}\!\! \hat{f}^{(n)}(s_n,\dots,s_1)\,dB^H_{s_n}\cdots dB^H_{s_1}}= 0 \cdot \E{1} = 0.
\end{multline}

The evaluation of the variance from experimental data is the most common topic in many works related to the optical properties of turbulence because it is directly related to the structure constant. Hence, we calculate it setting using property (\ref{eq:cov}),
\begin{align}
\E{Q^2(z)} &= c^2  \E{(B^H(\epsilon)\diamond X(z))^2}\nonumber\\
&= c^2 \left[\E{(D_{\Phi_{\chi_{[0,\epsilon)}}}X(z))^2} + \E{X^2(z)} \sabs{\chi_{[0,\epsilon)}}^2_{\phi}\right],
\end{align}
where we have set $X=Y$ and $f=g=\chi_{[0,\epsilon)}$. We have already evaluated $\E{X^2(z)}$ in the latter section. The fractional Malliavin derivative appearing at the right-hand side demands elaboration, property~(\ref{eq:dbrow1}) implies
\begin{equation}
D_{\Phi_{\chi_{[0,\epsilon)}}} X(z) = \int_{\Re_+} D^{\phi}_s X(z)\, \chi_{[0,\epsilon)}(s)\,ds.
\end{equation}
Since the $\phi$-differential is linear we have 
\begin{equation}
D^{\phi}_s X(z)= z \sum^{\infty}_{n=1}  \tilde{g}^n  D^{\phi}_s\!\left[ \int_{\Re^n_+}\hat{f}^{(n)}(s_n,\dots,s_1)\, dB^H_{s_n}\cdots dB^H_{s_1}\right].\label{eq:d-series}
\end{equation}
We are going to compute these derivatives now: let us fix $n\geq 2$, from the first theorem (\ref{thm:first}) we can commute the stochastic integral and $\phi$-differential,
\begin{multline}
D^{\phi}_s\!\!\left[ \int_{\Re^n_+}\hat{f}^{(n)}(s_n,\dots,s_1)\, dB^H_{s_n}\cdots dB^H_{s_1}\right]\\
= \int_{\Re_+}\!\! D^{\phi}_s\!\left[\int_{\Re^{n-1}_+}\!\hat{f}^{(n)}\; dB^H_{s_n}\cdots dB^H_{s_2}\right]dB^H_{s_1}+\int_{\Re^n_+}\!\hat{f}^{(n)}\; dB^H_{s_n}\cdots dB^H_{s_2}\phi(s,s_1)ds_1.
\end{multline}
Now, we recursively commute the operators, the $\phi$-differential and the Wick integral. Each time we do so another integral as the last one on the right-hand side of the equation above is added. After $(n-1)$ iterations we reach the innermost integral, thus we evaluate
\begin{equation*}
D^\phi_s\!\!\left[\int_{\Re_+}\hat{f}^{(n)}(s_n,\dots,s_1)\,dB^H_{s_n}\right]=\int_{\Re_+}\hat{f}^{(n)}(s_n,\dots,s_1)\phi(s_n,s)\,ds_n,
\end{equation*}
with the aid of property (\ref{eq:dbrow2}). Finally, 
\begin{multline}
D^{\phi}_s\!\!\left[ \int_{\Re^n_+}\hat{f}^{(n)}(s_n,\dots,s_1) dB^H_{s_n}\cdots dB^H_{s_1}\right] =\int_{\Re^{n}_+}\!\hat{f}^{(n)}\; \phi(s,s_n)ds_n dB^H_{s_{n-1}}\cdots dB^H_{s_1}+\\
+\dots+\int_{\Re^{n}_+}\!\hat{f}^{(n)}\; dB^H_{s_{n}}\cdots\phi(s,s_k)\;ds_k \cdots dB^H_{s_1}+\dots+\int_{\Re^n_+}\!\hat{f}^{(n)}\; dB^H_{s_n}\cdots dB^H_{s_2}\phi(s,s_1)ds_1\\
= n \int_{\Re^{n-1}_+}\!\left[\int_{\Re_+}\hat{f}^{(n)}\; \phi(s,s_n)ds_n \right]dB^H_{s_{n-1}}\cdots dB^H_{s_1},
\end{multline}
to arrive to the last equality the symmetry of $\hat{f}^{(n)}$ was employed. Instead, for $n=1$ we just use property (\ref{eq:dbrow2}):
\begin{equation}
D^{\phi}_s\!\!\left[ \int_{\Re_+}\!\!\hat{f}^{(1)}(s_1)\,  dB^H_{s_1}\right]=\int_{\Re_+}\!\!f^{(1)}(s_1)\,  \phi(s,s_1)\,ds_1.
\end{equation}

Afterwards, we can build the fractional Malliavin derivative (\ref{eq:malliavin-derivative}) from the series (\ref{eq:d-series}),
\begin{align}
\!\!\!\!\!D_{\Phi_{\chi_{[0,\epsilon)}}}& X(z)\! =  z\tilde{g}\Bigg\{\int_{\Re^2_+}\!\!f^{(1)}(s')\,\chi_{[0,\epsilon)}(s) \phi(s,s')\,ds'\, ds+\nonumber\\
+\sum^{\infty}_{n=1}&\,(n+1)\,\tilde{g}^n\!\!\!
\int_{\Re^n_+}\!\!\left[\int_{\Re^2_+}\!\hat{f}^{(n+1)}(s',\dots,s_1)\;\chi_{[0,\epsilon)}(s) \phi(s,s')\,ds' ds \right]\!dB^H_{s_n}\cdots dB^H_{s_1}\Bigg\};
\label{eq:diff-var}
\end{align}
its second moment is
\begin{multline*}
\E{(D_{\Phi_{\chi_{[0,\epsilon)}}} X(z) )^2}= z^2\tilde{g}^2\left\{\left[\int_{\Re^2_+}\!\!f^{(1)}(s')\,\chi_{[0,\epsilon)}(s) \phi(s,s')\,ds'\, ds\right]^2+\right.\\
+\left.\sum^{\infty}_{n=1}\,(n+1)^2\tilde{g}^{2n}\abs{\int_{\Re^2_+}\!\hat{f}^{(n+1)}(s',\cdot)\;\chi_{[0,\epsilon)}(s) \phi(s,s')\,ds' ds\:}^2_{\phi}\right\}
\end{multline*}
---we used the orthogonal property of these integrals. This series converges, we apply the same procedure as before to find a bound for the integrals. What is more, each norm appearing in the series is bounded by the zero term,
\begin{multline}
\abs{\int_{\Re^2_+}\!\hat{f}^{(n+1)}(s',\cdot)\;\chi_{[0,\epsilon)}(s) \phi(s,s')\,ds' ds\:}^2_{\phi}\leq \left[\int^{\epsilon}_0\int^1_0\!\!(1-s')\phi(s,s')\,ds'\, ds\right]^2 =\\
=\left\{\frac{1}{2(2H+1)}\left[1-\epsilon^{2H+1}-(1-\epsilon)^{2H+1}\right]-\frac{H}{2H+1}\,\epsilon^{2H+1}+\frac{1}{2}\right\}^2\leq \frac{1}{4}.\label{eq:var-bound}
\end{multline}
Thus, the existence and uniqueness of (\ref{eq:diff-var}) is guaranteed. 
Finally, we need the norm
\begin{equation*}
\sabs{\chi_{[0,\epsilon)}}_{\phi}^2= H(2H-1)\int^{\epsilon}_0\!\!\int^{\epsilon}_0 \abs{u-s}^{2H-2}du\,ds= {\epsilon}^{2H},
\end{equation*}
to calculate the variance of the displacements,
\begin{multline}
\E{Q^2(z)}=z^2 c^2 \left\{ \left[ \epsilon^{2H} + \tilde{g}^2\left(\int_{\Re^2_+}\!f^{(1)}(s')\;\chi_{[0,\epsilon)}(s) \phi(s,s')\,ds' ds\:\right)^{\!\!2}\right] + \right.\\
+\left.\sum^{\infty}_{n=1} \,\tilde{g}^{2n}\!\! \left[\epsilon^{2H}\sabs{\hat{f}^{(n)}}^2_{\phi}+(n+1)^2\tilde{g}^2\abs{\int_{\Re^2_+}\!\hat{f}^{(n+1)}(s',\cdot)\;\chi_{[0,\epsilon)}(s) \phi(s,s')\,ds' ds\:}^2_{\phi}\right]\right\}.
\label{eq:nonzeroterm}
\end{multline}

Now, as the correlation distance goes to zero we recover the initial condition. While terms coming from the second moment of $X(z)$ banish (they are all bounded and multiplied by $\epsilon^{2H}$), it is not the case with those coming from the fractional derivative. We will not go through copious calculations since we are interested in a general outline of the solution; thereof, the solution can be expressed as
\begin{multline}
\E{Q^2(z)}=\frac{z^2 c^2 \tilde{g}^2 }{4} + \\
+ z^2 c^2 \tilde{g}^2 \sum^{\infty}_{n=1} \,\tilde{g}^{2n}(n+1)^2 \left.\Bigg|\int_{\Re^2_+}\!\hat{f}^{(n+1)}(s',\cdot)\;\chi_{[0,\epsilon)}(s) \phi(s,s')\,ds' ds\:\Bigg|^2_{\phi}\right|_{\epsilon\rightarrow 0}.
\end{multline}
We can estimate a bound for the second term:
\begin{align*}
F(\tilde{g}^2)=4\sum^{\infty}_{n=1} \,\tilde{g}^{2n}(n+1)^2 \Bigg|\int_{\Re^2_+}\!\hat{f}^{(n+1)}(s',\cdot)\;\chi_{[0,\epsilon)}(s) \phi(s,s')&\left.\,ds' ds\:\Bigg|^2_{\phi}\right|_{\epsilon\rightarrow 0} \\
\leq & 4\sum^{\infty}_{n=1} \,\tilde{g}^{2n}(n+1)^2 \left(\frac{1}{4}\right)\\
\leq &\frac{1}{\tilde{g}^2}\sum^{\infty}_{n=2} n^2 \tilde{g}^{2n}.
\end{align*}
Now, because $x/(1-x)^2=\sum_{n=1}^\infty n x^{2n}$ it is
\begin{equation*}
\frac{1}{\tilde{g}^2} \sum^\infty_{n=2} n^2 \tilde{g}^{2n} = \frac{1}{\tilde{g}}\sum_{n=1}^\infty n^2 \tilde{g}^{2n-1}- 1 =\frac{1}{2\tilde{g}}\frac{d}{d\tilde{g}}\!\!\left[\frac{\tilde{g}^2}{(1-\tilde{g}^2)^2}\right] -1= \frac{(1+\tilde{g}^2)}{(1-\tilde{g}^2)^3}-1
\end{equation*}
and then $F(\tilde{g}^2)\leq \tilde{g}^2/(1-\tilde{g}^2)^3$, whenever $\tilde{g}<1$. Finally, replacing the values for $\tilde{g}$, we have
\begin{equation}
\E{Q^2(z)}=\frac{c^2}{4}\,  A\, z^{2H+2} l_0^{2/3-2H} \left[1 + F\left( A\, z^{2H} l_0^{2/3-2H}\right)\right].\label{example-solution}
\end{equation}
Furthermore, for the range of validity given in the past sections, the contribution of the function $F$ is less than $10^{-6}$. Thus, the first contribution to Malliavin derivative of $X$ completely characterize the variance once the interface's properties are defined. So, determining the behavior of the interface is crucial for the present model.
\addcontentsline{toc}{chapter}{\protect\numberline{Conclusions }{}}
\thispagestyle{fancy}
\renewcommand{\chaptermark}[1]{\markboth{#1}{}}
\renewcommand{\sectionmark}[1]{\markright{\thesection\ #1}}
\fancyhf{} 
\fancyhead[RO]{\bfseries\thepage}
\fancyhead[LO]{\bfseries Conclusions }
\renewcommand{\headrulewidth}{0.5pt}
\renewcommand{\footrulewidth}{0pt}
\addtolength{\headheight}{0.5pt} 
\fancypagestyle{plain}{%
\fancyhead{} 
\renewcommand{\headrulewidth}{0pt} 
}
\chapter*{Conclusions}
\setcounter{equation}{0}
\setcounter{footnote}{0}
\renewcommand{\theequation}{\textsc{Conclusions}.\arabic{equation}}

We started Chapter \ref{chp:index} making a revision of the up-to-date Passive Scalar Fields properties. Also, we have shown the refractive index is among them: this is well-known in Atmospheric Optics. Nevertheless, the progress made in Fluid Dynamics on scalar turbulence has hardly impacted turbulent propagation. Later on, we compared the properties a fair model should comply against those followed by actual optical models. Afterwards, we formulated the properties that make the family of \textit{isotropic} fractional Brownian motion a good candidate to simulate the turbulent refractive index:
\begin{itemize}
\item The Structure Function asociated to the index $\mu$, a scalar field, obeys the power law $\sim \norm{\vect}^{2H}$ with $0<H<1$. The value of the (Hurst) parameter $H$ depends on the state of the turbulence: $H>1/2$ for highly anisotropic scalar turbulence, and $H<1/2$, almost always near $1/3$, whenever the forces that generate the turbulence are not relevant.

\item The Structure Function dependence in $\vect$ induces a variance corresponding to a non-differentiable process.

\item It is assumed a Gaussian process. This is an \textit{ad-hoc} supposition widely used among the literature: it is specially applied when the process plays the role of a source in a fluid equation. This approximation is good whenever we are interested in the low moments associated to the stochastic process.
\end{itemize}

We have proved our proposed model (\ref{eq:fBm-permitivity}) fulfills all these conditions. Moreover, we obtained its fractal dimension, equation \eqref{eq:fractal-scalar-dimension}, matches the estimated by \citep{paper:constantin} for passive scalar: $\dim \epsilon^{-1} = 3-H$. Therefore, the exponent $H$ determines the state of the turbulence.

Finally, we must stress this model give us a local structure function for the refractive index---as suggested by \citet{paper:dario-garavaglia} and some preliminar experimental measures.

On Chapter \ref{chp:classic} we have shown under what conditions the wave-equation bring to us the paraxial approximation. Then, following \citet{paper:charnotskii} we have written its Green function using a path integral velocity representation.

All over this chapter the Markovian approximation is used. It has dominated the Atmospheric Optics scenario among the classic models. As it was noted, this model discriminates the direction of propagation, $z$, from the remaining coordinates. Implying a Brownian motion governs the behavior in that direction, that is, 
\begin{equation*}
\epsilon(\brho,z) \propto W(z).
\end{equation*}

Thus, we used this model to calculate the effects of the turbulence over a system of grids \citep{paper:perez-garavagia}. First, we have analized the image formation with and without turbulence. We observe the grids arrangement naturaly selects certains positions where the visibility is different from zero; that is, the formation of auto-images. After the introduction of the turbulence this property remains unchanged.

On the other hand, the quality of the image is degradated. It depends on the geometry of the grids, represented by $d$ and $L$, as it is shown in figures (\ref{fig:comp-turbulent-irradiance-pattern}) and (\ref{fig:turbulent-irradiance-pattern}). In the particular case $d\rightarrow\infty$, the visibility behaves as if the turbulence were absent in coincidence with \cite{paper:zavorotny}. 

Since the turbulent medium produces a cut-off in Fourier series for the irradiance pattern  introduces a method to evaluate the structure constant $C^2_\varepsilon$ as we showed. 

Finally, with the tools exposed in the third chapter we can advance to Chapter \ref{chp:stochastic} and solve the ray-equation coming from the Geometric Optics in the turbulent case.

At the introduction to this chapter we have shown substantial differences between our model and the Markovian approximation. We also proved that in the markovian case it is admisible to commutate derivatives and averages---this is assumed true in Optics not caring about the kind of process at hand. Also, this approximation has fractal dimension equal to $2\tfrac{1}{2}$, and thus it is not capable of determine the state of the turbulence. There are other models like thise. For example a set of fractal screens equispaced has dimension less than 2, and therefore it completely falls out of the foretold range for scalar fields. 

Next, we gave an alternative demonstration to \citet{synge}'s to find ray-equations for the (singular) optical lagrangian. The equations for ray light trayectories coming from this lagrangian are nonlinear, and then we proceed to linearize them.

We specifically studied the $H>1/2$ problem. The motivations for such a choice are various. From the mathematical point of view, we were able to define a composition of stochastic processes. Afterwards, we have shown the first order ray-equation corresponds to a Stochastic Volterra Equation. Moreover, we have shown that a unique analitical solution exists. This solution was expressed as kernel convolutions can be rewritten by means of a chaos expansion; thus, turning it into a manageable expression.

This analysis covers \textit{a priori} only those cases where average temperature gradients are relevant, that is, introduce strong anisotrophies. This behavior is likely to be found at the laboratory. Usually, these experiences disregard the process of turbulence making. It is considered that aligning a row of heaters along the ray trajectory (eventually using fans) and taking measures at a couple of meters high above them \citep[e.g.][]{paper:consortini-2} is enough to produce a completly developed turbulence. This asumption is at least ingenuous. As we have seen the conditions for isotropy and homogeneity are difficult to obtain. First, it must be known for certain the non-existence of a convective turbulence; that is, we must observe small Rayleigh numbers for the system (p. \pageref{eq:rayleighnr}). Also, an inertial tubulence does not necesarly produces an isotropic and homogeneous scalar turbulence. As was shown by \citet{paper:villermaux}, true isotropic and homogeneous scalars fields are obtained making the turbulent flow  circulate through some particular grid arrangement.
 
The validity range for our solution contains all the possible distances at the laboratory---$L\ll 10^6$m. Therefore, our problem is completly determined by the initial conditions; our election of the incoming angle as a fBm (\ref{eq:initialangle}) is the right choice given the behavior of the scalar quantities. Since this condition is related to short-range correlations we should find the constant $c$ depends on the inner scale and the structure constant.

Afterwards, when we use the solution (\ref{example-solution}) to estimate the variance of a laser beam going through the turbulence over a distance $L$. We obtain:
\begin{equation*}
\text{Var}[Q(L)]\simeq \frac{c^2}{4} A\, l^{2/3-2H}_0 L^{2+2H},
\label{eq:variance}
\end{equation*}
where the correction to this result is of order $\mathcal{O}(F)\sim 10^{-5}$. Moreover, this term comes from the Malliavin derivative. That is, the constant term, of order zero, does not contribute to the variance---see (\ref{eq:nonzeroterm}). We must stress that the anisotropy introduced by the mean flux should be observed in different constants $c$ at each axis.

Now, making  $H^+\rightarrow 1/2$ the displacements variance approachs to $A\, l^{-1/3}_0 L^{3}$. This is the behavior found by \citeauthor{consortini3}. It does not correspond to the Kolmogorov isotropic model, in accordance to the properties identified at the beginning, but to a brownian motion ($H=1/2$). That is, given a gaussian process with structure function like in (\ref{prop:zeta-correlation}) the result from Consortini does not hold. This can be achieved when $\zeta_n\searrow 1/3$; there exists anisotropy o convective turbulence.

Nevertheless, this result is coherent with the markovian model since the stochastic integrals exactly introduce such a dependence with the distance\footnote{
Observe \begin{equation*}
\E{Q^2(L)}\propto\E{\left(\int_0^L B(s) ds\right)^2}\sim L^2\times L^{2\cdot\frac{1}{2}},\end{equation*} 
according to the definition given at the beginning of this section.
}. The very same happens in our case. On the other hand, supposing the extension for $H\in[1/3,1/2)$ suggest a similar dependence. It should be changed  $l_0$ by $L$, and thus in that case $\text{Var } Q \sim L^{2.66}$ independently from $H$. Here we lack the knowledge to establish a value for the remaining quantities since the conditions on $c$ are undefined. Although, it is clear the power-law difference between this result and the markovian case is relatively small.

These results have been presented in \textsc{XIII Meeting on Nonequilibrium Statistical Mechanics and Nonlinear Physics (MEDYFINOL'02)}, December 9-13, 2002. Another version \citep{paper:dario} has been sent to be published.
\thispagestyle{fancy}
\renewcommand{\chaptermark}[1]{\markboth{#1}{}}
\renewcommand{\sectionmark}[1]{\markright{\thesection\ #1}}
\fancyhf{} 
\fancyhead[RO]{\bfseries\thepage}
\fancyhead[LO]{\bfseries Appendices}
\renewcommand{\headrulewidth}{0.5pt}
\renewcommand{\footrulewidth}{0pt}
\addtolength{\headheight}{0.5pt} 
\fancypagestyle{plain}{%
\fancyhead{} 
\renewcommand{\headrulewidth}{0pt} 
}
\appendix
\addtocontents{toc}{\contentsline {chapter}{\numberline{Appendices}}{}}
\addcontentsline{toc}{chapter}{\protect\numberline{A}{}}
\chapter*{{\appendixname} A}
\label{appxA}
\setcounter{footnote}{0}
\setcounter{equation}{0}
\renewcommand{\theequation}{A.\arabic{equation}}

\section*{Fractional Brownian motions}

Before introduce these processes, let us review some basic notions. To build a stochastic process a probability space $[\Omega,\mathcal{F},\mathbb{P}]$ must be provided, where $(\Omega,\mathcal{F})$ is a measurable space with measure $\mathbb{P}$ such that $\mathbb{P}(\Omega)=1$---it is the probability measure. The space $\Omega$ is an abstract space, whose characteristics are irrelevant for the present discussion. Now, let $(Y,\mathcal{Y})$ be another mensurable space and $T$ a parameter set (e.g., $\N$, \Re, etc.); thus, any given map $X:T\times \Omega\rightarrow Y$ is a \textit{stochastic process} if $\forall t\in T$
\begin{equation}
X^{-1}_t(B)=\{\omega:\, X(t,\omega)\in B\}\in \mathcal{F}, \text{ for any }B\in\mathcal{Y}.
\label{prop:stoch-process}\end{equation}
Here it will be only necessary to consider $Y=\Re$ and $\mathcal{Y}= \mathcal{B}(\Re)$ the Borel $\sigma$-algebra. 

There is an alternative definition: the canonical representation. We assign to each element $\omega\in\Omega$ a function $X(\omega)\in \Re^T$, where it is defined $\Re^T=\{ f:\; f(t): T\rightarrow\Re\}$. It is called \textit{realization} of the process. Also, we must provide a $\sigma$-algebra so within this space the property (\ref{prop:stoch-process}) is preserved: the \textit{Kolmogorov $\sigma$-algebra} $\mathcal{B}(\Re^T)$. It is generated by the cylinder sets
\begin{equation*}
\mathcal{Z}_t(B):= \{f\in \Re^T: f(t)\in B\},\text{ for }B\in \mathcal{B}(\Re).
\end{equation*}
Finally, from the original probability we can derive \textit{the distribution law of} $X$ over $(\Re^T,\mathcal{B}(\Re^T))$, 
\begin{equation*}
\mathbb{P}_X=\mathbb{P}\{\omega: X(\omega)\in A\}, \quad A\in \mathcal{B}(\Re^T).
\end{equation*}
Therefore, the triad $[\Re^T,\mathcal{B}(\Re^T),\mathbb{P}_X]$ constitutes the canonical probability space. The original abstract probability space $[\Omega,\mathcal{F},\mathbb{P}]$ is irrelevant if the distribution law of $X$ is given. That is, let us take 
\begin{equation*}
\mathcal{Z}_{t_1,\dots,t_n}(B_1\times\cdots\times B_n)=\{f\in \Re^T: (f(t_1),\dots,f(t_n))\in B_1\times\cdots\times B_n\}\in \mathcal{B}(\Re^T),
\end{equation*}
and thus define the $n$-dimensional distribution of the process $X$ as $P_{t_1,\dots,t_n}(B_1\times\cdots\times B_n):=\mathbb{P}_X(\mathcal{Z}_{t_1,\dots,t_n}(B_1\times\cdots\times B_n))$. Conversely, given  these finite distributions for all $n$ the probability law $\mathbb{P}_X$ can be recovered---Kolmogorov's Theorem \citep[p. 244]{book:shiryayev}. 

Henceforth, a Gaussian process can be build from the finite dimensional distributions, which are normal distributions; that is,  $\forall t_1,\dots,t_n\in T$ the random vector $(X_{t_1},\dots,X_{t_n})$ has distribution
\begin{equation}
\begin{aligned}
\mathbb{P}_{t_1,\dots,t_n}(x_1\leq X_{t_1}\leq x_1+ dx_1,\dots,x_n\leq &\, X_{t_n}\leq x_n+ dx_n)=\\
=\frac{dx_1\cdots dx_n}{\sqrt{(2\pi)^n\det \mathbf{V}}}\,& \exp\left\{-\frac{1}{2}(\mathbf{x}-\boldsymbol{\mu})^t\mathbf{V}^{-1}(\mathbf{x}-\boldsymbol{\mu})\right\},
\end{aligned}
\label{def:gaussian}
\end{equation}
where $\mathbf{x}=(x_1,\dots,x_n),\boldsymbol{\mu}\in\Re^n$ and $\mathbf{V}\in\Re^n\times\Re^n$ is a definite positive matrix. It is straightforward to find that $(\boldsymbol{\mu})_i=\E{X_{t_i}}$ is the mean value at times $t_1,\dots,t_n$, and $(\mathbf{V})_{i,j}=\text{Cov}(X_{t_i},X_{t_j})=\E{(X_{t_i}-\mu_i)(X_{t_j}-\mu_j)}$ is the associated covariance matrix, where $\mathbb{E}[\,\cdot\,]$ is the average calculated with $\mathbb{P}_X$. 
\label{page:stationary}
Finally, we can formally introduce stationarity for processes and its increments. The shift operator $\tau_s$ is defined $(\tau_s\circ f)(t)=f(s+t)$. A process is called \textit{stationary} if 
\begin{equation*}
\mathbb{P}_X\circ\tau^{-1}_s =\mathbb{P}_X.
\end{equation*}
It can be translated in terms of the finite distributions as  $P_{t_1+s,\dots,t_n+s}=P_{t_1,\dots,t_n}$ for any $n$. Moreover, for Gaussian process this is equivalent to 
\begin{equation*}
\text{Cov}(X_{t_i},X_{t_j}) =\text{Cov}(X_{\abs{t_i-t_j}},X_0),
\end{equation*}
 for any $(t_1,\dots,t_n)$. On the other hand, a process possess \textit{stationary increments} if the sets $\tau_{t+s}\circ X-X_{t+s}$ and $\tau_s\circ X- X_s$ has the same distribution. This implies its variance has the property
 \begin{equation}
 \E{(X_t-X_s)^2}=\E{(X_{\abs{t-s}}-X_0)^2}.
 \end{equation}

We are in conditions now to define the $1$-dimensional fractional Brownian motion. It is a Gaussian process with the following properties \citep{paper:mandelbrot-2}:
\begin{align}
B^H(s) &=0,\quad \text{almost surely,}
\label{eq:as-zero}\\
\E{B^H(s)}&=0,
\label{eq:mean-zero}\\
\E{B^H(s)B^H(t)}&= \frac{1}{2}\left[\abs{s}^{2H}+\abs{t}^{2H}-\abs{s-t}^{2H}\right],
\label{eq:fBm-covariance}\end{align}
for $s,t\in\Re$ and $0<H<1$. The exponent $H$ is called \textit{Hurst parameter}, because it was Hurst \citeyearpar{paper:hurst} who found Nile river's cumulated water flows vary proportional to $t^H$ ($t$ is the time) with $1/2<H<1$. In fact, the family of fBm processes should be separated in three subfamilies. When $H=1/2$ we recover the standard Brownian motion with covariance
\begin{equation}
\E{B^{1/2}(s)B^{1/2}(t)}= \min\{s,t\}:= s \wedge t.
\label{eq:Bm-covariance}\end{equation}
Now, given two dependent Gaussian random variables we have the property
\begin{equation}
\frac{\mathbb{E}(A|B)}{B}=\frac{\mathbb{E}(A B)}{\mathbb{Eb}(B^2)}. 
\label{eq:conditioned}\end{equation}
From this and the former equation whenever $s\geq t$, it is $\E{B^{1/2}(s)|B^{1/2}(t)}=B^{1/2}(t).$
This is a martingale, which has no long-memory and its intervals are not correlated. 

On the other hand, the case $1/2<H<1$ is the representative case of a long-memory process. That is, using equation (\ref{eq:conditioned}) again the conditioned average yields
\begin{equation}
\E{B^H(s)|B^H(t)}=\frac{1}{2}\left[\left(\frac{s}{t}\right)^{2H}+1-\left(\frac{s}{t}-1\right)^{2H}\right]B^H(t),
\label{eq:H-conditioned}\end{equation}
i.e., it is not a martingale\footnote{
If a process $X$ is a martingale it has the property $\E{X(s)|X(t)}=X(t)$ for $s\geq t$. 
}. As $s$ grows the conditioned mean behaves 
\begin{equation*}
\E{B^H(s)|B^H(t)}\simeq H \left(\frac{s}{t}\right)^{2H-1} B^H(t),													 \end{equation*}
and diverges at infinity. The long-range dependence is also represented by the divergence of the series $\sum^{\infty}_{n=1}\mathbb{E}[B^H(1)(B^H(n)-B^H(1))]=\infty$.

Finally, the case $0<H<1/2$ is left. In the very same way equation (\ref{eq:H-conditioned}) is valid for this range. We observe that $\E{B^H(s)|B^H(t)}\simeq \frac{1}{2} B^H(t)$: on the long-range it behaves like a martingale, since it posses short-memory. That is, the correlation of the increments is finite $0<\sum^{\infty}_{n=1}\mathbb{E}[B^H(1)(B^H(n)-B^H(1))]<\infty$ as the time goes to infinity. It is only zero for the Brownian motion, its increments are uncorrelated: it has no memory at all.

The fBm processes have stationary increments. We can evaluate the covariance for them, 
\begin{equation}
\begin{aligned}
\Eb{[B^H(t_4)-B^H(t_3)][B^H(t_2)-B^H(t_1)]}=& \\
=\frac{1}{2}\left[\abs{t_4-t_1}^{2H}+\abs{t_3-t_2}^{2H}\right.&\left.-\abs{t_4-t_2}^{2H}-\abs{t_3-t_1}^{2H}\right].
\end{aligned}
\label{eq:incr-covariance}\end{equation}
When $0\leq t_1=t_3< t_2=t_4$ we just have $\E{(B^H(t_2)-B^H(t_1))^2}=\abs{t_2-t_1}^{2H}$, and so the stationarity is accomplished for the increments. Moreover, if we pick in particular $t_4=t+h, t_3=t_2=t,$ and $t_1=0$, the covariance of the increments according to (\ref{eq:incr-covariance}) is 
\begin{equation*}
(t+h)^{2H}-t^{2H}-h^{2H}=\left\{\begin{aligned}
>0,\quad\text{if }H>\frac{1}{2}\\
=0,\quad\text{if }H=\frac{1}{2}\\
<0,\quad\text{if }H<\frac{1}{2}
\end{aligned}\right..
\end{equation*}
We observe that in the case $H>1/2$ consecutive increments tend to have the same sign, they are \textit{persistent}. For the Brownian motion these are as likely to have the same sign as the opposite. While in the last case $H<1/2$ the increments are more likely to have opposite signs, and so we call them \textit{anti-persitent}.

As we did with the translation, we define the operator $\phi_{\alpha}$ such that $(f\circ \phi_{\alpha})(t)=f({\alpha} t)$. We say then the process $X$ is scalar invariant if both $X\circ \phi_{\alpha}$ and $\alpha^{H}X$ have the same probability distribution for any $\alpha$ and $H$. For $0<H<1$ the fractional Brownian motion is scalar-invariant:
\begin{equation}
B^H(\alpha s) \overset{d}{=}\alpha^H B^H(s), \quad {\rm for\; any\, }\alpha,
\label{prop:selfsimilar}\end{equation}
where $\overset{d}{=}$ means they share the same probability law. Usually scalar-invariant processes are called \textit{self-similar} if they have stationary increments.

It is worth mentioning that given the change of variable $\Phi:T\rightarrow T'$, with $\Phi$ an invective transformation, the redefined stochastic process $Z_t=X_{\Phi(t)}$ is also a Gaussian process over $T$ with mean $\mu(t)=\mu'(\Phi(t))$ and covariance 
\begin{equation}
v(s,t)=v'(\Phi(s),\Phi(t)),
\label{prop:change}
\end{equation}
where $\mu$ and $\Phi$ are those defined for the original process $X$.

The fractional Brownian motion processes are not differentiable with probability 1\footnote{
It is said that a process is differentiable with probability 1 if $\mathbb{P}\{\xi_n\nrightarrow\xi\}=0$ as $n$ goes to infinity, and that a process converges in mean square if $\E{(\xi_n-\xi)^2}$ as $n\rightarrow \infty$. Both properties are equivalent when the process is Gaussian. \label{foot:conv}
}. This result can be proven from the following lemma found in  \citet[\textbf{\S 9.4}]{book:cramer}: \textit{If a Gaussian process $X$ is differentiable in $t$ with prob. 1 then $\exists(\partial^2 v/\partial s\,\partial t)(t,t)$}, $v(s,t)$ is the covariance function of the process. Therefore, the covariance of the fractional Brownian motion (\ref{eq:fBm-covariance}) implies
\begin{equation}
\frac{\partial v}{\partial s}(s,t)= \left\{\begin{aligned}
H [s^{2H-1} - (s-t)^{2H-1}],\quad s>t \\
H[s^{2H-1} + (t-s)^{2H-1}],\quad s<t
\end{aligned}\right..
\end{equation}
Then its diagonal is not derivable and so the second derivative does not exist.

We have introduced the 1-dimensional fractional Brownian motion process, its $n$-dimensional counterpart can be alternatively constructed through the covariance:
\begin{equation}
v(\mathbf{x},\mathbf{x'})= \frac{1}{2^n}\prod^n_{i=1} \left[\abs{x_i}^{2H}+\abs{x'_i}^{2H}-\abs{x_i-x'_i}^{2H}\right],
\label{eq:nfBm}
\end{equation}
for $\mathbf{x},\mathbf{x'}\in\Re^n$. 

\section*{Fractal Dimension}

The fractal dimension or Hausdorff dimension is defined through the \textit{Hausdorff measure} as follows.

Given a set $F$ first define a $\delta$-cover as the countable collection of sets $\{U_i\}$ covering $F$, each one with diameter not greater than $\delta$. Henceforth,
\begin{equation*}
\Hf{\delta}{s}(F)=\left\{\sum^{\infty}_{i=1}\text{diam}(U_i)^s: \{U_i\} \text{ is a $\delta$-cover of }F \right\},
\end{equation*}
where $\text{diam}(U) = \sup_{x,y\in U}\abs{x-y}$. Then the Hausdorff measure is defined as $\Hfm{s}(F) = \lim_{\delta\rightarrow 0}\Hf{\delta}{s}(F)$. Since this measure is either zero or infinity, the Hausdorff dimension of $F$ is univocally defined as
\begin{equation}
\dim_H F = \inf\{s:\Hfm{s}(F)=0\}=\sup\{s:\Hfm{s}(F)=\infty\},
\label{eq:hausdorff-dim}
\end{equation}
and thus,
\begin{equation}
\Hfm{s}(F)=\left\{\begin{aligned}\infty &\quad\text{if }s< \dim_H F\\
				 0 &\quad\text{if }s>\dim_H F\end{aligned}\right.
\label{prop:bound}
\end{equation}

The direct calculation of this dimension is almost impossible. It is usually done through some auxiliar theorems which provides us with upper an lower bounds for the Hausdorff dimension. 

In particular lower bounds to the Hausdorff dimension of a set $F$ can be found using the \textit{potential theory}. It is known \citep[\textbf{Theorem 4.13}, p. 64]{book:falconer} that given a mass distribution $\mu$ on $F$ such that
\begin{equation*}
\int\int \frac{\mu(dx)\mu(dy)}{\abs{x-y}^s}<\infty,
\end{equation*}
it is $\dim_H F\leq s$. The latter integral is known as \textit{$s$-potential}. Also, there are two other theorems from \citeauthor{book:falconer}'s book we would like to mention here without proof:

\textbf{Theorem 7.3:} \emph{For any sets $E\subset\Re^n$ and $F\subset\Re^m$}
\begin{equation}
\dim_H(E\times F)-n \leq \dim_H E + \overline{\dim}_B F.
\label{prop:cross-dim}
\end{equation}
Where $\overline{\dim}_B F$ is the \textit{upper box-counting dimension} defined as,
\begin{equation*}
\overline{\dim}_B F = \overline{\lim_{\delta\rightarrow 0}}\;\frac{\log N_\delta(F)}{-\log \delta}.
\label{prop:box-dim}
\end{equation*}
where $N_\delta(F)$ is the smallest number of cubes of side $\delta$ that cover $F$.

\textbf{Theorem 8.1:} \emph{If $E,F$ are Borel subsets of $\Re^n$ then}
\begin{equation}
\dim_H(E\cap(F+x))\leq \max\{0,\dim_H(E\times F)-n\}
\label{prop:intersect-dim}
\end{equation}
\textit{for almost all $x\in\Re^n$.}

\textbf{Theorem 8.2:} \emph{If $E,F\subset\Re^n$ be Borel subsets, and let $G$ be a group of transformations on $\Re^n$. Then}
\begin{equation}
\dim_H(E\cap\sigma(F))\geq \dim_H E +\dim_H F -n
\label{prop:intersect2-dim}
\end{equation}
\textit{for a set of motions $\sigma\in G$ of positive measure in the following cases:}
\renewcommand{\theenumi}{\emph{(\alph{enumi})}}
\begin{enumerate}
\item \emph{$G$ is the group of similarities and $E$ and $F$ are arbitrary sets.}
\item \emph{$G$ is the group of rigid motions, $E$ is arbitrary and $F$ is a rectificable curve, surface, or manifold.}
\item \emph{$G$ is the group of rigid motions and $E$ and $F$ are arbitrary, with either $\dim_H E> \frac{1}{2} (n+1)$ or $\dim_H F> \frac{1}{2}(n+1)$.}
\end{enumerate}
\addcontentsline{toc}{chapter}{\protect\numberline{B}{}}
\chapter*{{\appendixname} B}
\label{appx}
\setcounter{equation}{0}
\renewcommand{\theequation}{B.\arabic{equation}}

\section*{Markovian Model for the Turbulent Refractive Index}

The markovian model we introduced in Chapter \ref{chp:index} determines a preferred direction of propagation, let us say the $z$-axis, and thus the behavior across this direction is different from those perpendicular to it. That is, its increments the are independent, so they do not have memory of their past. This property, as we mentioned earlier, describes a martingale or markovian process. 

Here we will show how the function $A$ in equation (\ref{eq:markovian}) can be built from the original structure function. Let us begin with a locally homogeneous process $X$ having as structure function the following:
\begin{equation*}
D_X(\vect)=\langle \abs{X(\vect+\vect')-X(\vect')}^2\rangle.
\end{equation*}
Moreover, if we assume it is stationary and Gaussian, as discussed on page \pageref{page:stationary},  its correlation function has a spectral representation (\citet{book:shiryayev}, p. 387)
\begin{equation}
B_X(r):=\langle X(\vect+\vect')X(\vect')\rangle= \int_{\Re^3}\!\! d^3k\, F_X(\wn)\, e^{i \vect\cdot\wn}.
\label{eq:correlation-representation}\end{equation}
Because both functions are related by the equation $D_X(\vect)= 2B_X(\vect)-B_X(0)-B^{\ast}_X(0)$ we turn the former into
\begin{equation}
D_X(\vect)=2\int_{\Re^3}\!\! d^3k\, F_X(\wn) [1- \cos (\vect\cdot\wn)],
\label{eq:spectral-representation}\end{equation}
when $X$ is the turbulent refractive index the spectrum is the one discussed earlier in Section \ref{section:kraichnan-model}. 

Now, taking the inverse transform of (\ref{eq:correlation-representation}) and using (\ref{eq:markovian}) we find:
\begin{equation*}
A(\brho)= 2\pi \int_{\Re^2}\!\!d^2\kappa\; F_X(\Bkappa,0)\; e^{i\Bkappa\cdot\brho};
\end{equation*}
also, it is
\begin{equation*}
A(\brho)=\int_{\Re}\!\!dz B_X(\brho,z).
\end{equation*}
Besides, when the process is isotropic the spectrum only depends on the absolute value of the wavenumber and thus
\begin{equation}
A(\brho)= 4\pi^2 \int^{\infty}_0\!\!\kappa d\kappa\; F_X(\kappa,0)\; J_0(\kappa\norm{\brho}).
\label{eq:2-iso-spectral-representation}\end{equation}
Comparing equations (\ref{eq:spectral-representation}) and (\ref{eq:2-iso-spectral-representation})  we define the structure function over the $(x,y)$-plane as 
\begin{equation}
H(\brho)=\frac{1}{\pi}\left[A(0)-A(\brho)\right].
\end{equation}
In particular, suppose the power spectra has the form:
\begin{equation}
F_X(\Bkappa,0;z)= \frac{\Gamma(p+2)}{4\pi^2}\sin \frac{\pi p}{2}\; C^2_{\epsilon}(z)\norm{\Bkappa}^{-p-3},
\label{eq:local-power-spectra}\end{equation}
then we find 
\begin{equation}
H(\brho,z)=\frac{\Gamma(p+2)}{\Gamma\!\left[(p+3)/2\right]^2}\sin \frac{\pi p}{2}\; C^2_{\epsilon}(z)\norm{\brho}^{p+1}.
\label{eq:2-structure-function}\end{equation}

\section*{Validity Range of the Path Integral Representation}

To study the validity range of the Feymann's path integral representation we will look at the energy flux of a point source radiation through a pupil. If the pupil's transfer function is $O(\Vect)$ then we have
\begin{equation*}
P\propto \int_{\Re^2}\!\!  d^2R\; O(\Vect) \abs{G(\mathbf{0},L;\Vect,0)}^2.
\end{equation*}
The reciprocity principle implies that this is equivalent to the irradiance $I$ evaluated at the point $(\mathbf{0},L)$ provided the initial irradiance distribution function is the same as the transfer function.

The flux $P$ is a stochastic variable, so we can evaluate its normalized variance or \textit{scintillation index}, that is,
\begin{equation*}
\sigma^2_P = \frac{\mean{P^2}-\mean{P}^2}{\mean{P}^2}.
\end{equation*}
This quantity is an indicator of the type of approximation needed to solve a given propagation problem. Since the turbulent refractive index is a Gaussian process the mean and free-propagation fluxes coincide, that is, $\mean{P}=P_0\propto \Sigma k^2/ 4\pi^2 L^2$ with $\Sigma= \int_{\Re^2}\!d^2 r A(\vect)$ the effective area of the pupil. 

Therefore, we can compare the energy flux of the free propagating wave against the flux in the turbulent case. Combining equations (\ref{eq:inhomogeneous-green-2}), (\ref{eq:inhomogeneous-term}), and  (\ref{eq:irradiance}) evaluated at $\Vect=0$ we obtain
\begin{multline}
\frac{\mean{P^2}}{\mean{P}^2}=\sigma^2_P+1 = \frac{4\pi^2 L^2}{\Sigma k^2} \int_{\Re^2}\!\! d^2 r\;C_A(\vect) \int\!\!\!\int \mathcal{D}^2v_1(\zeta)\mathcal{D}^2v_2(\zeta)\, \exp\!\left[ik\int^L_0\!\!d\zeta\; \mathbf{v}_1(\zeta)\cdot\mathbf{v}_2(\zeta)\right]\\
\times \exp\!\left\{-\Phi\left[L,\vect_1(\zeta),\vect_2(\zeta)\right]\right\}\delta^{(2)}\!\left[\int^L_0\!\!d\zeta\,\mathbf{v}_1(\zeta)\right]\delta^{(2)}\!\left[\int^L_0\!\!d\zeta\, \mathbf{v}_2(\zeta)\right],
\label{eq:regime}
\end{multline}
where 
\begin{equation*}
C_A(\vect)=\frac{1}{\Sigma}\int d^2R\;A\!\left(\Vect +\frac{\vect}{2}\right)A\!\left(\Vect -\frac{\vect}{2}\right),
\end{equation*}
and
\begin{align*}
\Phi\left[L,\vect_1(\zeta),\vect_2(\zeta)\right]&=\frac{\pi k^2}{4}\int^L_0\!\!dz \left\{2H[\vect_1(z),z]-2H[\vect_2(z),z]-H[\vect_1(z)+\vect_2(z),z]\right.\nonumber\\
&\left.-H[\vect_1(z)-\vect_2(z),z]\right\}
\end{align*}
with $\vect_1(z)=\int^L_z\!\!d\zeta\, \mathbf{v}_1(\zeta)$,\, $\vect_2(z)=\int^L_z\!\!d\zeta\, \mathbf{v}_2(\zeta)+(1-z/L)\,\vect$, and  
the function $H$ is defined as in the former section. Because a more general situation is studied in Chapter 2, we have chosen not to give a detailed description for the calculations that lead to equation (\ref{eq:regime}).

We observe the strength of the turbulence is measured by the exponential factor in the latter equation. Since its arguments have no dimensions, we can show that $\sigma^2_P$ depends on two dimensionless parameters: the \textit{Fresnel number} $\Omega=k a^2/L$ corresponding to the pupil effective aperture size $a$, and $q=k\rho^2_0/L$ obtained from the \textit{spherical wave coherent radius} condition $D(\rho_0,L)=1$ \citep[p. 228]{paper:charnotskii-2}, where:
\begin{equation*}
D(\vect,L)=\frac{\pi k^2}{2}\int^L_0\!\!dz\,H\left[\left(1-\frac{z}{L}\right)\vect,z\right]\simeq\frac{\pi k^2}{2}\, r^{5/3}\int^L_0\!\!dz\;C^2_{\epsilon}(z)(1-z/L)^{5/3}
\end{equation*}
---we used (\ref{eq:2-structure-function}) for $p=2/3$. 

\begin{figure}
\begin{center}
\includegraphics[width=0.65\textwidth]{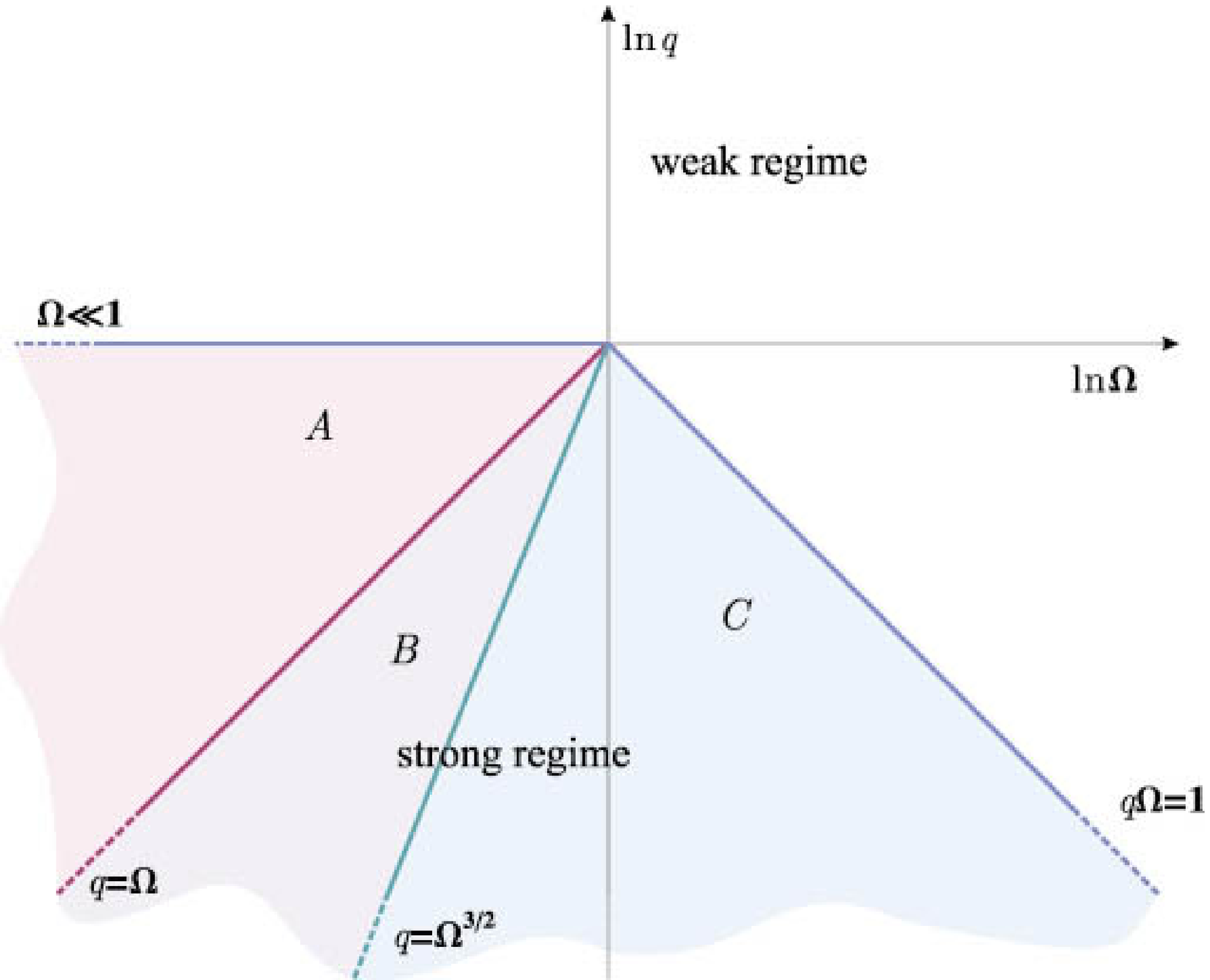}
\caption{The graphic displays the weak and strong regime regions. The latter is also divided into three subregions: within the region $A$ ($\Omega\ll q\ll 1$) the scintillation index is asymptotically equal to one, in $B$ ($q\ll\Omega\ll q^{2/3}$) is $\sigma^2_P=\mathcal{O}(q/\Omega)$, while in the last region ($q^{2/3}\ll\Omega\ll q^{-1}$), $C$, is $\sigma^2_P=\mathcal{O}(q^{1/3})$.
\label{fig:channels}}
\end{center}
\end{figure}

Thus, we define the \textit{weak scintillation} regime as the set of points $(\Omega,q)$ where $\sigma_P$ is asymptotically close to the first term of the Taylor expansion of $\exp(-\Phi).$ It is found \citep{paper:dashen, paper:tatarskii-1, paper:tatarskii-2} that this condition is reached when $q\gg 1$ and $\Omega\ll 1$, or $q\gg \Omega^{-1}$ and $\Omega\gg 1.$ Otherwise, the complement to this region corresponds to the \textit{strong scintillation} regime (Figure \ref{fig:channels}).
Defined as the region where $\sigma^2_P$ is asymptotically close to the coherent channel expansion. That is produced from two contributions: the \textit{main} channel expansion 
\begin{multline}
\exp\left(-\Phi\left[L,\vect_1(\zeta),\vect_2(\zeta)\right]\right) \\
=\exp\left\{-\frac{\pi k^2}{2}\int^L_0\!\!dz\,H[z,\vect_1(z)]\right\}\left\{1-Q[L,\vect_1(z),\vect_2(z)]+\dots\right\},
\label{eq:main-channel}\end{multline}
 where,
\begin{multline}
Q[L,\vect_1(z),\vect_2(z)]\\
=\frac{\pi k^2}{4}\int^L_0\!\!dz \left\{2H[z,\vect_2(z)]-H[z,\vect_1(z)+\vect_2(z)]-H[z,\vect_1(z)-\vect_2(z)]\right\},
\end{multline}
and the \textit{additional} coherence channel expansion obtained from $\vect_1$ and $\vect_2$ interchanging positions in (\ref{eq:main-channel}). The main idea behind is that the function $\Phi$ is less than unity in one of two regions $\abs{\vect_1}\sim\rho_0$ and $\abs{\vect_2}\sim\rho_0$. The coherent channel expansion is thus the sum of these two contributions into the scintillation definition
\begin{equation}
\sigma^2_P= M_2+M_3+\dots+N_1+N_2+\dots
\end{equation}
where $M_i$ corresponds to the contribution of the main channel and $N_i$ to that of the additional one. By completeness we give here the first terms of these expansions:
\begin{align}
N_1&=\frac{1}{\Sigma}\int_{\Re^2}\!\!d^2r\;C_A(\vect)\;\exp\left[ -D(\vect,z)\right]\\
N_2&=\frac{2\pi k^2}{\Sigma}\int^L_0\!\!dz\!\int_{\Re^2}\!\!d^2r\!\int_{\Re^2}\!\!d^2\kappa\; F_{\epsilon}(\Bkappa,0;z)\, C_A(\vect)\nonumber\\
&\times\sin^2\!\left[\left(1-\frac{z}{L}\right)\frac{\vect\cdot\Bkappa}{2}-p\!\left(\frac{z}{L},\frac{z}{L}\right)\frac{L}{2 k}\kappa^2\right]\nonumber\\
&\times\exp\!\left\{-\frac{\pi k^2}{2}\int^L_0\!\!dz'\;H\!\left[\left(1-\frac{z'}{L}\right)\vect-p\!\left(\frac{z'}{L},\frac{z}{L}\right)\frac{L}{k}\Bkappa,z'\right]\right\}
\end{align}
and
\begin{multline}
M_2=2\pi k^2\int^L_0\!\!dz\!\int_{\Re^2}\!\!d^2\kappa\,F_{\epsilon}\!(\Bkappa,0;z)\,\abs{\hat{A}\left[\left(1-\frac{z}{L}\right)\Bkappa\right]}^2\sin^2\!\left[p\!\left(\frac{z}{L},\frac{z}{L}\right)\frac{L}{2k}\kappa^2\right]\\
\times\exp\!\left\{-\frac{\pi k^2}{2}\int^L_0\!\!dz'\,H\!\left[p\!\left(\frac{z'}{L},\frac{z}{L}\right)\frac{L}{k}\Bkappa,z'\right]\right\};
\end{multline}
where $\hat{A}$ is the Fourier transform of the intensity distribution $A$,  $F_{\epsilon}$ is the structure function for the index fluctuations, and $p(x,y)=\min\{x,y\}(1-\max\{x,y\})$. Further terms can be obtained repeating the procedure outlined above.

This very same procedure can be extended to more complex propagation problems. We have seen the scintillation does not only depend on the propagation path but also on the aperture size; moreover, the condition $\Omega\ll q\ll 1$, the aperture size being larger than the coherence radius, is likely to occur in many situations, even for short propagation path, so the only viable tool is the strong-scintillation approach \citep{paper:charnotskii3}. Thus, the Feymann's path integral approach let us calculate the effects of the inhomogeneous media over an irradiance pattern generated by complex objects in every possible situation.
\addcontentsline{toc}{chapter}{\protect\numberline{C}{}}
\chapter*{{\appendixname} C}
\setcounter{equation}{0}
\renewcommand{\theequation}{C.\arabic{equation}}

This Appendix is meant to cover the inequalities shown on page \pageref{eq:kernel-al}. Let $\norm{\cdot}_{H,-q}$ be the norm defined in (\ref{eq:hida-exp}), $F,G, k^H\in \mathcal{S}_H^\ast$ and $J=(0,L]$. Using the H\"older inequality we have
\begin{multline}
\int_J\!\int_J \norm{G(s)k^H(s,t) F(t)}_{H,-q} ds\,dt\\
\leq \left[\int_J \norm{G(s)}_{H,-q}\left(\int_J\norm{k^H(s,t)}_{H,-q}^{p'}\,ds\right)^\frac{1}{p'}\,dt\right] \norm{F}_{H,-q,L^p(J)}\\
\leq \left[\int_J \left(\int_J\norm{k^H(s,t)}_{H,-q}^{p'}\,ds\right)^\frac{p}{p'}\,dt\right]^\frac{1}{p} \norm{F}_{H,-q,L^p(J)}\norm{G}_{H,-q,L^{p'}(J)}.
\end{multline}
for $1/p + 1/p'=1$.We apply this very same procedure but beggining with $G$, thus the inequality only has $p$ and $p'$ interchanged. So taking the supremum at both sides yields
\begin{multline}
\norm{k^H}_{L^{p.p'}(J),-q}\leq \left\{\left[\int_J\left(\int_J\norm{k^H(t,s)}^p_{H,-q}ds\right)^\frac{p'}{p} dt\right]^\frac{1}{p'}\!\!,\right.\\
\left.\left[\int_J\left(\int_J\norm{k^H(t,s)}^{p'}_{H,-q}dt\right)^\frac{p}{p'} ds\right]^\frac{1}{p}\right\} 
\end{multline}

Now, we can estimate a bound for the kernel given the above property. Since, we know
\begin{equation*}
\norm{k^H(z,s)}_{H,-q}\leq g \tilde{M}\; \chi_{[0,z]}(s)s^{-1}\abs{z-s}.
\end{equation*}
Thus, we evaluate:
\begin{equation}
\int^L_0 \left[\int^L_0 \left(\chi_{[0,t]}(s) \frac{\abs{t-s}}{s}\right)^{p'}\,ds\right]^\frac{p}{p'}\!dt=\int^L_0 \left(\frac{\pi p'}{\sin \pi p'}\; t\right)^\frac{p}{p'}\!dt=  \left(\frac{\pi p'}{\sin \pi p'}\right)^\frac{p}{p'} \frac{L^p}{p}
\end{equation}
and 
\begin{multline}
\int^L_0 \left[\int^L_0 \left(\chi_{[0,t]}(s) \frac{\abs{t-s}}{s}\right)^{p}\,dt\right]^\frac{p'}{p}\!ds\\
=\int^L_0 \left[\frac{(L-s)^{p+1}}{(p+1) s^p}\right]^\frac{p'}{p}\!ds=  \left(\frac{\pi p'}{\sin \pi p'}\;\frac{\Gamma(2p')}{\Gamma(p'+1)\Gamma(p')}\right)\frac{L^{p'}}{(p+1)^{p'/p}}.
\end{multline}
We just compare both terms to realize that equation (\ref{eq:gral-bound}) holds.

\thispagestyle{fancy}
\renewcommand{\chaptermark}[1]{\markboth{#1}{}}
\renewcommand{\sectionmark}[1]{\markright{\thesection\ #1}}
\fancyhf{} 
\fancyhead[RO]{\bfseries\thepage}
\fancyhead[LO]{\bfseries\rightmark}
\renewcommand{\headrulewidth}{0.5pt}
\renewcommand{\footrulewidth}{0pt}
\addtolength{\headheight}{0.5pt} 
\fancypagestyle{plain}{%
\fancyhead{} 
\renewcommand{\headrulewidth}{0pt} 
}
\label{bib}
\bibliography{thesis}
\bibliographystyle{jmr}
\backmatter
\end{document}